\documentstyle[preprint,aps,eqsecnum]{revtex}
\begin{document}
\draft
\tightenlines

\title{A Canonical Decomposition in Collective and Relative Variables
of a Klein-Gordon Field in the
Rest-Frame Wigner-Covariant Instant Form.}

\author{Luca Lusanna}

\address
{Sezione INFN di Firenze\\
L.go E.Fermi 2 (Arcetri)\\
50125 Firenze, Italy\\
E-mail: LUSANNA@FI.INFN.IT}

\author{and}

\author{Massimo Materassi}

\address
{Dipartimento di Fisica\\
Universita' di Perugia\\
Via Elce di Sotto,\\
06100 Perugia, Italy\\
E-mail: MATERASSI@PG.INFN.IT\\
MATERASSI@FI.INFN.IT}

\maketitle

\begin{abstract}

The canonical decomposition of a real Klein-Gordon field in collective and
relative variables proposed by Longhi and Materassi is reformulated on
spacelike hypersurfaces. This allows to obtain the complete canonical reduction
of the system on Wigner hyperplanes, namely in the rest-frame Wigner-covariant
instant form of dynamics. From the study of Dixon's multipoles for the
energy-momentum tensor on the Wigner
hyperplanes we derive the definition of the canonical center-of-mass
variable for a Klein-Gordon field configuration: it turns out that the
Longhi-Materassi global variable should be interpreted as a center of phase
of the field configuration. A detailed study of the kinematical ``external"
and ``internal" properties of the field configuration on the Wigner hyperplanes
is done. The construction is then extended to charged
Klein-Gordon fields: the centers of phase of the two real components can be
combined to define a global center of phase and a collective relative variable
describing the action-reaction between the two Feshbach-Villars components of
the field with definite sign of energy and charge. The
Dixon multipoles for both the energy-momentum and the electromagnetic current
are given. Also the  coupling of the Klein-Gordon field
to scalar relativistic particles is studied and it is shown that in the
reduced phase space, besides the particle and field relative variables, there
is also a collective relative variable describing the relative motion of the
particle subsytem with respect to the field one.

\vskip 1truecm

\date{\TeXButton{today}{\today}}

\vskip 1truecm

\end{abstract}
\vfill\eject

\section{Introduction}

Recently a new instant form of dynamics, the rest-frame Wigner-covariant
instant form, was defined in Ref.\cite{lus} as an important tool for the
program of obtaining a unified description of the four interactions with a
canonical reduction to the independent degrees of freedom (Dirac's observables
for a generalized Coulomb gauge; see Ref.\cite{dubna} for a review). This
tool emerged from the study of the reformulation of ordinary theories in
Minkowski spacetime on arbitrary spacelike hypersurfaces as a prerequisite to
their description in curved spacetimes. Since this formulation is the
classical background of the Tomonaga-Schwinger description of quantum field
theory, this new instant form will also open the possibility to obtain a
Wigner-covariant equal time quantization of Minkowski theories, to define a set
of Tomonaga-Schwinger asymptotic states and to open new possibilities for the
description of relativistic bound states, whose formulation in the framework
of the standard Fock approach is still problematic (see Ref.\cite{weinberg}
and the spurious solutions of the Bethe-Salpeter equation \cite{naka}).

In the rest-frame instant form of every isolated system there is a 3+1 splitting
of Minkowski spacetime with a foliation whose leaves, labelled by a scalar time
parameter $\tau$, are the spacelike Wigner hyperplanes $\Sigma_{W\tau}$
orthogonal to the four-momentum of the isolated system (assumed timelike). In
an arbitrary Lorentz frame the points of the Wigner hyperplane $\Sigma_{W\tau}$
have coordinates $z^{\mu}(\tau ,\vec \sigma )=x^{\mu}_s(\tau )+\epsilon^{\mu}_r
(u(p_s)) \sigma^r$: the point $x^{\mu}_s(\tau )$ is an arbitrary origin for the
3-coordinates $\vec \sigma$ on $\Sigma_{W\tau}$ and $p^{\mu}_s$, its conjugate
momentum, is orthogonal to $\Sigma_{W\tau}$ [$\epsilon^{\mu}_r(u(p_s))$ are
three suitable spacelike four-vectors tangent to $\Sigma_{W\tau}$]. After the
canonical reduction of the isolated system to the Wigner hyperplanes we remain
with four universal first class constraints\cite{lus}:\hfill\break
i) one, $p^2_s\approx M^2_{sys}$, identifying the invariant mass $M_{sys}$ of
the isolated system as the Hamiltonian for the evolution in the rest-frame
time;\hfill\break
ii) three others defining the rest frame by the requirement that the (Wigner
spin 1) total 3-momentum of the isolated system is zero, ${\vec P}_{sys}
\approx 0$ [so that $p^{\mu}_s\approx M_{sys} u^{\mu}(p_s)$, with the unit
four-vector $U^{\mu}(p_s)$ describing the orientation of the Wigner hyperplanes
with respect to the arbitrary Lorentz frame].

Among the physical canonical degrees of freedom there are
the coordinates of a point of the Wigner hyperplane, different from its
covariant origin $x^{\mu}_s(\tau )$ of the internal 3-coordinates, with
canonical noncovariant coordinates ${\tilde x}^{\mu}_s(\tau )$. This point, the
classical analogue of the Newton-Wigner position operator, is decoupled from
the system, describes its ``external" center-of-mass variable and may be
interpreted as a decoupled point particle observer with his clock measuring the
rest-frame time. Instead, on
the Wigner hyperplane there are only physical canonical relative degrees of
freedom, whose identification requires the addition of three gauge fixing
constraints ${\vec X}_{sys}\approx 0$, so that the constraints ${\vec X}_{sys}
\approx 0$ and ${\vec P}_{sys}\approx 0$ are second class and may be eliminated.
These constraints ${\vec X}_{sys}\approx 0$ may be interpreted as the
elimination of an ``internal" center-of-mass-like variable of the system,
which is the gauge
variable conjugated to the first class constraints ${\vec P}_{sys}\approx 0$,
so not to have a double counting of the center-of-mass degrees of freedom. The
constraints ${\vec X}_{sys}\approx 0$ eliminate this ``internal"
center-of-mass-like variable
from the physical degrees of freedom forcing it to coincide with the arbitrary
origin $x^{\mu}_s(\tau )=z^{\mu}(\tau ,\vec \sigma ={\vec X}_{sys}=0)$ of the
Wigner hyperplane.

An open problem is the identification of these constraints ${\vec X}_{sys}
\approx 0$ for the various isolated systems. In a future paper\cite{iten}
it will be studied for a system of scalar particles. Instead, in this
paper we will study this probem for the real Klein-Gordon field, as a
particular aspect of the more general problem of finding canonical collective
and relative variables for its configurations. In Refs.\cite{lon,lon1} Longhi
and Materassi found a set of such variables in the standard Lorentz
covariant approach. In this paper we shall reformulate their solution on the
Wigner hyperplanes starting from the formulation of the Klein-Gordon field on
arbitrary spacelike hypersurfaces in Minkowski spacetime\cite{albad}. Then,
after having expressed the energy-momentum tensor of the Klein-Gordon field
on the Wigner hyperplanes, where it assumes a form similar to the
energy-momentum tensor of a relativistic perfect fluid,
we shall study Dixon's multipoles\cite{dixon}
of the field. This will open the way to the identification of the ``internal"
center of mass ${\vec X}_{\phi}={\vec q}_{\phi}$ of the Klein-Gordon field and
to the realization that the Longhi-Materassi collective variable is not the
center of mass but a ``center of phase" of the field configuration.

By considering a Klein-Gordon field configuration as a relativistic extended
object, we can make a detailed study of the kinematical properties of the
description of such an object on Wigner hyperplanes.

Then we shall apply the same approach to the system of N scalar particles
interacting with a real Klein-Gordon field and to the system of a complex
charged Klein-Gordon field plus the electromagnetic field. In the latter
case its two components with definite sign of energy and charge (identified by
means of the Feshbach-Villars formalism in absence of interaction) generate two
centers of phase, so that, besides the global center of phase of the complex
field, there is a second collective variable describing the global
action-reaction between the two components. A similar collective variable
exists also for the system of N particles interacting with a real Klein-Gordon
field.

In Section II the real Klein-Gordon field is reformulated on spacelike
hypersurfaces in Minkowski spacetime, following Ref.\cite{albad}, and then
restricted to Wigner hyperplanes.

In Section III its modulus-phase variables on the Wigner hyperplane are defined.

In Section IV there is the definition of the collective and relative
canonical variables and the inverse canonical transformation to
express the original real Klein-Gordon field in terms of them. It is
shown that each constant energy surface of the Klein-Gordon field is
the disjoint union of symplectic manifolds. Also a set of canonical
multipoles is defined for a certain class of field configurations and
there are some comments on the self-interactions of the field.

In Section V the energy-momentum tensor of a configuration of the real
Klein-Gordon field is restricted to the Wigner hyperplanes and there is a study
of Dixon's multipoles on it. This suggests a definition for the relativistic
center of mass of the field configuration.

In Section VI, by using the group-theoretical methods of Ref.\cite{pauri}, there
is a complete discussion of the concepts of ``external" and ``internal" (with
respect to the Wigner hyperplane) centers of mass of the field configuration.
The Longhi-Materassi collective variable is interpreted as a ``center of phase"
of the field configuration.

In Section VII the isolated system of N scalar particles interacting with a real
Klein-Gordon field is studied to visualize its collective variables and
their interpretation.

In Section VIII the previous analysis is extended to the charged Klein-Gordon
field by defining the collective and relative canonical variables for its
positive and negative frequency parts with the help of the Feshbach-Villars
formalism\cite{fv,cs,gross}. The two collective variables can be combined to
give the overall ``center of phase" of the field configuration plus a
collective relative variable. Then there is a study of the coupling to the
electromagnetic field and of the Dixon multipoles.

In the Conclusions there are some comments about the utility of the multipolar
expansions of the Klein-Gordon field in general relativity and about the
problems existing for the quantization of the collective and relative variables.

In Appendix A there are some notations for spacelike hypersurfaces.

In Appendix B there is a review of Longhi-Materassi papers.

In Appendix C there is a review of the Feshbach-Villars formalism and the
definition of the Fourier coefficients of the Klein-Gordon field corresponding
to the solutions of the square-root Klein-Gordon equation with both positive
(or negative) energy and electric charge in the free case.

In Appendix D there is the expression of the invariant mass for the charged
Klein-Gordon field plus the electromagnetic field on the Wigner hyperplanes in
terms of the Fourier coefficients defined in Appendix C.

\vfill\eject

\section{The real klein-Gordon field on spacelike hypersurfaces.}

In Ref.\cite{albad} there is the description of scalar electrodynamics on
spacelike hypersurfaces. In this Section we shall review this description
restricting ourselves to a real Klein-Gordon field
with a self-interaction $V(\phi )$ following the scheme of Ref.\cite{lus}
[see Appendix A for the notations; $\dot \phi (\tau ,\vec \sigma )=\partial
\phi (\tau ,\vec \sigma )/\partial \tau$]. The action of a scalar
Klein-Gordon field reads

\begin{eqnarray}
S&=& \int d\tau d^3\sigma N(\tau ,\vec \sigma )\sqrt{\gamma (\tau ,\vec \sigma )
}\nonumber \\
&&{1\over 2} \Big[ g^{\tau\tau} {\dot \phi}^2+2 g^{\tau \check r} \dot \phi
\partial_{\check r}\phi + g^{\check r\check s}\partial_{\check r}\phi
\partial_{\check s}\phi -m^2\phi^2 -2V(\phi ) \Big]
(\tau ,\vec \sigma )=\nonumber \\
&=&\int d\tau d^3\sigma \sqrt{\gamma (\tau ,\vec \sigma )} {1\over 2}\Big[
{1\over N}[\partial_{\tau} -N^{\check r}\partial_{\check r}]\phi
[\partial_{\tau} -N^{\check s}\partial_{\check s}]\phi + \nonumber \\
&+&N [\gamma^{\check r\check s}\partial_{\check r}\phi
\partial_{\check s}\phi -m^2\phi^2-2V(\phi ) ]
\, \Big] (\tau ,\vec \sigma ).
\label{II1}
\end{eqnarray}

\noindent where the configuration variables are $z^{\mu}(\tau ,\vec \sigma )$,
$\phi (\tau ,\vec \sigma )={\tilde \phi}(z(\tau ,\vec \sigma ))$. The
main difference between the standard theory and the one on spacelike
hypersurfaces is that now the configuration variables are the fields
$\phi (\tau ,\vec \sigma )=\tilde \phi (z(\tau ,\vec \sigma ))=(\tilde
\phi \circ z) (\tau ,\vec \sigma )$, with $\tilde \phi (x)$ solution
of the Klein-Gordon equation. These new fields $\phi =\tilde \phi
\circ z$ contain the nonlocal information about the 3+1 splitting of
Minkowski spacetime $M^4$ with a foliation of spacelike hypersurfaces
$\Sigma_{\tau}$, obtained through an embedding $R\times \Sigma
\rightarrow M^4$, $(\tau ,\vec \sigma ) \mapsto z^{\mu}(\tau ,\vec
\sigma )$ [$\Sigma$ is an abstract 3-surface diffeomorphic to $R^3$].
The fields $\phi (\tau
,\vec \sigma )$ have a built-in definition of equal time associated
with the Lorentz-scalar time parameter $\tau$ which labels the leaves
of the foliation.

Since $z^{\mu}_{\tau}=Nl^{\mu}+N^{\check r}z^{\mu}_{\check r}$, with the lapse
and shift functions $N$, $N
^{\check r}$ functionals of $z^{\mu}(\tau ,\vec \sigma )$ through the metric
$g_{AB}(\tau ,\vec \sigma )$, one has
${{\partial}\over {\partial z^{\mu}_{\tau}}}=l_{\mu}{{\partial}\over {\partial
N}}+z_{\check s\mu}\gamma^{\check s\check r}{{\partial}\over {\partial N^{\check
r}}}$.

The canonical momenta are

\begin{eqnarray}
\pi(\tau ,\vec \sigma )&=&{{\partial L}\over {\partial \partial_{\tau}
\phi (\tau ,\vec \sigma )}}={{\sqrt{\gamma}(\tau ,\vec \sigma )}\over
{N(\tau ,\vec \sigma )}} \Big[ \dot \phi -N^{\check r} \partial_{\check r}\phi
\Big] (\tau ,\vec \sigma ),\nonumber \\
&&\Rightarrow \, \dot \phi (\tau ,\vec \sigma )=\Big[ {N\over {\sqrt{\gamma}}}
\pi +N^{\check r}\partial_{\check r}\phi \Big] (\tau ,\vec \sigma ),\nonumber \\
\rho_{\mu}(\tau ,\vec \sigma )&=&-{{\partial L}\over {\partial \partial_{\tau}
z^{\mu}(\tau ,\vec \sigma )}}=\nonumber \\
&=&l_{\mu}(\tau ,\vec \sigma ) \Big[ {{\sqrt{\gamma}}\over 2}
[{1\over {N^2}}(\dot \phi -N^{\check r}\partial_{\check r}\phi )^2-
\gamma^{\check r\check s}\partial_{\check r}\phi
\partial_{\check s}\phi +m^2\phi^2+2V(\phi )] \Big] (\tau ,\vec \sigma )
+\nonumber \\
&+&z_{\check s\mu}(\tau ,\vec \sigma )\gamma^{\check s\check r}(\tau ,\vec
\sigma ) \Big[ {{ \sqrt{\gamma}}\over N}\partial_{\check r}\phi (\dot \phi -
N^{\check u}\partial_{\check u}\phi ) \Big] (\tau ,\vec \sigma ).
\label{II2}
\end{eqnarray}

We have the following primary constraints

\begin{eqnarray}
{\cal H}_{\mu}(\tau ,\vec \sigma )&=& \rho_{\mu}(\tau ,\vec \sigma )-
\nonumber \\
&&-l_{\mu}(\tau ,\vec \sigma ) \Big[ {{\pi^2}\over {2\sqrt{\gamma}}}-
{{\sqrt{\gamma}}\over 2} [\gamma^{\check r\check s}\partial_{\check r}\phi
\partial_{\check s}\phi -m^2\phi^2-2V(\phi )] \Big] (\tau ,\vec \sigma )-
\nonumber \\
&-&z_{\check s\mu}(\tau ,\vec \sigma )\gamma^{\check r\check s}(\tau ,\vec
\sigma ) [ \pi \partial_{\check r}\phi ](\tau ,\vec \sigma ) \approx 0,
\label{II3}
\end{eqnarray}

\noindent and the following Dirac Hamiltonian [$\lambda^{\mu}(\tau ,\vec
\sigma )$ are Dirac multiplier]

\begin{equation}
H_D=\int d^3\sigma \lambda^{\mu}
(\tau ,\vec \sigma ){\cal H}_{\mu}(\tau ,\vec \sigma ).
\label{II4}
\end{equation}

By using the Poisson brackets

\begin{eqnarray}
\{ z^{\mu}(\tau ,\vec \sigma ),\rho_{\nu}(\tau ,{\vec \sigma}^{'}) \} &=&\eta
^{\mu}_{\nu} \delta^3(\vec \sigma -{\vec \sigma}^{'}),\nonumber \\
\{ \phi (\tau ,\vec \sigma ),\pi(\tau ,{\vec \sigma}^{'}) \} &=&
\delta^3(\vec \sigma -{\vec \sigma}^{'}),
\label{II5}
\end{eqnarray}

\noindent one finds that the time constancy of the primary constraints does not
imply the existence of  new secondary constraints. The constraints are first
class with the algebra  $\{ {\cal H}_{\mu}(\tau ,\vec \sigma ),{\cal H}_{\nu}
(\tau ,\vec \sigma ) \} =0$.

The conserved Poincare' generators are

\begin{eqnarray}
p_s^\mu \left( \tau \right) &=&\int \rho ^\mu \left( \tau ,\vec
\sigma \right) d^3\sigma ,\nonumber \\
J_s^{\mu \nu }\left( \tau \right) &=&\int \left[ z^\mu \left( \tau
,\vec \sigma \right) \rho ^\nu \left( \tau ,\vec \sigma \right) -z^\nu
\left( \tau ,\vec \sigma \right) \rho ^\mu \left( \tau ,\vec \sigma \right)
\right] d^3\sigma .
\label{II6}
\end{eqnarray}

Following Ref.\cite{lus}, we can restrict ourselves to spacelike hyperplanes
$z^{\mu}(\tau ,\vec \sigma )=x_s^{\mu}(\tau )+b^{\mu}_{\check r}(\tau )
\sigma^{\check r}$, where the normal $l^{\mu}=\epsilon^{\mu}
_{\alpha\beta\gamma}b_1^{\alpha}(\tau )b_2^{\beta}(\tau )b_3^{\gamma}(\tau )$
is $\tau$-independent [with $b^{\mu}_{\tau}=l^{\mu}$; $b^{\mu}_{\check
A}(\tau )=\Big( b^{\mu}_{\tau}, b^{\mu}_{\check r}(\tau ) \Big)$ is an
orthonormal
 tetrad]. Using the results
of that paper we find that $\{ x^{\mu}_s,p^{\nu}_s \} {}^{*}=-\eta^{\mu\nu}$,
$J^{\mu\nu}_s=x_s^{\mu}p_s^{\nu}-x_s^{\nu}p_s^{\mu}+S^{\mu\nu}_s$ and that the
constraints are reduced to the following ten ones
at the level of Dirac brackets [now $\gamma^{\check r\check s}=-\delta
^{\check r\check s}$]

\begin{eqnarray}
{\tilde {\cal H}}^{\mu}(\tau )&=& \int d^3\sigma {\cal H}^{\mu}(\tau ,\vec
\sigma )=\nonumber \\
&=&p^{\mu}_s-b^{\mu}_{\check A}(\tau ){\tilde P}^{\check A}_{\phi}=p^{\mu}_s-l
^{\mu}{\tilde P}^{\tau}_{\phi}-b^{\mu}_{\check r}(\tau ) P^{\check r}_{\phi}=
\nonumber \\
&=&p^{\mu}_s-l^{\mu}  {1\over 2} \int d^3\sigma
\Big[ \pi^2+(\vec \partial \phi )^2
+m^2\phi^2 +2 V(\phi )\Big] (\tau ,\vec \sigma ) -\nonumber \\
&-&b^{\mu}_{\check r}(\tau )  \int d^3\sigma [\pi \partial_{\check r}\phi ]
(\tau ,\vec \sigma ) \approx 0,\nonumber \\
{\tilde {\cal H}}^{\mu\nu}(\tau )&=&b^{\mu}_{\check r}(\tau ) \int d^3\sigma
\sigma^{\check r} {\cal H}^{\nu}(\tau ,\vec \sigma )-b^{\nu}_{\check r}(\tau )
\int d^3\sigma \sigma^{\check r} {\cal H}^{\mu}(\tau ,\vec \sigma )=\nonumber \\
&=&S^{\mu\nu}_s-\Big( b^{\mu}_{\check r}(\tau )l^{\nu}-b^{\nu}_{\check r}(\tau )
l^{\mu}\Big)\nonumber \\
&& {1\over 2}\int d^3\sigma \sigma^{\check r} \Big[ \pi^2+(\vec \partial
\phi )^2+m^2\phi^2+2 V(\phi )\Big] (\tau ,\vec \sigma ) -\nonumber \\
&-&\Big( b^{\mu}_{\check r}(\tau )b^{\nu}_{\check s}(\tau )-b^{\nu}
_{\check r}(\tau )b^{\mu}_{\check s}(\tau )\Big)
 \int d^3\sigma \sigma^{\check r} [\pi \partial^{\check s}\phi ]
(\tau ,\vec \sigma ) =\nonumber \\
&=&S^{\mu\nu}_s-b^{\mu}_{\check r}(\tau )b^{\nu}_{\check s}(\tau )S_{\phi}
^{\check r\check s} - \Big( b^{\mu}_{\check r}(\tau )l^{\nu}-b^{\nu}_{\check
r}(\tau )l^{\mu}\Big) S^{\check r\tau}_{\phi} \approx 0,
\label{II7}
\end{eqnarray}

\noindent where ${\tilde P}^{\check A}_{\phi}=({\tilde P}^{\tau}_{\phi}; P
^{\check r}_{\phi})$ is the (Lorentz-scalar) 4-momentum of the field
configuration and $S^{\check r\check s}_{\phi}=J^{\check r\check
s}_{\phi} {|}_{ {\vec P}_{\phi}=0}$, $S^{\tau \check
r}_{\phi}=J^{o\check r}_{\phi} {|}_{ {\vec P}_{\phi}=0}$ its spin
tensor [see Appendix B].

The configuration variables are reduced from $z^{\mu}(\tau ,\vec \sigma )$,
$\phi (\tau ,\vec \sigma )$, to $x_s^{\mu}(\tau )$, to the six independent
degrees of freedom hidden in the orthonormal tetrad $b^{\mu}_{\check A}(\tau )
$, to $\phi (\tau ,\vec \sigma )$, and to the associated momenta: $p^{\mu}_s$
conjugate to $x^{\mu}_s$; six degrees of freedom hidden in $S^{\mu\nu}_s$ as
the momenta conjugate to the variables hidden in the orthonormal tetrad
$b^{\mu}_{\check A}$ [see Ref.\cite{lus} for the associated Hanson-Regge Dirac
brackets for these degrees of freedom]; $\pi (\tau ,\vec \sigma )$ conjugate to
$\phi (\tau ,\vec \sigma )$.

If we select all the configurations of the system with timelike total momentum
[$p^2_s > 0$], we can restrict ourselves to the special Wigner hyperplanes
$\Sigma_{W\tau}$
orthogonal to $p^{\mu}_s$. The procedure for doing this canonical
reduction implies\cite{lus} to boost at rest the variables $b^{\mu}_{\check A}$,
$S_s^{\mu\nu}$, with the standard Wigner boost $L^{\mu}{}_{\nu}(p_s,{\buildrel
\circ \over p}_s)$ for timelike Poincar\'e orbits [${\buildrel \circ \over
p}_s^{\mu}
=(\epsilon_s=\eta_s\sqrt{p^2_s}; \vec 0)$; $\eta_s=sign\, p^o_s$] and then
to add the gauge-fixings $b^{\mu}_{\check A}-L^{\mu}{}_{\nu =\check A}(p_s,
{\buildrel \circ \over p}_s)=b^{\mu}_{\check A}-\epsilon^{\mu}_{\check A}(u(p
_s))\approx 0$ [$u^{\mu}(p_s)=p^{\mu}_s/\epsilon_s$]. These gauge-fixings
[only six of them are independent], together with ${\tilde {\cal H}}^{\mu\nu}
(\tau )\approx 0$, form six pairs of second class constraints. The Dirac
brackets with respect to these second class constraints [implying ${\tilde {\cal
H}}^{\mu\nu}(\tau )\equiv 0$] admit the following canonical basis: $\phi (\tau
,\vec \sigma )$, $\pi (\tau ,\vec \sigma )$, ${\tilde x}^{\mu}_s(\tau )$ [it is
not a 4-vector, but has only the covariance of the little group of timelike
Poincar\'e orbits like the Newton-Wigner position operator], $p^{\mu}_s$.
As shown in Ref.\cite{lus}, the indices $\check A=(\tau ,\check r)$ are
replaced by $A=(\tau ,r)$; all the quantities $B^{\tau}$ are Lorentz scalars,
while the quantities $\vec B=\{ B^r \}$ are spin-1 Wigner 3-vectors; instead
$p^{\mu}_s$ is a Minkowski 4-vector giving the orientation of the Wigner
hyperplane with respect to a given Lorentz frame.
We get $J^{\mu\nu}_s={\tilde x}^{\mu}_sp^{\nu}_s-{\tilde x}^{\nu}_sp^{\mu}_s+
{\tilde S}_s^{\mu\nu}$ with the spin tensor ${\tilde S}^{\mu\nu}_s$ given in
Eqs.(59) of Ref.\cite{lus}.

Therefore, the final effect of these gauge-fixings is a
canonical reduction to a phase space spanned only by the variables ${\tilde x}
^{\mu}_s(\tau )$, $p^{\mu}_s$, $\phi (\tau ,\vec \sigma )$, $\pi(\tau ,\vec
\sigma )$, with standard Dirac brackets. Instead of ${\tilde x}^{\mu}_s$,
$p^{\mu}_s$, describing a decoupled observer,
one can use the canonical variables\cite{lus}: $\epsilon_s$, $T_s=
p_s\cdot {\tilde x}_s/\epsilon_s=p_s\cdot x_s/\epsilon_s$, ${\vec k}_s={\vec
p}_s/\epsilon_s$, ${\vec z}_s=\epsilon_s[{\vec {\tilde x}}_s-{{{\vec p}_s}\over
{p^o_s}}{\tilde x}^o_s]$ [${\vec z}_s$ is the noncovariant 3-coordinate
corresponding to the Newton-Wigner position operator].

The only surviving four constraints are [now ${\vec P}_{\phi}=\{ P^r_{\phi} \}$
is a spin-1 Wigner 3-vector]

\begin{eqnarray}
{\cal H}(\tau )&=&\epsilon_s -{\tilde P}^{\tau}_{\phi}=\nonumber \\
&=&\epsilon_s- {1\over 2} \int d^3\sigma \Big[ \pi^2+(\vec \partial
\phi )^2+m^2\phi^2+2 V(\phi )\Big] (\tau ,\vec \sigma )  \approx 0,\nonumber \\
{\vec {\cal H}}_p(\tau )&=& {\vec P}_{\phi}=
\int d^3\sigma [\pi \vec \partial \phi ](\tau ,\vec \sigma ) \approx 0.
\label{II8}
\end{eqnarray}

\noindent where ${\tilde P}^A_{\phi}=({\tilde P}^{\tau}_{\phi}; {\vec P}
_{\phi})$, ${\tilde P}^{\tau}_{\phi}=P^{\tau}_{\phi}+\int d^3\sigma V(\phi )
(\tau ,\vec \sigma )$,  is the 4-momentum of the field configuration.
The Dirac Hamiltonian is now $H_D=\lambda (\tau ){\cal H} +\vec
\lambda (\tau )\cdot {\vec {\cal H}}_p$.

By defining ${\bar S}^{AB}_s=\epsilon^A_{\mu}(u(p_s))\epsilon^B_{\nu}(u(p_s))S
^{\mu\nu}_s$, we can showRef.\cite{lus} that on the Wigner
hyperplanes we have ${\bar S}^{AB}_s\equiv J^{AB}_{\phi}{|}_{P^D_{\phi}=0}=
S^{AB}_{\phi}$ [$J^{AB}_{\phi}$ is the angular momentum of the field
configuration] and ${\tilde S}^{ij}=\delta^{ir}\delta^{js}{\bar S}^{rs}_s\equiv
\delta^{ir}\delta^{js}{\bar S}^{rs}_{\phi}$, ${\tilde S}^{oi}_s=-\delta^{ir}
{\bar S}^{rs}_s\delta^{sj}p^j_s/(p^o_s+\epsilon_s)\equiv -\delta^{ir}
{\bar S}^{rs}_{\phi}\delta^{sj}p^j_s/(p^o_s+\epsilon_s)$, so that
$J^{\mu\nu}_s$ becomes independent from the boosts $S_{\phi}^{\tau
r}$. Therefore, the generators of the realization of the Poincar\'e
group in the rest-frame Wigner-covariant instant form of dynamics
(``external" Poincar\'e algebra) are $p_s^{\mu}$ [or $\epsilon_s$,
${\vec k}_s$] and the Lorentz generators

\begin{eqnarray}
J^{ij}_s&=&{\tilde x}_s^ip_s^j-{\tilde x}_s^jp_s^i+\delta^{ir}\delta^{js}
{\bar S}_s^{rs},\nonumber \\
J^{oi}_s&=&{\tilde x}_s^op_s^i-{\tilde x}^i_sp^o_s-{{\delta^{ir}{\bar S}_s^{rs}
p_s^s}\over {p^o_s+\epsilon_s}},\nonumber \\
&&{}\nonumber \\
{\bar S}_s^{rs}&\equiv& S^{rs}_{\phi}=J^{rs}_{\phi}{|}_{{\vec P}_{\phi}=0}=
\int d^3\sigma \{ \sigma^r[\pi \partial^s \phi ]
(\tau ,\vec \sigma )-(r \leftrightarrow s) \} {|}_{{\vec P}_{\phi}=0}.
\label{II9}
\end{eqnarray}

With the gauge fixing $\chi =T_s-\tau \approx 0$, we can eliminate the variables
$\epsilon_s$, $T_s$, and find that
the $\tau$-evolution (in the Lorentz scalar rest-frame
time $T_s\equiv \tau$) is governed by the Hamiltonian

\begin{eqnarray}
H_R&=&M_{\phi}-\vec \lambda (\tau )\cdot {\vec {\cal H}}_p(\tau ),\nonumber \\
&&{}\nonumber \\
M_{\phi}&=&{\tilde P}^{\tau}_{\phi}=
{1\over 2}\int d^3\sigma \Big[ \pi^2+(\vec \partial \phi )^2+m^2\phi^2
+2 V(\phi )\Big] (\tau ,\vec \sigma ),
\label{II10}
\end{eqnarray}

\noindent where $M_{\phi}$ is the invariant mass of the field configuration
[it replaces the non relativistic Hamiltonian $H_{rel}$ appearing in the
separation of the center-of-mass motion $H={{ {\vec P}^2}\over {2M}}+H_{rel}$].

In the gauge $\vec \lambda (\tau )=0$, the Hamilton equations are [$\triangle =-
{\vec \partial}^2$]

\begin{eqnarray}
\partial_{\tau} \phi (\tau ,\vec \sigma )\, &{\buildrel \circ \over =}\,&
\pi (\tau ,\vec \sigma ),\nonumber \\
\partial_{\tau}\pi (\tau ,\vec \sigma )\, &{\buildrel \circ
\over =}\,& [-\triangle -m^2] \phi (\tau ,\vec \sigma )-{{\partial V(\phi )}
\over {\partial \phi}}(\tau ,\vec \sigma ),\nonumber \\
&&\Rightarrow [\partial_{\tau}^2+\triangle +m^2] \phi (\tau ,\vec \sigma )\,
{\buildrel \circ \over =}\, -{{\partial V(\phi )}\over {\partial \phi}}(\tau
,\vec \sigma ).
\label{II11}
\end{eqnarray}

We got a description in which the noncovariant canonical ``external"
center-of-mass 3-variables ${\vec z}_s$, ${\vec k}_s$, move freely and
are decoupled from the Klein-Gordon field variables $\phi (\tau ,\vec
\sigma )$, $\pi (\tau ,\vec
\sigma )$, living on the Wigner hyperplane and restricted by ${\vec P}_{\phi}
\approx 0$. To reduce the field variables only to relative degrees of freedom
one has to find an ``internal" collective center-of-mass-like variable ${\vec X}
_{\phi}[\phi ,\pi]$ conjugate to ${\vec P}_{\phi}$, such that the gauge fixings
${\vec X}_{\phi}\approx 0$ force the position of the field ``internal"
collective variable $\vec \sigma ={\vec
X}_{\phi}$ to coincide with the origin $x^{\mu}_s(\tau )=z^{\mu}(\tau ,
\vec \sigma =0)$ of the Wigner hyperplane. In this way we get a decoupled point
particle observer (the ``external" center of mass ${\tilde x}^{\mu}_s$
of the isolated system) whose
scalar time $T_s\equiv \tau$ labels the evolution of the relative field
variables on the Wigner hyperplanes, defined by the field configuration itself,
foliating Minkowski spacetime.

\vfill\eject

\section{Fourier transform and modulus-phase variables on the Wigner
hyperplane.}

 In the rest-frame Wigner-covariant instant form on Wigner
hyperplanes [in which one considers only field configurations with a
timelike 4-momentum], we can define the Fourier coefficients of the fields
$\phi (\tau ,\vec \sigma )$, $\pi (\tau ,\vec \sigma )$

\begin{equation}
\left\{
\begin{array}{l}
a\left( \tau ,\vec q\right) =\int d^3\sigma \, \left[ \omega \left(
q\right) \phi \left( \tau ,\vec \sigma \right) +i\pi \left( \tau ,\vec
\sigma \right) \right] e^{i\left( \omega \left(  q\right) \tau -\vec
q\cdot \vec \sigma \right) }, \\
a^{*}\left( \tau ,\vec q\right) =\int d^3\sigma  \, \left[ \omega \left(
q\right) \phi \left( \tau ,\vec \sigma \right) -i\pi \left( \tau ,\vec
\sigma \right) \right] e^{-i\left( \omega \left(  q\right) \tau -\vec
q\cdot \vec \sigma \right) }, \\
\\
\phi \left( \tau ,\vec \sigma \right) =\int d\tilde q \, \left[ a\left( \tau
,\vec q\right) e^{-i\left( \omega \left(  q\right) \tau -\vec q\cdot
\vec \sigma \right) }+a^{*}\left( \tau ,\vec q\right) e^{+i\left( \omega
\left(  q\right) \tau -\vec q\cdot \vec \sigma \right) }\right] , \\
\pi \left( \tau ,\vec \sigma \right) =-i\int d\tilde q \, \omega \left(
q\right) \left[ a\left( \tau ,\vec q\right) e^{-i\left( \omega \left(
q\right) \tau -\vec q\cdot \vec \sigma \right) }-a^{*}\left( \tau ,\vec
q\right) e^{+i\left( \omega \left( \vec q\right) \tau -\vec q\cdot \vec
\sigma \right) }\right] ,
\end{array}
\right.
\label{III1}
\end{equation}

\noindent with

\begin{equation}
\begin{array}{ccc}
\omega \left(  q\right) =\sqrt{m^2+ {\vec q}^2}, & d\tilde q=%
\frac{d^3q}{\Omega \left(  q\right) }, & \Omega \left( q\right)
=(2\pi )^3\,2\omega \left(  q\right) ,\quad q=|\vec q|=\sqrt{{\vec q}^2}.
\end{array}
\label{III2}
\end{equation}

\noindent Here $q_A\sigma^A=\omega ( q)\tau -\vec q\cdot \vec \sigma$ is
defined by using $q^A=(q^{\tau}=\omega ( q);\, q^r)$ with $q^{\tau}$
Lorentz scalar and $\vec q$ a spin-1 Wigner 3-vector like $\vec \sigma$
[on arbitrary hyperplanes all the quantities $\tau$ ,$\vec \sigma$ ,$q^{\tau}$,
$\vec q$, would be Lorentz scalars]. The Fourier coefficients are
$\tau$-dependent only when there is a non null self interaction $V(\phi )$; they
and their gradients are assumed to belong to the space $L_2(d\tilde q)$.

From the Poisson brackets $\left\{ \phi \left( \tau ,\vec \sigma \right) ,\pi
\left( \tau ,\vec \sigma^{\prime }\right) \right\} =\delta ^3\left( \vec
\sigma -\vec \sigma^{\prime }\right) $, we get [leaving the factor $\Omega(
q)$ to agree with the standard notations]

\begin{equation}
\left\{ a\left( \tau ,\vec q\right) ,a^{*}\left( \tau ,\vec k
\right) \right\} =-i\Omega \left(  q\right) \delta ^3\left( \vec q-\vec
k\right) ,
\label{III3}
\end{equation}

The 4-momentum and angular momentum of the field configuration are
[${\cal V}\left[ a,a^{*}\right] $ being the contribution of the
potential $V\left( \phi \right) $ when present]

\begin{eqnarray}
{\tilde P}_{\phi}^\tau &=&P^{\tau}_{\phi}+{\cal V}\left[ a,a^{*}\right] =
{1\over 2} \int d^3\sigma [ \pi^2+(\vec \partial \phi )^2+m^2\phi^2+2V(\phi )]
(\tau ,\vec \sigma )=\nonumber \\
&=&\int d\tilde q \, \omega \left(  q\right) a^{*}\left( \tau ,\vec q\right)
a\left( \tau ,\vec q\right) +{\cal V}\left[ a,a^{*}\right] ,\nonumber \\
{\vec P}_{\phi}&=&\int d^3\sigma [\pi \vec \partial \phi ](\tau ,\vec \sigma )
=\int d\tilde q \, \vec q\, a^{*}\left( \tau ,\vec q\right)
a\left( \tau ,\vec q\right) ,
\label{III4}
\end{eqnarray}

\begin{eqnarray}
J_{\phi}^{rs}&=& \int d^3\sigma [\pi (\sigma^r\partial^s-\sigma^s\partial^r)\phi
](\tau ,\vec \sigma )=\nonumber \\
&=&-i\int d\tilde q \,
a^{*}\left( \tau ,\vec q\right) \left( q^r\frac
\partial {\partial q^s}-q^s\frac \partial {\partial q^r}\right) a\left(
\tau ,\vec q\right) , \nonumber \\
J^{\tau r}_{\phi}&=&-\tau P^r_{\phi}+{1\over 2} \int d^3\sigma \, \sigma^r\,\,
[\pi^2+(\vec \partial \phi )^2+m^2\phi^2](\tau ,\vec \sigma )=\nonumber \\
&=&-
\tau P_{\phi}^r+i\int d\tilde q \, \omega \left( q\right) a^{*}\left( \tau
,\vec q\right) \frac \partial {\partial q^r}a\left( \tau ,\vec q\right) .
\label{III5}
\end{eqnarray}

We want to define four variables $X_{\phi}^A[\phi ,\pi ]=(X_{\phi}^\tau ;{\vec
X}_{\phi})$ canonically
conjugated to $P_{\phi}^A[\phi ,\pi ]=(P_{\phi}^\tau ;{\vec P}_{\phi})$. First
of all we make a canonical transformation to modulus-phase canonical variables

\begin{equation}
\left\{
\begin{array}{l}
a\left( \tau ,\vec q\right) =\sqrt{I\left( \tau ,\vec q\right) }\, e^{\left[
i\varphi \left( \tau ,\vec q\right) \right]} , \\
a^{*}\left( \tau ,\vec q\right) =\sqrt{I\left( \tau ,\vec q\right) }\, e^{\left[
 -i\varphi \left( \tau ,\vec q\right) \right]} , \\
\\
I\left( \tau ,\vec q\right) =a^{*}\left( \tau ,\vec q\right) a\left( \tau
,\vec q\right) , \\
\varphi \left( \tau ,\vec q\right) =\frac 1{2i}\ln \left[ \frac{a\left( \tau ,
\vec q\right) }{a^{*}\left( \tau ,\vec q\right) }\right] ,
\end{array}
\right.
\label{III6}
\end{equation}

\begin{equation}
\left\{ I\left( \tau ,\vec q\right) ,\varphi \left( \tau ,\vec q^{\prime
}\right) \right\} =\Omega \left(  q\right) \delta ^3\left( \vec q-\vec
q^{\prime }\right) .
\label{III7}
\end{equation}

In terms of the original canonical variables $\phi$, $\pi$, we have

\begin{eqnarray}
I( \tau ,\vec q) &=&\int d^3\sigma \int d^3\sigma ^{\prime
}\, e^{i\vec q\cdot ( \vec \sigma -\vec \sigma ^{\prime }) }[
\omega (  q) \phi ( \tau ,\vec \sigma ) -i\pi
( \tau ,\vec \sigma ) ] [ \omega (  q)
\phi ( \tau ,\vec \sigma ^{\prime }) +i\pi ( \tau ,\vec
\sigma ^{\prime }) ] , \nonumber \\
\varphi ( \tau ,\vec q) &=&\frac 1{2i}\ln\, \Big[ \frac{\int
d^3\sigma \, [ \omega (  q) \phi ( \tau ,\vec \sigma
) +i\pi ( \tau ,\vec \sigma ) ] e^{i( \omega
(  q) \tau -\vec q\cdot \vec \sigma ) }}{\int
d^3\sigma ^{\prime }\, [ \omega (  q) \phi ( \tau
,\vec \sigma ^{\prime }) -i\pi ( \tau ,\vec \sigma ^{\prime
}) ] e^{-i( \omega (  q) \tau -\vec q\cdot
\vec \sigma ^{\prime }) }}\Big] =\nonumber \\
&=&\omega (q) \tau + \frac 1{2i}\ln\, \Big[ \frac{\int
d^3\sigma \, [ \omega (  q) \phi ( \tau ,\vec \sigma
) +i\pi ( \tau ,\vec \sigma ) ] e^{-i\vec q\cdot \vec \sigma  }}{\int
d^3\sigma ^{\prime }\, [ \omega (  q) \phi ( \tau
,\vec \sigma ^{\prime }) -i\pi ( \tau ,\vec \sigma ^{\prime
}) ] e^{ i\vec q\cdot \vec \sigma ^{\prime } }}\Big] .
\label{III8}
\end{eqnarray}

\begin{equation}
\left\{
\begin{array}{l}
\phi \left( \tau ,\vec \sigma \right) =\int d\tilde q\, \sqrt{I\left( \tau
,\vec q\right) }\left[ e^{i\varphi \left( \tau ,\vec q\right) -i\left( \omega
\left(  q\right) \tau -\vec q\cdot \vec \sigma \right) }+e^{-i\varphi
\left( \tau ,\vec q\right) +i\left( \omega \left(  q\right) \tau -\vec
q\cdot \vec \sigma \right) }\right] , \\
\pi \left( \tau ,\vec \sigma \right) =-i\int d\tilde q\, \omega \left(
q\right) \sqrt{I\left( \tau ,\vec q\right) }\left[ e^{i\varphi \left( \tau
,\vec q\right) -i\left( \omega \left(  q\right) \tau -\vec q\cdot \vec
\sigma \right) }-e^{-i\varphi \left( \tau ,\vec q\right) +i\left( \omega \left(
q\right) \tau -\vec q\cdot \vec \sigma \right) }\right] .
\end{array}
\right.
\label{III9}
\end{equation}

The Poincar\'e charges of the field configuration take the form

\begin{equation}
\left\{
\begin{array}{l}
{\tilde P}_{\phi}^\tau =P^{\tau}_{\phi}+{\cal V}\left[ I,\varphi \right] =
\int d\tilde q\,  \omega \left(  q\right) I\left( \tau ,\vec q\right)
+{\cal V}\left[ I,\varphi \right] , \\
\\
{\vec P}_{\phi}=\int d\tilde q\,  \vec q\, I\left( \tau ,\vec q\right) ,
\end{array}
\right.
\label{III10}
\end{equation}

\begin{equation}
\left\{
\begin{array}{l}
J^{rs}_{\phi}=\int d\tilde q\,
I\left( \tau ,\vec q\right) \left( q^r\frac \partial
{\partial q^s}-q^s\frac \partial {\partial q^r}\right) \varphi \left( \tau
,\vec q\right) , \\
\\
J_{\phi}^{\tau r}=-\tau P_{\phi}^r-\int d\tilde q\,
\omega \left( q\right) I\left( \tau
,\vec q\right) \frac \partial {\partial q^r}\varphi \left( \tau ,\vec q\right).
\end{array}
\right.
\label{III11}
\end{equation}

The classical analogue of the occupation number is [$\triangle =-{\vec
\partial}^2$]

\begin{eqnarray}
N_{\phi}&=&\int d\tilde q\, a^{*}(\tau ,\vec q) a(\tau ,\vec q) =
\int d\tilde q\, I(\tau ,\vec q)=\nonumber \\
&=&{1\over 2} \int d^3\sigma [\pi {1\over {\sqrt{m^2+\triangle}}}+\phi
\sqrt{m^2+\triangle} \phi ](\tau ,\vec \sigma ).
\label{III12}
\end{eqnarray}

\section{Collective and relative canonical variables.}

A sequence of canonical transformations from the field variables $\phi (x)$,
$\pi (x)={{\partial \phi (x)}\over {\partial x^o}}$ to the Fourier
coefficients $a(\vec k)$, $a^{*}(\vec k)$ [$k^{\mu}=(k^o=\omega (k=|\vec k|)=
\sqrt{m^2+{\vec k}^2};\, \vec k)$], then to the modulus-phase
(or action-angle) variables
$I(\vec k)$, $\varphi (\vec k)$, and finally to a canonical basis $X^{\mu}
_{\phi}[F]$, $P^{\mu}_{\phi}$, ${\cal H}(\vec k | F]$, ${\cal K}(\vec k | F]$
[depending on an arbitrary normalized weight function $F(P_{\phi},k)$] was
found in Ref.\cite{lon} when the real Klein-Gordon field $\phi (x)$ satisfies
certain conditions. The main results of that paper are reviewed in Appendix B.

Here, we shall reformulate this approach on the Wigner hyperplane. Due to its
peculiar Wigner covariance properties\cite{lus}, the weight function
$F(P_{\phi},k)$ [with its generic non polynomial (for instance $F \approx
e^{-P_{\phi}\cdot k}$) dependence upon $P^{\mu}_{\phi}$ due to manifest
Lorentz covariance] may now be replaced by two separate scalar functions
${\tilde F}^{\tau}(P^{\tau}_{\phi},q)$, $\tilde F({\vec P}_{\phi},\vec q)$.
Moreover, if we accept that these two functions are singular at $\vec q=0$, we
can take them linear in $P^{\tau}_{\phi}$ and ${\vec P}_{\phi}$ respectively:
${\tilde F}^{\tau}(P^{\tau}_{\phi},q)=P^{\tau}_{\phi}\, F^{\tau}(q)$,
$\tilde F({\vec P}_{\phi}, \vec q)=-{\vec P}_{\phi}\cdot \vec q F(q)$ with
a suitable normalization for $F^{\tau}(q)$ and $F(q)$.

From Appendix B, we see that  we must require the following
behaviours also on the Wigner hyperplane

\begin{eqnarray}
i)\, q\, \rightarrow \, \infty &&,\quad\quad \sigma > 0,\nonumber \\
&&|a(\tau ,\vec q)| \rightarrow q^{-{3\over 2}-\sigma},\quad
|I(\tau ,\vec q)| \rightarrow q^{-3-\sigma},\quad |\varphi (\tau ,\vec q)|
\rightarrow q,\nonumber \\
ii)\, q\, \rightarrow \, 0 &&,\quad\quad \epsilon > 0,\quad \eta > -\epsilon ,
\nonumber \\
&&|a(\tau ,\vec q)| \rightarrow q^{-{3\over 2}+\epsilon},\quad
|I(\tau ,\vec q)| \rightarrow q^{-3+\epsilon},\quad |\varphi (\tau ,\vec q)|
\rightarrow q^{\eta}.
\label{IV1}
\end{eqnarray}

\noindent in order to have the Poincar\'e generators and the
occupation number of Section III finite.

Let us remark that the collective and relative canonical variables can be
defined in closed form only in absence of self-interactions of the field, so
that in this Section we shall put $V(\phi )={\cal V}[I,\varphi ]=0$.

\subsection{Definition of the collective variables.}

Let us define the four functionals of the phases

\begin{eqnarray}
&&X_{\phi}^\tau =\int d\tilde q \omega (q) F^\tau \left( q\right) \varphi
\left( \tau ,\vec q\right) , \nonumber \\
&&{\vec X}_{\phi}=\int d\tilde q  \vec q\, F\left( q\right) \varphi
\left( \tau ,\vec q\right) ,\nonumber \\
&&\Rightarrow \lbrace X^r_{\phi},X^s_{\phi} \rbrace =0,\quad\quad \lbrace
X_{\phi}^\tau ,X_{\phi}^r \rbrace =0.
\label{IV2}
\end{eqnarray}

\noindent depending on two Lorentz scalar functions $F^{\tau}(q)$, $F(q)$,
whose form will be restricted by the following requirements implying that $X^A
_{\phi}$ and $P^A_{\phi}$ are canonical variables:

\begin{equation}
\begin{array}{cccc}
\left\{ P_{\phi}^\tau ,X_{\phi}^\tau \right\} =1, & \left\{ P^r_{\phi},X_{\phi}
^s \right\} =-\delta ^{rs}, & \left\{ P_{\phi}^r,X_{\phi}^\tau \right\}
=0, & \left\{ P_{\phi}^\tau ,X_{\phi}^r\right\} =0,
\end{array}
\label{IV3}
\end{equation}

Since $\left\{ P_{\phi}^\tau ,X_{\phi}^\tau \right\} =\int d\tilde q \omega^2
\left( q\right) F^\tau \left( q\right)$ and
$\left\{ P_{\phi}^r,X_{\phi}^s\right\} =\int d\tilde q
\, q^rq^s\, F\left( q \right) $,
we must require the following normalizations for $F^{\tau }(q)$, $F(q)$

\begin{equation}
\int d\tilde q \omega^2(q) F^\tau \left( q\right) =1
\label{IV4}
\end{equation}

\begin{equation}
\int d\tilde q\, q^rq^s
F\left( q\right) =-\delta ^{rs}.
\label{IV5}
\end{equation}

Moreover, $\left\{ P_{\phi}^r,X_{\phi}^\tau \right\} =\int d\tilde q
\omega (q) q^r F^\tau \left( q\right)$ and $\left\{ P_{\phi}^\tau ,X_{\phi}
^r\right\} =\int d\tilde q \omega (q)\, q^r F\left( q \right)$ ,
imply the following conditions

\begin{equation}
\int d\tilde q \omega (q)\, q^r F^\tau \left( q\right) =0,
\label{IV6}
\end{equation}

\begin{equation}
\int d\tilde q \omega (q)\, q^r F\left( q\right) =0.
\label{IV7}
\end{equation}

\noindent which are automatically satisfied because $F^{\tau}(q)$, $F(q)$,
$q=|\vec q|$, are even under $q^r\, \rightarrow -q^r$.

A solution of Eqs.(\ref{IV4}) and (\ref{IV5}) is

\begin{eqnarray}
F^{\tau}(q) &=& {{16 \pi^2}\over {m\, q^2\, \sqrt{m^2+q^2} }} e^{-{{4\pi}\over
{m^2}}\, q^2},\nonumber \\
&&{}\nonumber \\
F(q) &=& - {{48 \pi^2}\over {m\, q^4}} \sqrt{m^2+q^2}\, e^{-{{4\pi}\over
{m^2}} q^2}.
\label{IV8}
\end{eqnarray}

The singularity in $\vec q =0$ requires $\varphi (\tau ,\vec q)\,
{\rightarrow}_{q\rightarrow 0}\, q^{\eta}$, $\eta > 0$, [and not $\eta > -
\epsilon$ as in Eq.(\ref{IV1})] for the
existence of $X^{\tau}_{\phi}$, ${\vec X}_{\phi}$.

The analogue of the function $F(P_{\phi},k)$ of Ref.\cite{lon} is now

\begin{eqnarray}
{\cal F}\left( P_{\phi},\vec q\right) &=&F^\tau \left( q\right) \omega \left(
q\right) P^\tau_{\phi} -F\left( q\right) \vec q\cdot {\vec P}_{\phi},
\nonumber \\
&&{}\nonumber \\
&&\int d\tilde q\, {\cal F}\left( P_{\phi},\vec q\right) \omega \left( q
\right) =P_{\phi}^\tau , \nonumber \\
&& \int d\tilde q\,
{\cal F}\left( P_{\phi},\vec q\right) q^r=P_{\phi}^r.
\label{IV9}
\end{eqnarray}

Let us remark that for field configurations $\phi (\tau ,\vec \sigma )$ such
that the Fourier transform $\hat \phi (\tau ,\vec q)$ has compact support in a
sphere centered at $\vec q=0$ of volume V, we get $X^{\tau}_{\phi}=-{1\over V}
\int {{d^3q}\over {\omega (q)}} \varphi (\tau ,\vec q)$, ${\vec X}_{\phi}=
{1\over V}\int d^3q {{3\vec q}\over {{\vec q}^2}} \varphi (\tau ,\vec q)$.

\subsection{Auxiliary relative variables.}

As in Ref.\cite{lon}, let us define an auxiliary relative action variable
and an auxiliary relative phase variable

\begin{equation}
\hat I\left( \tau ,\vec q\right) =I\left( \tau ,\vec q\right) -F^\tau \left(
q\right) P_{\phi}^\tau \omega \left( q\right) +F\left( q\right)
\vec q\cdot {\vec P}_{\phi},
\label{IV10}
\end{equation}

\begin{equation}
\hat \varphi \left( \tau ,\vec q\right) =\varphi \left( \tau ,\vec q\right)
-\omega \left( q\right) X_{\phi}^\tau +\vec q\cdot {\vec X}_{\phi}.
\label{IV11}
\end{equation}

The previous canonicity conditions on $F^{\tau}(q)$, $F(q)$, imply

\begin{eqnarray}
&&\int d\tilde q\, \omega \left( q\right) \hat I\left( \tau ,\vec q\right) =0,
\nonumber \\
&&\int d\tilde q\, q^i\, \hat I\left( \tau ,\vec q\right) =0.
\label{IV12}
\end{eqnarray}

\begin{eqnarray}
&&\int d\tilde q F^\tau \left( \mathrm{q}\right) \omega \left( q\right) \hat
\varphi \left( \tau ,\vec q\right) d\tilde q=0,\nonumber \\
&&\int d\tilde q F\left( q\right) q^i\, \hat \varphi \left(
\tau ,\vec q\right)=0.
\label{IV13}
\end{eqnarray}

These auxiliary  variables have the following nonzero Poisson bracket

\begin{equation}
\left\{ \hat I\left( \tau ,\vec k\right) ,\hat \varphi \left( \tau ,\vec q
\right) \right\}
=\Delta \left( \vec k,\vec q\right) ,
\label{IV14}
\end{equation}

\noindent with

\begin{equation}
\Delta \left( \vec k,\vec q\right) =\Omega \left( k\right) \delta
^3\left( \vec k-\vec q\right) -F^\tau \left( k\right) \omega \left(
k\right) \omega \left( q\right) +F\left(
k\right) \vec k\cdot \vec q.
\label{IV15}
\end{equation}

The distribution $\Delta \left( \vec k,\vec q\right) $ has the
semigroup property

\begin{equation}
\int d\tilde q\Delta \left( \vec k,\vec q\right) \Delta \left( \vec q,\vec
k^{\prime }\right) =\Delta \left( \vec k,\vec k^{\prime }\right) ,
\label{IV16}
\end{equation}

\noindent and satisfies the constraints:

\begin{equation}
\begin{array}{cc}
\int d\tilde q\, \omega \left( q\right) \Delta \left( \vec q,\vec k\right)
=0, & \int d\tilde q\, q^r\, \Delta \left( \vec q,\vec k\right)=0,
\end{array}
\label{IV17}
\end{equation}

\begin{equation}
\begin{array}{cc}
\int d\tilde q\, F^\tau \left( q\right) \omega \left( q\right) \Delta
\left( \vec k,\vec q\right) =0, & \int d\tilde q\, q^r\,
F\left( q\right) \Delta \left( \vec k,\vec q\right) =0.
\end{array}
\label{IV18}
\end{equation}

At this stage the canonical variables $I(\tau ,\vec q)$, $\varphi (\tau ,\vec
q)$ for the Klein-Gordon field are replaced by the noncanonical set
$X_{\phi}^\tau$ ,  $P_{\phi}^\tau$ , ${\vec X}_{\phi}$, ${\vec P}_{\phi}$,
$\hat I(\tau ,\vec q)$, $\hat \varphi (\tau ,\vec q)$ with
the Poisson brackets

\begin{equation}
\left\{
\begin{array}{l}
\begin{array}{cccc}
\left\{ P_{\phi}^\tau ,X_{\phi}^\tau \right\} =1, & \left\{ P_{\phi}^r,
X_{\phi}^s\right\} =-\delta ^{rs}, & \left\{ P_{\phi}^r,X_{\phi}^\tau \right\}
=0, & \left\{ P_{\phi}^\tau ,X_{\phi}^r\right\} =0,
\end{array}
\\
\\
\begin{array}{ccc}
\left\{ X_{\phi}^r,X_{\phi}^s\right\} =0, & \left\{ X_{\phi}^\tau ,X_{\phi}^r
\right\} =0, & \left\{ P^A_{\phi}, P^B_{\phi} \right\} =0,\quad A,B=(\tau ,
r),
\end{array}
\\
\\
\begin{array}{cccc}
\left\{ P_{\phi}^\tau ,\hat I\left( \tau ,\vec q\right) \right\} =0,
& \left\{ P_{\phi}^r,\hat I\left( \tau ,\vec q\right) \right\} =0,
& \left\{ \hat I\left( \tau ,\vec q\right) ,X_{\phi}^\tau \right\} =0,
& \left\{ \hat I\left( \tau ,\vec q\right) ,X_{\phi}^r\right\} =0,
\end{array}
\\
\\
\begin{array}{cccc}
\left\{ X_{\phi}^r,\hat \varphi \left( \tau ,\vec q\right) \right\} =0,
& \left\{ X_{\phi}^\tau ,\hat \varphi \left( \tau ,\vec q\right) \right\} =0,
& \left\{ P^r_{\phi},\hat \varphi \left( \tau ,\vec q\right) \right\} =0,
& \left\{ P_{\phi}^\tau ,\hat \varphi \left( \tau ,\vec q\right) \right\} =0,
\end{array}
\\
\\
\left\{ \hat I\left( \tau ,\vec k\right) ,\hat \varphi \left( \tau ,\vec q
\right) \right\} =\Omega \left( k\right) \delta ^3\left( \vec k-\vec q\right)
-F^\tau \left( k\right) \omega \left( k\right) \omega
\left( q\right) +F\left( k\right) \vec k\cdot \vec q .
\end{array}
\right.
\label{IV19}
\end{equation}

The generators of Lorentz group are already decomposed into two
parts, the collective and the relative ones, each satisfying the Lorentz
algebra and having vanishing mutual Poisson brackets

\begin{equation}
\left\{
\begin{array}{l}
J^{rs}_{\phi}=L_{\phi}^{rs}+{\hat S}_{\phi}^{rs}, \\
\\
L_{\phi}^{rs}=X_{\phi}^rP_{\phi}^s-X_{\phi}^sP_{\phi}^r,
\\
\\
{\hat S}_{\phi}^{rs}=\int d\tilde q\,
\hat I\left( \tau ,\vec q\right) \left( q^r
\frac \partial {\partial q^s}-q^s\frac
\partial {\partial q^r}\right) \hat \varphi \left( \tau ,\vec q\right) ,
\end{array}
\right.
\label{IV20}
\end{equation}

\begin{equation}
\left\{
\begin{array}{l}
J^{\tau r}_{\phi}=L^{\tau r}_{\phi}+{\hat S}^{\tau r}_{\phi}, \\
\\
L_{\phi}^{\tau r}=[X_{\phi}^\tau -\tau ] P_{\phi}^r-X_{\phi}^r
P_{\phi}^\tau , \\
\\
{\hat S}_{\phi}^{\tau r}=-\int d\tilde q\, \omega \left( q\right) \hat
I\left( \tau ,\vec q\right) \frac \partial {\partial q^r}\hat \varphi
\left( \tau ,\vec q\right) .
\end{array}
\right.
\label{IV21}
\end{equation}

\subsection{Canonical relative variables.}

We must now find the canonical relative variables hidden inside the auxiliary
ones, which are not free but satisfy Eqs.(\ref{IV12}) and (\ref{IV13}).

As in Ref.\cite{lon}, let us introduce the following differential operator
[$\triangle_{LB}$ is the Laplace-Beltrami operator of the mass shell
submanifold $H^1_3$ (see Appendix B and Ref.\cite{lon,lon1})]

\begin{eqnarray}
{\cal D}_{\vec q}&=&3-m^2 \triangle_{LB}=\nonumber \\
&=&3-m^2\left[ \sum\limits_{i=1}^3\left( \frac \partial
{\partial q^i}\right) ^2+\frac 2{m^2}\sum\limits_{i=1}^3q^i\,
\frac \partial {\partial q^i}+\frac
1{m^2}\left( \sum\limits_{i=1}^3q^i\, \frac \partial
{\partial q^i}\right) ^2\right] ,
\label{IV22}
\end{eqnarray}

\noindent which is a scalar on the Wigner hyperplane since it is invariant
under Wigner's rotations.

Since $\omega \left( \mathrm{q}\right) $ and $\vec q$ are null modes of this
operator\cite{lon},  we can put

\begin{eqnarray}
\hat I\left( \tau ,\vec q\right) &=&{\cal D}_{\vec q}{\bf H}( \tau ,\vec q) ,
\nonumber \\
{\bf H}\left( \tau ,\vec q\right) &=&\int d\tilde k\, {\cal G}\left( \vec q,\vec
k\right) \hat I\left( \tau ,\vec k\right) ,
\label{IV23}
\end{eqnarray}

\noindent with  ${\cal G}\left( \vec q,\vec k\right) $ being the Green
function of  ${\cal D}_{\vec q}$ [see Refs.\cite{lon,lon1} for its expression]

\begin{equation}
{\cal D}_{\vec q}{\cal G}\left( \vec q,\vec k\right) =\Omega \left( k
\right) \delta ^3\left( \vec k-\vec q\right) .
\label{IV24}
\end{equation}

Like in Ref.\cite{lon}, for each zero mode $f_o(\vec q)$ of ${\cal D}_{\vec q}$
[${\cal D}_{\vec q}\, f_o(\vec q)=0$] for which one has $| \int d\tilde q\,
f_o(\vec q) \hat I(\tau ,\vec q) | < \infty$, we have by integration by parts

\begin{eqnarray}
\int d\tilde q && f_o(\vec q) \hat I(\tau ,\vec q) = \int d\tilde q\, f_o(\vec
q) {\cal D}_{\vec q} {\bf H}(\tau ,\vec q)=\nonumber \\
&=&-{1\over {2(2\pi )^3}} \int d^3q\, {{\partial}\over {\partial q^r}} \Big(
{{m^2\delta^{rs}+q^rq^s}\over {\omega (q)}} \Big[ f_o(\vec q) {{\partial}\over
{\partial q^s}} {\bf H}(\tau ,\vec q)-{\bf H}(\tau ,\vec q) {{\partial}\over
{\partial q^s}} f_o(\vec q)\Big] \Big) .
\label{IV25}
\end{eqnarray}

The boundary conditions (ensuring finite Poincar\'e generators)

\begin{eqnarray}
{\bf H}(\tau ,\vec q)\, &{\rightarrow}_{q \rightarrow 0}\,& q^{-1+\epsilon},
\quad\quad \epsilon > 0,\nonumber \\
{\bf H}(\tau ,\vec q)\, &{\rightarrow}_{q \rightarrow \infty}\,& q^{-3-\sigma},
\quad\quad \sigma > 0,
\label{IV26}
\end{eqnarray}

\noindent imply $\int d\tilde q\, f_o(\vec q) \hat I(\tau ,\vec q)=0$, or

\begin{eqnarray}
\int d\tilde q\, f_o(\vec q) I(\tau ,\vec q) &=& P^{\tau}_{\phi}\, \int
d\tilde q\, \omega (q) f_o(\vec q) F^{\tau}(q)-\nonumber \\
&-&{\vec P}_{\phi}\, \cdot \, \int d\tilde q\, \vec q\, f_o(\vec q) F(q).
\label{IV27}
\end{eqnarray}

We shall restrict ourselves to field configurations
for which $I(\tau ,\vec q)\, {\rightarrow}_{q\rightarrow 0}\, q^{-3+\eta}$
with  $\eta \in (0,1]$ to avoid the extra condition connected with the zero
modes $f_o(\vec q)=v^{(o)}_{2,-3,lm}(\vec q)$ [see the $Q_{lm}$ of Appendix B].

Instead,
for the zero modes $f_o(\vec q)= v^{(o)}_{1,-3,lm}(\vec q)= q^l\,\, {}_2F_1
({{l-1}\over 2}, {{l+3}\over 2}; l+{3\over 2}; -q^2)\, Y_{lm}(\alpha ,\beta )$,
[$\vec q=q (cos\, \alpha cos\, \beta, cos\, \alpha sin\, \beta , sin\,
\alpha )$] of Ref.\cite{lon} we get from Eq.(\ref{IV27}): i) for l=0,1, the
identities $P^{\tau}_{\phi}=P^{\tau}_{\phi}$, ${\vec P}_{\phi}={\vec P}_{\phi}$
[$v^{(o)}_{1,-3,00}$ and $v^{(o)}_{1,-3,1m}$ give $f_o(\vec q)=\omega (q),
\vec q$]; ii) for $l \geq 2$ the conditions

\begin{eqnarray}
P_{lm}&=& \int d\tilde q\, v^{(o)}_{1,-3,lm}(\vec q) I(\tau ,\vec q) =
\nonumber \\
&=& const. \int d\tilde q\, q^l\, {}_2F_1
({{l-1}\over 2}, {{l+3}\over 2}; l+{3\over 2}; -q^2)\, Y_{lm}(\theta ,\varphi )
\nonumber \\
&&\int d^3\sigma \int d^3\sigma^{'} e^{i\vec q\cdot (\vec \sigma -{\vec \sigma}
^{'})} [\omega (q)\phi (\tau ,\vec \sigma )-i\pi (\tau ,\vec \sigma )][\omega
(q)\phi (\tau ,{\vec \sigma}^{'})+i\pi(\tau ,{\vec \sigma}^{'})]=0.
\label{IV28}
\end{eqnarray}

\noindent
Since $\vec q\cdot (\vec \sigma -{\vec \sigma}^{'})=q cos\, \theta
|\vec \sigma -{\vec \sigma}^{'}|$, one can check that $P_{l\, m\not= 0}$ is
automatically zero. Moreover, in $P_{l0}$ the term $\omega (q) [\phi (\tau
,\vec \sigma )\pi(\tau ,{\vec \sigma}^{'})-\phi (\tau ,{\vec \sigma}^{'})
\pi (\tau ,\vec \sigma )]$ does not contribute for symmetry reasons. Therefore,
the final restriction is

\begin{eqnarray}
P_{l0}&=&const. \int d\tilde q\, q^l\, {}_2F_1
({{l-1}\over 2}, {{l+3}\over 2}; l+{3\over 2}; -q^2)\, Y_{l0}(\theta ,\varphi )
\nonumber \\
&&\int d^3\sigma \int d^3\sigma^{'} e^{i\vec q\cdot (\vec \sigma -{\vec \sigma}
^{'})} \Big[ (m^2+q^2)\phi (\tau ,\vec \sigma )\phi (\tau ,{\vec \sigma}^{'})+
\pi (\tau ,\vec \sigma )\pi (\tau ,{\vec \sigma}^{'})\Big] =0.
\label{IV29}
\end{eqnarray}

These conditions on $\phi (\tau ,\vec \sigma )$ and $\pi (\tau ,\vec \sigma )
=\partial_{\tau} \phi (\tau ,\vec \sigma )$ identify the class of
configurations of the Klein-Gordon field for which one can define the previous
canonical transformation and for which there is no ambiguity in defining a
unique realization of the Poincar\'e group (the BMS algebra degenerates in the
Poincar\'e algebra in this case).

We can satisfy the constraints on $\hat \varphi (\tau ,\vec q)$ with the
definition [${\cal D}_{\vec q} \omega (q)={\cal D}_{\vec q} \vec q =0$]

\begin{eqnarray}
\hat \varphi (\tau ,\vec q) &=&\int d\tilde k\int d\tilde k^{\prime }
{\bf K}(\tau ,\vec k) {\cal G}(\vec k,\vec k^{\prime}) \Delta (\vec k^{\prime},
\vec q) ,\nonumber \\
{\bf K}(\tau ,\vec q) &=&{\cal D}_{\vec q}\hat \varphi (\tau ,\vec q) =
{\cal D}_{\vec q}\varphi (\tau ,\vec q),\nonumber \\
&&{}\nonumber \\
&&{\rightarrow}_{q\rightarrow \infty}\, q^{1-\epsilon},\quad \epsilon > 0,
\quad\quad {\rightarrow}_{q\rightarrow 0}\, q^{\eta -2},\quad \eta > 0,
\label{IV30}
\end{eqnarray}

\noindent which also implies

\begin{equation}
\left\{
\begin{array}{l}
\begin{array}{cc}
\left\{ {\bf H}\left( \tau ,\vec q\right) ,X_{\phi}^\tau \right\} =0, & \left\{
{\bf H}\left( \tau ,\vec q\right) ,P_{\phi}^\tau \right\} =0, \\
 \left\{
{\bf H}\left( \tau ,\vec q\right) ,X_{\phi}^r\right\} =0, & \left\{ {\bf H}
\left( \tau ,\vec q\right) ,P_{\phi}^r\right\} =0,
\end{array}
\\
\\
\begin{array}{cc}
\left\{ {\bf K}\left( \tau ,\vec q\right) ,X_{\phi}^\tau \right\} =0, & \left\{
{\bf K}\left( \tau ,\vec q\right) ,P_{\phi}^\tau \right\} =0, \\ \left\{
{\bf K}\left( \tau ,\vec q\right) ,X_{\phi}^r\right\} =0, & \left\{ {\bf K}
\left( \tau ,\vec q\right) ,P_{\phi}^r\right\} =0,
\end{array}
\\
\\
\left\{ {\bf H}\left( \tau ,\vec q\right) ,\mathbf{K}\left( \tau ,\vec
q^{\prime }\right) \right\} =\Omega \left(  q\right) \delta ^3\left(
\vec q-\vec q^{\prime }\right) .
\end{array}
\right.
\label{IV31}
\end{equation}

The final decomposition of Lorentz generators is

\begin{equation}
\left\{
\begin{array}{l}
J_{\phi}^{rs}=L_{\phi}^{rs}+S_{\phi}^{rs}, \\
\\
L_{\phi}^{rs}=X_{\phi}^rP_{\phi}^s-X_{\phi}^sP_{\phi}^r,
\\
\\
S_{\phi}^{rs}=\int d\tilde k{\bf H}\left( \tau ,\vec k\right) \left( k^r
\frac \partial {\partial k^s}-k^s\frac
\partial {\partial k^r}\right) {\bf K}\left( \tau ,\vec k\right) ,
\end{array}
\right)
\label{IV32}
\end{equation}

\begin{equation}
\left\{
\begin{array}{l}
J_{\phi}^{\tau r}=L_{\phi}^{\tau r}+S_{\phi}^{\tau r}, \\
\\
L_{\phi}^{\tau r}=(X_{\phi}^\tau -\tau )P_{\phi}^r-X_{\phi}^rP_{\phi}^\tau , \\
\\
S_{\phi}^{\tau r}=-\int d\tilde q\omega \left( q\right) {\bf H}
\left( \tau ,\vec q\right) \frac \partial {\partial q^r}{\bf K}
\left( \tau ,\vec q\right) .
\end{array}
\right)
\label{IV33}
\end{equation}

\subsection{Field variables in terms of collective-relative variables.}

We have found the canonical transformation

\begin{eqnarray}
I\left( \tau ,\vec q\right) &=&
F^{\tau}(q)\omega (q) P^{\tau}_{\phi}-F(q)\vec q\cdot {\vec P}_{\phi}+
{\cal D}_{\vec q} {\bf H}(\tau ,\vec q),\nonumber \\
\varphi (\tau ,\vec q)&=&\int d\tilde k\int d\tilde k^{\prime } {\bf K}
( \tau ,\vec k) {\cal G}\left( \vec k,\vec k^{\prime }\right) \Delta
\left( \vec k^{\prime },\vec q\right) +\omega \left( q\right)
X_{\phi}^\tau -\vec q\cdot {\vec X}_{\phi},\nonumber \\
&&{}\nonumber \\
N_{\phi}&=&P^{\tau}_{\phi} \int d\tilde q\, \omega (q) F^{\tau}(q)-{\vec P}
_{\phi}\cdot \,\, \int d\tilde q\, \vec q F(q) +\int d\tilde q\,
{\cal D}_{\vec q} {\bf H}(\tau ,\vec q)=\nonumber \\
&=& \tilde c {{ P^{\tau}_{\phi}}\over m}+\int d\tilde q\,
{\cal D}_{\vec q} {\bf H}(\tau ,\vec q),\nonumber \\
&&\tilde c = m \int d\tilde q \omega (q) F^{\tau}(q)=2\int_0^{\infty}{{dq}\over
{\sqrt{m^2+q^2}}}e^{-{{4\pi}\over {m^2}}q^2}=2 e^{4\pi} \int^{\infty}_m
{{dx}\over {\sqrt{x^2-m^2}}}e^{-{{4\pi}\over {m^2}}x^2},\nonumber \\
&&{}
\label{IV34}
\end{eqnarray}

\noindent with the two functions $F^{\tau}(q)$, $F(q)$ given in
Eqs.(\ref{IV8}). Its inverse is

\begin{eqnarray}
P^{\tau}_{\phi}&=&\int d\tilde q \omega (q) I(\tau ,\vec q)={1\over 2}\int
d^3\sigma \Big[ \pi^2 +(\vec \partial \phi )^2+m^2\phi^2\Big] (\tau ,\vec
\sigma ),\nonumber \\
{\vec P}_{\phi}&=&\int d\tilde q \vec q\, I(\tau ,\vec q)=\int d^3\sigma
\Big[ \pi \vec \partial \phi \Big] (\tau ,\vec \sigma ),\nonumber \\
X^{\tau}_{\phi}&=&\int d\tilde q \omega (q) F^{\tau}(q) \varphi (\tau ,\vec q)=
\tau +\nonumber \\
&+&{1\over {2\pi i m}} \int d^3q {{e^{-{{4\pi}\over {m^2}} q^2} }\over
{q^2\, \omega (q)}} ln\, \Big[ {{\omega (q) \int d^3\sigma e^{i\vec q\cdot \vec
\sigma} \phi (\tau ,\vec \sigma ) +i\int d^3\sigma e^{i\vec q\cdot \vec \sigma}
\pi (\tau ,\vec \sigma )}\over {\omega (q) \int d^3\sigma e^{-i\vec q\cdot \vec
\sigma} \phi (\tau ,\vec \sigma ) -i \int d^3\sigma e^{-i\vec q\cdot \vec
\sigma} \pi (\tau ,\vec \sigma )}}\Big] =\nonumber \\
&{\buildrel {def} \over =}\,& \tau + {\tilde X}^{\tau}_{\phi},\quad\quad
\Rightarrow \quad L^{\tau r}_{\phi}={\tilde X}^{\tau}_{\phi}P^r_{\phi}-
X^r_{\phi}P^{\tau}_{\phi},\nonumber \\
&&{}\nonumber \\
{\vec X}_{\phi}&=&\int d\tilde q \vec q\, F(q) \varphi (\tau ,\vec q)=
\nonumber \\
&=&{{2i}\over {\pi m}} \int d^3q\, {{q^i}\over {q^4}}\, e^{-{{4\pi}\over {m^2}}
q^2}  ln\, \Big[ {{\sqrt{m^2+q^2} \int d^3\sigma e^{i\vec q\cdot \vec \sigma}
\phi (\tau ,\vec \sigma ) +i \int d^3\sigma e^{i\vec q\cdot \vec \sigma}
\pi (\tau ,\vec \sigma )}\over {\sqrt{m^2+q^2} \int d^3\sigma e^{-i\vec q\cdot
\vec \sigma} \phi (\tau ,\vec \sigma ) -i \int d^3\sigma e^{-i\vec q\cdot \vec
\sigma} \pi (\tau ,\vec \sigma )}} \Big] ,\nonumber \\
&&{}\nonumber \\
{\bf H}(\tau ,\vec q)&=&\int d\tilde k {\cal G}(\vec q,\vec k) [I(\tau ,\vec k)
-F^{\tau}(k)\omega (k) \int d{\tilde q}_1 \omega (q_1) I(\tau ,{\vec q}_1)+
\nonumber \\
&+&F(k) \vec k \cdot \, \int d{\tilde q}_1 {\vec q}_1\, I(\tau ,{\vec q}_1) ]=
\nonumber \\
&=&\int d^3\sigma_1d^3\sigma_2 \Big[ \pi (\tau ,{\vec \sigma}_1) \pi (\tau
,{\vec \sigma}_2) \int d\tilde k {\cal G}(\vec q,\vec k) \int d{\tilde k}_1
\triangle (\vec k,{\vec k}_1)e^{i{\vec k}_1\cdot ({\vec \sigma}_1-{\vec
\sigma}_2)}+\nonumber \\
&+&\phi (\tau ,{\vec \sigma}_1) \phi (\tau ,{\vec \sigma}_2) \int d\tilde k
{\cal G}(\vec q,\vec k) \int d{\tilde k}_1 \omega^2(k_1)\triangle (\vec k,{\vec
k}_1) e^{i{\vec k}_1\cdot ({\vec \sigma}_1-{\vec \sigma}_2)}-\nonumber \\
&-&i\Big( \pi (\tau ,{\vec \sigma}_1) \phi (\tau ,{\vec \sigma}_2)+\pi (\tau
,{\vec \sigma}_2) \phi (\tau ,{\vec \sigma}_1)\Big) \nonumber \\
&& \int d\tilde k {\cal G}(\vec q,\vec k) \int d{\tilde k}_1 \omega (k_1)
\triangle (\vec k.{\vec k}_1) e^{i{\vec k}_1\cdot ({\vec \sigma}_1-{\vec
\sigma}_2)} \Big] ,\nonumber \\
&&{}\nonumber \\
{\bf K}(\tau ,\vec q)&=&{\cal D}_{\vec q} \hat \varphi (\tau ,\vec q)={\cal D}
_{\vec q}\varphi (\tau ,\vec q)=\nonumber \\
&=&{1\over {2i}} {\cal D}_{\vec q}
\ln \left[ \frac{\int
d^3\sigma \, \left[ \omega \left(  q\right) \phi \left( \tau ,\vec \sigma
\right) +i\pi \left( \tau ,\vec \sigma \right) \right]
e^{-i\vec q\cdot \vec \sigma  }}{\int
d^3\sigma ^{\prime }\, \left[ \omega \left(  q\right) \phi \left( \tau
,\vec \sigma ^{\prime }\right) -i\pi \left( \tau ,\vec \sigma ^{\prime
}\right) \right] e^{i\vec q\cdot
\vec \sigma ^{\prime } }}\right] .
\label{IV35}
\end{eqnarray}

\noindent We get the following expression of the other canonical variables
$a(\tau ,\vec q)$, $\phi (\tau ,\vec \sigma )$, $\pi (\tau ,\vec \sigma )$,
in terms of the final ones

\begin{eqnarray}
a(\tau ,\vec q)&=&\sqrt{F^{\tau}(q)\omega (q)P^{\tau}
_{\phi}-F(q)\vec q\cdot {\vec P}_{\phi}+{\cal D}_{\vec q}
{\bf H}(\tau , \vec q) }\nonumber \\
&&e^{i[\omega (q) X_{\phi}^\tau -\vec q\cdot {\vec X}_{\phi}]
+i\int d\tilde k\int d\tilde k^{\prime }\, {\bf K}( \tau ,\vec k)
{\cal G}\left( \vec k,\vec k^{\prime }\right) \Delta \left( \vec k^{\prime }
,\vec q\right)  },\nonumber \\
&&{}\nonumber \\
N_{\phi}&=& \tilde c {{P^{\tau}_{\phi}}\over m}-
+\int d\tilde k\, {\cal D}_{\vec k} {\bf H}(\tau ,\vec k),
\label{IV36}
\end{eqnarray}

\begin{eqnarray}
\phi (\tau ,\vec \sigma ) &=&  \int d\tilde q\sqrt{F^{\tau}(q)\omega (q)P^{\tau}
_{\phi}-F(q)\vec q\cdot {\vec P}_{\phi}+{\cal D}_{\vec q}
{\bf H}(\tau , \vec q) }\nonumber \\
&&\left[ e^{-i[\omega (q) \left( \tau -X_{\phi}^\tau \right) -
\vec q\cdot \left(\vec \sigma -{\vec X}_{\phi}\right)]
+i\int d\tilde k\int d\tilde k^{\prime }\, {\bf K}( \tau ,\vec k)
{\cal G}\left( \vec k,\vec k^{\prime }\right) \Delta \left( \vec k^{\prime }
,\vec q\right)  }\right. + \nonumber \\
&&+\left. e^{i[\omega \left( q\right) \left( \tau
-X_{\phi}^\tau \right) -\vec q\cdot \left( \vec \sigma -{\vec X}_{\phi}\right) ]
-i\int d\tilde k\int d\tilde k^{\prime }\, {\bf K}\left( \tau ,\vec
k\right) {\cal G}\left( \vec k,\vec k^{\prime }\right) \Delta \left( \vec
k^{\prime },\vec q\right) }\right] =\nonumber \\
&=&2 \int d\tilde q {\bf A}_{\vec q}(\tau ;P^A_{\phi},{\bf H}]\, cos\,
\Big[ \vec q\cdot \vec \sigma +{\bf B}_{\vec q}(\tau ;X^A_{\phi},{\bf K}]
\Big] ,\nonumber \\
&&{}\nonumber \\
\pi (\tau ,\vec \sigma )&=&-i \int d\tilde q \omega (q)\sqrt{F^{\tau}(q)
\omega (q)P^{\tau}_{\phi}-F(q)\vec q\cdot {\vec P}_{\phi}+{\cal D}_{\vec q}
{\bf H}(\tau , \vec q) }\nonumber \\
&&\left[ e^{-i[\omega (q) \left( \tau -X_{\phi}^\tau \right)-
\vec q\cdot \left(\vec \sigma -{\vec X}_{\phi}\right)]
+i\int d\tilde k\int d\tilde k^{\prime }\, {\bf K}( \tau ,\vec k)
{\cal G}\left( \vec k,\vec k^{\prime }\right) \Delta \left( \vec k^{\prime }
,\vec q\right)  }\right. - \nonumber \\
&&-\left. e^{+i[\omega \left( q\right) \left( \tau
-X_{\phi}^\tau \right) -\vec q\cdot \left( \vec \sigma -{\vec X}_{\phi}\right)
]-i\int d\tilde k\int d\tilde k^{\prime }\, {\bf K}\left( \tau ,\vec
k\right) {\cal G}\left( \vec k,\vec k^{\prime }\right) \Delta \left( \vec
k^{\prime },\vec q\right) }\right] =\nonumber \\
&=&-2\int d\tilde q \omega (q) {\bf A}_{\vec q}(\tau ;P^A_{\phi},{\bf H}]\,
sin\, \Big[ \vec q\cdot \vec \sigma +{\bf B}_{\vec q}(\tau ;X^A_{\phi},{\bf K}]
\Big] ,\nonumber \\
&&{}\nonumber \\
{\bf A}_{\vec q}(\tau ;P^A_{\phi},{\bf H}]&=&\sqrt{F^{\tau}(q)\omega (q)
P^{\tau}_{\phi}-F(q)\vec q\cdot {\vec P}_{\phi}+{\cal D}_{\vec q}{\bf H}(\tau
,\vec q)}=\sqrt{I(\tau ,\vec q)},\nonumber \\
{\bf B}_{\vec q}(\tau ;X^A_{\phi},{\bf K}]&=&-\vec q\cdot {\vec X}_{\phi}-
\omega (q)(\tau -X^{\tau}_{\phi})+\int d\tilde kd{\tilde k}^{'} {\bf K}(\tau
,\vec k){\cal G}(\vec k,{\vec k}^{'})\triangle ({\vec k}^{'},\vec q)=
\nonumber \\
&=&\varphi (\tau ,\vec q)-\omega (q) \tau .
\label{IV37}
\end{eqnarray}

The Klein-Gordon field configuration is described by:\hfill\break
i) its energy $P^{\tau}
_{\phi}$ and the conjugate field time-variable $X^{\tau}_{\phi}$, which is
equal to $\tau$ plus some kind of internal time ${\tilde X}^{\tau}_{\phi}$;
\hfill\break
ii) the conjugate reduced canonical variables of a free point ${\vec
X}_{\phi}$, ${\vec P}_{\phi}$;\hfill\break iii) an infinite set of
canonically conjugate relative variables ${\bf H}(\tau ,\vec q)$,
${\bf K}(\tau ,\vec q)$. \hfill\break While the sets i) and ii)
describe a ``monopole" field configuration see Section V), which
depends only on 8 degrees of freedom like a scalar particle at rest
[${\vec P}_{\phi}\approx 0$] and with mass $\epsilon_s\approx
\sqrt{(P^{\tau}_{\phi})^2-{\vec P}_{\phi}^2}\approx P^{\tau}_{\phi}$,
corresponding to the decoupled collective variables of the field
configuration, the set iii) describes an infinite set of ``canonical relative
variables" with respect to the relativistic collective variables
of the sets i) and ii).

The conditions ${\bf H}(\tau ,\vec q)={\bf K}(\tau ,\vec q)=0$ select
the class of field configurations, solutions of the Klein-Gordon
equation, which are of the ``monopole" type on the Wigner hyperplanes

\begin{eqnarray}
\phi_{mon}(\tau ,\vec \sigma )&=&2 \int d\tilde q \sqrt{F^{\tau}(q)\omega (q)
P^{\tau}_{\phi}-F(q)\vec q\cdot {\vec P}_{\phi}} cos\, \Big[ \vec q\cdot (\vec
\sigma -{\vec X}_{\phi})-\omega (q)(\tau -X^{\tau}_{\phi})\Big] \approx
\nonumber \\
&\approx& 2 \sqrt{P^{\tau}_{\phi}} \int d\tilde q \sqrt{F^{\tau}(q)\omega (q)}
cos\, \Big[ \vec q\cdot (\vec
\sigma -{\vec X}_{\phi})-\omega (q)(\tau -X^{\tau}_{\phi})\Big] ,\nonumber \\
\pi_{mon}(\tau ,\vec \sigma )&=& -2 \int d\tilde q
\sqrt{F^{\tau}(q)\omega (q)P^{\tau}_{\phi}-F(q)\vec q\cdot {\vec P}_{\phi}}
sin\, \Big[ \vec q\cdot (\vec
\sigma -{\vec X}_{\phi})-\omega (q)(\tau -X^{\tau}_{\phi})\Big] \approx
\nonumber \\
&\approx& -2 \sqrt{P^{\tau}_{\phi}} \int d\tilde q \sqrt{F^{\tau}(q)\omega (q)}
sin\, \Big[ \vec q\cdot (\vec
\sigma -{\vec X}_{\phi})-\omega (q)(\tau -X^{\tau}_{\phi})\Big] .
\label{IV38}
\end{eqnarray}

If we add the gauge-fixings ${\vec X}_{\phi}\approx 0$ to ${\vec P}_{\phi}
\approx 0$ [this implies $\vec \lambda (\tau )=0$ in Eq.(\ref{II10})]
and go to Dirac brackets, the rest-frame instant-form
Klein-Gordon canonical variables in the gauge $\tau \equiv T_s=p_s\cdot x_s/
\epsilon_s$ (see the end of Section II) are [in the following formulas one has
$T_s-X^{\tau}_{\phi}=-{\tilde X}^{\tau}_{\phi}$]

\begin{eqnarray}
a(T_s ,\vec q)&=&\sqrt{F^{\tau}(q)\omega (q)P^{\tau}_{\phi}
+{\cal D}_{\vec q} {\bf H}(T_s , \vec q) }\nonumber \\
&&e^{i[\omega (q) {\tilde X}_{\phi}^\tau  +
\vec q\cdot \vec \sigma ]+
i\int d\tilde k\int d\tilde k^{\prime }\, {\bf K}(T_s ,\vec k)
{\cal G}\left( \vec k,\vec k^{\prime }\right) \Delta \left( \vec k^{\prime }
,\vec q\right)   },\nonumber \\
&&{}\nonumber \\
N_{\phi}&=&\tilde c {{P^{\tau}_{\phi}}\over m} +\int d\tilde q
{\cal D}_{\vec q} {\bf H}(T_s,\vec q),\nonumber \\
&&{}\nonumber \\
\phi (T_s ,\vec \sigma ) &=&  \int d\tilde q\sqrt{F^{\tau}(q)\omega (q)P^{\tau}
_{\phi}+{\cal D}_{\vec q} {\bf H}(T_s , \vec q) }
\nonumber \\
&&\left[ e^{i[\omega (q) {\tilde X}_{\phi}^\tau +
\vec q\cdot \vec \sigma ]+
i\int d\tilde k\int d\tilde k^{\prime }\, {\bf K}( T_s ,\vec k)
{\cal G}\left( \vec k,\vec k^{\prime }\right) \Delta \left( \vec k^{\prime }
,\vec q\right)  }\right. + \nonumber \\
&&+\left. e^{-i[\omega \left( q\right)
{\tilde X}_{\phi}^\tau +\vec q\cdot \vec \sigma ]
-i\int d\tilde k\int d\tilde k^{\prime }\, {\bf K}\left( T_s ,\vec
k\right) {\cal G}\left( \vec k,\vec k^{\prime }\right) \Delta \left( \vec
k^{\prime },\vec q\right) }\right] =\nonumber \\
&=&2 \int d\tilde q\, {\bf A}_{\vec q}(T_s;P^{\tau}
_{\phi},{\bf H}] \, cos\, \Big[ \vec q\cdot \vec \sigma +{\bf B}_{\vec q}(T_s;
{\tilde X}^{\tau}_{\phi},{\bf K}]\, \Big] ,\nonumber \\
&&{}\nonumber \\
\pi (T_s ,\vec \sigma )&=&-i \int d\tilde q \omega (q)\sqrt{F^{\tau}(q)
\omega (q)P^{\tau}_{\phi}+{\cal D}_{\vec q} {\bf H}(T_s , \vec q) }\nonumber \\
&&\left[ e^{i[\omega (q) {\tilde X}_{\phi}^\tau +
\vec q\cdot\vec \sigma ]
+i\int d\tilde k\int d\tilde k^{\prime }\, {\bf K}( T_s ,\vec k)
{\cal G}\left( \vec k,\vec k^{\prime }\right) \Delta \left( \vec k^{\prime }
,\vec q\right)  }\right. - \nonumber \\
&&-\left. e^{-i[\omega \left( q\right)
{\tilde X}_{\phi}^\tau +\vec q\cdot \vec \sigma ]
-i\int d\tilde k\int d\tilde k^{\prime }\, {\bf K}\left( T_s ,\vec
k\right) {\cal G}\left( \vec k,\vec k^{\prime }\right) \Delta \left( \vec
k^{\prime },\vec q\right) }\right] =\nonumber \\
&=&-2 \int d\tilde q\, \omega (q) {\bf A}_{\vec q}(T_s;P^{\tau}
_{\phi},{\bf H}] \, sin\, \Big[ \vec q\cdot \vec \sigma +{\bf B}_{\vec q}(T_s;
{\tilde X}^{\tau}_{\phi},{\bf K}]\, \Big] ,\nonumber \\
&&{}\nonumber \\
{\bf A}_{\vec q}(T_s;P^{\tau}_{\phi},{\bf H}]&=& \sqrt{ F^{\tau}(q)
\omega (q) P^{\tau}_{\phi}+ {\cal D}_{\vec q} {\bf H}(T_s,\vec q)}=
\sqrt{I(T_s,\vec q)},\nonumber \\
{\bf B}_{\vec q}(T_s;X^{\tau}_{\phi},{\bf K}]&=& \int d\tilde k
d\tilde k^{'}\, {\bf K}(T_s,\vec k){\cal G}(\vec k,{\vec k}^{'}) \triangle
({\vec k}^{'},\vec q) +\omega (q) {\tilde X}^{\tau}_{\phi}=\nonumber \\
&=&\varphi (T_s,\vec q)-\omega (q) T_s,\nonumber \\
&&\Downarrow \nonumber \\
{{\partial \phi (T_s,\vec \sigma )}\over {\partial P^{\tau}_{\phi}}}&=&
\int d\tilde q F^{\tau}(q)\omega (q) {{cos\, \Big[ \vec q\cdot \vec \sigma +
{\bf B}_{\vec q}(T_s;{\tilde X}^{\tau}_{\phi},{\bf K}] \Big]}\over {{\bf A}
_{\vec q}(T_s;P^{\tau}_{\phi},{\bf K}] }},\nonumber \\
{{\partial \phi (T_s,\vec \sigma )}\over {\partial X^{\tau}_{\phi}}}&=&
2\int d\tilde q \omega(q) {\bf A}_{\vec q}(T_s;P^{\tau}_{\phi},{\bf H}]
sin\, \Big[ \vec q\cdot \vec \sigma +
{\bf B}_{\vec q}(T_s;{\tilde X}^{\tau}_{\phi},{\bf K}] \Big],\nonumber \\
{{\delta \phi (T_s,\vec \sigma )}\over {\delta {\bf K}(T_s,\vec q)}}&=&
2 \int d\tilde k\, {\bf A}_{\vec k}(T_s;P^{\tau}_{\phi},{\bf H}]\nonumber \\
&& sin\, \Big[ \vec k\cdot \vec \sigma +{\bf B}_{\vec k}(T_s;{\tilde X}
^{\tau}_{\phi},{\bf K}] \Big] \int d{\tilde k}^{'} {\cal G}(\vec q,{\vec k}
^{'})\triangle ({\vec k}^{'},\vec k),\nonumber \\
{{\delta \phi (T_s,\vec \sigma )}\over {\delta {\bf H}(T_s,\vec q)}}&=&
\int d\tilde k \Big( {{cos\, \Big[ \vec k\cdot \vec \sigma +
{\bf B}_{\vec k}(T_s;{\tilde X}^{\tau}
_{\phi},{\bf K}] \Big]}\over {{\bf A}_{\vec k}(T_s;P^{\tau}_{\phi},{\bf H}]}}
\Big) {\cal D}_{\vec k}\delta^3(\vec q-\vec k)=\nonumber \\
&=&-{\cal D}_{\vec q} \Big( {{cos\, \Big[ \vec q\cdot \vec \sigma +
{\bf B}_{\vec q}(T_s;{\tilde X}^{\tau}
_{\phi},{\bf K}] \Big]}\over {{\bf A}_{\vec q}(T_s;P^{\tau}_{\phi},{\bf H}]}}
\Big) .
\label{IV39}
\end{eqnarray}

From Eq.(\ref{II10}), the Hamiltonian is now $M_{\phi}=P^{\tau}_{\phi}$: it
generates the following evolution in $T_s$

\begin{eqnarray}
&&{{\partial}\over {\partial T_s}}\, X^{\tau}_{\phi}\, {\buildrel \circ \over
=}\, \{ X^{\tau}_{\phi},P^{\tau}_{\phi} \} =-1,\quad\quad
\Rightarrow \quad X^{\tau}_{\phi}\, {\buildrel \circ \over =}\, -T_s,
\nonumber \\
&&{{\partial}\over {\partial T_s}}\, P^{\tau}_{\phi}\, {\buildrel \circ \over
=}\, 0,\nonumber \\
&&{{\partial}\over {\partial T_s}}\, {\bf H}(T_s,\vec q)\, {\buildrel \circ
\over =}\, 0,\nonumber \\
&&{{\partial}\over {\partial T_s}}\, {\bf K}(T_s,\vec q)\, {\buildrel \circ
\over =}\, 0,\nonumber \\
&&\Rightarrow \,\, {{\partial}\over {\partial T_s}}{\bf A}_{\vec q}(T_s;P
^{\tau}_{\phi},{\bf H}]={{\partial}\over {\partial T_s}}{\bf B}_{\vec
q}(T_s;{\tilde X}^{\tau}_{\phi},{\bf K}]\, {\buildrel \circ \over =}\, 0,
\nonumber \\
&&{}\nonumber \\
&&{{\partial}\over {\partial T_s}}\, \phi (T_s,\vec \sigma )\, {\buildrel
\circ \over =}\, -{{\partial}\over {\partial X^{\tau}_{\phi}}}\, \phi (T_s,
\vec \sigma )=\pi (T_s,\vec \sigma ),\nonumber \\
&&{{\partial}\over {\partial T_s}}\, \pi (T_s,\vec \sigma )\, {\buildrel \circ
\over =}\, -{{\partial}\over {\partial X^{\tau}_{\phi}}}\, \pi (T_s,\vec
\sigma )=-[\triangle +m^2]\phi (T_s,\vec \sigma ),\nonumber \\
\Rightarrow&& ({{\partial^2}\over {\partial T_s^2}}-{{\partial^2}\over
{\partial {\vec \sigma}^2}}+m^2) \phi (T_s,\vec \sigma )\, {\buildrel \circ
\over =}\, 0.
\label{IV40}
\end{eqnarray}

Therefore, in the free case ${\bf H}(T_s,\vec q)$, ${\bf K}(T_s,\vec q)$ are
constants of the motion [complete integrability and Liouville theorem for the
Klein-Gordon field]. Since the canonical variable $P^{\tau}_{\phi}$ is the
Hamiltonian for the evolution in $T_s\equiv \tau$, we need the ``internal"
variable $X^{\tau}_{\phi}=\tau +{\tilde X}^{\tau}_{\phi}$ [i.e. the ``internal
time variable" ${\tilde X}^{\tau}_{\phi}$] to write Hamilton's equations
${{\partial}\over {\partial T_s}} F\, {\buildrel \circ \over =}\, \{ F,P^{\tau}
_{\phi} \} =-{{\partial F}\over {\partial X^{\tau}_{\phi}}}=
-{{\partial F}\over {\partial {\tilde X}^{\tau}_{\phi}}}$; in the free case we
have ${{\partial}\over {\partial T_s}}\, {\buildrel \circ \over =}\,
-{{\partial}\over {\partial X^{\tau}_{\phi}}}$ on $\phi (T_s,\vec \sigma )[X
^{\tau}_{\phi},P^{\tau}_{\phi},{\bf H},{\bf K}]$ and $\pi (T_s,\vec \sigma )[X
^{\tau}_{\phi},P^{\tau}_{\phi},{\bf H},{\bf K}]$, so that the evolution in the
time $X^{\tau}_{\phi}=T_s+{\tilde X}^{\tau}_{\phi}$ [which takes place
inside the Wigner hyperplane and which can be interpreted as an evolution
in the internal time ${\tilde X}^{\tau}_{\phi}$]
is equal and opposite to the evolution in the rest-frame
time $T_s$ from a Wigner hyperplane to the next one in the free case.

By adding the two second class constraints $X^{\tau}_{\phi}-T_s ={\tilde X}
^{\tau}_{\phi} \approx 0$, $P^{\tau}_{\phi}-const. \approx 0$, and by going to
Dirac brackets, we get the rest-frame Hamilton-Jacobi formulation corresponding
to the given constant value of the total energy [the field $\phi (T_s,\vec
\sigma )$, which is $T_s$-independent since it depends only on the internal
time ${\tilde X}^{\tau}_{\phi}$, now becomes also ${\tilde X}^{\tau}_{\phi}$-
independent]. In this way we find a symplectic subspace (spanned by the
canonical variables ${\bf H}$, ${\bf K}$) of each constant energy [$P^{\tau}
_{\phi}=const.$] surface of the Klein-Gordon field. Each constant energy
surface is not a symplectic manifold, but in this way it turns out to be the
disjoint union (over ${\tilde X}^{\tau}_{\phi}$) of the symplectic manifolds
determined by ${\tilde X}^{\tau}_{\phi}=const.$, $P^{\tau}_{\phi}=const.$

\subsection{A family of canonical multipoles.}

From Eqs.(\ref{IV39}) we get [assuming that we can interchange the
sums with the integrals]

\begin{eqnarray}
\phi (T_s,\vec \sigma )&=&2 \int d\tilde q {\bf A}_{\vec q}(T_s;P^{\tau}_{\phi},
{\bf H}] \Big( cos\, \vec q\cdot \vec \sigma \,\, cos\, {\bf B}_{\vec q}(T_s;
{\tilde X}^{\tau}_{\phi},{\bf K}]-\nonumber \\
&-&sin\, \vec q\cdot \vec \sigma \,\, sin\, {\bf B}
_{\vec q}(T_s;{\tilde X}^{\tau}_{\phi},{\bf K}] \Big) =\nonumber \\
&=& \sum_{k=0}^{\infty} (-)^k \Big[ {{\sigma^{r_1}...\sigma^{r_{2k+1}}}\over
{(2k+1)!}} {\cal T}_{\phi ,O}^{r_1...r_{2k+1}}-
{{\sigma^{r_1}...\sigma^{r_{2k}}}\over {(2k)!}} {\cal T}_{\phi ,E}
^{r_1..r_{2k}} \Big] ,\nonumber \\
&&{}\nonumber \\
\pi (T_s,\vec \sigma )&=&-2 \int d\tilde q\omega (q)
{\bf A}_{\vec q}(T_s;P^{\tau}_{\phi},
{\bf H}] \Big( sin\, \vec q\cdot \vec \sigma \,\, cos\, {\bf B}_{\vec q}(T_s;
{\tilde X}^{\tau}_{\phi},{\bf K}]+\nonumber \\
&+&cos\, \vec q\cdot \vec \sigma \,\, sin\, {\bf B}
_{\vec q}(T_s;{\tilde X}^{\tau}_{\phi},{\bf K}] \Big) =\nonumber \\
&=& -\sum_{k=0}^{\infty} (-)^k \Big[ {{\sigma^{r_1}...\sigma^{r_{2k+1}}}\over
{(2k+1)!}} {\cal T}_{\pi ,O}^{r_1...r_{2k+1}}-
{{\sigma^{r_1}...\sigma^{r_{2k}}}\over {(2k)!}} {\cal T}_{\pi ,E}
^{r_1..r_{2k}} \Big] ,\nonumber \\
&&{}\nonumber \\
{\cal T}_{\phi ,O}^{r_1...r_{2k+1}}&=&
2\int d\tilde q\, q^{r_1}...q^{r_{2k+1}}\,
{\bf A}_{\vec q}(T_s;P^{\tau}_{\phi},{\bf H}]
cos\, {\bf B}_{\vec q}(T_s;{\tilde X}^{\tau}_{\phi},{\bf K}],\nonumber \\
{\cal T}_{\phi ,E}^{r_1..r_{2k}}&=&
2\int d\tilde q\, q^{r_1}...q^{r_{2k}}\,
{\bf A}_{\vec q}(T_s;P^{\tau}_{\phi},{\bf H}]
sin\, {\bf B}_{\vec q}(T_s;{\tilde X}^{\tau}_{\phi},{\bf K}],\nonumber \\
{\cal T}_{\pi ,O}^{r_1...r_{2k+1}}&=&
2\int d\tilde q\omega (q)\, q^{r_1}...q^{r_{2k+1}}\,
{\bf A}_{\vec q}(T_s;P^{\tau}_{\phi},{\bf H}]
sin\, {\bf B}_{\vec q}(T_s;{\tilde X}^{\tau}_{\phi},{\bf K}],\nonumber \\
{\cal T}_{\pi ,E}^{r_1..r_{2k}}&=&
2\int d\tilde q\omega (q)\, q^{r_1}...q^{r_{2k}}\,
{\bf A}_{\vec q}(T_s;P^{\tau}_{\phi},{\bf H}]
cos\, {\bf B}_{\vec q}(T_s;{\tilde X}^{\tau}_{\phi},{\bf K}] .
\label{IV41}
\end{eqnarray}

In this way we have defined a set of canonical multipoles of the fields $\phi
(T_s,\vec \sigma )$, $\pi (T_s,\vec \sigma )$, with respect to $\vec \sigma =0$
, i.e. with respect to the origin $x^{\mu}_s(\tau )$ of the Wigner hyperplanes
(a noncanonical covariant centroid) and they are expressed in
terms of the final canonical basis for the Klein-Gordon field.

When the fields have a compact support W in momentum space, it can be shown
that the only nonvanishing Poisson brackets of these multipoles are

\begin{eqnarray}
\{ {\cal T}_{\pi ,O}^{r_1...r_{2k+1}},{\cal T}_{\phi ,O}^{s_1...s_{2h+1}}\} &=&
{\cal I}^{r_1...r_{2k+1},s_1...s_{2h+1}},\nonumber \\
\{ {\cal T}_{\pi ,E}^{r_1...r_{2k}},{\cal T}_{\phi ,E}^{s_1...s_{2h}}\} &=&-
{\cal I}^{r_1...r_{2k},s_1...s_{2h}},\nonumber \\
&&{}\nonumber \\
{\cal I}^{r_1...r_m,s_1...s_n}&=&2 \int_Wd\tilde q \omega^2(q) q^{r_1}\cdots
q^{r_m}q^{s_1}\cdots q^{s_n}.
\label{IV42}
\end{eqnarray}

That is they form a closed algebra with a generalized Kronecker symbol, which
could be quantized instead of the Fourier coefficients.

\subsection{Effects of self interactions $V(\phi )$.}

When there is an interaction term $V(\phi )$, the Hamiltonian becomes $M_{\phi}=
{\tilde P}^{\tau}_{\phi}=P^{\tau}_{\phi}+\int d^3\sigma V(\phi )(T_s,\vec
\sigma )$ and one has [see the difference between the rest-frame time evolution
and the internal time one in presence of interactions]

\begin{eqnarray}
{{\partial}\over {\partial T_s}}\, X^{\tau}_{\phi}\, &{\buildrel \circ \over
=}\,& -1-{{\partial}\over {\partial P^{\tau}_{\phi}}} \int d^3\sigma V(\phi
(T_s,\vec \sigma ))=\nonumber \\
&=&-1-\int d\tilde k {{F^{\tau}(k)\omega (k)}\over { {\bf A}_{\vec k}(T_s;P
^{\tau}_{\phi},{\bf H}]}} \int d^3\sigma V^{'}(\phi (T_s,\vec \sigma ))
cos\, [\vec k\cdot \vec \sigma +{\bf B}_{\vec k}(T_s;X^{\tau}_{\phi},{\bf K}]
\, ],\nonumber \\
{{\partial}\over {\partial T_s}}\, {\tilde P}^{\tau}_{\phi}\, &{\buildrel \circ
\over =}\,& 0,\nonumber \\
{{\partial}\over {\partial T_s}}\, P^{\tau}_{\phi}\, &{\buildrel \circ \over
=}\,& {{\partial}\over {\partial X^{\tau}_{\phi}}}\, \int d^3\sigma V(\phi (T_s
,\vec \sigma ))=\nonumber \\
&=&-2 \int d\tilde k \omega (k) {\bf A}_{\vec k}(T_s;P^{\tau}_{\phi},{\bf H}]
\int d^3\sigma V^{'}(\phi (T_s,\vec \sigma )) sin\, [\vec k\cdot \vec \sigma +
{\bf B}(T_s;X^{\tau}_{\phi},{\bf K}]\, ],\nonumber \\
{{\partial}\over {\partial T_s}}\, {\bf H}(T_s,\vec q)\, &{\buildrel \circ
\over =}\,& {{\delta}\over {\delta {\bf K}(T_s,\vec q)}} \int d^3\sigma
V(\phi (T_s ,\vec \sigma ))=\nonumber \\
&=&-2 \int d^3\sigma V^{'}(\phi (T_s,\vec \sigma )) \int d\tilde k\, {\bf A}
_{\vec k}(T_s;P^{\tau}_{\phi},{\bf H}]\nonumber \\
&&sin\, [\vec k\cdot \vec \sigma +{\bf B}_{\vec k}(T_s;X^{\tau}_{\phi},
{\bf K}]\, ] \int d\tilde k^{'}\, {\cal G}(\vec q,{\vec k}^{'})\triangle
({\vec k}^{'},\vec k) ,\nonumber \\
{{\partial}\over {\partial T_s}}\, {\bf K}(T_s,\vec q)\, &{\buildrel \circ
\over =}\,& -{{\delta}\over {\delta {\bf H}(T_s,\vec q)}} \int d^3\sigma
V(\phi (T_s,\vec \sigma ))=\nonumber \\
&=&-\int d^3\sigma V^{'}(\phi (T_s,\vec \sigma ))\, {\cal D}_{\vec q}
\Big( {{cos\, \Big[ \vec q\cdot \vec \sigma +{\bf B}_{\vec q}(T_s;X^{\tau}
_{\phi},{\bf K}] \Big]}\over {{\bf A}_{\vec q}(T_s;P^{\tau}_{\phi},{\bf H}]}}
\Big),
\label{IV43}
\end{eqnarray}

\noindent with $V^{'}(\phi )=\partial V(\phi )/\partial \phi$. In particular
we get

\begin{eqnarray}
{{\partial}\over {\partial T_s}}{\bf A}_{\vec q}(T_s;P^{\tau}_{\phi},{\bf H}]
&=&{1\over {2{\bf A}_{\vec q}(T_s;P^{\tau}_{\phi},{\bf H}]}} \Big[ F^{\tau}(q)
\omega (q) {{\partial P^{\tau}_{\phi}}\over {\partial T_s}}+{\cal D}_{\vec q}
{{\partial {\bf H}(T_s,\vec q)}\over {\partial T_s}} \Big] \,
{\buildrel \circ \over =}\nonumber \\
&{\buildrel \circ \over =}\,& -{1\over {{\bf A}_{\vec q}(T_s;P^{\tau}_{\phi},
{\bf H}]}} \int d^3\sigma V^{'}(\phi (T_s,\vec \sigma ))
\int d\tilde k \Big[ F^{\tau}(q)\omega (q)\omega (k)+\triangle (\vec q,
\vec k)\Big] \nonumber \\
&& {\bf A}_{\vec k}(T_s;P^{\tau}_{\phi},{\bf H}] sin\, \Big[ \vec
k\cdot \vec \sigma +{\bf B}_{\vec k}(T_s;X^{\tau}_{\phi},{\bf K}]\Big] =
\nonumber \\
&=&-{1\over {{\bf A}_{\vec q}(T_s;P^{\tau}_{\phi},{\bf H}]}} \int d\tilde k
{\bf A}_{\vec k}(T_s;P^{\tau}_{\phi},{\bf H}]
\int d^3\sigma V^{'}(\phi (T_s,\vec \sigma ))\nonumber \\
&& sin\, \Big[ \vec
k\cdot \vec \sigma +{\bf B}_{\vec k}(T_s;X^{\tau}_{\phi},{\bf K}]\Big]
\Big[ \Omega (k)\delta^3(\vec q-\vec k) +F(q) \vec q\cdot \vec k\Big]
,\nonumber \\
{{\partial}\over {\partial T_s}}{\bf B}_{\vec q}(T_s;X^{\tau}_{\phi},{\bf K}]
&=&-\omega (q)\Big( 1-{{\partial X^{\tau}_{\phi}}\over {\partial T_s}}\Big) +
\int d\tilde k d{\tilde k}^{'} {{\partial {\bf K}(T_s,\vec k)}\over
{\partial T_s}} {\cal G}(\vec k,{\vec k}^{'}) \triangle ({\vec k}^{'},\vec q)\,
{\buildrel \circ \over =}\nonumber \\
&{\buildrel \circ \over =}\,& -2\omega (q)-\int d^3\sigma V^{'}(\phi (T_s,\vec
\sigma ))\nonumber \\
&&\int d\tilde k \Big[ \omega (q)F^{\tau}(k)\omega (k)+\triangle (\vec k,\vec q)
\Big] {{ cos\, \Big[ \vec k\cdot \vec \sigma +{\bf B}_{\vec k}(T_s;X^{\tau}
_{\phi},{\bf K}]\Big] }\over {{\bf A}_{\vec k}(T_s;P^{\tau}_{\phi},{\bf H}]}}
=\nonumber \\
&=&-2\omega (q)-\int d\tilde k {{\Omega (k)\delta^3(\vec q-\vec k)+F(k)\vec
q\cdot \vec k}\over {{\bf A}(T_s;P^{\tau}_{\phi},{\bf H}]}}\nonumber \\
&& \int d^3\sigma
V^{'}(\phi (T_s,\vec \sigma )) cos\, \Big[ \vec k\cdot \vec \sigma +{\bf B}
_{\vec k}(T_s;X^{\tau}_{\phi},{\bf K}]\Big] .
\label{IV44}
\end{eqnarray}

For $V(\phi )={1\over {n+1}} \xi \phi^{n+1}$ we have:\hfill\break
\hfill\break
$V^{'}(\phi (T_s,\vec \sigma ))=
\xi \phi^n(T_s,\vec \sigma )= 2^n \xi \int d{\tilde q}_1..d{\tilde q}_n
{\bf A}_{{\vec q}_1}(T_s;P^{\tau}_{\phi},{\bf H}]....{\bf A}_{{\vec q}_n}(T_s;
P^{\tau}_{\phi},{\bf H}] cos\, \Big( {\vec q}_1\cdot \vec \sigma +{\bf B}_{{\vec
q}_1}(T_s;{\tilde X}^{\tau}_{\phi},{\bf K}]\Big) ... cos\,  \Big( {\vec q}_n
\cdot \vec \sigma +{\bf B}_{{\vec q}_n}(T_s;{\tilde X}^{\tau}_{\phi},{\bf K}]
\Big)$.

For the sine-Gordon case we have $V(\phi )={{\mu^4}\over {\lambda}}[cos\,
({{\sqrt{\lambda}}\over {\mu}}-1]$ and $V^{'}(\phi (T_s,\vec \sigma ))=$
\hfill\break
\hfill\break
${{\mu^3}\over {\sqrt{\lambda}}} sin\, \Big({{\sqrt{\lambda}}\over {\mu}} \phi
(T_s,\vec \sigma )\Big) ={{\mu^3}\over {\sqrt{\lambda}}} sin\,
\Big( 2{{\sqrt{\lambda}}\over {\mu}} \int d\tilde q {\bf A}_{\vec q}(T_s;P
^{\tau}_{\phi},{\bf H}] cos\, \Big[ \vec q\cdot \vec \sigma +{\bf B}_{\vec
q}(T_s;{\tilde X}^{\tau}_{\phi},{\bf K}]\Big] \Big)$.

With a completely integrable interaction, it should exist a new canonical
basis ${\hat {\tilde X}}^{\tau}_{\phi}$, ${\hat P}^{\tau}_{\phi}$, ${\hat {\bf
H}}(T_s,\vec q)$, ${\hat {\bf K}}(T_s,\vec q)$ with ${\hat {\tilde X}}
^{\tau}_{\phi}$ the real ``internal" time like in the free case.

For instance, in the new canonical basis of Eqs.(\ref{IV35})
one can define a family of completely
integrable interactions associated with the Hamiltonian $M_{\phi}=P^{\tau}
_{\phi}+\int d\tilde q [terms\, quadratic\, in\, {\bf H}(T_s,\vec q)\, and\,
{\bf K}(T_s,\vec q)]$. However, their associated Lagrangian density as
functionals
of $\phi (\tau ,\vec \sigma )$ would be completely non local first because the
canonical transformation $\phi , \pi \mapsto X^A_{\phi}, P^A_{\phi},
{\bf H}, {\bf K}$ is non local, and
second because one would have to solve an integral equation to get the
momenta $\pi (\tau ,\vec \sigma )$ in terms of the velocities $\dot \phi (\tau
,\vec \sigma )$ [${{\partial}\over {\partial T_s}} \phi (T_s,\vec \sigma )\,
{\buildrel \circ \over =}\, \{ \phi (T_s,\vec \sigma ),M_{\phi} \}$].

\vfill\eject

\section{The energy-momentum tensor on Wigner hyperplanes, Dixon's multipoles
and the center of mass of a field configuration.}

Let us now look at other properties of the Klein-Gordon field on the Wigner
hyperplanes. In particular we are interested in identifying which kind of
collective variables describe the center of mass of a field configuration and
which is their relation, if any, with the previous collective variables. In so
doing, we shall consider a field configuration as a relativistic extended body
and we shall study its Dixon multipoles\cite{dixon}.

The Euler-Lagrange equations from the action (\ref{II1}) are

\begin{eqnarray}
&&\Big( {{\partial {\cal L}}\over {\partial z^{\mu}}}-\partial_A
{{\partial {\cal L}}\over {\partial z^{\mu}_A}}\Big) (\tau ,\vec
\sigma )=\eta_{\mu\nu}\partial_A[\sqrt{g} T^{AB}[\phi ] z_B^{\nu}](\tau ,\vec
\sigma )\, {\buildrel \circ \over =}\, 0,\nonumber \\
&&\Big( {{\partial {\cal L}}\over {\partial
\phi}}-\partial_A{{\partial {\cal L}}\over {\partial \partial_A\phi}}\Big)
(\tau ,\vec \sigma )\, {\buildrel \circ \over =}\, 0,
\label{V00}
\end{eqnarray}

\noindent where we introduced the energy-momentum tensor

\begin{eqnarray}
T^{AB}(\tau ,\vec \sigma )[\phi ]&=&-\Big[ {2\over {\sqrt{g}}}{{\delta
S}\over {\delta g_{AB}}}\Big] (\tau ,\vec \sigma )=\nonumber \\
 &=&[\partial^A\phi \partial^B\phi -{1\over 2}g^{AB}(\pi^2-(\vec \partial
 \phi )^2-m^2\phi^2)](\tau ,\vec \sigma ).
\label{V0}
\end{eqnarray}

When $\partial_A[\sqrt{g} z^{\mu}_B]=0$, as it happens on the Wigner
hyperplanes in the gauge $T_s-\tau \approx 0$, $\vec \lambda (\tau
)=0$, we get the conservation of the energy-momentum tensor $T^{AB}$,
i.e. $\partial_AT^{AB}\, {\buildrel \circ \over =}\, 0$. Otherwise,
there is compensation coming from the dynamics of the surface.

The conserved, manifestly Lorentz covariant, energy-momentum tensor of the
Klein-Gordon field $T^{\mu\nu}(x)[\tilde \phi ]=-{1\over 2}
\eta^{\mu\nu}[\partial_{\alpha}\tilde \phi (x)\partial^{\alpha}\tilde \phi (x)
-m^2{\tilde \phi}^2(x)]+\partial^{\mu}\tilde \phi (x)\partial^{\nu}\tilde \phi
(x)$ becomes $T^{AB}(\tau ,\vec \sigma )[\phi ]$ in  coordinates
adapted to the hypersurface $\Sigma_{\tau}$ [$\sigma^A=(\tau ,\vec
\sigma )$, $q^A=(\omega (q),\vec q)$], where

\begin{eqnarray}
T^{AB}(\tau ,\vec \sigma )[\phi ]&=&z^A_{\mu}(\tau ,\vec \sigma )z^B_{\nu}(\tau
,\vec \sigma )T^{\mu\nu}(z(\tau ,\vec \sigma ))[\tilde \phi ]=\nonumber \\
&=&z^A_{\mu}(\tau ,\vec \sigma )z^B_{\nu}(\tau
,\vec \sigma )T^{\mu\nu}(\tau ,\vec \sigma )[\phi =\tilde \phi \circ z]=
\nonumber \\
&=&-{1\over 2}g^{AB}(\tau ,\vec \sigma )\Big[ \pi^2 -
(\vec \partial \phi )^2-m^2\phi^2 \Big] (\tau ,\vec \sigma )+\partial^A\phi
(\tau ,\vec \sigma ) \partial^B\phi (\tau ,\vec \sigma ).
\label{V1}
\end{eqnarray}

By using the results in Ref.\cite{lus}, we have the following expression
for the energy-momentum tensor

1) On arbitrary spacelike hypersurfaces we get

\begin{eqnarray}
T^{\mu\nu}(\tau ,\vec \sigma )[\phi ]&=&-{1\over 2}\eta^{\mu\nu}\Big[ \partial
_A\phi (\tau ,\vec \sigma )\partial^A\phi (\tau ,\vec \sigma )-m^2\phi^2(\tau
,\vec \sigma )\Big]+\nonumber \\
&+&z^{\mu}_A(\tau ,\vec \sigma )z^{\nu}_B(\tau ,\vec \sigma )
\partial^A\phi (\tau ,\vec \sigma )\partial^B\phi (\tau ,\vec \sigma )=
\nonumber \\
&=&\Big[ -{1\over 2}\eta^{\mu\nu} g^{\tau\tau}+(z^{\mu}_{\tau}g^{\tau\tau}+
z^{\mu}_ug^{u\tau})(z^{\nu}_{\tau}g^{\tau\tau}+z^{\nu}_vg^{v\tau})\Big] (\tau
,\vec \sigma ) \pi^2(\tau ,\vec \sigma )+\nonumber \\
&+&\Big[ -\eta^{\mu\nu}g^{\tau r}+(z^{\mu}_{\tau}g^{\tau\tau}+z^{\mu}_ug
^{u\tau})(z^{\nu}_{\tau}g^{\tau r}+z^{\nu}_vg^{vr})+\nonumber \\
&+&(z^{\mu}_{\tau}g
^{\tau r}+z^{\mu}_ug^{ur})(z^{\nu}_{\tau}g^{\tau\tau}+z^{\nu}_vg^{v\tau})\Big]
(\tau ,\vec \sigma ) [\pi \, \partial_r\phi ](\tau ,\vec \sigma )+\nonumber \\
&+&\Big[-{1\over 2}\eta^{\mu\nu}g^{rs}+(z^{\mu}_{\tau}g^{\tau r}+z^{\mu}_ug
^{ur})(z^{\nu}_{\tau}g^{\tau s}+z^{\nu}_vg^{vs})\Big] (\tau ,\vec \sigma )
[\partial_r\phi \partial_s\phi ](\tau ,\vec \sigma )+\nonumber \\
&+&{1\over 2}\eta^{\mu\nu} m^2 \phi^2(\tau ,\vec \sigma ).
\label{V2}
\end{eqnarray}

2) On arbitrary spacelike hyperplanes, where

\begin{eqnarray}
z^{\mu}(\tau ,\vec \sigma )&=&x^{\mu}_s(\tau )+b^{\mu}_u(\tau
)\sigma^u,\nonumber \\
 z^{\mu}_r(\tau ,\vec \sigma
)&=&b^{\mu}_r(\tau ),\quad\quad z^{\mu}_{\tau}(\tau ,\vec \sigma
)={\dot x}_s^{\mu}(\tau )+{\dot b}^{\mu}_u (\tau
)\sigma^u=l^{\mu}/\sqrt{g^{\tau\tau}}-g_{\tau r}z^{\mu}_r,\nonumber \\
 &&{}\nonumber \\
g_{\tau \tau}&=& {\dot x}_s\cdot {\dot b}_r\sigma^r]^2,\quad\quad
g_{\tau r}=b_{r\mu}[{\dot x}_s^{\mu}+ {\dot b}^{\mu}_s\sigma^s],
\nonumber \\
g_{rs}&=&-\delta_{rs},\quad\quad \gamma^{rs}=-\delta^{rs},\quad\quad
\gamma =1,\nonumber \\
g&=&g_{\tau\tau}+\sum_rg^2_{\tau r}, \nonumber \\
  g^{\tau\tau}&=&1/[l_{\mu}({\dot x}_s^{\mu}+{\dot b}^{\mu}_u\sigma^u)]^2,
\quad\quad g^{\tau r}
=g^{\tau\tau}g_{\tau r}=b_{r\mu}({\dot x}_s^{\mu}+{\dot b}^{\mu}_u\sigma^u)/
[l_{\mu}({\dot x}_s^{\mu}+{\dot b}^{\mu}_u\sigma^u)]^2,\nonumber \\
g^{rs}&=&-\delta^{rs} +g^{\tau\tau}g_{\tau r}g_{\tau s}=-\delta^{rs}+
b_{r\mu}({\dot x}_s^{\mu}+{\dot b}^{\mu}_u\sigma^u) b_{r\nu}({\dot
x}_s^{\nu}+{\dot b}^{\nu}_v\sigma^v)
/[l_{\mu}({\dot x}_s^{\mu}+{\dot b}^{\mu}_u\sigma^u)]^2,
\label{V3}
\end{eqnarray}

\noindent there is no relevant simplification for $T^{\mu\nu}(x^{\beta}_s(\tau
)+b^{\beta}_u(\tau )\sigma
^u)[\phi]$.

3) On Wigner hyperplanes, where

\begin{eqnarray}
z^{\mu}(\tau ,\vec \sigma )&=&x^{\mu}_s(\tau
)+\epsilon^{\mu}_u(u(p_s))\sigma^u,\nonumber \\
 &&{}\nonumber \\
z^{\mu}_r&=&\epsilon^{\mu}_r(u(p_s),\quad\quad
l^{\mu}=u^{\mu}(p_s),\quad\quad z^{\mu}_{\tau}={\dot x}^{\mu}_s(\tau
),\nonumber \\
 g&=&[{\dot x}_s(\tau )\cdot u(p_s)],\quad\quad g_{\tau\tau}={\dot x}^2_s,
 \quad\quad g_{\tau r}={\dot x}_{s\mu} \epsilon^{\mu}_r(u(p_s)),\quad\quad
 g_{rs}=-\delta_{rs},\nonumber \\
g^{\tau\tau}&=&1/[{\dot x}_{s\mu}u^{\mu}(p_s)]^2,\quad\quad g^{\tau
r}={\dot x}_{s\mu}\epsilon^{\mu}
_r(u(p_s))/[{\dot x}_{s\mu}u^{\mu}(p_s)]^2,\nonumber \\
g^{rs}&=&-\delta^{rs}+{\dot x}
_{s\mu}\epsilon^{\mu}_r(u(p_s)) {\dot x}_{s\nu}\epsilon^{\nu}_s(u(p_s))
/[{\dot x}_{s\mu}u^{\mu}(p_s)]^2,
\label{V4}
\end{eqnarray}

\noindent we get

\begin{eqnarray}
T^{\mu\nu}[x^{\mu}_s(\tau )&+&\epsilon^{\mu}_u(u(p_s))\sigma^u][\phi ]=
{1\over {[{\dot x}_s(\tau )\cdot u(p_s)]^2}}\Big[ -{1\over 2}\eta^{\mu\nu}+
\nonumber \\
&+&[{\dot x}^{\mu}_s(\tau )+{\dot x}_{s\rho}(\tau )\epsilon^{\rho}_u(u(p_s))
\epsilon^{\mu}_u(u(p_s))][{\dot x}^{\nu}_s(\tau )+{\dot x}_{s\sigma}(\tau )
\epsilon^{\sigma}_u(u(p_s))\epsilon^{\nu}_u(u(p_s))]\Big] \pi^2(\tau ,\vec
\sigma )+\nonumber \\
&+&{1\over {[{\dot x}_s(\tau )\cdot u(p_s)]^2}}
\Big[ {1\over 2}\eta^{\mu\nu}\Big( [{\dot x}_s(\tau )\cdot u(p_s)]^2\delta
_{rs}-{\dot x}_{s\alpha}(\tau )\epsilon^{\alpha}_r(u(p_s)){\dot x}
_{s\beta}(\tau )\epsilon^{\beta}_s(u(p_s))\big)+\nonumber \\
&+&[{\dot x}_s(\tau )\cdot u(p_s)]^2\Big( \epsilon^{\mu}_r(u(p_s))-
{{{\dot x}_{s\beta}(\tau )\epsilon^{\beta}_r(u(p_s))}\over {[{\dot x}_s(\tau )
\cdot u(p_s)]^2}}[{\dot x}^{\mu}_s(\tau )+{\dot x}_{s\alpha}(\tau )
\epsilon^{\alpha}_u(u(p_s))\epsilon^{\mu}_u(u(p_s))]\Big)\nonumber \\
&&\Big( \epsilon^{\nu}_r(u(p_s))-
{{{\dot x}_{s\beta}(\tau )\epsilon^{\beta}_r(u(p_s))}\over {[{\dot x}_s(\tau )
\cdot u(p_s)]^2}}[{\dot x}^{\nu}_s(\tau )+{\dot x}_{s\alpha}(\tau )
\epsilon^{\alpha}_v(u(p_s))\epsilon^{\nu}_v(u(p_s))]\Big) \Big] \nonumber \\
&& [\partial_r\phi \partial_s\phi ](\tau ,\vec \sigma )+
{1\over 2}\eta^{\mu\nu} m^2 \phi^2(\tau ,\vec \sigma )+\nonumber \\
&+&{1\over {[{\dot x}_s(\tau )\cdot u(p_s)]^2}} \Big[ -\eta^{\mu\nu}{\dot x}
_{s\rho}(\tau )\epsilon^{\rho}_r(u(p_s))-\nonumber \\
&-&[{\dot x}^{\mu}_s(\tau )+{\dot x}_{s\rho}(\tau )\epsilon^{\rho}_u(u(p_s))
\epsilon^{\mu}_u(u(p_s))]\nonumber \\
&&\Big( \epsilon^{\nu}_r(u(p_s))-
{{{\dot x}_{s\beta}(\tau )\epsilon^{\beta}_r(u(p_s))}\over {[{\dot x}_s(\tau )
\cdot u(p_s)]^2}}[{\dot x}^{\nu}_s(\tau )+{\dot x}_{s\alpha}(\tau )
\epsilon^{\alpha}_v(u(p_s))\epsilon^{\nu}_v(u(p_s))]\Big)-\nonumber \\
&-&\Big( \epsilon^{\mu}_r(u(p_s))-
{{{\dot x}_{s\beta}(\tau )\epsilon^{\beta}_r(u(p_s))}\over {[{\dot x}_s(\tau )
\cdot u(p_s)]^2}}[{\dot x}^{\mu}_s(\tau )+{\dot x}_{s\alpha}(\tau )
\epsilon^{\alpha}_u(u(p_s))\epsilon^{\mu}_u(u(p_s))]\Big) \nonumber \\
&&[{\dot x}^{\nu}_s(\tau )+{\dot x}_{s\rho}(\tau )\epsilon^{\rho}_v(u(p_s))
\epsilon^{\nu}_v(u(p_s))] [\pi \partial_r\phi ](\tau ,\vec \sigma ) .
\label{V5}
\end{eqnarray}

Since we have

\begin{eqnarray}
{\dot x}_s^{\mu}(\tau )&=&-\lambda^{\mu}(\tau
)=[u^{\mu}(p_s)u^{\nu}(p_s)-\epsilon^{\mu}_r(u(p_s))\epsilon^{\nu}_r(u(p_s))]
{\dot x}_{s \nu}(\tau )=\nonumber \\
&=&- u^{\mu}(p_s) \lambda (\tau )+
\epsilon^{\mu}_r(u(p_s)) \lambda_r(\tau ),\nonumber \\
{\dot x}^2_s(\tau )&=& \lambda^2(\tau )-{\vec \lambda}(\tau ) > 0,\nonumber \\
U^{\mu}_s(\tau )&=& {{ {\dot x}^{\mu}_s(\tau )}\over { \sqrt{{\dot x}^2
_s(\tau )} }}={{-\lambda (\tau )u^{\mu}(p_s)+\lambda_r(\tau )\epsilon^{\mu}_r
(u(p_s))}\over {\sqrt{\lambda^2(\tau )-{\vec \lambda}^2(\tau )} }},
\label{V6}
\end{eqnarray}

\noindent the timelike worldline described by the origin of the Wigner
hyperplane is arbitrary (i.e. gauge dependent): $x^{\mu}_s(\tau )$ may
be any covariant noncanonical centroid [the real ``external" center of
mass is the canonical noncovariant ${\tilde x}^{\mu}_s(T_s)= x^{\mu}
_s(T_s)-{1\over {\epsilon_s(p^o_s+\epsilon_s)}}\Big[
p_{s\nu}S^{\nu\mu}_s+\epsilon_s(S^{o\mu}_s+S^{o\nu}_s {{p_{s\nu}p^{\mu}_s}\over
{\epsilon_s^2}}) \Big]$: it describes a decoupled point particle observer ;
see Section VI].

In the gauge $T_s-\tau \approx 0$, ${\vec X}_{\phi}\approx 0$,
implying $\lambda (\tau )=-1$, $\vec \lambda (\tau )=0$ [$g_{\tau\tau}=1$,
$g_{\tau r}=0$], we get ${\dot x}^{\mu}_s(T_s)=u^{\mu}(p_s)$. Therefore, in this
gauge, we have the centroid $x^{\mu}_s(T_s)=x_s^{({\vec X}_{\phi})\mu }(T_s)=
x^{\mu}_s(0)+T_s u^{\mu}(p_s)$, which carries the Klein-Gordon ``internal"
collective variable ${\vec \sigma}_{\phi}={\vec X}_{\phi}\approx 0$ (see
Section VI).

In this gauge we get the following form of the energy-momentum tensor
[$\eta^{\mu\nu}=u^{\mu}(p_s)u^{\nu}(p_s)-\sum_{r=1}^3\epsilon^{\mu}_r(u(p_s))
\epsilon^{\nu}_r(u(p_s))$]

\begin{eqnarray}
T^{\mu\nu}[x^{\mu}_s(T_s)+\epsilon^{\mu}_u(u(p_s))\sigma^u][\phi ]&=&
T^{\mu\nu}[x^{\mu}_s(T_s)+\epsilon^{\mu}_u(u(p_s))\sigma^u][X^A_{\phi},
P^A_{\phi},{\bf H},{\bf K}]=\nonumber \\
&=&{1\over 2}u^{\mu}(p_s)u^{\nu}(p_s)[\pi^2+(\vec \partial \phi )^2+
m^2\phi^2](T_s,\vec \sigma )+\nonumber \\
&+&\epsilon^{\mu}_r(u(p_s))\epsilon^{\nu}_s(u(p_s))[-{1\over 2}\delta_{rs}
[\pi^2-(\vec \partial \phi )^2-m^2\phi^2]+\nonumber \\
&+&\partial_r\phi\partial_s\phi ](T_s,
\vec \sigma )-\nonumber \\
&-&[u^{\mu}(p_s)\epsilon^{\nu}_r(u(p_s))+u^{\nu}(p_s)\epsilon^{\mu}_r(u(p_s))]
[\pi \partial_r\phi ](T_s,\vec \sigma )=\nonumber \\
&=& \Big[ \rho [\phi ,\pi ] u^{\mu}(p_s)u^{\nu}(p_s)+{\cal P}[\phi ,\pi ] [\eta
^{\mu\nu}-u^{\mu}(p_s)u^{\nu}(p_s)]+\nonumber \\
&+&u^{\mu}(p_s)q^{\nu}[\phi ,\pi ]+
u^{\nu}(p_s)q^{\mu}[\phi ,\pi ]+\nonumber \\
&+&T^{rs}_{an\, stress}[\phi ,\pi ] \epsilon^{\mu}_r(u(p_s))\epsilon^{\nu}
_s(u(p_s))\Big] (T_s,\vec \sigma ),\nonumber \\
 &&{}\nonumber \\
 \rho [\phi ,\pi ]&=&{1\over 2}[\pi^2+(\vec \partial \phi
)^2+m^2\phi^2],\nonumber \\
 {\cal P}[\phi ,\pi ]&=&{1\over 2}
[\pi^2-{5\over 3}(\vec \partial \phi )^2- m^2\phi^2],\nonumber \\
 q^{\mu}[\phi ,\pi ]&=&-\pi \partial_r\phi
\epsilon^{\mu}_r(u(p_s)),\nonumber \\
 T^{rs}_{an\, stress}[\phi ,\pi
]&=&-[\partial^r\phi \partial^s\phi -{1\over 3}
\delta^{rs}(\vec \partial \phi )^2],\nonumber \\
 &&{}{}{} \delta_{uv}T^{uv}_{an\,
stress}[\phi ,\pi ]=0,\nonumber \\
&&{}\nonumber \\
T^{rs}_{stress}(T_s,\vec \sigma )[\phi ]&=&\epsilon^r_{\mu}(u(p_s))\epsilon^s
_{\nu}(u(p_s))T^{\mu\nu}[x^{\mu}_s(T_s)+\epsilon^{\mu}
_u(u(p_s))\sigma^u][\phi ]=\nonumber \\
&=&[\partial^r\phi \partial^s\phi ](T_s,\vec \sigma )-
{1\over 2}\delta^{rs}
[\pi^2-(\vec \partial \phi )^2-m^2\phi^2](T_s,\vec \sigma ),\nonumber \\
&&{}\nonumber \\
T^{\mu}{}_{\mu}[x^{\mu}_s(T_s)+\epsilon^{\mu}_u(u(p_s))\sigma^u][\phi ]&=& 2
[\pi^2-(\vec \partial \phi )^2-m^2 \phi^2](T_s, \vec \sigma ),\nonumber \\
&&{}\nonumber \\
T^{\mu\nu}[x^{\mu}_s(T_s)+\epsilon^{\mu}_u(u(p_s))\sigma^u][\phi ]\,\,
u_{\nu}(p_s)&=&{1\over 2}[\pi^2+(\vec \partial \phi )^2+m^2\phi^2](T_s,\vec
\sigma ) u^{\mu}(p_s)+\nonumber \\
&+&[\pi \partial^r\phi ](T_s,\vec \sigma )\epsilon^{\mu}
_r(u(p_s)),\nonumber \\
&&{}\nonumber \\
P^{\mu}_T[\phi]&=&\int d^3\sigma T^{\mu\nu}[x^{\mu}_s(T_s)+\epsilon^{\mu}
_u(u(p_s))\sigma^u][\phi ]u_{\nu}(p_s)=\nonumber \\
&=&P^{\tau}_{\phi} u^{\mu}(p_s)+P^r_{\phi}\epsilon^{\mu}_r(u(p_s))\approx
P^{\tau}_{\phi} u^{\mu}(p_s) \approx p^{\mu}_s,\nonumber \\
M_{\phi}&=&P^{\mu}_T[\phi ] u_{\mu}(p_s) =P^{\tau}_{\phi}.\nonumber \\
\label{V7}
\end{eqnarray}

\noindent While the stress tensor of the Klein-Gordon field on the Wigner
hyperplanes is $T^{rs}_{stress}(T_s,\vec \sigma )[\phi ]$, from the last line
of the expression of the energy-momentum tensor we see that it acquires a form
reminiscent of the energy-momentum tensor of an ideal relativistic fluid as
seen from a local observer at rest (Eckart decomposition; see Ref.\cite{israel}
):
i) the constant normal $u^{\mu}(p_s)$ to the Wigner hyperplanes replaces the
hydrodynamic velocity field of the fluid; ii) $\rho [\phi ,\pi ](T_s,\vec
\sigma ]$ is the energy density; iii) ${\cal P}[\phi ,\pi ](T_s,\vec \sigma )$
is the analogue of the pressure (sum of the thermodynamical pressure and of the
non-equilibrium bulk stress or viscous pressure); iv) $q^{\mu}[\phi ,\pi ](T_s,
\vec \sigma )$ is the analogue of the heat flow; v) $T^{rs}_{an\, stress}[\phi
,\pi ](T_s,\vec \sigma )$ is the shear (or anisotropic) stress tensor.

We can now study the manifestly Lorentz covariant Dixon multipoles
\cite{dixon} for the free real Klein-Gordon field on the Wigner
hyperplanes in the gauge $\lambda (\tau )=-1$, $\vec \lambda (\tau
)=0$ [so that ${\dot x}_s^{\mu}(T_s )=u^{\mu}(p_s)$, ${\ddot
x}^{\mu}_s(T_s)=0$, $x^{\mu}_s(T_s)=u^{\mu}(p_s) T_s+x^{\mu}_s(0)$]
with respect to the origin $x^{\mu}_s(T_s)$ [$\delta x^{\mu}_s(\vec
\sigma )=\epsilon^{\mu}_u(u(p_s)) \sigma^u$; $(\mu_1..\mu_n)$ means
symmetrization, while $[\mu_1..\mu_n]$ means antisymmetrization]

\begin{eqnarray}
t_T^{\mu_1...\mu_n\mu\nu}(T_s)&=&t_T^{(\mu_1...\mu_n)(\mu\nu)}(T_s)=
\nonumber \\
&=&\int d^3\sigma \delta x^{\mu_1}_s(\vec \sigma )...\delta x^{\mu_n}_s(\vec
\sigma ) T^{\mu\nu}[x^{\mu}_s(T_s)+\epsilon^{\mu}_u(u(p_s))\sigma^u][\phi ]
=\nonumber \\
&=&\epsilon^{\mu_1}_{r_1}(u(p_s))...\epsilon^{\mu_n}_{r_n}(u(p_s))\nonumber \\
&&\Big[ u^{\mu}(p_s) u^{\nu}(p_s) {1\over 2}\int d^3\sigma \sigma^{r_1}...
\sigma^{r_n}[\pi^2+(\vec \partial \phi )^2+m^2\phi^2](T_s,\vec \sigma )+
\nonumber \\
&+&\epsilon^{\mu}_r(u(p_s))\epsilon^{\nu}_s(u(p_s))\int d^3\sigma \sigma^{r_1}
...\sigma^{r_n}[-{1\over 2}\delta_{rs}
[\pi^2-(\vec \partial \phi )^2-m^2\phi^2]+\nonumber \\
&+&\partial_r\phi\partial_s\phi ](T_s,
\vec \sigma )+\nonumber \\
&+&[u^{\mu}(p_s)\epsilon^{\nu}_r(u(p_s))+u^{\nu}(p_s)\epsilon^{\mu}_r(u(p_s))]
\int d^3\sigma \sigma^{r_1}...\sigma^{r_n}[\pi \partial^r\phi ](T_s,\vec
\sigma ) \Big] =\nonumber \\
&=&\epsilon^{\mu_1}_{r_1}(u(p_s))...\epsilon^{\mu_n}_{r_n}(u(p_s))
\epsilon^{\mu}_A(u(p_s))\epsilon^{\nu}_B(u(p_s)) I_T^{r_1..r_nAB}(T_s)=\nonumber \\
&=&\epsilon^{\mu_1}_{r_1}(u(p_s))...\epsilon^{\mu_n}_{r_n}(u(p_s))
\Big[ u^{\mu}(p_s) u^{\nu}(p_s) I_T^{r_1...r_n\tau\tau}(T_s)+\nonumber \\
&+&\epsilon^{\mu}_r(u(p_s))\epsilon^{\nu}_s(u(p_s))
I_T^{r_1...r_nrs}(T_s)+\nonumber \\
&+&[u^{\mu}(p_s)\epsilon^{\nu}_r(u(p_s))+u^{\nu}(p_s)\epsilon^{\mu}_r(u(p_s))]
I_T^{r_1...r_nr\tau}(T_s) \Big] ,\nonumber \\
 &&{}\nonumber \\
u_{\mu_1}(p_s)&& t_T^{\mu_1...\mu_n\mu\nu}(T_s)=0,\nonumber \\
 &&{}\nonumber \\
 For && n=0\, (monopole)\quad I^{\tau\tau}_T(T_s)=P^{\tau}_{\phi},\quad\quad
 I_T^{r\tau}(T_s)=P^r_{\phi},\nonumber \\
 &&{}\nonumber \\
t_T^{\mu_1...\mu_n\mu}{}_{\mu}(T_s)&=&\int d^3\sigma \delta
x^{\mu_1}_s(\vec
\sigma )...\delta x^{\mu_n}_s(\vec \sigma ) T^{\mu}{}_{\mu}
[x^{\mu}_s(T_s)+\epsilon^{\mu}_u(u(p_s))\sigma^u][\phi ]
=\nonumber \\
&=&\epsilon^{\mu_1}_{r_1}(u(p_s))...\epsilon^{\mu_n}_{r_n}(u(p_s))\nonumber \\
&&\int d^3\sigma \sigma^{r_1}...\sigma^{r_n} \Big[ 2[\pi^2-(\vec \partial \phi
)^2]-m^2\phi^2\Big] (T_s,\vec \sigma )=\nonumber \\
&{\buildrel {def} \over =}&\epsilon^{\mu_1}_{r_1}(u(p_s))...\epsilon^{\mu_n}
_{r_n}(u(p_s))I_T^{r_1...r_nA}{}_A(T_s)\nonumber \\
&&{}\nonumber \\
 I_T^{r_1r_2A}{}_A(T_s)&=&{\check I}_T^{r_1r_2A}{}_A(T_s)-{1\over 3}\delta^{r_1r_2}\delta
_{uv}I_T^{uvA}{}_A(T_s)=i^{r_1r_2}_T(T_s)-{1\over 2}\delta^{r_1r_2}\delta_{uv}i_T
^{uv}(T_s),\nonumber \\
{\check I}_T^{r_1r_2A}{}_A(T_s)&=&i_T^{r_1r_2}(T_s)-{1\over
3}\delta^{r_1r_2}\delta
_{uv}i_T^{uv}(T_s),\quad\quad \delta_{uv}{\check I}_T^{uvA}{}_A(T_s)=0,\nonumber \\
i^{r_1r_2}_T(T_s)&=&I_T^{r_1r_2A}{}_A(T_s)-\delta^{r_1r_2}\delta_{uv}I_T^{uvA}{}_A(T_s)=
\nonumber \\
&=&\int d^3\sigma (\sigma^{r_1}\sigma^{r_2}-\delta
^{r_1r_2}{\vec \sigma}^2) \Big[ 2[\pi^2-(\vec \partial \phi )^2]-m^2\phi^2\Big]
(T_s,\vec \sigma ),\nonumber \\
 &&{}\nonumber \\
 {\tilde t}_T^{\mu_1...\mu_n}(T_s)&=& t_T^{\mu_1...\mu_n\mu\nu}(T_s)
u_{\mu}(p_s)u_{\nu}(p_s)=\nonumber \\
&=&\epsilon^{\mu_1}_{r_1}(u(p_s))...\epsilon^{\mu_n}_{r_n}(u(p_s))
I_T^{r_1..r_n\tau\tau}(T_s),\nonumber \\
 {\tilde t}_T^{\mu_1}(T_s)&=&
\epsilon^{\mu_1}_{r_1}(u(p_s)) {1\over 2}
\int d^3\sigma \sigma^{r_1}
[\pi^2+(\vec \partial \phi )^2+m^2\phi^2](T_s ,\vec \sigma )=
\nonumber \\
 &=&\epsilon^{\mu_1}_{r_1}(u(p_s)) I_T^{r_1\tau\tau}(T_s)=-P^{\tau}_{\phi}
\epsilon^{\mu_1}_{r_1}(u(p_s)) r^{r_1}_{\phi},\nonumber \\
 &&{}\nonumber \\
{\tilde t}_T^{\mu_1\mu_2}(T_s)&=&
\epsilon^{\mu_1}_{r_1}(u(p_s))\epsilon^{\mu_2}_{r_2}(u(p
_s))I_T^{r_1r_2\tau\tau}(T_s),\nonumber \\
 &&{}\nonumber \\
I^{r_1r_2\tau\tau}_T(T_s)&=&{\hat I}^{r_1r_2\tau\tau}_T(T_s)-{1\over
3}\delta^{r_1r_2}
\delta_{uv}I^{uv\tau\tau}_T(T_s)={\tilde i}^{r_1r_2}_T(T_s)-{1\over 2}\delta
^{r_1r_2}\delta_{uv}{\tilde i}^{uv}_T(T_s),\nonumber \\
{\hat I}^{r_1r_2\tau\tau}_T(T_s)&=&{\tilde i}^{r_1r_2}_T(T_s)-{1\over
3}\delta^{r_1r_2}
\delta_{uv}{\tilde i}^{uv}_T(T_s),\quad\quad \delta_{uv}{\hat I}^{uv\tau\tau}_T(T_s)=0,
\nonumber \\
{\tilde
i}_T^{r_1r_2}(T_s)&=&I_T^{r_1r_2\tau\tau}(T_s)-\delta^{r_1r_2}\delta
_{uv}I_T^{uv\tau\tau}(T_s).
\label{V8}
\end{eqnarray}

The Wigner covariant multipoles $I_T^{r_1..r_n\tau\tau}(T_s)$,
$I_T^{r_1..r_nrs}(T_s)$, $I_T^{r_1..r_nr\tau}(T_s)$ are the mass,
stress and momentum multipoles respectively.

The quantities ${\check I}_T^{r_1r_2A}{}_A(T_s)$ and $i^{r_1r_2}_T
(T_s)$ are the traceless quadrupole moment and the inertia tensor
defined by Thorne in Ref.\cite{thorne}.

The quantities $I_T^{r_1r_2\tau\tau}(T_s)$ and ${\tilde i}_T^{r_1r_2}
(T_s)$ are Dixon's definitions of quadrupole moment and of tensor of
inertia respectively.

Moreover, Dixon's definition of ``center of mass" of an extended
object is ${\tilde t}^{\mu_1}_T(T_s)=0$ or
$I_T^{r\tau\tau}(T_s)=-P^{\tau}_{\phi} r^r_{\phi}=0$: therefore the
quantity ${\vec r}_{\phi} $ defined in the previous equation is a
noncanonical [$\{ r^r_{\phi},r^s_{\phi}
 \} =S^{rs}_{\phi}$] candidate for the ``internal" center of mass of the
field configuration: its vanishing is a gauge fixing for ${\vec
P}_{\phi}\approx 0$ and implies $x^{\mu}_s(T_s)=x^{({\vec
r}_{\phi})\mu}_s(T_s)=x^{\mu}_s(0)+u^{\mu}(p_s) T_s$ (see next
Section).

When $I^{r\tau\tau}_T(T_s)=0$, the equations
$0={{dI^{r\tau\tau}_T(T_s)}\over {dT_s}}=-P^{\tau}_{\phi}
{{dr^r_{\phi}}\over {dT_s}}= {d\over {dT_s}} \int d^3\sigma \,
\sigma^r[\pi^2+(\vec \partial \phi )^2+m^2\phi^2](T_s,\vec \sigma )\,
{\buildrel \circ \over =}\, -P^r_{\phi}$ implies the correct
momentum-velocity relation ${{{\vec P}_{\phi}}\over
{P^{\tau}_{\phi}}}\, {\buildrel \circ \over =}\, {{d{\vec
r}_{\phi}}\over {dT_s}} \approx 0$>

These multipoles,  whose Poisson brackets are non-trivial, are related
to the previous ones of Eqs.(\ref{IV41}), with the Poisson brackets
given in Eqs.(\ref{IV42}),in the following way (remember that these
other multipoles exist only if the Klein-Gordon fields have compact
support V in momentum space)

\begin{eqnarray}
I_T^{r_1...r_n\tau\tau}(T_s)&=&{1\over 2}\int d^3\sigma
\sigma^{r_1}...
\sigma^{r_n}[\pi^2+(\vec \partial \phi )^2+m^2\phi^2](T_s,\vec \sigma )=
\nonumber \\
&=&2 \int d{\tilde q}_1d{\tilde q}_2 {\bf A}_{{\vec q}_1}(T_s;P^{\tau}_{\phi},
{\bf H}] {\bf A}_{{\vec q}_2}(T_s;P^{\tau}_{\phi},{\bf H}] \int d^3\sigma
\sigma^{r_1}...\sigma^{r_n} \nonumber \\
&&\Big( [\omega (q_1)\omega (q_2) +{\vec q}_1\cdot {\vec q}_2] sin\,
({\vec q}_1\cdot \vec \sigma +{\bf B}_{{\vec q}_1}(T_s;{\tilde X}^{\tau}_{\phi},
{\bf K}]) sin\,
({\vec q}_2\cdot \vec \sigma +{\bf B}_{{\vec q}_2}(T_s;{\tilde X}^{\tau}_{\phi},
{\bf K}]) +\nonumber \\
&+&m^2 cos\,
({\vec q}_1\cdot \vec \sigma +{\bf B}_{{\vec q}_1}(T_s;{\tilde X}^{\tau}_{\phi},
{\bf K}]) cos\,
({\vec q}_2\cdot \vec \sigma +{\bf B}_{{\vec q}_2}(T_s;{\tilde X}^{\tau}_{\phi},
{\bf K}]) \Big) =\nonumber \\
&=&{1\over 2} \sum_{h,k=0}^{\infty} (-)^{h+k} \nonumber \\
&&\Big( {1\over {(2h+1)!(2k+1)!}}
\Big[ V_1^{r_1...r_nu_1...u_{2h+1}v_1...v_{2k+1}} \nonumber \\
&&[{\cal T}_{\pi ,O}^{u_1...u
_{2h+1}}{\cal T}_{\pi ,O}^{v_1...v_{2k+1}}+m^2{\cal T}_{\phi ,O}^{u_1...u
_{2h+1}}{\cal T}_{\phi ,O}^{v_1...v_{2k+1}}]+\nonumber \\
&+&V_2^{r_1...r_nu_1...u_{2h+1}v_1...v_{2k+1}}{\cal T}_{\phi ,O}^{u_1...u
_{2h+1}}{\cal T}_{\phi ,O}^{v_1...v_{2k+1}} \Big] -\nonumber \\
&-&{1\over {(2h+1)!(2k)!}}
\Big[ V_1^{r_1...r_nu_1...u_{2h+1}v_1...v_{2k}} [{\cal T}_{\pi ,O}^{u_1...u
_{2h+1}}{\cal T}_{\pi ,E}^{v_1...v_{2k}}+m^2{\cal T}_{\phi ,O}^{u_1...u
_{2h+1}}{\cal T}_{\phi ,E}^{v_1...v_{2k}}]+\nonumber \\
&+&V_2^{r_1...r_nu_1...u_{2h+1}v_1...v_{2k}}{\cal T}_{\phi ,O}^{u_1...u_{2h+1}}
{\cal T}_{\phi ,E}^{v_1...v_{2k}} \Big] -\nonumber \\
&-&{1\over {(2h)!(2k+1)!}}
\Big[ V_1^{r_1...r_nu_1...u_{2h}v_1...v_{2k+1}} [{\cal T}_{\pi ,E}^{u_1...u
_{2h}}{\cal T}_{\pi ,O}^{v_1...v_{2k+1}}+m^2{\cal T}_{\phi ,E}^{u_1...u
_{2h}}{\cal T}_{\phi ,O}^{v_1...v_{2k+1}}]+\nonumber \\
&+&V_2^{r_1...r_nu_1...u_{2h}v_1...v_{2k+1}}{\cal T}_{\phi ,E}^{u_1...u_{2h}}
{\cal T}_{\phi ,O}^{v_1...v_{2k+1}} \Big] +\nonumber \\
&+&{1\over {(2h)!(2k)!}}
\Big[ V_1^{r_1...r_nu_1...u_{2h}v_1...v_{2k}} [{\cal T}_{\pi ,E}^{u_1...u
_{2h}}{\cal T}_{\pi ,E}^{v_1...v_{2k}}+m^2{\cal T}_{\phi ,E}^{u_1...u
_{2h}}{\cal T}_{\phi ,E}^{v_1...v_{2k}}]+\nonumber \\
&+&V_2^{r_1...r_nu_1...u_{2h}v_1...v_{2k}}{\cal T}_{\phi
,E}^{u_1...u_{2h}} {\cal T}_{\phi ,E}^{v_1...v_{2k}} \Big] \Big)
,\nonumber \\
 I_T^{r_1...r_nrs}(T_s)&=&\int d^3\sigma
\sigma^{r_1}...\sigma^{r_n}[-{1\over 2}
\delta_{rs}[\pi^2-(\vec \partial \phi )^2-m^2\phi^2]+\partial_r\phi\partial_s
\phi ](T_s,\vec \sigma )=\nonumber \\
&=&4\int d{\tilde q}_1d{\tilde q}_2 {\bf A}_{{\vec q}_1}(T_s;P^{\tau}_{\phi},
{\bf H}] {\bf A}_{{\vec q}_2}(T_s;P^{\tau}_{\phi},{\bf H}] \int d^3\sigma
\sigma^{r_1}...\sigma^{r_n}\nonumber \\
&&\Big( -{1\over 2}\delta^{rs}\Big[ [\omega (q_1)\omega (q_2) -{\vec q}_1\cdot
{\vec q}_2] \nonumber \\
&&sin\,
({\vec q}_1\cdot \vec \sigma +{\bf B}_{{\vec q}_1}(T_s;{\tilde X}^{\tau}_{\phi},
{\bf K}]) sin\,
({\vec q}_2\cdot \vec \sigma +{\bf B}_{{\vec q}_2}(T_s;{\tilde X}^{\tau}_{\phi},
{\bf K}])-\nonumber \\
&-&m^2 cos\,
({\vec q}_1\cdot \vec \sigma +{\bf B}_{{\vec q}_1}(T_s;{\tilde X}^{\tau}_{\phi},
{\bf K}]) cos\,
({\vec q}_2\cdot \vec \sigma +{\bf B}_{{\vec q}_2}(T_s;{\tilde X}^{\tau}_{\phi},
{\bf K}]) \Big] +\nonumber \\
&+& q^r_1 q^s_2 sin\,
({\vec q}_1\cdot \vec \sigma +{\bf B}_{{\vec q}_1}(T_s;{\tilde X}^{\tau}_{\phi},
{\bf K}]) sin\,
({\vec q}_2\cdot \vec \sigma +{\bf B}_{{\vec q}_2}(T_s;{\tilde X}^{\tau}_{\phi},
{\bf K}]) \Big) =\nonumber \\
&=&\sum_{h,k=0}^{\infty} (-)^{h+k}\nonumber \\
&&\Big( {1\over {(2h+1)!(2k+1)!}}
\Big[-{1\over 2}\delta^{rs} V_1^{r_1...r_nu_1...u_{2h+1}v_1...v_{2k+1}}
\nonumber \\
&&[{\cal T}_{\pi ,O}^{u_1...u
_{2h+1}}{\cal T}_{\pi ,O}^{v_1...v_{2k+1}}-m^2{\cal T}_{\phi ,O}^{u_1...u
_{2h+1}}{\cal T}_{\phi ,O}^{v_1...v_{2k+1}}]+\nonumber \\
&+&({1\over 2}\delta^{rs}
V_2^{r_1...r_nu_1...u_{2h+1}v_1...v_{2k+1}}+V_3^{r_1...r_nu_1...u_{2h+1}v_1...
v_{2k+1}rs})
{\cal T}_{\phi ,O}^{u_1...u_{2h+1}}
{\cal T}_{\phi ,O}^{v_1...v_{2k+1}} \Big] -\nonumber \\
&-&{1\over {(2h+1)!(2k)!}}
\Big[ -{1\over 2}\delta^{rs}V_1^{r_1...r_nu_1...u_{2h+1}v_1...v_{2k}}
\nonumber \\
&&[{\cal T}_{\pi ,O}^{u_1...u
_{2h+1}}{\cal T}_{\pi ,E}^{v_1...v_{2k}}-m^2{\cal T}_{\phi ,O}^{u_1...u
_{2h+1}}{\cal T}_{\phi ,E}^{v_1...v_{2k}}]+\nonumber \\
&+&({1\over 2}\delta^{rs}V_2^{r_1...r_nu_1...u_{2h+1}v_1...v_{2k}}+V_3^{r_1...
r_nu_1...u_{2h+1}v_1...v_{2k}rs}){\cal T}_{\phi ,O}^{u_1...u_{2h+1}}
{\cal T}_{\phi ,E}^{v_1...v_{2k}} \Big] -\nonumber \\
&-&{1\over {(2h)!(2k+1)!}}
\Big[ -{1\over 2}\delta^{rs}V_1^{r_1...r_nu_1...u_{2h}v_1...v_{2k+1}}
\nonumber \\
&&[{\cal T}_{\pi ,E}^{u_1...u
_{2h}}{\cal T}_{\pi ,O}^{v_1...v_{2k+1}}-m^2{\cal T}_{\phi ,E}^{u_1...u
_{2h}}{\cal T}_{\phi ,O}^{v_1...v_{2k+1}}]+\nonumber \\
&+&({1\over 2}\delta^{rs}V_2^{r_1...r_nu_1...u_{2h}v_1...v_{2k+1}}+V_3^{r_1...
r_nu_1...u_{2h}v_1...v_{2k+1}rs}){\cal T}_{\phi ,E}^{u_1...u_{2h}}
{\cal T}_{\phi ,O}^{v_1...v_{2k+1}} \Big] +\nonumber \\
&+&{1\over {(2h)!(2k)!}}
\Big[ -{1\over 2}\delta^{rs}
V_1^{r_1...r_nu_1...u_{2h}v_1...v_{2k}} [{\cal T}_{\pi ,E}^{u_1...u
_{2h}}{\cal T}_{\pi ,E}^{v_1...v_{2k}}-m^2{\cal T}_{\phi ,E}^{u_1...u
_{2h}}{\cal T}_{\phi ,E}^{v_1...v_{2k}}]+\nonumber \\
&+&({1\over
2}\delta^{rs}V_2^{r_1...r_nu_1...u_{2h}v_1...v_{2k}}+V_3^{r_1...
r_nu_1...u_{2h}v_1...v_{2k}rs}){\cal T}_{\phi ,E}^{u_1...u_{2h}} {\cal
T}_{\phi ,E}^{v_1...v_{2k}} \Big] \Big) ,\nonumber \\
I_T^{r_1...r_nr\tau}(T_s)&=&\int d^3\sigma
\sigma^{r_1}...\sigma^{r_n}[\pi
\partial^r
\phi ](T_s,\vec \sigma )=\nonumber \\
&=&4 \int d{\tilde q}_1d{\tilde q}_2 {\bf A}_{{\vec q}_1}(T_s;P^{\tau}_{\phi},
{\bf H}] {\bf A}_{{\vec q}_2}(T_s;P^{\tau}_{\phi},{\bf H}] \omega (q_1)
q^r_2 \int d^3\sigma \sigma^{r_1}...\sigma^{r_n}\nonumber \\
&&sin\,
({\vec q}_1\cdot \vec \sigma +{\bf B}_{{\vec q}_1}(T_s;{\tilde X}^{\tau}_{\phi},
{\bf K}]) sin\,
({\vec q}_2\cdot \vec \sigma +{\bf B}_{{\vec q}_2}(T_s;{\tilde X}^{\tau}_{\phi},
{\bf K}]) =\nonumber \\
&=&\sum_{h,k=0}^{\infty} (-)^{h+k}\nonumber \\
&&\Big( {1\over {(2h+1)!(2k+1)!}}
V_4^{r_1...r_nu_1...u_{2h+1}v_1...v_{2k+1}} {\cal T}_{\pi ,O}^{u_1...u
_{2h+1}}{\cal T}_{\phi ,O}^{v_1...v_{2k+1}}-\nonumber \\
&-&{1\over {(2h+1)!(2k)!}}
V_4^{r_1...r_nu_1...u_{2h+1}v_1...v_{2k}} {\cal T}_{\pi ,O}^{u_1...u
_{2h+1}}{\cal T}_{\phi ,E}^{v_1...v_{2k}}-\nonumber \\
&-&{1\over {(2h)!(2k+1)!}}
V_4^{r_1...r_nu_1...u_{2h}v_1...v_{2k+1}} {\cal T}_{\pi ,E}^{u_1...u
_{2h}}{\cal T}_{\phi ,O}^{v_1...v_{2k+1}}+\nonumber \\
&+&{1\over {(2h)!(2k)!}}
V_4^{r_1...r_nu_1...u_{2h}v_1...v_{2k}} {\cal T}_{\pi ,E}^{u_1...u
_{2h}}{\cal T}_{\phi ,E}^{v_1...v_{2k}}\Big) ,\nonumber \\
&&{}\nonumber \\
 &&where\nonumber \\
  &&{}\nonumber \\
&&V_1^{r_1...r_nu_1...u_hv_1...v_k}=\int_V d^3\sigma \sigma^{r_1}...\sigma^{r_n}
\sigma^{u_1}...\sigma^{u_h}\sigma^{v_1}...\sigma^{v_k},\nonumber \\
&&V_2^{r_1...r_nu_1...u_hv_1...v_k}=\int_V d^3\sigma \sigma^{r_1}...\sigma^{r_n}
\vec \partial (\sigma^{u_1}...\sigma^{u_h}) \cdot \vec \partial (\sigma^{v_1}
...\sigma^{v_k}),\nonumber \\
&&V_3^{r_1...r_nu_1...u_hv_1...v_krs}=\int_V d^3\sigma \sigma^{r_1}...\sigma
^{r_n} \partial^r (\sigma^{u_1}...\sigma^{u_h})\, \partial^s (\sigma^{v_1}...
\sigma^{v_k}),\nonumber \\
&&V_4^{r_1...r_nu_1...u_hv_1...v_kr}=\int_V d^3\sigma \sigma^{r_1}...\sigma
^{r_n}\sigma^{u_1}...\sigma^{u_h}\, \partial^r (\sigma^{v_1}...\sigma^{v_k}).
\label{V9}
\end{eqnarray}

Then there are the related Dixon multipoles

\begin{eqnarray}
p^{\mu_1...\mu_n\mu}_T(T_s)&=&t_T^{\mu_1...\mu_n\mu\nu}(T_s)
u_{\nu}(p_s)= p_T^{(\mu_1...\mu_n)\mu}(T_s)=\nonumber \\
 &=&\epsilon^{\mu_1}_{r_1}(u(p_s))...\epsilon^{\mu_n}_{r_n}(u(p_s))
\epsilon^{\mu}_A(u(p_s)) I_T^{r_1..r_nA\tau}(T_s)=\nonumber \\
&=&\epsilon^{\mu_1}_{r_1}(u(p_s))...\epsilon^{\mu_n}_{r_n}(u(p_s)) \int
d^3\sigma \sigma^{r_1}...\sigma^{r_n} \Big[
{1\over 2}[\pi^2+(\vec \partial \phi )^2+m^2\phi^2](T_s ,\vec
\sigma ) u^{\mu}(p_s)+\nonumber \\
&+&[\pi \partial^r\phi ](T_s ,\vec \sigma )\epsilon^{\mu}
_r(u(p_s)) \Big] =\nonumber \\
&=&2 \epsilon^{\mu_1}_{r_1}(u(p_s))...\epsilon^{\mu_n}_{r_n}(u(p_s))
\nonumber \\
&&\int d{\tilde q}_1 d{\tilde q}_2 {\bf A}_{{\vec q}_1}(T_s;P^{\tau}_{\phi},
{\bf H}] {\bf A}_{{\vec q}_2}(T_s;P^{\tau}_{\phi},{\bf H}] \int d^3\sigma
\sigma^{r_1}...\sigma^{r_n}\nonumber \\
&&\Big[ u^{\mu}(p_s) \Big( (\omega (q_1) \omega (q_2) +{\vec q}_1\cdot {\vec
q}_2) \nonumber \\
&&sin\, ({\vec q}_1\cdot \vec \sigma +{\bf B}_{{\vec q}_1}(T_s;{\tilde X}
^{\tau}_{\phi},{\bf K}])
sin\, ({\vec q}_2\cdot \vec \sigma +{\bf B}_{{\vec q}_2}(T_s;{\tilde X}
^{\tau}_{\phi},{\bf K}])+\nonumber \\
&+& m^2 cos\, ({\vec q}_1\cdot \vec \sigma +{\bf B}_{{\vec q}_1}(T_s;{\tilde X}
^{\tau}_{\phi},{\bf K}])
cos\, ({\vec q}_1\cdot \vec \sigma +{\bf B}_{{\vec q}_1}(T_s;{\tilde X}
^{\tau}_{\phi},{\bf K}]) \Big) -\nonumber \\
&-&\epsilon^{\mu}_r(u(p_s))\,\, q^r_2\,\,
sin({\vec q}_1\cdot \vec \sigma +{\bf B}_{{\vec q}_1}(T_s;{\tilde X}
^{\tau}_{\phi},{\bf K}])
sin({\vec q}_2\cdot \vec \sigma +{\bf B}_{{\vec q}_2}(T_s;{\tilde X}
^{\tau}_{\phi},{\bf K}]) \Big] ,\nonumber \\
 &&{}\nonumber \\
u_{\mu_1}(p_s)&& p_T^{\mu_1...\mu_n\mu}(T_s)=0,\nonumber \\
 &&{}\nonumber \\
 n=0&&\Rightarrow p^{\mu}_T(T_s)=P^{\mu}_T[\phi ]=\epsilon^{\mu}_A(u(p_s))
 P^A_{\phi}\approx p^{\mu}_s,\nonumber \\
 &&{}\nonumber \\
 p_T^{\mu_1..\mu_n\mu}(T_s)u_{\mu}(p_s)&=&{\tilde
t}_T^{\mu_1...\mu_n}(T_s)
=\epsilon^{\mu_1}_{r_1}(u(p_s))...\epsilon^{\mu_n}_{r_n}(u(p_s))
I_T^{r_1..r_n\tau\tau}(T_s).
\label{V10}
\end{eqnarray}

The spin dipole is defined as

\begin{eqnarray}
S^{\mu\nu}_T(T_s)[\phi ]&=&2 p_T^{[\mu\nu ]}(T_s)=2
\epsilon^{[\mu}_r(u(p_s))\epsilon^{\nu ]}_A(u(p_s)) I_T^{rA\tau}(T_s)=\nonumber \\
 &=&[\epsilon^{\mu}_r(u(p_s)) u^{\nu}(p_s)-\epsilon^{\nu}_r(u(p_s)) u^{\mu}
(p_s)] {1\over 2} \int d^3\sigma \sigma^r [\pi^2+(\vec \partial \phi )^2
+m^2\phi^2](T_s ,\vec \sigma )+\nonumber \\
&+&[\epsilon^{\mu}_r(u(p_s)) \epsilon^{\nu}_s(u(p_s))-
\epsilon^{\nu}_r(u(p_s)) \epsilon^{\mu}_s(u(p_s))] \int d^3\sigma \sigma^r
[\pi \partial^s\phi ](T_s,\vec \sigma )=\nonumber \\
&=&S^{\mu\nu}_s=\epsilon^{\mu}_r(u(p_s))\epsilon^{\nu}_s(u(p_s))S^{rs}_{\phi}+
[\epsilon^{\mu}_r(u(p_s)) u^{\nu}(p_s)-\epsilon^{\nu}_r(u(p_s)) u^{\mu}
(p_s)]S^{\tau r}_{\phi}=\nonumber \\
&=&\epsilon^{\mu}_r(u(p_s))\epsilon^{\nu}_s(u(p_s)) \int d\tilde q
{\bf H}(T_s,\vec q) \Big(q^r{{\partial}\over {\partial q^s}}-q^s{{\partial}
\over {\partial q^r}}\Big) {\bf K}(T_s,\vec q)-\nonumber \\
&-&[\epsilon^{\mu}_r(u(p_s)) u^{\nu}(p_s)-\epsilon^{\nu}_r(u(p_s)) u^{\mu}
(p_s)]\int d\tilde q\, \omega (q) {\bf H}(T_s,\vec q) {{\partial}\over
{\partial q^r}} {\bf K}(T_s,\vec q),\nonumber \\
 &&{}\nonumber \\
 &&u_{\mu}(p_s) S^{\mu\nu}_T(T_s)[\phi ]=-\epsilon^{\nu}_r(u(p_s))
S^{\tau r}_{\phi}=-{\tilde
t}^{\nu}_T(T_s)=P^{\tau}_{\phi}\epsilon^{\nu}_r(u(p_s)) r^r_{\phi},
\label{V11}
\end{eqnarray}

\noindent with $u_{\mu}(p_s)S^{\mu\nu}_T(T_s)[\phi ]=0$
when ${\tilde t}_T^{\mu_1}(T_s)=0$ and this condition can be taken as
a definition of center of mass equivalent to Dixon's one. When this
condition holds, the barycentrice spin dipole is
$S^{\mu\nu}_T(T_s)[\phi ]=2 \epsilon^{[\mu}_r(u(p_s)) \epsilon^{\nu
]}_s(u(p_s)) I_T^{rs\tau}(T_s)$, so that
$I^{[rs]\tau}_T(T_s)=\epsilon^r_{\mu}(u(p_s)) \epsilon^s_{\nu}(u(p_s))
S^{\mu\nu}_T(T_s)[\phi ]$.

As shown in Ref.\cite{dixon}, if the Klein-Gordon field has a compact support
W on the Wigner hyperplanes $\Sigma_{W\tau}$ and if $f(x)$ is a $C^{\infty}$
complex-valued scalar function on Minkowski spacetime with compact support [so
that its Fourier transform $\tilde f(k)=\int d^4x f(x) e^{ik\cdot x}$ is a
slowly increasing entire analytic function on Minkowski spacetime ($|(x^o+iy^o)
^{q_o}...(x^3+iy^3)^{q_3}f(x^{\mu}+iy^{\mu})| < C_{q_o...q_3} e^{a_o|y^o|+...+
a_3|y^3|}$, $a_{\mu} > 0$, $q_{\mu}$ positive integers for every $\mu$ and
$C_{q_o...q_3} > 0$), whose inverse is $f(x)=\int {{d^4k}\over {(2\pi )^4}}
\tilde f(k)e^{-ik\cdot x}$], we have

\begin{eqnarray}
< T^{\mu\nu},f >&=& \int d^4x T^{\mu\nu}(x) f(x)=\nonumber \\
&=&\int dT_s\int d^3\sigma f(x_s+\delta x_s)T^{\mu\nu}[x_s(T_s)+\delta x_s(\vec
\sigma )][\phi ]=\nonumber \\
&=&\int dT_s \int d^3\sigma \int {{d^4k}\over {(2\pi )^4}} \tilde f(k)
e^{-ik\cdot [x_s(T_s)+\delta x_s(\vec \sigma )]}T^{\mu\nu}
[x_s(T_s)+\delta x_s(\vec \sigma )][\phi ]=\nonumber \\
&=&\int dT_s\int {{d^4k}\over {(2\pi )^4}} \tilde f(k) e^{-ik\cdot x_s(T_s)}
\int d^3\sigma T^{\mu\nu}[x_s(T_s)+\delta x_s(\vec \sigma )][\phi ]\nonumber \\
&&\sum_{n=0}^{\infty} {{(-i)^n}\over {n!}} [k_{\mu}\epsilon^{\mu}_u(u(p_s))
\sigma^u]^n=\nonumber \\
&=&\int dT_s\int {{d^4k}\over {(2\pi )^4}} \tilde f(k) e^{-ik\cdot
x_s(T_s)}
\sum_{n=0}^{\infty} {{(-i)^n}\over {n!}} k_{\mu_1}...k_{\mu_n} t_T^{\mu_1...
\mu_n\mu\nu}(T_s),
\label{V12}
\end{eqnarray}

\noindent and, but only for $f(x)$ analytic in W \cite{dixon}, we get

\begin{eqnarray}
< T^{\mu\nu},f >&=& \int dT_s \sum_{n=0}^{\infty}{1\over {n!}} t_T^{\mu_1...
\mu_n\mu\nu}(T_s) {{\partial^nf(x)}\over {\partial x^{\mu_1}...\partial x^{\mu
_n}}}{|}_{x=x_s(T_s)},\nonumber \\
&&\Downarrow \nonumber \\
T^{\mu\nu}(x)[\phi ]&=& \sum_{n=0}^{\infty}{{(-1)^n}\over {n!}}
{{\partial^n}\over {\partial x^{\mu_1}...\partial x^{\mu_n}}} \int dT_s
\delta^4(x-x_s(T_s)) t_T^{\mu_1...\mu_n\mu\nu}(T_s).
\label{V13}
\end{eqnarray}

For a non analytic $f(x)$ we have

\begin{eqnarray}
< T^{\mu\nu},f > &=& \int dT_s \sum^N_{n=0} {1\over {n!}} t_T^{\mu_1...\mu_n
\mu\nu}(T_s) {{\partial^nf(x)}\over {\partial x^{\mu_1}...\partial x^{\mu
_n}}}{|}_{x=x_s(T_s)}+\nonumber \\
&+&\int dT_s \int {{d^4k}\over {(2\pi )^4}} \tilde f(k) e^{-ik\cdot x_s(T_s)}
\sum_{n=N+1}^{\infty} {{(-i)^n}\over {n!}} k_{\mu_1}...k_{\mu_n} t_T^{\mu_1...
\mu_n\mu\nu}(T_s),
\label{V14}
\end{eqnarray}

\noindent and, as shown in Ref.\cite{dixon}, from the knowledge of the moments
$t_T^{\mu_1...\mu_n\mu}(T_s)$ for all $n > N$ we can get $T^{\mu\nu}(x)$ and,
thus, all the moments with $n\leq N$.

In Appendix C  other types of Dixon's multipoles are analyzed. From
this study it turns out that the multipolar expansion(\ref{V13}) may
be rearranged with the help of the Hamilton equations implying
$\partial_{\mu}T^{\mu\nu}\, {\buildrel \circ \over =}\, 0$, so that
for analytic Klein-Gordon configurations from Eq.(\ref{c5}) we get

\begin{eqnarray}
T^{\mu\nu}(x)[\phi]\, &{\buildrel \circ \over =}\,& u^{(\mu}(p_s)
\epsilon^{\nu )}_A(u(p_s)) \int dT_s\, \delta^4(x-x_s(T_s))\,
P^A_{\phi}+\nonumber \\
 &+&{1\over 2} {{\partial}\over {\partial x^{\rho}}} \int dT_s\,
 \delta^4(x-x_s(T_s))\, S_T^{\rho (\mu}(T_s)[\phi] u^{\nu )}(p_s)+\nonumber \\
 &+&\sum_{n=2}^{\infty}{{(-1)^n}\over {n!}} {{\partial^n}\over
 {\partial x^{\mu_1}...\partial x^{\mu_n}}} \int dT_s\, \delta^4(x-x_s(T_s))
 {\cal I}_T^{\mu_1..\mu_n\mu\nu}(T_s),
\label{V15}
\end{eqnarray}

\noindent where for $n\geq 2$ $\quad {\cal I}_T^{\mu_1..\mu_n\mu\nu}(T_s)
={{4(n-1)}\over {n+1}} J_T^{(\mu_1..\mu_{n-1} | \mu | \mu_n)\nu}(T_s)$,
with the quantities $J_T^{\mu_1..\mu_n\mu\nu\rho\sigma}(T_s)$ being
the Dixon $2^{2+n}$-pole inertial moment tensors given in
Eqs.(\ref{c7}) [the quadrupole and related inertia tensor are
proportional to $I_T^{r_1r_2\tau\tau}(T_s)$].

The equations $\partial_{\mu}T^{\mu\nu}\, {\buildrel \circ \over =}\,
0$ imply the Papapetrou-Dixon-Souriau equations for the `pole-dipole'
system $P_T^{\mu}(T_s)$ and $S^{\mu\nu}_T(T_s)[\phi ]$ [see
Eqs.(\ref{c1}) and (\ref{c4})]

\begin{eqnarray}
{{d P^{\mu}_T(T_s)}\over {dT_s}}\, &{\buildrel \circ \over =}\,&
0,\nonumber \\
 {{d S^{\mu\nu}_T(T_s)[\phi ]}\over {dT_s}}\,
&{\buildrel \circ \over =}\,& 2 P^{[\mu}_T(T_s) u^{\nu ]}(p_s)=2
P^r_{\phi} \epsilon^{[\mu}_r(u(p_s)) u^{\nu ]}(p_s) \approx 0.
\label{V16}
\end{eqnarray}

\vfill\eject

\section{External and internal canonical center of mass, Moller's center of
energy and Fokker-Pryce center of inertia}

Let us now consider the problem of the definition of the relativistic
center of mass of a Klein-Gordon field configuration, after having
seen in the previous Section, Eq(\ref{V10}), Dixon's definition of
this concept in the multipolar approach. Let us remark that in the
approach leading to the rest-frame instant form of dynamics on
Wigner's hyperplanes there is a splitting of this concept in an
``external" and an ``internal" one. One can either look at the
isolated system from an arbitrary Lorentz frame or put himself inside
the Wigner hyperplane.

From outside one finds after the canonical reduction to Wigner hyperplane that
there is an origin $x^{\mu}_s(\tau )$ for these hyperplanes (a covariant
noncanonical centroid) and a noncovariant canonical coordinate ${\tilde x}
^{\mu}_s(\tau )$ describing an ``external" decoupled point particle observer
with a clock measuring the rest-frame time $T_s$. Associated with them
there is the ``external" realization (\ref{II9}) of the Poincar\'e
group.

Instead, all the degrees of freedom of the isolated system (here the
Klein-Gordon field configuration) are described by canonical variables on the
Wigner hyperplane restricted by the rest-frame condition ${\vec P}_{\phi}
\approx 0$, implying that an ``internal" collective variable ${\vec X}_{\phi}$
is a gauge variable and that only relative variables are physical
degrees of freedom (a form of weak Mach principle). Inside the Wigner
hyperplane at $\tau =0$ there is another realization of the Poincar\'e
group given by Eqs. (\ref{III4}), (\ref{III5}) with generators
$P^{\tau}_{\phi}$, $P^r_{\phi}$, $J^{rs}_{\phi}$, $K^r_{\phi}=J^{\tau
r}_{\phi}{|}_{\tau =0}=S^{\tau r}_{\phi}$, the ``internal" Poincar\' e
algebra. By using the methods of Ref.\cite{pauri} (where there is a
complete discussion of many definitions of relativistic
center-of-mass-like variables) we can build the following three
'internal" (that is inside the Wigner hyperplane) Wigner 3-vectors
corresponding to the 3-vectors 'canonical center of mass' ${\vec
q}_{\phi}$, 'Moller center of energy' ${\vec r}_{\phi}$ and
'Fokker-Pryce center of inertia' ${\vec y}_{\phi}$ (the analogous
concepts for the relativistic N-body problem are under study
\cite{iten}).

The noncanonical ``internal" M\o ller center of energy and the
associated spin 3-vector are

\begin{eqnarray}
{\vec r}_{\phi}&=& - {{\vec K}_{\phi}\over {P^{\tau}_{\phi}}} =
-{1\over {2P^{\tau}_{\phi}}} \int d^3\sigma
\,\, \vec \sigma \, [\pi^2+(\vec \partial \phi )^2+m^2\phi^2](\tau ,\vec
\sigma )=\nonumber \\
&=&-{2\over {P^{\tau}_{\phi}}}
\int d{\tilde q}_1d{\tilde q}_2 {\bf A}_{{\vec q}_1}(\tau ;P^A_{\phi},{\bf H}]
{\bf A}_{{\vec q}_2}(\tau ;P^A_{\phi},{\bf H}] \int d^3\sigma \, \vec \sigma
\nonumber \\
&&\Big([\omega (q_1)\omega (q_2)+{\vec q}_1\cdot {\vec q}_2] sin\, ({\vec q}_1
\cdot \vec \sigma +{\bf B}_{{\vec q}_1}(\tau ;X^A_{\phi},{\bf K}]) sin\,
({\vec q}_2 \cdot \vec \sigma +{\bf B}_{{\vec q}_2}(\tau ;X^A_{\phi},{\bf K}])
+\nonumber \\
&+&m^2 cos\, ({\vec q}_1 \cdot \vec \sigma +{\bf B}_{{\vec q}_1}(\tau ;X^A
_{\phi},{\bf K}]) cos\,
({\vec q}_2 \cdot \vec \sigma +{\bf B}_{{\vec q}_2}(\tau ;X^A_{\phi},{\bf K}])
\Big), \nonumber \\
&&{}\nonumber \\
{\vec \Omega}_{\phi} &=& {\vec J}_{\phi} -{\vec r}_{\phi}\times {\vec P}
_{\phi},\nonumber \\
&&\{ r^r_{\phi},P^s_{\phi} \} =\delta^{rs},\quad\quad \{ r^r_{\phi},P^{\tau}
_{\phi} \} ={{P^r_{\phi}}\over {P^{\tau}_{\phi}}},\nonumber \\
&&\{ r^r_{\phi},r^s_{\phi} \} =-{1\over {(P^{\tau}_{\phi})^2}} \epsilon^{rsu}
\Omega^u_{\phi},\nonumber \\
&&\{ \Omega^r_{\phi},\Omega^s_{\phi} \} =\epsilon^{rsu}(\Omega^u_{\phi}-{1\over
{(P^{\tau}_{\phi})^2}}({\vec \Omega}_{\phi} \cdot {\vec P}_{\phi})\,\, P_{\phi}
^u),\quad\quad \{ \Omega^r_{\phi},P^{\tau}_{\phi} \} =0.
\label{VI1}
\end{eqnarray}

\noindent We see that ${\vec r}_{\phi}$ coincides with Dixon's definition of
Eq.(\ref{V10}).

The canonical ``internal"
center of mass [$\{ q^r_{\phi},q^s_{\phi} \} =0$, $\{ q^r_{\phi},P^s_{\phi} \} =
\delta^{rs}$, $\{ J^r_{\phi},q^s_{\phi} \} =\epsilon^{rsu}q^u_{\phi}$] is

\begin{eqnarray}
{\vec q}_{\phi}&=& {\vec r}_{\phi}- {{{\vec J}_{\phi}\times {\vec \Omega}
_{\phi}}\over {\sqrt{(P^{\tau}_{\phi})^2-{\vec P}^2_{\phi}}(P^{\tau}_{\phi}+
\sqrt{(P^{\tau}_{\phi})^2-{\vec P}^2_{\phi}})}}=
\nonumber \\
&=&-{{{\vec K}_{\phi}}\over {\sqrt{(P^{\tau}_{\phi})^2-{\vec P}^2_{\phi}}}}+
{{{\vec J}_{\phi}\times {\vec P}_{\phi}}\over {\sqrt{(P^{\tau}_{\phi})^2-{\vec
P}^2_{\phi}}(P^{\tau}_{\phi}+\sqrt{(P^{\tau}_{\phi})^2
-{\vec P}^2_{\phi}})}}+\nonumber \\
&+&{{({\vec K}_{\phi}\cdot {\vec P}_{\phi})\,\, {\vec P}_{\phi}}\over {P^{\tau}
_{\phi}\sqrt{(P^{\tau}_{\phi})^2-{\vec P}^2_{\phi}}\Big( P^{\tau}_{\phi}+
\sqrt{(P^{\tau}_{\phi})^2-{\vec P}^2_{\phi}}\Big) }},\nonumber \\
&&\approx {\vec r}_{\phi}\quad for\quad {\vec P}_{\phi}\approx 0;\quad\quad
\{ {\vec q}_{\phi},P^{\tau}_{\phi} \} ={{{\vec P}_{\phi}}\over {P^{\tau}
_{\phi}}}\approx 0,\nonumber \\
&&{}\nonumber \\
{\vec S}_{q \phi} &=&{\vec J}_{\phi}-{\vec q}_{\phi}\times {\vec P}_{\phi}=
\nonumber \\
&=& {{P^{\tau}_{\phi}{\vec J}_{\phi}}\over {\sqrt{(P^{\tau}_{\phi})^2-{\vec P}
^2_{\phi}}}}+{{{\vec K}_{\phi}\times {\vec P}_{\phi}}\over {\sqrt{(P^{\tau}
_{\phi})^2-{\vec P}^2_{\phi}}}}-{{({\vec J}_{\phi}\cdot {\vec P}_{\phi})\,\,
{\vec P}_{\phi}}\over {\sqrt{(P^{\tau}_{\phi})^2-{\vec P}^2_{\phi}}\Big(
P^{\tau}_{\phi}+\sqrt{(P^{\tau}_{\phi})^2-{\vec P}^2_{\phi}}\Big) }}\approx
\nonumber \\
&\approx& {\vec S}_{\phi}, \quad for\quad {\vec P}_{\phi}\approx 0,\nonumber \\
&&\{ {\vec S}_{q \phi},{\vec P}_{\phi} \} =\{ {\vec S}_{q \phi},{\vec q}_{\phi}
\} =0,\quad\quad \{ S^r_{q \phi},S^s_{q \phi} \} =\epsilon^{rsu}S^u_{q \phi}.
\label{VI2}
\end{eqnarray}

The ``internal" noncanonical Fokker-Pryce center of inertia' ${\vec
y}_{\phi}$ is

\begin{eqnarray}
{\vec y}_{\phi}&=& {\vec q}_{\phi}+{{{\vec S}_{q \phi}\times {\vec P}_{\phi}}
\over {\sqrt{(P^{\tau}_{\phi})^2-{\vec P}_{\phi}^2} (P^{\tau}_{\phi}+
\sqrt{(P^{\tau}_{\phi})^2-{\vec P}_{\phi}^2})}}
={\vec r}_{\phi}+{{{\vec S}_{q \phi}\times {\vec P}_{\phi}}\over {P^{\tau}
_{\phi}\sqrt{(P^{\tau}_{\phi})^2-{\vec P}_{\phi}^2}}},\nonumber \\
&&{}\nonumber \\
{\vec q}_{\phi}&=&{\vec r}_{\phi}+{{{\vec S}_{q \phi}\times {\vec P}_{\phi}}
\over {P^{\tau}_{\phi}(P^{\tau}_{\phi}+\sqrt{(P^{\tau}_{\phi})^2-{\vec P}
_{\phi}^2})}} = {{P^{\tau}_{\phi}{\vec r}_{\phi}+\sqrt{(P^{\tau}_{\phi})^2-
{\vec P}_{\phi}^2} {\vec y}_{\phi}}\over {P^{\tau}_{\phi}+\sqrt{(P^{\tau}
_{\phi})^2-{\vec P}_{\phi}^2}}},\nonumber \\
&&\{ y^r_{\phi},y^s_{\phi} \} ={1\over {P^{\tau}_{\phi}\sqrt{(P^{\tau}_{\phi})
^2-{\vec P}_{\phi}^2} }}\epsilon^{rsu}\Big[ S^u_{q \phi}+{{ ({\vec S}_{q \phi}
\cdot {\vec P}_{\phi})\, P^u_{\phi}}\over {\sqrt{(P^{\tau}_{\phi})^2-{\vec P}
_{\phi}^2}(P^{\tau}_{\phi}+\sqrt{(P^{\tau}_{\phi})^2-{\vec P}_{\phi}^2})}}
\Big] ,\nonumber \\
&&{}\nonumber \\
{\vec P}_{\phi}\approx 0 &\Rightarrow& {\vec q}_{\phi}\approx {\vec r}_{\phi}
\approx {\vec y}_{\phi}.
\label{VI3}
\end{eqnarray}

The Wigner 3-vector ${\vec q}_{\phi}$ is therefore the canonical 3-center of
mass of the Klein-Gordon field [since ${\vec q}_{\phi}\approx {\vec r}_{\phi}$,
it also describe that point $z^{\mu}(\tau ,{\vec q}_{\phi})=x^{\mu}_s(\tau )+
q^r_{\phi} \epsilon^{\mu}_r(u(p_s))$ where the energy of the field
configuration is concentrated]. Instead, the previous variable ${\vec X}_{\phi}$
should be better named the 'center of phase' of the field configuration.
There should exist a canonical transformation from the canonical basis
${\tilde X}^{\tau}_{\phi}$, $P^{\tau}_{\phi}$, ${\vec X}_{\phi}$, ${\vec P}
_{\phi}$, ${\bf H}(\tau ,\vec k)$, ${\bf K}(\tau ,\vec k)$, to a new basis
$q^{\tau}_{\phi}$, $P^{\tau \, '}_{\phi} =\sqrt{ (P^{\tau}_{\phi})^2-{\vec P}^2
_{\phi}}$ [since $\{ {\vec
q}_{\phi},P^{\tau}_{\phi} \} = {\vec P}_{\phi}/P^{\tau}_{\phi}$ only
weakly zero], ${\vec q}_{\phi}$, ${\vec P}_{\phi}$, ${\bf H}^{'}(\tau
,\vec k)$, ${\bf K}^{'}(\tau ,\vec k)$ containing relative variables
with respect to the true center of mass of the field configuration.
However it does not seem easy to identify this final canonical basis;
in particular it is not clear how to find the `time' variable
$q^{\tau}_{\phi}$. Maybe the methods used in Ref.\cite{iten} can be
extended from particles to fields.

The gauge fixing ${\vec q}_{\phi}\approx 0$ [it also implies $\vec
\lambda (\tau )=0$ like ${\vec X}_{\phi}\approx 0$] forces all three
internal center-of-mass variables to coincide with the origin
$x^{\mu}_s$ of the Wigner hyperplane. We shall denote $x_s^{({\vec
q}_{\phi})\mu}(\tau )=x^{\mu}_s(0)+\tau u^{\mu}(p_s)$ the origin in
this gauge, because, as we shall see, in this case it enjoys of
properties not present in the ${\vec X}_{\phi}\approx 0$ gauge [where
we have that $x^{({\vec X}_{\phi})\mu}_s(\tau )=x
^{\mu}_s(0)+\tau u^{\mu}_s(p_s)$ is formally equal to $x^{({\vec q}_{\phi})\mu}
_s(\tau )$, but, since ${\tilde t}_T^{\mu}(T_s)\not= 0$,
$u_{\mu}(p_s) S_T^{\mu\nu}(T_s)[\phi ]\not= 0$, it has to be
interpreted as a different centroid].

In the gauge ${\vec q}_{\phi}\approx 0$ the origin becomes the Dixon's
center of mass which coincide with the internal Moller center of
energy as shown in Eq.(\ref{V10}) [and then Eq.(\ref{V14}) implies
that the origin is also the Tulczyjew\cite{tul} and the Pirani
\cite{pirani} centroid (see Ref.\cite{bini} for a review of these
concepts in relation with the Papapetrou-Dixon-Souriau pole-dipole
approximation of an extended object), because they are defined by
$p_{s\mu}S^{\mu\nu}_T(T_s)[\phi ]\approx 0$ and by $u_{\mu}(p_s)
S_T^{\mu\nu}(T_s)[\phi ]\approx 0$ respectively and we have $p^{\mu}_s
=\epsilon_s u^{\mu}(p_s)$]. Therefore, the worldline $x^{({\vec q}_{\phi})\mu}
_s$ is the unique center-of-mass worldline of special relativity in the sense
of Refs.\cite{bei}.

The rest-frame instant form ``external" realization of the Poincar\'e
algebra of Eq. (\ref{II9}) has the generators $p^{\mu}_s$,
$J^{ij}_s={\tilde x}^i_sp^j_s-{\tilde x}^j_sp^i_s+\delta^{ir}
\delta^{js} S^{rs}_{\phi}$, $K^i_s=J^{oi}_s={\tilde x}^o_sp^i_s-
{\tilde x}^i_sp^o_s-{{\delta^{ir} S^{rs}_{\phi}\, p^s_s}\over {p^o_s+
\epsilon_s}}={\tilde x}^o_sp^i_s-{\tilde x}^i_sp^o_s+\delta^{ir}{{({\vec S}
_{\phi}\times {\vec p}_s)^r}\over {p^o_s+\epsilon_s}}$ [for ${\tilde
x}^o_s=0$ this is the Newton-Wigner decomposition of $J^{\mu\nu}_s$].
As already said
the canonical variables ${\tilde x}^{\mu}_s$, $p^{\mu}_s$, may be replaced by
the canonical pairs $\epsilon_s=\pm \sqrt{p^2_s}$, $T_s=p_s\cdot {\tilde x}_s/
\epsilon_s$ [to be gauge fixed with $T_s-\tau \approx 0$]; ${\vec k}_s={\vec
p}_s/\epsilon_s=\vec u(p_s)$, ${\vec z}_s= \epsilon_s ({\vec {\tilde x}}_s-
{{{\vec p}_s}\over {p^o_s}} {\tilde x}^o_s)\equiv \epsilon_s {\vec q}_s$.

One can build three ``external" 3-variables, the
canonical ${\vec Q}_s$, the Moller ${\vec R}_s$ and the Fokker-Pryce ${\vec Y}
_s$ by using this rest-frame ``external" realization of the Poincar\'e algebra

\begin{eqnarray}
{\vec R}_s&=& -{1\over {p^o_s}}{\vec K}_s=({\vec {\tilde x}}_s-{{{\vec
p}_s}\over {p^o_s}} {\tilde x}^o_s)-{{{\vec S}_{\phi}\times {\vec p}_s}
\over {p^o_s(p^o_s+\epsilon_s)}},\nonumber \\
{\vec Q}_s&=&{\vec {\tilde x}}_s-{{{\vec p}_s}\over {p^o_s}}{\tilde x}^o_s=
{{{\vec z}_s}\over {\epsilon_s}}=
{\vec R}_s+{{ {\vec S}_{\phi}\times {\vec p}_s}\over {p^o_s(p^o_s+
\epsilon_s)}}={{p^o_s {\vec R}_s+\epsilon_s {\vec Y}_s}\over {p^o_s+\epsilon_s}}
,\nonumber \\
{\vec Y}_s&=&{\vec Q}_s+{{ {\vec S}_{\phi}\times {\vec p}_s}\over
{\epsilon_s (p^o_s+\epsilon_s)}}={\vec R}_s+{{ {\vec S}_{\phi}\times
{\vec p}_s}\over {p^o_s\epsilon_s}},\nonumber \\
&&\{ R^r_s,R^s_s \} =-{1\over {(p^o_s)^2}}\epsilon^{rsu}\Omega^u_s,
\quad\quad {\vec \Omega}_s={\vec J}_s-{\vec R}_s\times {\vec p}_s,\nonumber \\
&&\{ Y^r_s,Y^s_s \} ={1\over {\epsilon_sp^o_s}}\epsilon^{rsu}\Big[ S^u_{\phi}
+{{ ({\vec S}_{\phi}\cdot {\vec p}_s)\, p^u_s}\over {\epsilon_s(p^o_s+
\epsilon_s)}}\Big] ,\nonumber \\
&&{}\nonumber \\
&&{\vec p}_s\cdot {\vec Q}_s={\vec p}_s\cdot {\vec R}_s={\vec p}_s\cdot {\vec
Y}_s={\vec k}_s\cdot {\vec z}_s,\nonumber \\
{\vec p}_s=0 &\Rightarrow& {\vec Q}_s={\vec Y}_s={\vec R}_s,
\label{VI4}
\end{eqnarray}

\noindent with the same velocity and coinciding in the Lorentz rest frame where
${\buildrel \circ \over p}^{\mu}_s=\epsilon_s (1;\vec 0)$

In Ref.\cite{pauri} in a one-time framework without constraints and at a fixed
time, it is shown that the 3-vector ${\vec Y}_s$ [but not ${\vec Q}_s$ and
${\vec R}_s$] satisfies the condition $\{ K^r_s,Y^s_s \} =
Y^r_s\, \{ Y^s_s,p^o_s \}$ for being the space component of a 4-vector
$Y^{\mu}_s$. In the enlarged canonical treatment including time variables, it is
not clear which are the time components to be added to ${\vec Q}_s$, ${\vec R}
_s$, ${\vec Y}_s$, to rebuild 4-dimesnional quantities ${\tilde x}^{\mu}_s$,
$R^{\mu}_s$, $Y^{\mu}_s$, in an arbitrary Lorentz frame $\Gamma$, in which the
origin of the Wigner hyperplane is the 4-vector $x^{\mu}_s = (x^o_s; {\vec x}
_s)$. We have

\begin{eqnarray}
{\tilde x}^{\mu}_s(\tau )&=&
({\tilde x}^o_s(\tau ); {\vec {\tilde x}}_s(\tau ) )=
x^{\mu}_s-{1\over {\epsilon_s(p^o_s+\epsilon_s)}}\Big[
p_{s\nu}S_s^{\nu\mu}+\epsilon_s(S^{o\mu}_s-S^{o\nu}_s{{p_{s\nu}p_s^{\mu}}\over
{\epsilon^2_s}}) \Big],\quad\quad p^{\mu}_s,\nonumber \\
{\tilde x}^o_s&=&\sqrt{1+{\vec k}_s^2} (T_s+{{{\vec k}_s\cdot {\vec z}_s}\over
{\epsilon_s}})=\sqrt{1+{\vec k}^2_s}(T_s+{\vec k}_s\cdot {\vec q}_s)\not=
x^0_s,\quad\quad p^o_s=\epsilon_s\sqrt{1+{\vec k}_s^2},\nonumber \\
{\vec {\tilde x}}_s&=&{{ {\vec z}_s}\over {\epsilon_s}}+(T_s+{{{\vec k}_s\cdot
{\vec z}_s}\over {\epsilon_s}}) {\vec k}_s={\vec q}_s+(T_s+{\vec k}_s\cdot
{\vec q}_s){\vec k}_s,\quad\quad {\vec p}_s=\epsilon_s {\vec k}_s.
\label{VI5}
\end{eqnarray}

\noindent for the non-covariant (frame-dependent) canonical center of mass and
its conjugate momentum.

Each Wigner hyperplane intersects the worldline of the arbitrary origin
4-vector $x^{\mu}_s(\tau )=z^{\mu}(\tau ,\vec 0)$ in $\vec \sigma =0$, the
pseudo worldline of ${\tilde x}^{\mu}_s(\tau )=z^{\mu}(\tau ,{\tilde {\vec
\sigma}})$ in some ${\tilde {\vec \sigma}}$ and the worldline of the
 Fokker-Pryce 4-vector $Y^{\mu}_s(\tau )=z^{\mu}(\tau ,{\vec \sigma}_Y)$ in
some ${\vec \sigma}_Y$ [on this worldline one can put the ``internal
center of mass" with the gauge fixing ${\vec q}_{\phi}\approx 0$
(${\vec q}_{\phi}\approx {\vec r}_{\phi}\approx {\vec y}_{\phi}$ due
to ${\vec P}_{\phi}\approx 0$)]; one also has $R^{\mu}_s=z^{\mu}(\tau
,{\vec \sigma}_R)$. Since we have $T_s=u(p_s)\cdot x_s=u(p_s)\cdot
{\tilde x}_s\equiv \tau$ on the Wigner hyperplane labelled by $\tau$,
we require that also $Y^{\mu}_s$, $R^{\mu}_s$ have time components
such that they too satisfy $u(p_s)\cdot Y_s=u(p_s)\cdot R_s=T_s\equiv
\tau$. Therefore, it is reasonable to assume that ${\tilde
x}^{\mu}_s$, $Y^{\mu}_s$ and $R^{\mu}_s$ satisfy the following
equations consistently with Eqs.(\ref{VI1}), (\ref{VI2}) when
$T_s\equiv
\tau$ and ${\vec q}_{\phi}\approx 0$

\begin{eqnarray}
{\tilde x}^{\mu}_s&=&( {\tilde x}^o_s; {\vec {\tilde x}}_s)=({\tilde x}^o_s;
{{{\vec z}_s}\over {\epsilon_s}}+(T_s+{{{\vec k}_s\cdot {\vec z}_s}\over
{\epsilon_s}}){\vec k}_s )
=x^{({\vec q}_{\phi})\mu}_s+\epsilon^{\mu}_u(u(p_s)) {\tilde \sigma}^u,
\nonumber \\
Y^{\mu}_s&=&({\tilde x}^o_s;\, {1\over {\epsilon_s}}[{\vec z}_s+{{{\vec
S}_{\phi}\times {\vec p}_s}\over {\epsilon_s[1+u^o(p_s)]}}]+(T_s+
{{{\vec k}_s\cdot {\vec z}_s}\over {\epsilon_s}}){\vec k}_s\, )=\nonumber \\
&=&{\tilde x}^{\mu}_s+\eta^{\mu}_r{{({\vec
S}_{\phi}\times {\vec p}_s)^r}\over {\epsilon_s[1+u^o(p_s)]}}=\nonumber \\
&=&x^{({\vec q}_{\phi})\mu}_s+\epsilon^{\mu}_u(u(p_s)) \sigma^u_Y,\nonumber \\
R^{\mu}_s&=&( {\tilde x}^o_s;\, {1\over {\epsilon_s}}[{\vec z}_s-
{{{\vec S}_{\phi}\times {\vec p}_s}\over {\epsilon_s u^o(p_s)
[1+u^o(p_s)]}}]+(T_s+
{{{\vec k}_s\cdot {\vec z}_s}\over {\epsilon_s}}){\vec k}_s\, )=\nonumber \\
&=&{\tilde x}^{\mu}_s-\eta^{\mu}_r{{({\vec S}_{\phi}\times {\vec p}_s)^r}
\over {\epsilon_su^o(p_s)[1+u^o(p_s)]}}=\nonumber \\
&=&x^{({\vec q}_{\phi})\mu}_s+\epsilon^{\mu}_u(u(p_s)) \sigma^u_R,\nonumber \\
&&{}\nonumber \\
T_s&=&u(p_s)\cdot x_s^{({\vec q}_{\phi})}=u(p_s)\cdot {\tilde x}_s=u(p_s)\cdot
Y_s=u(p_s)\cdot R_s,\nonumber \\
&&{}\nonumber \\
{\tilde \sigma}^r&=&\epsilon_{r\mu}(u(p_s))
[x^{({\vec q}_{\phi})\mu}_s-{\tilde x}^{\mu}_s]=
{{ \epsilon_{r\mu}(u(p_s)) [u_{\nu}(p_s)S^{\nu\mu}_s+S^{o\mu}_s]}\over
{[1+u^o(p_s)]}}=\nonumber \\
&=&-S_{\phi}^{\tau r}+{{S_{\phi}^{rs}p^s_s}\over {\epsilon_s[1+u^o(p_s)]}}
=\epsilon_s r^r_{\phi}+{{S_{\phi}^{rs}u^s(p_s)}\over {1+u^o(p_s)}}
\approx \nonumber \\
&\approx& \epsilon_s q^r_{\phi}+{{S_{\phi}^{rs}u^s(p_s)}\over {1+
u^o(p_s)}}\approx {{S_{\phi}^{rs}u^s(p_s)}\over {1+u^o(p_s)}}
,\nonumber \\
\sigma^r_Y&=&\epsilon_{r\mu}(u(p_s))[x^{({\vec q}_{\phi})\mu}_s-Y^{\mu}_s]=
{\tilde \sigma}^r-\epsilon_{ru}(u(p_s)){{({\vec
S}_{\phi}\times {\vec p}_s)^u}\over {\epsilon_s[1+u^o(p_s)]}}=\nonumber \\
&=&{\tilde \sigma}^r+{{S^{rs}_{\phi}u^s(p_s)}\over {1+u^o(p_s)}}=
\epsilon_s r^r_{\phi} \approx \epsilon_s q^r_{\phi} \approx 0,\nonumber \\
\sigma^r_R&=&\epsilon_{r\mu}(u(p_s))
[x^{({\vec q}_{\phi})\mu}_s-R^{\mu}_s]={\tilde \sigma}^r+
\epsilon_{ru}(u(p_s)) {{({\vec S}_{\phi}\times {\vec p}_s)^u}
\over {\epsilon_su^o(p_s)[1+u^o(p_s)]}}=\nonumber \\
&=&{\tilde \sigma}^r-{{S_{\phi}^{rs}u^s(p_s)}\over {u^o(p_s)[1+
u^o(p_s)]}}=\epsilon_sr^r_{\phi}+{{[1-u^o(p_s)]S^{rs}_{\phi}u^s(p_s)}\over
{u^o(p_s)[1+u^o(p_s)]}}\approx \nonumber \\
&\approx& {{[1-u^o(p_s)]S^{rs}_{\phi}u^s(p_s)}\over
{u^o(p_s)[1+u^o(p_s)]}},\nonumber \\
&&{}\nonumber \\
&\Rightarrow& x^{({\vec q}_{\phi})\mu}_s(\tau ) = Y^{\mu}_s,\quad for\quad
{\vec q}_{\phi}\approx 0,
\label{VI6}
\end{eqnarray}

\noindent namely in the gauge ${\vec q}_{\phi}\approx 0$
the external Fokker-Pryce non canonical center of inertia
coincides with the origin $x^{({\vec q}_{\phi})\mu}_s(\tau )$ carrying the
``internal" center of mass (coinciding with the ``internal" M\"oller center of
energy and with the ``internal" Fokker-Pryce center of inertia)
and also being the Pirani centroid and the Tulczyjew centroid.

Therefore, if we would find the canonical basis $q^{\tau}_{\phi}$,
$P^{\tau \, {'}}_{\phi}=
\sqrt{(P^{\tau}_{\phi})^2-{\vec P}^2_{\phi}}$, ${\vec q}_{\phi}$, ${\vec P}
_{\phi}$, ${\bf H}^{'}(\tau ,\vec q)$, ${\bf K}^{'}(\tau ,\vec q)$, then,
in the gauge ${\vec q}_{\phi}\approx 0$ and $T_s \approx \tau$, the
Klein-Gordon field configurations would have the four-momentum density
peaked on the worldline $x^{({\vec q}_{\phi})\mu}_s(T_s)$; the
canonical variables ${\bf H}^{'}(\tau ,\vec q)$, ${\bf K}^{'}(\tau
,\vec q)$ would characterize the relative motions with respect to the
``monopole" configuration describing the center of mass of the field
configuration. The ``monopole" solutions of the Klein-Gordon equation
would be identified by the conditions ${\bf H}
^{'}(\tau ,\vec q)={\bf K}^{'}(\tau ,\vec q)=0$ [formally they are given by
Eqs.(\ref{IV38}) with $X^{\tau}_{\phi}$, ${\vec X}_{\phi}$ and
$P^{\tau}
_{\phi}$ replaced by $q^{\tau}_{\phi}$, ${\vec q}_{\phi}$ and $\sqrt{(P^{\tau}
_{\phi})^2-{\vec P}^2_{\phi}}$]: these field configurations
have the same independent degrees of freedom of a free scalar particle at rest
with mass $P^{\tau \, {'}}_{\phi}\approx P^{\tau}_{\phi}$ [its conjugate
``time" $q^{\tau}_{\phi}$ would satisfy $\partial /\partial T_s\, {\buildrel
\circ \over =}\, \partial /\partial q^{\tau}_{\phi}$ in the free case, see
Subsection D of Section IV].

Remember that the canonical center of mass  lies in between the Moller center
of energy and the Fokker-Pryce center of inertia and that the noncovariance
region around the Fokker-Pryce 4-vector extends to a worldtube with
radius (the Moller radius) $|{\vec S}_{\phi}| / P^{\tau}_{\phi}$.

\vfill\eject

\section{Coupling to scalar particles.}

In this Section we shall consider a Yukawa-type  coupling of the Klein-Gordon
field to the masses of scalar relativistic particles\cite{lus} to find which
are the Dirac observables spanning the reduced phase space. The action is

\begin{eqnarray}
S&=& \int d\tau L(\tau )=
\int d\tau d^3\sigma \Big( N(\tau ,\vec \sigma )
\sqrt{\gamma (\tau ,\vec \sigma )}\nonumber \\
&&{1\over 2} \Big[ g^{\tau\tau} {\dot \phi}^2+2 g^{\tau \check r} \dot \phi
\partial_{\check r}\phi + g^{\check r\check s}\partial_{\check r}\phi
\partial_{\check s}\phi -m^2\phi^2  \Big]
(\tau ,\vec \sigma )-\nonumber \\
&-&\sum_{i=1}^N\delta^3(\vec \sigma -{\vec \eta}_i(\tau )) \Big[\eta_im_i+G\phi
(\tau ,\vec \sigma )]\sqrt{[g_{\tau\tau}+2g_{\tau \check r}{\dot \eta}_i
^{\check r}(\tau )+g_{\check r\check s}{\dot \eta}_i^{\check r}(\tau ){\dot
\eta}_i^{\check s}(\tau )\Big] (\tau ,\vec \sigma )} \Big)=\nonumber \\
&=&\int d\tau d^3\sigma \Big( \sqrt{\gamma (\tau ,\vec \sigma )} {1\over 2}
\Big[ {1\over N}[\partial_{\tau} -N^{\check r}\partial_{\check r}]\phi \,
[\partial_{\tau} -N^{\check s}\partial_{\check s}]\phi + \nonumber \\
&+&N [\gamma^{\check r\check s}\partial_{\check r}\phi
\partial_{\check s}\phi -m^2\phi^2 ]\, \Big] (\tau ,\vec \sigma )
-\sum_{i=1}^N\delta^3(\vec \sigma -{\vec \eta}_i(\tau
)) [\eta_im_i+G\phi (\tau ,\vec \sigma )]\nonumber \\
&&\sqrt{ [N^2+g_{\check r\check s}(N^{\check r}+{\dot \eta}_i^{\check r}(\tau ))
(N^{\check s}+{\dot \eta}_i^{\check s}(\tau ))](\tau ,\vec \sigma )} \Big) ,
\nonumber \\
&&{}
\label{VII1}
\end{eqnarray}

\noindent and the canonical momenta are

\begin{eqnarray}
\kappa_{i\check r}(\tau )&=&-{{\partial L(\tau )}\over {\partial {\dot \eta}
_i^{\check r}}}= \Big[ \eta_im_i+G\phi (\tau ,{\vec \eta}_i(\tau ))\Big]
\nonumber \\
&&{{ g_{\check r\check s} (N^{\check s}+{\dot \eta}_i^{\check s}(\tau ))}\over
{\sqrt{N^2+g_{\check r\check s}(N^{\check r}+{\dot \eta}_i^{\check r}(\tau ))
(N^{\check s}+{\dot \eta}_i^{\check s}(\tau ))} }}(\tau ,{\vec \eta}_i(\tau )),
\nonumber \\
\pi(\tau ,\vec \sigma )&=&{{\partial L}\over {\partial \partial_{\tau}
\phi (\tau ,\vec \sigma )}}=\nonumber \\
&=&{{\sqrt{\gamma}(\tau ,\vec \sigma )}\over
{N(\tau ,\vec \sigma )}}\Big[ \dot \phi -N^{\check r}\partial_{\check r}\phi
\Big] (\tau ,\vec \sigma ),\nonumber \\
\rho_{\mu}(\tau ,\vec \sigma )&=&-{{\partial L}\over {\partial \partial_{\tau}
z^{\mu}(\tau ,\vec \sigma )}}=\nonumber \\
&=&l_{\mu}(\tau ,\vec \sigma ) \Big[\, \Big( {{\sqrt{\gamma}}\over 2}
[{1\over {N^2}}(\dot \phi -N^{\check r}\partial_{\check r}\phi )^2-
\gamma^{\check r\check s}\partial_{\check r}\phi
\partial_{\check s}\phi +m^2\phi^2]\, \Big) (\tau ,\vec \sigma )
+\nonumber \\
&+&\sum_{i=1}^N\delta^3(\vec \sigma -{\vec \eta}_i(\tau )) {{[\eta_im_i+G\phi
(\tau ,\vec \sigma )]\, N(\tau ,\vec \sigma )}\over {
\sqrt{[N^2+g_{\check r\check s}(N^{\check r}+{\dot \eta}_i^{\check r}(\tau ))
(N^{\check s}+{\dot \eta}_i^{\check s}(\tau ))](\tau ,\vec \sigma )} }}\,
\Big] +\nonumber \\
&+&z_{\check s\mu}(\tau ,\vec \sigma )\gamma^{\check s\check r}(\tau ,\vec
\sigma ) \Big[ \Big( {{ \sqrt{\gamma}}\over N}\partial_{\check r}\phi
(\dot \phi -N^{\check u}\partial_{\check u}\phi )\, \Big)
(\tau ,\vec \sigma )+\nonumber \\
&+&\sum_{i=1}^N\delta^3(\vec \sigma -{\vec \eta}_i(\tau )) {{[\eta_im_i+G\phi ]
g_{\check r\check s}(N^{\check u}+{\dot \eta}_i^{\check u}(\tau ))}\over
{\sqrt{N^2+g_{\check r\check s}(N^{\check r}+{\dot \eta}_i^{\check r}(\tau ))
(N^{\check s}+{\dot \eta}_i^{\check s}(\tau ))} }} (\tau ,\vec \sigma )\,
\Big] .
\label{VII2}
\end{eqnarray}

We get the first class constraints

\begin{eqnarray}
{\cal H}_{\mu}(\tau ,\vec \sigma )&=&\rho_{\mu}(\tau ,\vec \sigma )-
\nonumber \\
&&-l_{\mu}(\tau ,\vec \sigma ) \Big[ \Big( {{\pi^2}\over {2\sqrt{\gamma}}}-
{{\sqrt{\gamma}}\over 2} [\gamma^{\check r\check s}\partial_{\check r}\phi
\partial_{\check s}\phi -m^2\phi^2] \Big) (\tau ,\vec \sigma )+
\nonumber \\
&+&\sum_{i=1}^N\delta^3(\vec \sigma -{\vec \eta}_i(\tau ))\eta_i\sqrt{[\eta
_im_i+G\phi (\tau ,\vec \sigma )]^2-\gamma^{\check r\check s}(\tau ,\vec
\sigma )\kappa_{i\check r}(\tau )\kappa_{i\check s}(\tau )} \Big] -\nonumber \\
&-&z_{\check s\mu}(\tau ,\vec \sigma )\gamma^{\check r\check s}(\tau ,\vec
\sigma ) \Big[ (\pi \partial_{\check r}\phi )(\tau ,\vec \sigma ) +\sum_{i=1}^N
\delta^3(\vec \sigma -{\vec \eta}_i(\tau ))\kappa_{i\check r}(\tau )\Big]
\approx 0.
\label{VII3}
\end{eqnarray}

Following the procedure of Section II, one can arrive at the reduction
on the Wigner hyperplanes, where the remaining four first class
constraints and Dirac Hamiltonian are

\begin{eqnarray}
{\cal H}(\tau )&=&\epsilon_s - \Big[ \sum_{i=1}^N\eta_i\sqrt{[\eta_im_i+G\phi
(\tau ,{\vec \eta}_i(\tau ))]^2+{\vec \kappa}_i^2(\tau )}+\nonumber \\
&+& {1\over 2} \int d^3\sigma [\pi^2+(\vec \partial
\phi )^2+m^2\phi^2](\tau ,\vec \sigma ) \Big] =
\epsilon_s-M \approx 0,\nonumber \\
{\vec {\cal H}}_p(\tau )&=& \sum_{i=1}^N {\vec \kappa}_i(\tau )+
\int d^3\sigma [\pi \vec \partial \phi ](\tau ,\vec \sigma ) \approx 0,
\nonumber \\
 &&{}\nonumber \\
 H_D&=& \lambda (\tau ) {\cal H} + \vec \lambda (\tau )\cdot {\vec {\cal H}}_p.
\label{VII4}
\end{eqnarray}

In the gauge $T_s-\tau \approx 0$ the Hamiltonian is $H_R=M -\vec
\lambda (\tau )\cdot {\vec {\cal H}}_p$.

Let us make a canonical transformation from the canonical basis ${\vec \eta}_i
(\tau )$, ${\vec \kappa}_i(\tau )$, $\phi (\tau ,\vec \sigma )$, $\pi (\tau
,\vec \sigma )$, to the new basis ${\vec \eta}_{+}(\tau )$, ${\vec \kappa}
_{+}(\tau )$, ${\vec \rho}_a(\tau )$, ${\vec \pi}_a(\tau )$, $X^{\tau}_{\phi}$,
$P^{\tau}_{\phi}$, ${\vec X}_{\phi}$, ${\vec P}_{\phi}$, ${\bf H}(\tau ,\vec
q)$, ${\bf K}(\tau ,\vec q)$ [$\phi$, $\pi$, are assumed to satisfy
Eq.(\ref{IV29})] with the particle variables defined by

\begin{eqnarray}
{\vec \eta}_i&=&{\vec \eta}_{+}+{1\over {\sqrt{N}}} \sum_{a=1}^{N-1}{\hat
\gamma}_{ai}{\vec \rho}_a,\nonumber \\
{\vec \kappa}_i&=&{1\over N} {\vec \kappa}_{+}+\sqrt{N} \sum_{a=1}^{N-1}{\hat
\gamma}_{ai} {\vec \pi}_a,\nonumber \\
&&{}\nonumber \\
{\vec \eta}_{+}&=&{1\over N} \sum_{i=1}^N {\vec \eta}_i,\quad\quad
{\vec \rho}_a=\sqrt{N} \sum_{i=1}^N {\hat \gamma} {\vec \eta}_i,\nonumber \\
{\vec \kappa}_{+}&=&\sum_{i=1}^N {\vec \kappa}_i,\quad\quad
{\vec \pi}_a={1\over {\sqrt{N}}} \sum_{i=1}^N {\hat \gamma}_{ai} {\vec
\kappa}_i,\nonumber \\
&&{}\nonumber \\
&&\sum_{i=1}^N{\hat \gamma}_{ai}=0,\quad\quad \sum_{a=1}^{N-1}{\hat \gamma}_{ai}
{\hat \gamma}_{aj}=\delta_{ij}-{1\over N},\quad\quad
\sum_{i=1}^N{\hat \gamma}_{ai}{\hat \gamma}_{bi}=\delta_{ab}.
\label{VII5}
\end{eqnarray}

The variable ${\vec \eta}_{+}(\tau )$ is playing the role of a naive
``internal" center of mass for the particles on the Wigner hyperplane.
From the discussion of the previous Section, it is clear that the real
``internal" center of mass of the N particles is a ${\vec q}_{+}$
defined like ${\vec  q}_{\phi}$ of Eq.(\ref{VI2}) with ${\vec
r}_{+}=\sum_{i=1}^N
\sqrt{m^2_i+{\vec
\kappa}^2_i}\, {\vec \eta}_i
/ \sum_{k=1}^N\sqrt{m^2_k+{\vec \kappa}^2_k}$. But, since it is not known
(like it happens for the Klein-Gordon field)  the
canonical transformation ${\vec \eta}_i, {\vec \kappa}_i\, \mapsto {\vec q}_{+},
{\vec \kappa}_{+}, {\vec \rho}_{qa}, {\vec \pi}_{qa}$ identifying the real
relative variables ${\vec \rho}_{qa}$, ${\vec \pi}_{qa}$ [see however Ref.
\cite{iten}], we shall go on with
the previous naive canonical transformation in the following discussion to see
which kind of collective and relative variables emerge for the particles plus
the real Klein-Gordon field.

We  also remark that we use the field ``center of phase" ${\vec
X}_{\phi}$ and not the real internal center of mass ${\vec q}_{\phi}$,
because the knowledge of the canonical basis containing ${\vec
X}_{\phi}$ allows us to illustrate interpretational aspects which will
hold also with the canonical basis containing ${\vec q}_{\phi}$ when
it will be found.

The constraints become

\begin{eqnarray}
{\cal H}(\tau )&=& \epsilon_s-\Big[ P^{\tau}_{\phi}+\sum_{i=1}^N \eta_i
\Big( [\eta_im_i+G\phi (\tau,{\vec \eta}_i(\tau ))]^2+\nonumber \\
&+&[{1\over N}{\vec \kappa}_{+}(\tau )+\sqrt{N}\sum_{a=1}^{N-1}{\hat \gamma}
_{ai} {\vec \pi}_a(\tau )]^2 \Big)^{1/2} \approx 0,\nonumber \\
{\vec {\cal H}}_p(\tau )&=&{\vec \kappa}_{+}+{\vec P}_{\phi}\approx 0,
\label{VII6}
\end{eqnarray}

We can now replace the canonical variables ${\vec \eta}_{+}$, ${\vec \kappa}
_{+}$, ${\vec X}_{\phi}$, ${\vec P}_{\phi}$ with the following canonical ones

\begin{eqnarray}
\vec Y&=&{1\over 2} ({\vec \eta}_{+}+{\vec X}_{\phi}),\nonumber \\
{\vec {\cal H}}_p&=&{\vec \kappa}_{+}+{\vec P}_{\phi} \approx 0,\nonumber \\
\vec \zeta &=&{\vec \eta}_{+}-{\vec X}_{\phi},\nonumber \\
{\vec \pi}_{\zeta}&=&{1\over 2} ({\vec \kappa}_{+}-{\vec P}_{\phi}),\nonumber \\
&&{}\nonumber \\
{\vec \eta}_{+}&=&{1\over 2}\vec \zeta +\vec Y,\nonumber \\
{\vec X}_{\phi}&=&-{1\over 2}\vec \zeta +\vec Y,\nonumber \\
{\vec \kappa}_{+}&=&{1\over 2} {\vec {\cal H}}_p+{\vec \pi}_{\zeta}\approx
{\vec \pi}_{\zeta},\nonumber \\
{\vec P}_{\phi}&=&{1\over 2}{\vec {\cal H}}_p-{\vec \pi}_{\zeta}\approx
-{\vec \pi}_{\zeta}.
\label{VII7}
\end{eqnarray}

We see that $\vec Y$ is playing the role of the naive ``internal" center of
mass of the full ``particles+field" system: it is the gauge variable conjugate
to ${\vec {\cal H}}_p\approx 0$

The global relative variables $\vec \zeta (\tau )$, ${\vec \pi}_{\zeta}(\tau
)$,  describe the relative motion of the particle and the field naive centers
of mass and rule the action-reaction between the two subsystems.

With the natural gauge-fixings: $\vec Y\approx 0$, $T_s-\tau \approx 0$ [so
that ${\vec \eta}_{+}\approx {1\over 2}\vec \zeta \approx -{\vec X}_{\phi}$]
and the associated Dirac brackets, we get the reduced phase space ${\vec z}_s$,
 ${\vec k}_s$, ${\vec \zeta}$, ${\vec \pi}_{\zeta}$, ${\vec \rho}_a(\tau )$,
${\vec \pi}_a(\tau )$, $X^{\tau}_{\phi}$,
$P^{\tau}_{\phi}$, ${\bf H}(T_s,\vec q)$, ${\bf K}(T_s,\vec q)$ with the
evolution in $T_s\equiv \tau$ ruled by the Hamiltonian

\begin{eqnarray}
H&=&P^{\tau}_{\phi}+\sum_{i=1}^N \eta_i \Big( [\eta_im_i+G\phi (T_s,{1\over 2}
\vec \zeta (T_s)+{1\over {\sqrt{N}}}\sum_{a=1}^{N-1}{\hat \gamma}_{ai}{\vec
\rho}_a(T_s))]^2+\nonumber \\
&+&[{1\over N}{\vec \pi}_{\zeta}(T_s)+\sqrt{N}\sum_{a=1}^{N-1}{\hat \gamma}
_{ai} {\vec \pi}_a(T_s)]^2 \Big)^{1/2}=\nonumber \\
&{\buildrel {def} \over =}\,& P^{\tau}_{\phi} +\sum_{i=1}^N \eta_i M_i,
\nonumber \\
&&{}\nonumber \\
M^2_i&=&(\eta_im_i+G\phi )^2+[{{ {\vec \pi}_{\zeta}}\over N}+\sqrt{N}\sum_{a=1}
^{N-1}{\hat \gamma}_{ai} {\vec \pi}_a ]^2,\nonumber \\
&&{}\nonumber \\
\phi &=&\phi \Big( T_s,{1\over 2}
\vec \zeta (T_s)+{1\over {\sqrt{N}}}\sum_{a=1}^{N-1}{\hat \gamma}_{ai}{\vec
\rho}_a(T_s)\Big)=\nonumber \\
&=& 2 \int d\tilde q  \sqrt{F^{\tau}(q)\omega (q)P^{\tau}_{\phi}+
F(q)\vec q\cdot {\vec \pi}_{\zeta}(T_s)+{\cal D}_{\vec q}{\bf H}(T_s,\vec
q)}\nonumber \\
&&cos\, \Big[ \vec q\cdot \Big(
\vec \zeta (T_s)+{1\over {\sqrt{N}}}\sum_{a=1}^{N-1}{\hat \gamma}_{ai}{\vec
\rho}_a(T_s)\Big) -\nonumber \\
&-&\omega (q)(T_s-X^{\tau}_{\phi})+ \int d\tilde kd{\tilde k}^{'} {\bf K}(T_s,
\vec k){\cal G}(\vec k,{\vec k}^{'})\triangle ({\vec k}^{'},\vec q)
\Big].
\label{VII8}
\end{eqnarray}

The equations of motion for $X^{\tau}_{\phi}$, $P^{\tau}_{\phi}$, $\vec
\zeta (\tau )$, ${\vec \pi}_{\zeta}(\tau )$, ${\vec \rho}_a(\tau )$, ${\vec
\pi}_a(\tau )$, ${\bf H}(\tau ,\vec q)$, ${\bf K}(\tau ,\vec q)$ are
[remember that $T_s-X^{\tau}_{\phi}=-{\tilde X}^{\tau}_{\phi}$]

\begin{eqnarray}
{{d \vec \zeta (T_s)}\over {dT_s}}&\, {\buildrel \circ \over =}\,& \{ \vec
\zeta (T_s),H \} ={{\partial H}\over {\partial {\vec \pi}_{\zeta}(T_s)}}=
\nonumber \\
&=&\sum_{i=1}^N {1\over {\eta_iM_i}} \Big[ {{ {\vec \pi}_{\zeta}}\over {N^2}}+
{1\over {\sqrt{N}}} \sum_{a=1}^{N-1}{\hat \gamma}_{ai}{\vec \pi}_a +\nonumber \\
&+&G(\eta_im_i+G\phi ) \int d\tilde q {{F(q)\vec q}\over {\sqrt{F^{\tau}(q)
\omega (q) P^{\tau}_{\phi}+F(q)\vec q\cdot {\vec \pi}_{\zeta}(T_s)+{\cal D}
_{\vec q}{\bf H}(T_s,\vec q)}}} \nonumber \\
&&cos\, \Big( \vec q\cdot (\vec \zeta (T_s)+
{1\over {\sqrt{N}}}\sum_{a=1}^{N-1}{\hat \gamma}_{ai}{\vec \rho}_{ai}(T_s))-
\nonumber \\
&-&\omega (q)(T_s-X^{\tau}_{\phi})+\int d\tilde kd{\tilde k}^{'} {\bf K}(T_s,
\vec k) {\cal G}(\vec k,{\vec k}^{'}) \triangle ({\vec k}^{'},\vec q)\Big) \Big]
,\nonumber \\
{{d {\vec \pi}_{\zeta}(T_s)}\over {dT_s}}&\, {\buildrel \circ \over =}\,&
\{ {\vec \pi}_{\zeta}(T_s),H \} =-{{\partial H}\over {\partial \vec \zeta (T_s)
}}=\nonumber \\
&=&2G \sum_{i=1}^N {{\eta_im_i+G\phi }\over {\eta_iM_i}} \int d\tilde q \vec q
\sqrt{F^{\tau}(q)\omega (q) P^{\tau}_{\phi}+F(q)\vec q\cdot {\vec \pi}
_{\zeta}(T_s)+{\cal D}_{\vec q}{\bf H}(T_s,\vec q)}\nonumber \\
&&sin\, \Big( \vec q\cdot (\vec \zeta (T_s)+
{1\over {\sqrt{N}}}\sum_{a=1}^{N-1}{\hat \gamma}_{ai}{\vec \rho}_{ai}(T_s))-
\nonumber \\
&-&\omega (q)(T_s-X^{\tau}_{\phi})+\int d\tilde kd{\tilde k}^{'} {\bf K}(T_s,
\vec k) {\cal G}(\vec k,{\vec k}^{'}) \triangle ({\vec k}^{'},\vec q)\Big)
,\nonumber \\
&&{}\nonumber \\
{{d {\vec \rho}_a(T_s)}\over {dT_s}}&\, {\buildrel \circ \over =}\,&
\{ {\vec \rho}_a(T_s),H \} = {{\partial H}\over {\partial {\vec \pi}_a(T_s)}}=
\nonumber \\
&=&\sum_{i=1}^N {{ {\hat \gamma}_{ai}}\over {\eta_iM_i}}({{ {\vec \pi}
_{\zeta}}\over {\sqrt{N}}}+N\sum_{b=1}^{N-1} {\hat \gamma}_{bi} {\vec \pi}_b)
,\nonumber \\
{{d {\vec \pi}_a(T_s)}\over {dT_s}}&\, {\buildrel \circ \over =}\,&
\{ {\vec \pi}_a(T_s),H \} =-{{\partial H}\over {\partial {\vec \rho}_a(T_s)}}=
\nonumber \\
&=&{{2G}\over {\sqrt{N}}}\sum_{i=1}^N {\hat \gamma}_{ai} {{\eta_im_i+G\phi}\over
{\eta_iM_i}} \int d\tilde q \vec q\, \sqrt{F^{\tau}(q)
\omega (q) P^{\tau}_{\phi}+F(q)\vec q\cdot {\vec \pi}_{\zeta}(T_s)+{\cal D}
_{\vec q}{\bf H}(T_s,\vec q)}\nonumber \\
&&sin\, \Big( \vec q\cdot (\vec \zeta (T_s)+
{1\over {\sqrt{N}}}\sum_{a=1}^{N-1}{\hat \gamma}_{ai}{\vec \rho}_{ai}(T_s))-
\nonumber \\
&-&\omega (q)(T_s-X^{\tau}_{\phi})+\int d\tilde kd{\tilde k}^{'} {\bf K}(T_s,
\vec k) {\cal G}(\vec k,{\vec k}^{'}) \triangle ({\vec k}^{'},\vec q)\Big)
,\nonumber \\
&&{}\nonumber \\
{{d X^{\tau}_{\phi}}\over {dT_s}}&\, {\buildrel \circ \over =}\,&
\{ X^{\tau}_{\phi},H \} =- {{\partial H}\over {\partial P^{\tau}_{\phi}}}=
\nonumber \\
&=&-1+G\sum_{i=1}^N {{\eta_im_i+G\phi}\over {\eta_iM_i}} \int d\tilde q
{{F^{\tau}(q)\omega (q)}\over {\sqrt{F^{\tau}(q)
\omega (q) P^{\tau}_{\phi}+F(q)\vec q\cdot {\vec \pi}_{\zeta}(T_s)+{\cal D}
_{\vec q}{\bf H}(T_s,\vec q)} }}\nonumber \\
&&cos\, \Big( \vec q\cdot (\vec \zeta (T_s)+
{1\over {\sqrt{N}}}\sum_{a=1}^{N-1}{\hat \gamma}_{ai}{\vec \rho}_{ai}(T_s))-
\nonumber \\
&-&\omega (q)(T_s-X^{\tau}_{\phi})+\int d\tilde kd{\tilde k}^{'} {\bf K}(T_s,
\vec k) {\cal G}(\vec k,{\vec k}^{'}) \triangle ({\vec k}^{'},\vec q)\Big)
,\nonumber \\
{{d P^{\tau}_{\phi}}\over {dT_s}}&\, {\buildrel \circ \over =}\,&
\{ P^{\tau}_{\phi},H \} ={{\partial H}\over {\partial X^{\tau}_{\phi}}}=
\nonumber \\
&=&-2G\sum_{i=1}^N {{\eta_im_i+G\phi}\over {\eta_iM_i}} \int d\tilde q \omega
(q) \sqrt{F^{\tau}(q)
\omega (q) P^{\tau}_{\phi}+F(q)\vec q\cdot {\vec \pi}_{\zeta}(T_s)+{\cal D}
_{\vec q}{\bf H}(T_s,\vec q)}\nonumber \\
&&sin\, \Big( \vec q\cdot (\vec \zeta (T_s)+
{1\over {\sqrt{N}}}\sum_{a=1}^{N-1}{\hat \gamma}_{ai}{\vec \rho}_{ai}(T_s))-
\nonumber \\
&-&\omega (q)(T_s-X^{\tau}_{\phi})+\int d\tilde kd{\tilde k}^{'} {\bf K}(T_s,
\vec k) {\cal G}(\vec k,{\vec k}^{'}) \triangle ({\vec k}^{'},\vec q)\Big)
,\nonumber \\
&&{}\nonumber \\
{{\partial {\bf H}(T_s,\vec q)}\over {\partial T_s}}&\, {\buildrel \circ \over
=}\,&
\{ {\bf H}(T_s,\vec q),H \} = {{\delta H}\over {\delta {\bf K}(T_s,\vec q)}}=
\nonumber \\
&=&-2G \sum_{i=1}^N {{\eta_im_i+G\phi}\over {\eta_iM_i}} \int d\tilde kd{\tilde
k}^{'} {\cal G}(\vec q,{\vec k}^{'}) \triangle ({\vec k}^{'},\vec k)
\nonumber \\
&&\sqrt{F^{\tau}(q)
\omega (q) P^{\tau}_{\phi}+F(q)\vec q\cdot {\vec \pi}_{\zeta}(T_s)+{\cal D}
_{\vec q}{\bf H}(T_s,\vec q)}\nonumber \\
&&sin\, \Big( \vec k\cdot (\vec \zeta (T_s)+
{1\over {\sqrt{N}}}\sum_{a=1}^{N-1}{\hat \gamma}_{ai}{\vec \rho}_{ai}(T_s))-
\nonumber \\
&-&\omega (k)(T_s-X^{\tau}_{\phi})+\int d{\tilde k}_1d{\tilde k}_2 {\bf K}(T_s,
{\vec k}_1) {\cal G}({\vec k}_1,{\vec k}_2) \triangle ({\vec k}_2,\vec k)\Big)
,\nonumber \\
{{\partial {\bf K}(T_s,\vec q)}\over {\partial T_s}}&\, {\buildrel \circ \over
=}\,& \{ {\bf K}(T_s,\vec q),H \} =- {{\delta H}\over {\delta {\bf H}(T_s,\vec
q)}} =\nonumber \\
&=&-2G \sum_{i=1}^N {{\eta_im_i+G\phi}\over {\eta_iM_i}}\nonumber \\
&&{\cal D}_{\vec q} \Big[ cos\, \Big( \vec q\cdot (\vec \zeta (T_s)+
{1\over {\sqrt{N}}}\sum_{a=1}^{N-1}{\hat \gamma}_{ai}{\vec \rho}_{ai}(T_s))-
\nonumber \\
&-&\omega (q)(T_s-X^{\tau}_{\phi})+\int d\tilde kd{\tilde k}^{'} {\bf K}(T_s,
\vec k) {\cal G}(\vec k,{\vec k}^{'}) \triangle ({\vec k}^{'},\vec q)\Big)
\nonumber \\
&& \Big( \sqrt{F^{\tau}(q)
\omega (q) P^{\tau}_{\phi}+F(q)\vec q\cdot {\vec \pi}_{\zeta}(T_s)+{\cal D}
_{\vec q}{\bf H}(T_s,\vec q)} \Big) {}^{-1} \Big] .
\label{VII9}
\end{eqnarray}

When the decomposition into center-of-mass and relative variables will be
available for the transverse electromagnetic field in the Coulomb gauge,
one will be able to make a similar treatment of the isolated system consisting
of N charged particles (with Grassmann-valued electric charges) plus the
electromagnetic field \cite{lus,albad}.

\vfill\eject

\section{Charged Klein-Gordon Field.}

\subsection{On spacelike hypersurfaces.}

Following Ref.\cite{albad},
let us now consider a charged Klein-Gordon field: $\phi ={1\over {\sqrt{2}}}
(\phi_1+i\phi_2)$, $\phi^{*}={1\over {\sqrt{2}}}(\phi_1-i\phi_2)$ [$\phi_1=
{1\over {\sqrt{2}}}(\phi +\phi^{*})$, $\phi_2={{-i}\over {\sqrt{2}}}(\phi -
\phi^{*})$], whose action is

\begin{eqnarray}
S&=& \int d\tau d^3\sigma N(\tau ,\vec \sigma )\sqrt{\gamma (\tau ,\vec \sigma )
} \Big( g^{\tau\tau} {\dot \phi}^{*}\, \dot \phi +\nonumber \\
&&+g^{\tau \check r} \Big[ {\dot \phi}^{*}\, \partial
_{\check r}\phi +\partial_{\check r}\phi^{*}\, \dot \phi \Big] +
g^{\check r\check s}\partial_{\check r}\phi^{*}\, \partial
_{\check s}\phi - m^2 \phi^{*}\phi \Big) (\tau ,\vec \sigma )=\nonumber \\
&&=\int d\tau d^3\sigma \sqrt{\gamma (\tau ,\vec \sigma )} \Big( {1\over N}
[{\dot \phi}^{*} -N^{\check r}\partial_{\check r}\phi^{*}]\,
[\dot \phi -N^{\check s}\partial_{\check s}\phi ]+\nonumber \\
&+&N \Big[ \gamma^{\check r\check s}\partial_{\check r}\phi^{*}
\partial_{\check s}\phi -m^2\phi^{*}\phi \Big] \Big) (\tau ,\vec \sigma ).
\label{VIII1}
\end{eqnarray}

The canonical momenta are

\begin{eqnarray}
\pi_{\phi}(\tau ,\vec \sigma )&=&{{\partial L}\over {\partial \partial_{\tau}
\phi (\tau ,\vec \sigma )}}={1\over {\sqrt{2}}}(\pi_1-i\pi_2)(\tau ,\vec
\sigma )=\nonumber \\
&=&{{\sqrt{\gamma (\tau ,\vec \sigma )}}\over
{N(\tau ,\vec \sigma )}}\Big[ {\dot \phi}^{*}-N^{\check r}\partial
_{\check r} \phi^{*}\Big] (\tau ,\vec \sigma ),
\nonumber \\
\pi_{\phi^{*}}(\tau ,\vec \sigma )&=&{{\partial L}\over {\partial \partial
_{\tau}\phi^{*}(\tau ,\vec \sigma )}}={1\over {\sqrt{2}}}(\pi_1+i\pi_2)(\tau
,\vec \sigma )=\nonumber \\
&=&{{\sqrt{\gamma (\tau ,\vec \sigma )}}\over
{N(\tau ,\vec \sigma )}} \Big[ \dot \phi -N^{\check r}\partial
_{\check r}\phi \Big] (\tau ,\vec \sigma ),
\nonumber \\
\rho_{\mu}(\tau ,\vec \sigma )&=&-{{\partial L}\over {\partial \partial_{\tau}
z^{\mu}(\tau ,\vec \sigma )}}=\nonumber \\
&=&l_{\mu}(\tau ,\vec \sigma ) \Big( {{\pi_{\phi}
\pi_{\phi^{*}}}\over {\sqrt{\gamma}}}-\sqrt{\gamma} \Big[ \gamma
^{\check r\check s}\partial_{\check r}\phi^{*} \partial_{\check s}\phi -
m^2\phi^{*}\phi \Big] \Big) (\tau ,\vec \sigma )+\nonumber \\
&&+z_{\check s\mu}(\tau ,\vec \sigma )\gamma^{\check r\check s}(\tau ,\vec
\sigma ) \Big( \pi_{\phi^{*}}\partial_{\check r}\phi^{*}+
\pi_{\phi} \partial_{\check r}\phi \Big) (\tau ,\vec \sigma ).
\label{VIII2}
\end{eqnarray}

Therefore, we have the following primary constraints

\begin{eqnarray}
{\cal H}_{\mu}(\tau ,\vec \sigma )&=& \rho_{\mu}(\tau ,\vec \sigma )-
\nonumber \\
&&-l_{\mu}(\tau ,\vec \sigma ) \Big( {{\pi_{\phi}
\pi_{\phi^{*}}}\over {\sqrt{\gamma}}}-\sqrt{\gamma} [\gamma^{\check r\check s}
\partial_{\check r}\phi^{*} \partial_{\check s}\phi -m^2\phi^{*}\phi]
\Big) (\tau ,\vec \sigma )+\nonumber \\
&&+z_{\check s\mu}(\tau ,\vec \sigma )\gamma^{\check r\check s}(\tau ,\vec
\sigma ) \Big( \pi_{\phi^{*}}\partial_{\check r}\phi^{*}+
\pi_{\phi} \partial_{\check r}\phi \Big) (\tau ,\vec \sigma ) \approx 0,
\label{VIII3}
\end{eqnarray}

\noindent and the following Dirac Hamiltonian

\begin{equation}
H_D=\int d^3\sigma \lambda^{\mu}
(\tau ,\vec \sigma ){\cal H}_{\mu}(\tau ,\vec \sigma ).
\label{VIII4}
\end{equation}

By using the Poisson brackets

\begin{eqnarray}
\{ z^{\mu}(\tau ,\vec \sigma ),\rho_{\nu}(\tau ,{\vec \sigma}^{'}) \} &=&\eta
^{\mu}_{\nu} \delta^3(\vec \sigma -{\vec \sigma}^{'}),\nonumber \\
\{ \phi (\tau ,\vec \sigma ),\pi_{\phi}(\tau ,{\vec \sigma}^{'}) \} &=&
\{ \phi^{*}(\tau ,\vec \sigma ),\pi_{\phi^{*}}(\tau ,{\vec \sigma}^{'}) \} =
\delta^3(\vec \sigma -{\vec \sigma}^{'}),
\label{VIII5}
\end{eqnarray}

\noindent we find that
the time constancy of the primary constraints implies the existence of no
secondary constraint. The constraints turn out to be first class.

After the restriction  to spacelike hyperplanes
$z^{\mu}(\tau ,\vec \sigma )=x_s^{\mu}(\tau )+b^{\mu}_{\check r}(\tau )
\sigma^{\check r}$, the constraints are reduced to the following ones

\begin{eqnarray}
{\tilde {\cal H}}^{\mu}(\tau )&=& \int d^3\sigma {\cal H}^{\mu}(\tau ,\vec
\sigma )=\nonumber \\
&=&p^{\mu}_s-l^{\mu} \Big( \int d^3\sigma \Big[ \pi_{\phi^{*}} \pi_{\phi}+
\vec \partial \phi^{*} \cdot \vec \partial \phi +m^2\phi^{*}\phi \Big] (\tau
,\vec \sigma ) \Big) -\nonumber \\
&-&b^{\mu}_{\check r}(\tau ) \{ \int d^3\sigma [\pi_{\phi^{*}}\partial
_{\check r}\phi^{*}+\pi_{\phi} \partial_{\check r}\phi ](\tau ,\vec \sigma ) \}
\approx 0,\nonumber \\
{\tilde {\cal H}}^{\mu\nu}(\tau )&=&b^{\mu}_{\check r}(\tau ) \int d^3\sigma
\sigma^{\check r} {\cal H}^{\nu}(\tau ,\vec \sigma )-b^{\nu}_{\check r}(\tau )
\int d^3\sigma \sigma^{\check r} {\cal H}^{\mu}(\tau ,\vec \sigma )=\nonumber \\
&=&S^{\mu\nu}_s-\Big( b^{\mu}_{\check r}(\tau )l^{\nu}-b^{\nu}_{\check r}(\tau )
l^{\mu}\Big) \nonumber \\
&&\int d^3\sigma \sigma^{\check r}\Big[ \pi_{\phi^{*}}\pi_{\phi}+\vec \partial
\phi^{*}\cdot \vec \partial \phi +m^2\phi^{*}\phi \Big] (\tau ,
\vec \sigma ) +\nonumber \\
&+&\Big( b^{\mu}_{\check r}(\tau )b^{\nu}_{\check s}(\tau )-b^{\nu}
_{\check r}(\tau )b^{\mu}_{\check s}(\tau )\Big) \nonumber \\
&& \int d^3\sigma \sigma^{\check r} \Big[ \pi_{\phi^{*}}\partial_{\check s}
\phi^{*}+\pi_{\phi} \partial_{\check s}\phi \Big]
(\tau ,\vec \sigma )  \approx 0.
\label{VIII6}
\end{eqnarray}

The only surviving constraints on the Wigner hyperplanes [with the reduced
canonical variables ${\tilde x}^{\mu}_s(\tau )$, $p^{\mu}_s$,
$\phi (\tau ,\vec \sigma )$, $\pi_{\phi}(\tau ,\vec \sigma )={\dot \phi}^{*}
(\tau ,\vec \sigma )$, $\phi^{*}(\tau ,\vec \sigma )$, $\pi_{\phi^{*}}(\tau
,\vec \sigma )=\dot \phi (\tau ,\vec \sigma )$, satisfying
standard Dirac brackets] are

\begin{eqnarray}
{\cal H}(\tau )&=& \epsilon_s-
\int d^3\sigma \Big[ \pi_{\phi^{*}} \pi_{\phi}+\vec \partial
\phi^{*} \cdot \vec \partial \phi +m^2\phi^{*}\phi \Big] (\tau ,\vec
\sigma )  \approx 0,\nonumber \\
{\vec {\cal H}}_p(\tau )&=&
\int d^3\sigma \Big[ \pi_{\phi^{*}} \vec \partial \phi^{*} +\pi_{\phi}
\vec \partial \phi \Big] (\tau ,\vec \sigma ) \approx 0.
\label{VIII7}
\end{eqnarray}

The Lorentz generators have the form of Eq.(\ref{II9}) with the spin tensor

\begin{equation}
{\bar S}_s^{rs}\equiv S^{rs}_{\phi}=J^{rs}_{\phi}{|}_{{\vec P}_{\phi}=0}=
\int d^3\sigma \Big( \sigma^r\Big[ \pi_{\phi^{*}}\partial^s\phi^{*}+\pi
_{\phi}\partial^s\phi \Big] (\tau ,\vec \sigma )-(r \leftrightarrow s) \Big).
\label{VIII8}
\end{equation}

With the gauge fixing $\chi =T_s-\tau \approx 0$ the final Hamiltonian becomes

\begin{eqnarray}
H_R&=&M_{\phi}-\vec \lambda (T_s)\cdot {\vec {\cal H}}_p(T_s),\nonumber \\
M_{\phi}&=& \int d^3\sigma \Big[ \pi_{\phi^{*}} \pi_{\phi}+\vec \partial
\phi^{*} \cdot \vec \partial \phi +m^2\phi^{*}\phi \Big] (T_s ,\vec
\sigma )
\label{VIII9}
\end{eqnarray}

\subsection{Collective and relative variables.}

A priori we have the following two possibilities for the description  of the
complex Klein-Gordon field $\phi (\tau ,\vec \sigma )$ in terms of a
Fourier transform

\begin{eqnarray}
\phi ( \tau ,\vec \sigma ) &=&{1\over {\sqrt{2}}}[ \phi_1(\tau ,\vec \sigma )
+i\phi_2(\tau ,\vec \sigma )]=\nonumber \\
&=&{1\over {\sqrt{2}}} \int d\tilde q  \Big(\, [ a_1( \tau
,\vec q) e^{-i( \omega ( q) \tau -\vec q\cdot
\vec \sigma ) }+a_1^{*}(\tau ,\vec q) e^{+i( \omega
( q) \tau -\vec q\cdot \vec \sigma ) } ] + \nonumber \\
&+&i [ a_2( \tau
,\vec q) e^{-i( \omega (  q) \tau -\vec q\cdot
\vec \sigma ) }+a_2^{*}( \tau ,\vec q) e^{+i( \omega
(  q) \tau -\vec q\cdot \vec \sigma ) }] \, \Big) = \nonumber \\
&=&\int d\tilde q [ a ( \tau
,\vec q) e^{-i( \omega (  q) \tau -\vec q\cdot
\vec \sigma ) }+b^{*} ( \tau ,\vec q) e^{+i( \omega
(  q) \tau -\vec q\cdot \vec \sigma ) }] ,\nonumber \\
\phi^{*}(\tau ,\vec \sigma )&=&{1\over {\sqrt{2}}}[\phi_1(\tau ,\vec \sigma )-
i\phi_2(\tau ,\vec \sigma )]=\nonumber \\
&=&{1\over {\sqrt{2}}}
\int d\tilde q  \Big( \, [ a_1( \tau
,\vec q) e^{-i( \omega (  q) \tau -\vec q\cdot
\vec \sigma ) }+a_1^{*}( \tau ,\vec q) e^{i( \omega
(  q) \tau -\vec q\cdot \vec \sigma ) }] - \nonumber \\
&-&i [ a_2( \tau
,\vec q) e^{-i( \omega (  q) \tau -\vec q\cdot
\vec \sigma ) }+a_2^{*}( \tau ,\vec q) e^{+i( \omega
(  q) \tau -\vec q\cdot \vec \sigma ) }] \, \Big) =\nonumber \\
&=&\int d\tilde q [ b ( \tau ,\vec q) e^{-i( \omega
(  q) \tau -\vec q\cdot \vec \sigma ) } + a^{*} ( \tau
,\vec q) e^{+i( \omega (  q) \tau -\vec q\cdot
\vec \sigma ) } ] ,\nonumber \\
&&{}\nonumber \\
 \pi ( \tau ,\vec \sigma )& =&{1\over {\sqrt{2}}}[\pi_1(\tau ,\vec \sigma )-i
\pi_2(\tau ,\vec \sigma )]=\nonumber \\
&=&{{-i}\over {\sqrt{2}}} \int d\tilde q  \Big( \, [ a_1( \tau
,\vec q) e^{-i( \omega (  q) \tau -\vec q\cdot
\vec \sigma ) }-a_1^{*}( \tau ,\vec q) e^{+i( \omega
(  q) \tau -\vec q\cdot \vec \sigma ) }] - \nonumber \\
&-&i [ a_2( \tau
,\vec q) e^{-i( \omega (  q) \tau -\vec q\cdot
\vec \sigma ) }-a_2^{*}( \tau ,\vec q) e^{+i( \omega
(  q) \tau -\vec q\cdot \vec \sigma ) }] \, \Big) =\nonumber \\
&=&-i\int d\tilde q\omega (
q) [ b( \tau ,\vec q) e^{-i( \omega (
q) \tau -\vec q\cdot \vec \sigma ) }-a^{*}( \tau ,\vec
q) e^{+i( \omega ( \vec q) \tau -\vec q\cdot \vec
\sigma ) }] ,\nonumber \\
\pi^{*}(\tau ,\vec \sigma )&=&{1\over {\sqrt{2}}}[\pi_1(\tau ,\vec \sigma )+i
\pi_2(\tau ,\vec \sigma )]=\nonumber \\
&=&{{-i}\over {\sqrt{2}}} \int d\tilde q  \Big( \, [ a_1( \tau
,\vec q) e^{-i( \omega (  q) \tau -\vec q\cdot
\vec \sigma ) }-a_1^{*}( \tau ,\vec q) e^{+i( \omega
(  q) \tau -\vec q\cdot \vec \sigma ) }] + \nonumber \\
&+&i [ a_2( \tau
,\vec q) e^{-i( \omega (  q) \tau -\vec q\cdot
\vec \sigma ) }-a_2^{*}( \tau ,\vec q) e^{+i( \omega
(  q) \tau -\vec q\cdot \vec \sigma ) }] \, \Big) =\nonumber \\
&=&-i\int d\tilde q\omega (
q) [ a( \tau ,\vec
q) e^{-i( \omega ( \vec q) \tau -\vec q\cdot \vec
\sigma ) } - b^{*}( \tau ,\vec q) e^{+i( \omega (
q) \tau -\vec q\cdot \vec \sigma ) } ] ,\nonumber \\
&&{}\nonumber  \\
a_i(\tau ,\vec q)&=&\sqrt{I_i(\tau ,\vec q)} e^{i\varphi_i(\tau ,\vec q)},
\quad\quad i=1,2,\nonumber \\
a( \tau ,\vec q) &=&{1\over {\sqrt{2}}} [a_1(\tau ,\vec q)+ia_2(\tau
,\vec q) ]={1\over {\sqrt{2}}} [\sqrt{I_1}e^{i\varphi_1}+i\sqrt{I_2}
e^{i\varphi_2}](\tau ,\vec q)=\nonumber \\
&=&\int d^3\sigma [ \omega (
q) \phi ( \tau ,\vec \sigma ) +i\pi^{*} ( \tau ,\vec
\sigma ) ] e^{i( \omega (  q) \tau -\vec
q\cdot \vec \sigma ) }=\sqrt{I_a(\tau ,\vec q)}e^{i\varphi_a(\tau
,\vec q)}, \nonumber \\
b(\tau ,\vec q)&=&{1\over {\sqrt{2}}} [a_1(\tau ,\vec q)-ia_2(\tau ,\vec q)]=
{1\over {\sqrt{2}}}[\sqrt{I_1}e^{i\varphi_1}-i\sqrt{I_2}e^{i\varphi_2}](\tau
,\vec q)=\nonumber \\
&=&\int d^3\sigma [\omega (q) \phi^{*}(\tau ,\vec \sigma )+i \pi (\tau ,\vec
\sigma )] e^{i(\omega (q)\tau -\vec q\cdot \vec \sigma )}=\sqrt{I_b(\tau
,\vec q)}e^{i\varphi_b(\tau ,\vec q)},\nonumber \\
a^{*}( \tau ,\vec q)  &=&{1\over {\sqrt{2}}}[a_1^{*}(\tau ,\vec q)-ia
_2^{*}(\tau ,\vec q)]=\nonumber \\
&=&\int d^3\sigma [ \omega (
q) \phi^{*} ( \tau ,\vec \sigma ) -i\pi ( \tau ,\vec
\sigma ) ] e^{-i( \omega (  q) \tau -\vec
q\cdot \vec \sigma ) }, \nonumber \\
b^{*}(\tau ,\vec q)&=&{1\over {\sqrt{2}}}[a_1^{*}(\tau ,\vec q)+
ia_2^{*}(\tau ,\vec q)]=\nonumber \\
&=&\int d^3\sigma [\omega (q)\phi (\tau ,\vec \sigma )-i\pi^{*}(\tau ,\vec
\sigma )] e^{-i(\omega (q)\tau -\vec q\cdot \vec \sigma )},\nonumber \\
&&{}\nonumber \\
I_i(\tau ,\vec q)&=&a^{*}_i(\tau ,\vec q) a_i(\tau ,\vec q),\nonumber \\
I_a(\tau ,\vec q)&=&a^{*}(\tau ,\vec q)a(\tau ,\vec q)={1\over 2}[a^{*}_1a_1+
a^{*}_2a_2+i(a^{*}_1a_2-a^{*}_2a_1)](\tau ,\vec q),\nonumber \\
I_b(\tau ,\vec q)&=&b^{*}(\tau ,\vec q)b(\tau ,\vec q)={1\over 2}[a^{*}_1a_1+
a^{*}_2a_2-i(a^{*}_1a_2-a^{*}_2a_1)],\nonumber \\
(I_a+I_b)&&(\tau ,\vec q)=[I_1+I_2](\tau ,\vec q),\nonumber \\
(I_a-I_b)&&(\tau ,\vec q)=i[a^{*}_1a_2-a^{*}_2a_1](\tau ,\vec q),\nonumber \\
&&{}\nonumber \\
N_{i \phi}&=&\int d\tilde q a^{*}_i(\tau ,\vec q) a_i(\tau ,\vec q)= \int
d\tilde q I_i(\tau ,\vec \sigma ),\nonumber \\
N_{a \phi}&=&\int d\tilde q a^{*}(\tau ,\vec q)a(\tau ,\vec q)
={1\over 2}[N_{1 \phi}+N_{2 \phi}]+{i\over 2}\int d\tilde q [a^{*}_1a_2-
a^{*}_2a_1](\tau ,\vec q)=\nonumber \\
&=& \int d\tilde q I_a(\tau ,\vec q)=\nonumber \\
&=&{1\over 2} \int d^3\sigma [\pi_{\phi^{*}} {1\over {\sqrt{m^2+\triangle}}}
\pi +\phi^{*} \sqrt{m^2+\triangle} \phi +i(\pi_{\phi^{*}}\phi^{*}-\pi \phi )
](\tau ,\vec \sigma ),\nonumber \\
N_{b \phi}&=&\int d\tilde q b^{*}(\tau ,\vec q) b(\tau ,\vec q)
={1\over 2}[N_{1 \phi}+N_{2 \phi}]-{i\over 2}\int d\tilde q [a^{*}_1a_2-
a^{*}_2a_1](\tau ,\vec q)=\nonumber \\
&+&\int d\tilde q I_b(\tau ,\vec q)=\nonumber \\
&=&{1\over 2} \int d^3\sigma [\pi_{\phi^{*}} {1\over {\sqrt{m^2+\triangle}}}
\pi +\phi^{*} \sqrt{m^2+\triangle} \phi -i(\pi_{\phi^{*}}\phi^{*}-\pi \phi )
](\tau ,\vec \sigma ),\nonumber \\
&&{}\nonumber \\
N_{\phi}&=&N_{1 \phi}+N_{2 \phi} = N_{a \phi} + N_{b \phi}=\nonumber \\
&=&\int d^3\sigma [\pi_{\phi^{*}} {1\over {\sqrt{m^2+\triangle}}}
\pi +\phi^{*} \sqrt{m^2+\triangle} \phi ](\tau ,\vec \sigma )
\label{VIII10}
\end{eqnarray}

However, the description of the conserved electric charge
of the Klein-Gordon field privileges the use of
the Fourier coefficients $a(\tau ,\vec q)$, $b(\tau ,\vec q)$ rather than of
the $a_i(\tau ,\vec q)$'s, because we have

\begin{equation}
q_{\phi}=i\int d^3\sigma [\pi_{\phi^{*}}\phi^{*}-\pi \phi ](\tau ,\vec \sigma )=
N_{b \phi} - N_{a \phi},\quad {{dq_{\phi}}\over {d\tau}}\, {\buildrel \circ
\over =}\, 0;
\label{VIII11}
\end{equation}

\noindent on the Wigner hyperplanes the conservation $\partial_{\mu}J^{\mu}
_{\phi}\, {\buildrel \circ \over =}\, 0$ of the electromagnetic current
$J^{\mu}_{\phi}(x)=-i [\partial^{\mu} {\tilde\phi}^{*} \tilde \phi -{\tilde
\phi}^{*} \partial^{\mu} \tilde \phi ](x)$ is replaced by the existence of the
Hamiltonian constant of motion $q_{\phi}$.

The Fourier coefficients $a_i(\tau ,\vec \sigma )$, i=1,2, correspond
to the description of the field as two real Klein-Gordon fields
$\phi_i$. Instead, as shown in Appendix D, the Fourier coefficients
``a" and ``b" correspond to two Klein-Gordon fields $\phi_a$,
$\phi_b$, with positive ($\phi_a$) or negative ($\phi_b$)  energy and
electric charge, which are nonlocal combinations of $\phi$ and
$\pi_{\phi^{*}}$.

The total 4-momentum of the Klein-Gordon field may be described either in terms
of the ``a" and ``b" or of the ``$a_i$"

\begin{eqnarray}
P^{\tau}_{\phi}&=&\int d\tilde q \omega (q) [I_1+I_2](\tau ,\vec q)=P^{\tau}
_{1 \phi}(\tau )+P^{\tau}_{2 \phi}(\tau )=\nonumber \\
&=&\int d\tilde q \omega (q) [I_a+I_b](\tau ,\vec q)=P^{\tau}_{a \phi}(\tau )+
P^{\tau}_{b \phi}(\tau )=\nonumber \\
&=&\int d^3\sigma \Big[ \pi_{\phi^{*}} \pi_{\phi}+\vec \partial
\phi^{*} \cdot \vec \partial \phi +m^2\phi^{*}\phi \Big] (\tau ,\vec
\sigma ),\nonumber \\
{\vec P}_{\phi}&=&\int d\tilde q \vec q [I_1+I_2](\tau ,\vec q)={\vec P}
_{1 \phi}(\tau )+{\vec P}_{2 \phi}(\tau )=\nonumber \\
&=&\int d\tilde q \vec q [I_a+I_b](\tau ,\vec q)={\vec P}_{a \phi}(\tau )+
{\vec P}_{b \phi}(\tau )=\nonumber \\
&=&\int d^3\sigma \Big[ \pi_{\phi^{*}} \vec \partial \phi^{*} +\pi_{\phi}
\vec \partial \phi \Big] (\tau ,\vec \sigma ),
\label{VIII12}
\end{eqnarray}

\noindent but it is possible to
define a global ``relative 4-momentum" of the field
 only in terms of the ``a" and ``b"

\begin{eqnarray}
Q^{\tau}_{\phi}(\tau )&=&{1\over 2}\int d\tilde q \omega (q) [I_a-I_b](\tau
,\vec q)=P^{\tau}_{a \phi}(\tau )-P^{\tau}_{b \phi}(\tau )=\nonumber \\
&=&{i\over 2} \int d^3\sigma [\pi_{\phi^{*}} \sqrt{m^2+\triangle} \phi^{*}-
\pi \sqrt{m^2+\triangle} \phi ](\tau ,\vec \sigma )\not= \nonumber \\
&\not=& {1\over 2}\int d\tilde q \omega (q) [I_1-I_2](\tau ,\vec q)=
P^{\tau}_{1 \phi}(\tau )-P^{\tau}_{2 \phi}(\tau ),\nonumber \\
{\vec Q}_{\phi}(\tau )&=&{1\over 2}\int d\tilde q \vec q [I_a-I_b](\tau ,\vec
q)={\vec P}_{a \phi}(\tau )-{\vec P}_{b \phi}(\tau )=\nonumber \\
&=&-{i\over 2} \int d^3\sigma [\pi_{\phi^{*}} {{\vec \partial}\over {\sqrt{m^2
+\triangle}}} \pi -\phi^{*} \sqrt{m^2+\triangle} \vec \partial \phi ](\tau
,\vec \sigma )\not= \nonumber \\
&\not=&{1\over 2}\int d\tilde q \vec q [I_1-I_2](\tau ,\vec q)={\vec P}
_{1 \phi}(\tau )-{\vec P}_{2 \phi}(\tau ),\nonumber \\
&&{}\nonumber \\
P^{\tau}_{a \phi}&=&{1\over 2}(P^{\tau}_{\phi}+Q^{\tau}_{\phi}),\quad\quad
{\vec P}_{a \phi}={1\over 2}({\vec P}_{\phi}+{\vec Q}_{\phi}),\nonumber \\
P^{\tau}_{b \phi}&=&{1\over 2}(P^{\tau}_{\phi}-Q^{\tau}_{\phi}),\quad\quad
{\vec P}_{b \phi}={1\over 2}({\vec P}_{\phi}-{\vec Q}_{\phi}),
\label{VIII13}
\end{eqnarray}

The two partial centers of phase have coordinates

\begin{eqnarray}
X^{\tau}_{a\, \phi}&=& \int d\tilde q \omega (q) F^{\tau}(q) \varphi_a(\tau
,\vec q)={1\over {2i}} \int d\tilde q \omega (q) F^{\tau}(q) ln\, {{a(\tau
,\vec q)}\over {a^{*}(\tau ,\vec q)}}=\tau +\nonumber \\
&+&{1\over {2\pi i m}} \int d^3q {{e^{-{{4\pi}\over {m^2}} q^2} }\over
{q^2\, \omega (q)}} ln\, \Big[ {{\omega (q) \int d^3\sigma e^{i\vec q\cdot \vec
\sigma} \phi (\tau ,\vec \sigma ) +i\int d^3\sigma e^{i\vec q\cdot \vec \sigma}
\pi_{\phi^{*}}(\tau ,\vec \sigma )}\over
{\omega (q) \int d^3\sigma e^{-i\vec q\cdot \vec
\sigma} \phi^{*}(\tau ,\vec \sigma ) -i \int d^3\sigma e^{-i\vec q\cdot \vec
\sigma} \pi (\tau ,\vec \sigma )}}\Big]=\nonumber \\
&{\buildrel {def} \over =}& \tau +{\tilde X}^{\tau}_{a\, \phi},\nonumber \\
{\vec X}_{a\, \phi}&=& \int d\tilde q \, \vec q\, F(q) \varphi_a(\tau ,\vec q)
={1\over {2i}} \int d\tilde q\, \vec q\, F(q) ln\, {{a(\tau
,\vec q)}\over {a^{*}(\tau ,\vec q)}}=\nonumber \\
&=&{{2i}\over {\pi m}} \int d^3q\,
{{\vec q}\over {q^4}}\, e^{-{{4\pi}\over {m^2}}
q^2}  ln\, \Big[ {{\sqrt{m^2+q^2} \int d^3\sigma e^{i\vec q\cdot \vec \sigma}
\phi (\tau ,\vec \sigma ) +i \int d^3\sigma e^{i\vec q\cdot \vec \sigma}
\pi_{\phi^{*}}(\tau ,\vec \sigma )}\over
{\sqrt{m^2+q^2} \int d^3\sigma e^{-i\vec q\cdot
\vec \sigma} \phi^{*}(\tau ,\vec \sigma ) -i \int d^3\sigma e^{-i\vec q\cdot
\vec \sigma} \pi (\tau ,\vec \sigma )}} \Big] ,\nonumber \\
&&{}\nonumber \\
X^{\tau}_{b\, \phi}&=& \int d\tilde q \omega (q) F^{\tau}(q) \varphi_b(\tau
,\vec q)={1\over {2i}} \int d\tilde q \omega (q) F^{\tau}(q) ln\, {{b(\tau
,\vec q)}\over {b^{*}(\tau ,\vec q)}}=\tau +\nonumber \\
&+&{1\over {2\pi i m}} \int d^3q {{e^{-{{4\pi}\over {m^2}} q^2} }\over
{q^2\, \omega (q)}} ln\, \Big[ {{\omega (q) \int d^3\sigma e^{i\vec q\cdot \vec
\sigma} \phi^{*}(\tau ,\vec \sigma ) +i\int d^3\sigma e^{i\vec q\cdot \vec
\sigma}\pi (\tau ,\vec \sigma )}\over
{\omega (q) \int d^3\sigma e^{-i\vec q\cdot \vec
\sigma} \phi (\tau ,\vec \sigma ) -i \int d^3\sigma e^{-i\vec q\cdot \vec
\sigma} \pi_{\phi^{*}}(\tau ,\vec \sigma )}}\Big]=\nonumber \\
&{\buildrel {def} \over =}& \tau +{\tilde X}^{\tau}_{b\, \phi},\nonumber \\
{\vec X}_{b\, \phi}&=& \int d\tilde q \, \vec q\, F(q) \varphi_b(\tau ,\vec q)
={1\over {2i}} \int d\tilde q\, \vec q\, F(q) ln\, {{b(\tau
,\vec q)}\over {b^{*}(\tau ,\vec q)}}=\nonumber \\
&=&{{2i}\over {\pi m}} \int d^3q\,
{{\vec q}\over {q^4}}\, e^{-{{4\pi}\over {m^2}}
q^2}  ln\, \Big[ {{\sqrt{m^2+q^2} \int d^3\sigma e^{i\vec q\cdot \vec \sigma}
\phi^{*}(\tau ,\vec \sigma ) +i \int d^3\sigma e^{i\vec q\cdot \vec \sigma}
\pi (\tau ,\vec \sigma )}\over {\sqrt{m^2+q^2} \int d^3\sigma e^{-i\vec q\cdot
\vec \sigma} \phi (\tau ,\vec \sigma ) -i \int d^3\sigma e^{-i\vec q\cdot
\vec \sigma} \pi_{\phi^{*}}(\tau ,\vec \sigma )}} \Big] .
\label{VIII14}
\end{eqnarray}

Therefore, we shall define the canonical transformation of Section IV
separately to $a,a^{*}\mapsto$ $P^A_{a \phi}$, $X^A_{a \phi} (X^{\tau}_{a\phi}=
\tau +{\tilde X}^{\tau}_{a\phi})$, ${\bf H}_a$, ${\bf K}_a$ and to
$b,b^{*}\mapsto$ $P^A_{b \phi}$, $X^A_{b \phi} (X^{\tau}_{b\phi}=\tau +{\tilde
X}_{b\phi})$, ${\bf H}_b$, ${\bf K}_b$ and not to $a_i,a^{*}_i$,
i=1,2, $\mapsto P^A_{i \phi}, X^A_{i \phi}, {\bf H}_i, {\bf K}_i$
[assuming that the fields $\phi_a$, $\pi_a$, $\phi_b$, $\pi_b$ satisfy
Eq.(\ref{IV29})]. Then, we shall do the further canonical transformation from
the two sets of collective variables $X^A_{a\phi}$,
$P^A_{a\phi}$, $X^A_{b\phi}$, $P^A_{b\phi}$, to global center of phase variables
$X^A_{\phi}$, $P^A_{\phi}$, and global relative variables $R^A_{\phi}$, $Q^A
_{\phi}$ [$R^A_{\phi}$ describes the action-reaction between the centers of
phase of positive and negative energy field configutations]

\begin{eqnarray}
X^A_{\phi}&=&{1\over 2}(X^A_{a\, \phi}+X^A_{b\, \phi})\, [=
{1\over 2}(X^A_{1\phi}+X^A_{2\phi})],\quad\quad
X^{\tau}_{\phi}=\tau +{\tilde X}^{\tau}_{\phi},\nonumber \\
P^A_{\phi}&=&P^A_{a\, \phi}+P^A_{b\, \phi}[=P^A_{1\phi}+P^A_{2\phi}],
\nonumber \\
R^A_{\phi}&=&X^A_{a\, \phi}-X^A_{b\, \phi},\nonumber \\
Q^A_{\phi}&=&{1\over 2}(P^A_{a\, \phi}-P^A_{b\, \phi}).
\label{VIII15}
\end{eqnarray}

The Poincar\'e generators are $P^{\tau}_{\phi}$, ${\vec P}_{\phi}$ and

\begin{eqnarray}
J^{rs}_{\phi}&=&\int d^3\sigma [\pi_{\phi^{*}}(\sigma^r\partial^s-\sigma^s
\partial^r)\phi^{*}+\pi_{\phi}(\sigma^r\partial^s-\sigma^s\partial^r)
\phi](\tau ,\vec \sigma )=\nonumber \\
&=&-i \sum_{i=1}^2 \int d\tilde q a^{*}_i(\tau ,\vec q)(q^r{{\partial}\over
{\partial q^s}}-q^s{{\partial}\over {\partial q^r}}) a_i(\tau ,\vec q)=
\nonumber \\
&=&-i \int d\tilde q \Big[ a^{*}(\tau ,\vec q)(q^r{{\partial}\over {\partial
q^s}}-q^s{{\partial}\over {\partial q^r}})a(\tau ,\vec q)+b^{*}(\tau ,\vec q)
(q^r{{\partial}\over {\partial q^s}}-q^s{{\partial}\over {\partial q^r}})
b(\tau ,\vec q)\Big],\nonumber \\
J^{\tau r}_{\phi}&=&-\tau P^r_{\phi}+{1\over 2}\int d^3\sigma \, \sigma^r [\pi
_{\phi^{*}}\pi_{\phi}+\vec \partial \phi^{*}\cdot \vec \partial \phi +m^2
\phi^{*}\phi ](\tau ,\vec \sigma )=\nonumber \\
&=&-\tau P^r_{\phi}+i\sum_{i=1}^2 \int d\tilde q \omega (q) a^{*}_i(\tau ,\vec
q) {{\partial}\over {\partial q^r}} a_i(\tau ,\vec q)=\nonumber \\
&=&-\tau P^r_{\phi}+i \int d\tilde q \omega (q) \Big[ a^{*}(\tau ,\vec q)
{{\partial}\over {\partial q^r}}a(\tau ,\vec q)+b^{*}(\tau ,\vec q) {{\partial}
\over {\partial q^r}} b(\tau ,\vec q).
\label{VIII16}
\end{eqnarray}

To find the real center of mass ${\vec q}_{\phi}$ we must use
Eq.(\ref{VI1}) of Section VI, namely $q^r_{\phi} \approx
r^r_{\phi}=-J^{\tau r}_{\phi}/P^{\tau}
_{\phi}{|}_{\tau =0}$.

The original variables have the following expression

\begin{eqnarray}
a(\tau ,\vec k)&=&\sqrt{F^{\tau}(k)\omega (k)P^{\tau}
_{a \phi}-F(k)\vec k\cdot {\vec P}_{a \phi}(\tau )+{\cal D}_k {\bf H}_a(\tau ,
\vec k) }\nonumber \\
&&e^{i\int d\tilde q\int d\tilde q^{\prime }\, {\bf K}_a( \tau ,\vec q)
{\cal G}\left( \vec q,\vec q^{\prime }\right) \Delta \left( \vec q^{\prime }
,\vec k\right) +i\omega (k) X_{a \phi}^\tau (\tau ) -
i\vec k\cdot {\vec X}_{a \phi}(\tau ) },\nonumber \\
&&{}\nonumber \\
b(\tau ,\vec q)&=&\sqrt{F^{\tau}(k)\omega (k)P^{\tau}_{b \phi}(\tau )-
F(k)\vec k\cdot {\vec P}_{b \phi}(\tau )+{\cal D}_k {\bf H}_b(\tau , \vec k) }
\nonumber \\
&&e^{i\int d\tilde q\int d\tilde q^{\prime }\, {\bf K}_b( \tau ,\vec q)
{\cal G}\left( \vec q,\vec q^{\prime }\right) \Delta \left( \vec q^{\prime }
,\vec k\right) +i\omega (k) X_{b \phi}^\tau (\tau ) -
i\vec k\cdot {\vec X}_{b \phi}(\tau ) } ,\nonumber \\
&&{}\nonumber \\
\phi (\tau ,\sigma )&=& \int d\tilde k \Big(\, \sqrt{F^{\tau}(k)\omega (k)
P^{\tau}_{a \phi}(\tau )-F(k)\vec k\cdot {\vec P}_{a \phi}(\tau )+{\cal D}
_k {\bf H}_a(\tau , \vec k) }\nonumber \\
&&e^{i\int d\tilde q\int d\tilde q^{\prime }\, {\bf K}_a( \tau ,\vec q)
{\cal G}\left( \vec q,\vec q^{\prime }\right) \Delta \left( \vec q^{\prime }
,\vec k\right) +i\omega (k) {\tilde X}_{a \phi}^\tau (\tau ) +
i\vec k\cdot \left(\vec \sigma -{\vec X}_{a \phi}(\tau )\right) }+\nonumber \\
&&+\sqrt{F^{\tau}(k)\omega (k)P^{\tau}_{b \phi}(\tau )-
F(k)\vec k\cdot {\vec P}_{b \phi}(\tau )+{\cal D}_k {\bf H}_b(\tau , \vec k) }
\nonumber \\
&&e^{-i\int d\tilde q\int d\tilde q^{\prime }\, {\bf K}_b( \tau ,\vec q)
{\cal G}\left( \vec q,\vec q^{\prime }\right) \Delta \left( \vec q^{\prime }
,\vec k\right) -i\omega (k) {\tilde X}_{b \phi}^\tau (\tau ) -
i\vec k\cdot \left(\vec \sigma -{\vec X}_{b \phi}(\tau )\right) }
\Big)=\nonumber \\
&&=\int d\tilde k  \Big(\, \sqrt{F^{\tau}(k)\omega (k)[{1\over 2}P^{\tau}
_{\phi}+Q^{\tau}_{\phi}(\tau )]-F(k)\vec k\cdot [{1\over 2}{\vec P}_{\phi}+
{\vec Q}_{\phi}(\tau )]+{\cal D}_k {\bf H}_a(\tau ,
\vec k) }\nonumber \\
&&e^{+i\int d\tilde q\int d\tilde q^{\prime }\, {\bf K}_a( \tau ,\vec q)
{\cal G}\left( \vec q,\vec q^{\prime }\right) \Delta \left( \vec q^{\prime }
,\vec k\right) +i\omega (k) [{\tilde X}_{\phi}^\tau +{1\over 2}{\tilde R}^{\tau}
_{\phi}(\tau )] +i\vec k\cdot \left(\vec \sigma -[{\vec X}_{\phi}+{1\over 2}
{\vec R}_{\phi}(\tau )]\right) }+\nonumber \\
&&+\sqrt{F^{\tau}(k)\omega (k)[{1\over 2}P^{\tau}_{\phi}-Q^{\tau}_{\phi}(\tau )]
-F(k)\vec k\cdot [{1\over 2}{\vec P}_{\phi}-{\vec Q}_{\phi}(\tau )]+
{\cal D}_k {\bf H}_b(\tau , \vec k) }\nonumber \\
&&e^{-i\int d\tilde q\int d\tilde q^{\prime }\, {\bf K}_b( \tau ,\vec q)
{\cal G}\left( \vec q,\vec q^{\prime }\right) \Delta \left( \vec q^{\prime }
,\vec k\right) -i\omega (k) [{\tilde X}_{\phi}^\tau -{1\over 2}{\tilde R}^{\tau}
_{\phi}(\tau )]  -i\vec k\cdot \left(\vec \sigma -[{\vec X}_{\phi}-
{1\over 2}{\vec R}_{\phi}(\tau )]\right) } \Big)=\nonumber \\
&=&\int d\tilde k\Big[ {\bf A}_{a\vec k}(\tau ;{1\over 2}P^A_{\phi}+Q^A_{\phi},
{\bf H}_a] e^{i\Big( \vec k\cdot \vec \sigma +{\bf B}_{a\vec k}(\tau ;X^A
_{\phi}+{1\over 2}R^A_{\phi},{\bf K}_a]\Big)} +\nonumber \\
&+&{\bf A}_{b\vec k}(\tau ;{1\over 2}P^A_{\phi}-Q^A_{\phi},{\bf H}_b]
e^{-i\Big( \vec k\cdot \vec \sigma +{\bf B}_{b\vec k}(\tau ;X^A_{\phi}-{1\over
2}R^A_{\phi},{\bf K}_b]\Big)}\Big] ,\nonumber \\
&&{}\nonumber \\
\pi (\tau ,\vec \sigma )&=&-i \int d\tilde k \omega (k) \nonumber \\
&&\Big(
\sqrt{F^{\tau}(k)\omega (k)[{1\over 2}P^{\tau}_{\phi}-Q^{\tau}_{\phi}(\tau )]
-F(k)\vec k\cdot [{1\over 2}{\vec P}_{\phi}-{\vec Q}_{\phi}(\tau )]+
{\cal D}_k {\bf H}_b(\tau , \vec k) }\nonumber \\
&&e^{+i\int d\tilde q\int d\tilde q^{\prime }\, {\bf K}_b( \tau ,\vec q)
{\cal G}\left( \vec q,\vec q^{\prime }\right) \Delta \left( \vec q^{\prime }
,\vec k\right) +i\omega (k) [{\tilde X}_{\phi}^\tau -{1\over 2}{\tilde R}^{\tau}
_{\phi}(\tau )]  +i\vec k\cdot \left(\vec \sigma -[{\vec X}_{\phi}-
{1\over 2}{\vec R}_{\phi}(\tau )]\right) }-\nonumber \\
&&-\sqrt{F^{\tau}(k)\omega (k)[{1\over 2}P^{\tau}
_{\phi}+Q^{\tau}_{\phi}(\tau )]-F(k)\vec k\cdot [{1\over 2}{\vec P}_{\phi}+
{\vec Q}_{\phi}(\tau )]+{\cal D}_k {\bf H}_a(\tau ,
\vec k) }\nonumber \\
&&e^{-i\int d\tilde q\int d\tilde q^{\prime }\, {\bf K}_a( \tau ,\vec q)
{\cal G}\left( \vec q,\vec q^{\prime }\right) \Delta \left( \vec q^{\prime }
,\vec k\right) -i\omega (k) [{\tilde X}_{\phi}^\tau +{1\over 2}{\tilde R}^{\tau}
_{\phi}(\tau )]  -i\vec k\cdot \left(\vec \sigma -[{\vec X}_{\phi}+{1\over 2}
{\vec R}_{\phi}(\tau )]\right) } \Big)=\nonumber \\
&=&i \int d\tilde k \omega (k)\nonumber \\
&&\Big[ {\bf A}_{a\vec k}(\tau ;{1\over 2}P^A_{\phi}+Q^A_{\phi},
{\bf H}_a] e^{-i\Big( \vec k\cdot \vec \sigma +{\bf B}_{a\vec k}(\tau ;X^A
_{\phi}+{1\over 2}R^A_{\phi},{\bf K}_a]\Big)}-\nonumber \\
&-&{\bf A}_{b\vec k}(\tau ;{1\over 2}P^A_{\phi}-Q^A_{\phi},{\bf H}_b]
e^{i\Big( \vec k\cdot \vec \sigma +{\bf B}_{b\vec k}(\tau ;X^A_{\phi}-{1\over
2}R^A_{\phi},{\bf K}_b]\Big)} \Big],\nonumber \\
&&{}\nonumber \\
&&{}\nonumber \\
&&{\bf A}_{a\vec k}(\tau ;{1\over 2}P^A_{\phi}+Q^A_{\phi},{\bf H}_a] =
\nonumber \\
&&\sqrt{F^{\tau}(k)\omega (k)[{1\over 2}P^{\tau}
_{\phi}+Q^{\tau}_{\phi}(\tau )]-F(k)\vec k\cdot [{1\over 2}{\vec P}_{\phi}+
{\vec Q}_{\phi}(\tau )]+{\cal D}_{\vec k} {\bf H}_a(\tau ,\vec k) },\nonumber \\
&&{\bf A}_{b\vec k}(\tau ;{1\over 2}P^A_{\phi}-Q^A_{\phi},{\bf H}_b] =
\nonumber \\
&&\sqrt{F^{\tau}(k)\omega (k)[{1\over 2}P^{\tau}
_{\phi}-Q^{\tau}_{\phi}(\tau )]-F(k)\vec k\cdot [{1\over 2}{\vec P}_{\phi}-
{\vec Q}_{\phi}(\tau )]+{\cal D}_{\vec k} {\bf H}_b(\tau ,\vec k) },\nonumber \\
&&{\bf B}_{a\vec k}(\tau ;X^A_{\phi}+{1\over 2}R^A_{\phi},{\bf K}_a] =i\omega
(k)[{\tilde X}^{\tau}_{\phi}+{1\over 2}R^{\tau}_{\phi}(\tau )]-i\vec k\cdot
[{\vec X}_{\phi}+{1\over 2}{\vec R}_{\phi}(\tau )]+\nonumber \\
&+&\int d\tilde qd{\tilde q}^{'}\, {\bf K}_a(\tau ,\vec q) {\cal G}(\vec q
,{\vec q}^{'}) \triangle ({\vec q}^{'},\vec k),\nonumber \\
&&{\bf B}_{b\vec k}(\tau ;X^A_{\phi}-{1\over 2}R^A_{\phi},{\bf K}_a] =i\omega
(k)[{\tilde X}^{\tau}_{\phi}-{1\over 2}R^{\tau}_{\phi}(\tau )]-i\vec k\cdot
[{\vec X}_{\phi}-{1\over 2}{\vec R}_{\phi}(\tau )]+\nonumber \\
&+&\int d\tilde qd{\tilde q}^{'}\, {\bf K}_b(\tau ,\vec q) {\cal G}(\vec q
,{\vec q}^{'}) \triangle ({\vec q}^{'},\vec k),
\label{VIII17}
\end{eqnarray}

In the gauge $\chi =T_s-\tau \approx 0$ we get

\begin{eqnarray}
\phi (T_s,\vec \sigma )&=&
\int d\tilde k  \Big(\, \sqrt{F^{\tau}(k)\omega (k)[{1\over 2}P^{\tau}
_{\phi}+Q^{\tau}_{\phi}(T_s)]-F(k)\vec k\cdot
{\vec Q}_{\phi}(T_s)+{\cal D}_k {\bf H}_a(T_s ,
\vec k) }\nonumber \\
&&e^{i\int d\tilde q\int d\tilde q^{\prime }\, {\bf K}_a( T_s ,\vec q)
{\cal G}\left( \vec q,\vec q^{\prime }\right) \Delta \left( \vec q^{\prime }
,\vec k\right) +i\omega (k) [{\tilde X}_{\phi}^\tau +{1\over 2}{\tilde R}^{\tau}
_{\phi}(T_s)]  +i\vec k\cdot \left(\vec \sigma -{1\over 2}
{\vec R}_{\phi}(T_s)\right) }+\nonumber \\
&&+\sqrt{F^{\tau}(k)\omega (k)[{1\over 2}P^{\tau}_{\phi}-Q^{\tau}_{\phi}(T_s)]
+F(k)\vec k\cdot {\vec Q}_{\phi}(T_s)+
{\cal D}_k {\bf H}_b(T_s , \vec k) }\nonumber \\
&&e^{-i\int d\tilde q\int d\tilde q^{\prime }\, {\bf K}_b( T_s ,\vec q)
{\cal G}\left( \vec q,\vec q^{\prime }\right) \Delta \left( \vec q^{\prime }
,\vec k\right) -i\omega (k) [{\tilde X}_{\phi}^\tau -{1\over 2}{\tilde R}^{\tau}
_{\phi}(T_s)]  -i\vec k\cdot \left(\vec \sigma +
{1\over 2}{\vec R}_{\phi}(T_s)\right) } \Big),\nonumber \\
&&{}\nonumber \\
\pi (T_s ,\vec \sigma )&=&-i \int d\tilde k \omega (k)\nonumber \\
&& \Big(
\sqrt{F^{\tau}(k)\omega (k)[{1\over 2}P^{\tau}_{\phi}-Q^{\tau}_{\phi}(T_s)]
+F(k)\vec k\cdot {\vec Q}_{\phi}(T_s)+
{\cal D}_k {\bf H}_b(T_s , \vec k) }\nonumber \\
&&e^{+i\int d\tilde q\int d\tilde q^{\prime }\, {\bf K}_b( T_s ,\vec q)
{\cal G}\left( \vec q,\vec q^{\prime }\right) \Delta \left( \vec q^{\prime }
,\vec k\right) +i\omega (k) [{\tilde X}_{\phi}^\tau -{1\over 2}{\tilde R}^{\tau}
_{\phi}(T_s)]  +i\vec k\cdot \left(\vec \sigma +
{1\over 2}{\vec R}_{\phi}(T_s)\right) }-\nonumber \\
&&-\sqrt{F^{\tau}(k)\omega (k)[{1\over 2}P^{\tau}
_{\phi}+Q^{\tau}_{\phi}(T_s)]-F(k)\vec k\cdot
{\vec Q}_{\phi}(T_s)+{\cal D}_k {\bf H}_a(T_s , \vec k) }\nonumber \\
&&e^{-i\int d\tilde q\int d\tilde q^{\prime }\, {\bf K}_a( T_s ,\vec q)
{\cal G}\left( \vec q,\vec q^{\prime }\right) \Delta \left( \vec q^{\prime }
,\vec k\right) -i\omega (k) [{\tilde X}_{\phi}^\tau +{1\over 2}{\tilde R}^{\tau}
_{\phi}(T_s)]  -i\vec k\cdot \left(\vec \sigma -{1\over 2}
{\vec R}_{\phi}(T_s)\right) } ),\nonumber \\
&&{}\nonumber \\
N_{\phi} &=& P^{\tau}_{\phi} \int d\tilde k \omega (k) F^{\tau}(k) +\int
d\tilde k (-F(k) \vec k\cdot {\vec P}_{\phi} +{\cal D}_k[{\bf H}_a(T_s,\vec k)
+{\bf H}_b(T_s,\vec k)])\approx \nonumber \\
&&\approx P^{\tau}_{\phi} \int d\tilde k \omega (k) F^{\tau}(k) +\int
d\tilde k {\cal D}_k[{\bf H}_a(T_s,\vec k)+{\bf H}_b(T_s,\vec k)],\nonumber \\
q_{\phi}&=&2 Q^{\tau}_{\phi}(T_s) \int d\tilde k \omega (k) F^{\tau}(k) + \int
d\tilde k (-2 F(k) \vec k\cdot {\vec Q}_{\phi}(T_s)
+{\cal D}_k[{\bf H}_a(T_s,\vec k)-{\bf H}_b(T_s,\vec k)]).\nonumber \\
&&
\label{VIII18}
\end{eqnarray}

\subsection{The coupling to the electromagnetic field.}

Let us consider the action describing a charged Klein Gordon field interacting
with the electromagnetic field on spacelike hypersurfaces following the scheme
of Ref.\cite{albad}

\begin{eqnarray}
S&=& \int d\tau d^3\sigma N(\tau ,\vec \sigma )\sqrt{\gamma (\tau ,\vec \sigma )
}\nonumber \\
&&\{ g^{\tau\tau} (\partial_{\tau}+ieA_{\tau}) \phi^{*}\, (\partial_{\tau}-ieA
_{\tau}) \phi +\nonumber \\
&&+g^{\tau \check r} [(\partial_{\tau}+ieA_{\tau}) \phi^{*}\, (\partial
_{\check r}-ieA_{\check r}) \phi +(\partial_{\check r}+ieA_{\check r}) \phi^{*}\,
(\partial_{\tau}-ieA_{\tau}) \phi ] +\nonumber \\
&&+g^{\check r\check s}(\partial_{\check r}+ieA_{\check r}) \phi^{*}\, (\partial
_{\check s}-ieA_{\check s}) \phi - m^2 \phi^{*}\phi -{1\over 4}g^{\check A\check
C}g^{\check B\check D}F_{\check A\check B}F_{\check C\check D}\, \}
(\tau ,\vec \sigma )=\nonumber \\
&&=\int d\tau d^3\sigma \sqrt{\gamma (\tau ,\vec \sigma )} \{ {1\over N}
[\partial_{\tau}+ieA_{\tau} -N^{\check r}(\partial_{\check r}+ieA_{\check r})]
\phi^{*}\nonumber \\
&&[\partial_{\tau}-ieA_{\tau} -N^{\check s}(\partial_{\check s}-ieA_{\check s})]
\phi + N [\gamma^{\check r\check s}(\partial_{\check r}+ieA_{\check r})\phi^{*}
(\partial_{\check s}-ieA_{\check s})\phi -m^2\phi^{*}\phi ]-\nonumber \\
&&-{1\over {2N}}(F_{\tau\check r}-N^{\check u}F_{\check u\check r})\gamma
^{\check r\check s}(F_{\tau \check s}-N^{\check v}F_{\check v\check s})-{N\over
4} \gamma^{\check r\check s}\gamma^{\check u\check v}F_{\check r\check u}
F_{\check s\check v}\, \}(\tau ,\vec \sigma ).
\label{VIII19}
\end{eqnarray}

\noindent where the configuration variables are $z^{\mu}(\tau ,\vec \sigma )$,
$\phi (\tau ,\vec \sigma )={\tilde \phi}(z(\tau ,\vec \sigma ))$ and $A_{\check
A}(\tau ,\vec \sigma )=z^{\mu}_{\check A}(\tau ,\vec \sigma ){\tilde A}_{\mu}(z
(\tau ,\vec \sigma ))$ [$\tilde \phi (z)$ and
${\tilde A}_{\mu}(z)$ are the standard
Klein-Gordon field and electromagnetic potential, which do not know the
embedding of the spacelike hypersurface $\Sigma$ in Minkowski spacetime like
$\phi$ and $A_{\check A}$].

The canonical momenta are

\begin{eqnarray}
\pi^{\tau}(\tau ,\vec \sigma )&=&{{\partial L}\over {\partial \partial_{\tau}
A_{\tau}(\tau ,\vec \sigma )}}=0,\nonumber \\
\pi^{\check r}(\tau ,\vec \sigma )&=&{{\partial L}\over {\partial \partial
_{\tau}A_{\check r}(\tau ,\vec \sigma )}}=
-{{\sqrt{\gamma (\tau ,\vec \sigma )}}\over
{N(\tau ,\vec \sigma )}}\gamma^{\check r\check s}(\tau ,\vec \sigma )(F_{\tau
\check s}-N^{\check u}F_{\check u\check s})(\tau ,\vec \sigma ),\nonumber \\
\pi_{\phi}(\tau ,\vec \sigma )&=&{{\partial L}\over {\partial \partial_{\tau}
\phi (\tau ,\vec \sigma )}}=
{{\sqrt{\gamma (\tau ,\vec \sigma )}}\over
{N(\tau ,\vec \sigma )}}[\partial_{\tau}+ieA_{\tau}-N^{\check r}(\partial
_{\check r}+ieA_{\check r})](\tau ,\vec \sigma ) \phi^{*}(\tau ,\vec \sigma ),
\nonumber \\
\pi_{\phi^{*}}(\tau ,\vec \sigma )&=&{{\partial L}\over {\partial \partial
_{\tau}\phi^{*}(\tau ,\vec \sigma )}}=
{{\sqrt{\gamma (\tau ,\vec \sigma )}}\over
{N(\tau ,\vec \sigma )}} [\partial_{\tau}-ieA_{\tau}-N^{\check r}(\partial
_{\check r}-ieA_{\check r})](\tau ,\vec \sigma ) \phi (\tau ,\vec \sigma ),
\nonumber \\
\rho_{\mu}(\tau ,\vec \sigma )&=&-{{\partial L}\over {\partial \partial_{\tau}
z^{\mu}(\tau ,\vec \sigma )}}=\nonumber \\
&=&l_{\mu}(\tau ,\vec \sigma ) \{ {{\pi_{\phi}
\pi_{\phi^{*}}}\over {\sqrt{\gamma}}}-\sqrt{\gamma} [\gamma^{\check r\check s}
(\partial_{\check r}+ieA_{\check r})\phi^{*} (\partial_{\check s}-ieA_{\check
s})\phi -\nonumber \\
&&-m^2\phi^{*}\phi] +{1\over {2\sqrt{\gamma}}}\pi^{\check r}g_{\check r\check
s}\pi^{\check s}-{{\sqrt{\gamma}}\over 4}\gamma^{\check r\check s}\gamma
^{\check u\check v}F_{\check r\check u}F_{\check s\check v} \}(\tau ,\vec
\sigma )+\nonumber \\
&&+z_{\check s\mu}(\tau ,\vec \sigma )\gamma^{\check r\check s}(\tau ,\vec
\sigma ) \{ \pi_{\phi^{*}}(\partial_{\check r}+ieA_{\check r})\phi^{*}+
\pi_{\phi} (\partial_{\check r}-ieA_{\check r})\phi-F_{\check r\check u}
\pi^{\check u} \} (\tau ,\vec \sigma ).
\label{VIII20}
\end{eqnarray}

Therefore, we have the following primary constraints

\begin{eqnarray}
\pi^{\tau}(\tau ,\vec \sigma )&\approx& 0,\nonumber \\
{\cal H}_{\mu}(\tau ,\vec \sigma )&=& \rho_{\mu}(\tau ,\vec \sigma )-
\nonumber \\
&&-l_{\mu}(\tau ,\vec \sigma ) \{ {{\pi_{\phi}
\pi_{\phi^{*}}}\over {\sqrt{\gamma}}}-\sqrt{\gamma} [\gamma^{\check r\check s}
(\partial_{\check r}+ieA_{\check r})\phi^{*} (\partial_{\check s}-ieA_{\check
s})\phi -\nonumber \\
&&-m^2\phi^{*}\phi] +{1\over {2\sqrt{\gamma}}}\pi^{\check r}g_{\check r\check
s}\pi^{\check s}-{{\sqrt{\gamma}}\over 4}\gamma^{\check r\check s}\gamma
^{\check u\check v}F_{\check r\check u}F_{\check s\check v} \}(\tau ,\vec
\sigma )+\nonumber \\
&&+z_{\check s\mu}(\tau ,\vec \sigma )\gamma^{\check r\check s}(\tau ,\vec
\sigma ) \{ \pi_{\phi^{*}}(\partial_{\check r}+ieA_{\check r})\phi^{*}+
\pi_{\phi} (\partial_{\check r}-ieA_{\check r})\phi-F_{\check r\check u}
\pi^{\check u} \} (\tau ,\vec \sigma ) \approx 0,\nonumber \\
&&
\label{VIII21}
\end{eqnarray}

\noindent and the following Dirac Hamiltonian [$\lambda (\tau ,\vec \sigma )$
and $\lambda^{\mu}(\tau ,\vec \sigma )$ are Dirac multiplier]

\begin{equation}
H_D=\int d^3\sigma [-A_{\tau}(\tau ,\vec \sigma ) \Gamma (\tau ,\vec \sigma )+
\lambda (\tau ,\vec \sigma )\pi^{\tau}(\tau ,\vec \sigma )+\lambda^{\mu}
(\tau ,\vec \sigma ){\cal H}_{\mu}(\tau ,\vec \sigma )].
\label{VIII22}
\end{equation}

By using the Poisson brackets

\begin{eqnarray}
\{ z^{\mu}(\tau ,\vec \sigma ),\rho_{\nu}(\tau ,{\vec \sigma}^{'}) \} &=&\eta
^{\mu}_{\nu} \delta^3(\vec \sigma -{\vec \sigma}^{'}),\nonumber \\
\{ A_{\check A}(\tau ,\vec \sigma ),\pi^{\check B}(\tau ,{\vec \sigma}^{'}) \}
&=&\eta^{\check B}_{\check A} \delta^3(\vec \sigma -{\vec \sigma}^{'}),
\nonumber \\
\{ \phi (\tau ,\vec \sigma ),\pi_{\phi}(\tau ,{\vec \sigma}^{'}) \} &=&
\{ \phi^{*}(\tau ,\vec \sigma ),\pi_{\phi^{*}}(\tau ,{\vec \sigma}^{'}) \} =
\delta^3(\vec \sigma -{\vec \sigma}^{'}),
\label{VIII23}
\end{eqnarray}

\noindent we find that
the time constancy of the primary constraints implies the existence of only
one secondary constraint

\begin{equation}
\Gamma (\tau ,\vec \sigma ) =\partial_{\check r}\pi^{\check r}(\tau ,\vec
\sigma )+ie (\pi_{\phi^{*}} \phi^{*}-\pi_{\phi} \phi )(\tau ,\vec \sigma )
\approx 0.
\label{VIII24}
\end{equation}

We can verify that these constraints are first class with the algebra given
in Eqs.(125) of Ref.\cite{lus}.

The Poincar\'e generators are like in Eq.(\ref{II6}).

On spacelike hyperplanes $z^{\mu}(\tau ,\vec \sigma )=x_s^{\mu}(\tau )+
b^{\mu}_{\check r}(\tau )\sigma^{\check r}$,
the constraints are reduced to the following ones

\begin{eqnarray}
\pi^{\tau}(\tau ,\vec \sigma )&\approx& 0,\nonumber \\
\Gamma (\tau ,\vec \sigma )&=&-\vec \partial \vec \pi (\tau ,\vec \sigma )+
ie [\pi_{\phi^{*}}\phi^{*}-\pi_{\phi}\phi ](\tau ,\vec \sigma )\approx 0,
\nonumber \\
{\tilde {\cal H}}^{\mu}(\tau )&=& \int d^3\sigma {\cal H}^{\mu}(\tau ,\vec
\sigma )=\nonumber \\
&=&p^{\mu}_s-l^{\mu} \{ {1\over 2} \int d^3\sigma [{\vec \pi}^2+{\vec B}^2]
(\tau ,\vec \sigma )+\nonumber \\
&+&\int d^3\sigma [\pi_{\phi^{*}} \pi_{\phi}+(\vec \partial +ie\vec A)
\phi^{*} \cdot (\vec \partial -ie\vec A)\phi +m^2\phi^{*}\phi ](\tau ,\vec
\sigma ) \}-\nonumber \\
&-&b^{\mu}_{\check r}(\tau ) \{ \int d^3\sigma (\vec \pi \times \vec B)_{\check
r}(\tau ,\vec \sigma )+\int d^3\sigma [\pi_{\phi^{*}}(\partial_{\check r}+
ieA_{\check r})\phi^{*}+\nonumber \\
&+&\pi_{\phi} (\partial_{\check r}-ieA_{\check r}) \phi ](\tau ,\vec \sigma ) \}
\approx 0,\nonumber \\
{\tilde {\cal H}}^{\mu\nu}(\tau )&=&b^{\mu}_{\check r}(\tau ) \int d^3\sigma
\sigma^{\check r} {\cal H}^{\nu}(\tau ,\vec \sigma )-b^{\nu}_{\check r}(\tau )
\int d^3\sigma \sigma^{\check r} {\cal H}^{\mu}(\tau ,\vec \sigma )=\nonumber \\
&=&S^{\mu\nu}_s-(b^{\mu}_{\check r}(\tau )l^{\nu}-b^{\nu}_{\check r}(\tau )
l^{\mu}) [{1\over 2}\int d^3\sigma \sigma^{\check r} ({\vec \pi}^2+{\vec B}^2)
(\tau ,\vec \sigma )+\nonumber \\
&+&\int d^3\sigma \sigma^{\check r}[\pi_{\phi^{*}}\pi_{\phi}+(\vec \partial +
ie\vec A)\phi^{*}\cdot (\vec \partial -ie\vec A)\phi +m^2\phi^{*}\phi ](\tau ,
\vec \sigma ) \}+\nonumber \\
&+&(b^{\mu}_{\check r}(\tau )b^{\nu}_{\check s}(\tau )-b^{\nu}_{\check r}(\tau
)b^{\mu}_{\check s}(\tau )) \{ \int d^3\sigma \sigma^{\check r} (\vec \pi
\times \vec B)_{\check s}(\tau ,\vec \sigma )+\nonumber \\
&+&\int d^3\sigma \sigma^{\check r} [\pi_{\phi^{*}}(\partial_{\check s}+ieA
_{\check s})\phi^{*}+\pi_{\phi} (\partial_{\check s}-ieA_{\check s})\phi ]
(\tau ,\vec \sigma ) \} \approx 0.
\label{VIII25}
\end{eqnarray}

With the final restriction to Wigner hyperplanes, where the canonical variables
are ${\tilde x}
^{\mu}_s(\tau )$, $p^{\mu}_s$, $A_{\tau}(\tau ,\vec \sigma )$, $\pi^{\tau}
(\tau ,\vec \sigma )$, $\vec A(\tau ,\vec \sigma )$, $\vec \pi (\tau ,\vec
\sigma )$, $\phi (\tau ,\vec \sigma )$, $\pi_{\phi}(\tau ,\vec \sigma )=
[{\dot \phi}^{*}-ieA_{\tau}](\tau ,\vec \sigma )$,
$\phi^{*}(\tau ,\vec \sigma )$, $\pi_{\phi^{*}}(\tau ,\vec \sigma )=
[\dot \phi -ieA_{\tau}](\tau ,\vec \sigma )$,
the only surviving constraints are

\begin{eqnarray}
\pi^{\tau}(\tau ,\vec \sigma )&\approx& 0,\nonumber \\
\Gamma (\tau ,\vec \sigma )&=&-\vec \partial \vec \pi (\tau ,\vec \sigma )+
ie [\pi_{\phi^{*}}\phi^{*}-\pi_{\phi}\phi ](\tau ,\vec \sigma )\approx 0,
\nonumber \\
{\cal H}(\tau )&=& \epsilon_s-\{ {1\over 2} \int d^3\sigma ({\vec \pi}^2+{\vec
B}^2)(\tau ,\vec \sigma )+\nonumber \\
&+&\int d^3\sigma [\pi_{\phi^{*}}\phi^{*} \pi_{\phi}+(\vec \partial +ie\vec A)
\phi^{*} \cdot (\vec \partial -ie\vec A)\phi +m^2\phi^{*}\phi ](\tau ,\vec
\sigma ) \} \approx 0,\nonumber \\
{\vec {\cal H}}_p(\tau )&=& \int d^3\sigma (\vec \pi \times \vec B)(\tau ,\vec
\sigma )+\nonumber \\
&+&\int d^3\sigma [\pi_{\phi^{*}} (\vec \partial +ie\vec A)\phi^{*} +\pi_{\phi}
(\vec \partial -ie\vec A)\phi ](\tau ,\vec \sigma ) \approx 0.
\label{VIII26}
\end{eqnarray}

In the Lorentz generators of Eq.(\ref{II9}) the spin tensor has the  form

\begin{eqnarray}
{\bar S}_s^{rs}&\equiv& S^{rs}_{\phi}
=\int d^3\sigma \{ \sigma^r (\vec \pi \times \vec B)^s(\tau
,\vec \sigma )-\sigma^s (\vec \pi \times \vec B)^r(\tau ,\vec \sigma ) \}+
\nonumber \\
&+&\int d^3\sigma \{ \sigma^r[\pi_{\phi^{*}}(\partial^s+ieA^s)\phi^{*}+\pi
_{\phi}(\partial^s-ieA^s)\phi](\tau ,\vec \sigma )-(r \leftrightarrow s) \}.
\label{VIII27}
\end{eqnarray}

To make the reduction to Dirac's observables with respect to the electromagnetic
gauge transformations, let us recall \cite{lusa,val} that the electromagnetic
gauge degrees of freedom are described by the two pairs of conjugate variables
$A_{\tau}(\tau ,\vec \sigma )$, $\pi_{\tau}(\tau ,\vec \sigma )[\approx 0]$,
$\eta_{em}(\tau ,\vec \sigma )=-{1\over {\triangle}} {{\partial}\over {\partial
\vec \sigma}}\cdot \vec A(\tau ,\vec \sigma )$, $\Gamma (\tau ,\vec \sigma )
[\approx 0]$, so that we have the decompositions

\begin{eqnarray}
A^r(\tau ,\vec \sigma )&=&{{\partial}\over {\partial \sigma^r}} \eta_{em}(\tau
,\vec \sigma )+A^r_{\perp}(\tau ,\vec \sigma ),\nonumber \\
\pi^r(\tau ,\vec \sigma )&=&\pi^r_{\perp}(\tau ,\vec \sigma )+\nonumber \\
&+&{1\over {\triangle}} {{\partial}\over {\partial \sigma^r}} [-\Gamma (\tau
,\vec \sigma )+ie(\pi_{\phi^{*}}\phi^{*}-\pi_{\phi}\phi )(\tau ,\vec \sigma )]
\approx 0,\nonumber \\
&&{}\nonumber \\
\lbrace A^r_{\perp}(\tau ,\vec \sigma )&,&\pi^s_{\perp}(\tau ,{\vec \sigma}^{'})
\rbrace
=-P^{rs}_{\perp}(\vec \sigma ) \delta^3(\vec \sigma -{\vec \sigma}^{'}),
\label{VIII28}
\end{eqnarray}

\noindent where $P^{rs}_{\perp}(\vec \sigma )=\delta^{rs}+{{\partial^r
\partial^s}\over {\triangle}}$, $\triangle =-{\vec \partial}^2$. Then, we have

\begin{eqnarray}
\int d^3\sigma &&{\vec \pi}^2(\tau ,\vec \sigma )= \int d^3\sigma {\vec \pi}^2
_{\perp}(\tau ,\vec \sigma )-\nonumber \\
&-&{{e^2}\over {4\pi}} \int d^3\sigma_1d^3\sigma_2 {{i(\pi_{\phi^{*}}\phi^{*}-
\pi_{\phi}\phi )(\tau ,{\vec \sigma}_1)\, i(\pi_{\phi^{*}}\phi^{*}-
\pi_{\phi}\phi )(\tau ,{\vec \sigma}_2)}\over {|{\vec \sigma}_1-{\vec \sigma}
_2|}}.
\label{VIII29}
\end{eqnarray}

Since we have

\begin{eqnarray}
\lbrace \phi (\tau ,\vec \sigma ),\Gamma (\tau ,{\vec \sigma}^{'})\rbrace &=&
-ie \phi(\tau ,\vec \sigma ) \delta^3(\vec \sigma -{\vec \sigma}^{'}),
\nonumber \\
\lbrace \phi^{*}(\tau ,\vec \sigma ),\Gamma (\tau ,{\vec \sigma}^{'})\rbrace &=&
ie \phi^{*}(\tau ,\vec \sigma ) \delta^3(\vec \sigma -{\vec \sigma}^{'}),
\nonumber \\
\lbrace \pi_{\phi}(\tau ,\vec \sigma ),\Gamma (\tau ,{\vec \sigma}^{'})\rbrace
&=&ie \pi_{\phi}(\tau ,\vec \sigma ) \delta^3(\vec \sigma -{\vec \sigma}^{'}),
\nonumber \\
\lbrace \pi_{\phi^{*}}(\tau ,\vec \sigma ),\Gamma (\tau ,{\vec \sigma}^{'})
\rbrace&=&-ie \pi_{\phi^{*}}(\tau ,\vec \sigma ) \delta^3(\vec \sigma -{\vec
\sigma}^{'}),
\label{VIII30}
\end{eqnarray}

\noindent the Dirac observables for the Klein-Gordon field are

\begin{eqnarray}
\hat \phi (\tau ,\vec \sigma )&=&
e^{-ie\eta_{em}(\tau ,\vec \sigma )} \phi (\tau ,\vec \sigma ),\nonumber \\
{\hat \phi}^{*}(\tau ,\vec \sigma )&=&
e^{ie\eta_{em}(\tau ,\vec \sigma )} \phi^{*}(\tau ,\vec \sigma ),\nonumber \\
{\hat \pi}_{\phi}(\tau ,\vec \sigma )&=&e^{ie\eta_{em}(\tau ,\vec \sigma )}
\pi_{\phi}(\tau ,\vec \sigma ),\nonumber \\
{\hat \pi}_{\phi^{*}}(\tau ,\vec \sigma )&=&e^{-ie\eta_{em}(\tau ,\vec
\sigma )} \pi_{\phi^{*}}(\tau ,\vec \sigma ),\nonumber \\
&&{}\nonumber \\
\lbrace \hat \phi (\tau ,\vec \sigma )&,&\Gamma (\tau ,{\vec \sigma}^{'})\rbrace
=\lbrace {\hat \pi}_{\phi}(\tau ,\vec \sigma ),\Gamma (\tau ,{\vec \sigma}^{'})
\rbrace =0,\nonumber \\
\lbrace {\hat \phi}^{*}(\tau ,\vec \sigma )&,&\Gamma (\tau ,{\vec \sigma}^{'})
\rbrace =\lbrace {\hat \pi}_{\phi^{*}}(\tau ,\vec \sigma ),\Gamma (\tau
,{\vec \sigma}^{'})\rbrace =0,\nonumber \\
&&{}\nonumber \\
\lbrace \hat \phi (\tau ,\vec \sigma ),{\hat \pi}_{\phi}(\tau ,{\vec \sigma}
^{'}) \rbrace &=& \lbrace {\hat \phi}^{*}(\tau ,\vec \sigma ),{\hat \pi}
_{\phi^{*}}(\tau ,{\vec \sigma}^{'}) \rbrace = \delta^3(\vec \sigma ,{\vec
\sigma}^{'}).
\label{VIII31}
\end{eqnarray}

Therefore, in the generalized Wigner-covariant Coulomb gauge $A_{\tau}(\tau
,\vec \sigma )=\eta_{em}(\tau ,\vec \sigma )=0$, we have ${\hat \phi}^{*}
(\tau ,\vec \sigma )=[\hat \phi (\tau ,\vec \sigma )]^{*}$, ${\hat \pi}_{\phi
^{*}}(\tau ,\vec \sigma )=[{\hat \pi}_{\phi}(\tau ,\vec \sigma )]^{*}$ like
in the free case.

The constraints take the following form

\begin{eqnarray}
{\cal H}(\tau )&=&\epsilon_s- \{ \, {1\over 2}\int d^3\sigma ({\vec \pi}^2
_{\perp}+{\vec B}^2)(\tau ,\vec \sigma )+\nonumber \\
&+&\int d^3\sigma [{\hat \pi}_{\phi^{*}}{\hat \pi}_{\phi}+(\vec \partial +ie
{\vec A}_{\perp}){\hat \phi}^{*}\cdot (\vec \partial -ie{\vec A}_{\perp})\hat
\phi +m^2{\hat \phi}^{*}\hat \phi ](\tau ,\vec \sigma )-\nonumber \\
&-&{{e^2}\over {8\pi}}\int d^3\sigma_1d^3\sigma_2 {{i({\hat \pi}_{\phi^{*}}{\hat
\phi}^{*}-{\hat \pi}_{\phi}\hat \phi )(\tau ,{\vec \sigma}_1)\, i({\hat \pi}
_{\phi^{*}}{\hat \phi}^{*}-{\hat \pi}_{\phi}\hat \phi )(\tau ,{\vec \sigma_2)}}
\over {|{\vec \sigma}_1-{\vec \sigma}_2|}} \} ,\nonumber \\
&&{}\nonumber \\
{\vec {\cal H}}_p(\tau )&=&\int d^3\sigma ({\vec \pi}_{\perp}\times \vec B)
(\tau ,\vec \sigma )+\int d^3\sigma ({\hat \pi}_{\phi^{*}}\vec \partial {\hat
\phi}^{*}+{\hat \pi}_{\phi}\vec \partial \hat \phi )(\tau ,\vec \sigma )
\approx 0,
\label{VIII32}
\end{eqnarray}

\noindent where the Coulomb self-interaction appears in the invariant mass and
where the 3 constraints defining the rest frame do not depend on the interaction
since we are in an instant form of the dynamics. The final form of the
rest-frame spin tensor is

\begin{eqnarray}
{\bar S}^{rs}_s&=&
\int d^3\sigma \{ \sigma^r [({\vec \pi}_{\perp}\times \vec B)^s
+{\hat \pi}_{\phi^{*}}\partial^s{\hat \phi}^{*}+{\hat \pi}_{\phi}\partial^s
\hat \phi ] - (r\leftrightarrow s) \} (\tau ,\vec \sigma )=\nonumber \\
&=& S^{rs}_{\phi}(\tau )+\int d^3\sigma \{ \sigma^r [({\vec \pi}_{\perp}\times
 \vec B)^s- (r\leftrightarrow s) \} (\tau ,\vec \sigma ).
\label{VIII33}
\end{eqnarray}

If we go to the gauge $\chi =T_s-\tau \approx 0$, we can eliminate the variables
$\epsilon_s$, $T_s$, and the $\tau$-evolution (in the Lorentz scalar rest-frame
time) is governed by the Hamiltonian

\begin{eqnarray}
H_R&=&M-\vec \lambda (\tau )\cdot {\vec {\cal H}}_p(\tau ),\nonumber \\
&&{}\nonumber \\
M&=&{1\over 2}\int d^3\sigma ({\vec \pi}^2
_{\perp}+{\vec B}^2)(\tau ,\vec \sigma )+\nonumber \\
&+&\int d^3\sigma [{\hat \pi}_{\phi^{*}}{\hat \pi}_{\phi}+(\vec \partial +ie
{\vec A}_{\perp}){\hat \phi}^{*}\cdot (\vec \partial -ie{\vec A}_{\perp})\hat
\phi +m^2{\hat \phi}^{*}\hat \phi ](\tau ,\vec \sigma )-\nonumber \\
&-&{{e^2}\over {8\pi}}\int d^3\sigma_1d^3\sigma_2 {{i({\hat \pi}_{\phi^{*}}{\hat
\phi}^{*}-{\hat \pi}_{\phi}\hat \phi )(\tau ,{\vec \sigma}_1)\, i({\hat \pi}
_{\phi^{*}}{\hat \phi}^{*}-{\hat \pi}_{\phi}\hat \phi )(\tau ,{\vec \sigma)_2)}}
\over {|{\vec \sigma}_1-{\vec \sigma}_2|}}.
\label{VIII34}
\end{eqnarray}

In the gauge $\vec \lambda (\tau )=0$, the Hamilton equations are

\begin{eqnarray}
\partial_{\tau} \hat \phi (\tau ,\vec \sigma )\, &{\buildrel \circ \over =}\,&
{\hat \pi}_{\phi^{*}}(\tau ,\vec \sigma )+\nonumber \\
&+&{{ie^2}\over {4\pi}}\hat \phi (\tau ,\vec \sigma ) \int d^3\bar \sigma
{{i({\hat \pi}_{\phi^{*}}{\hat \phi}^{*}-{\hat \pi}_{\phi}\hat \phi )(\tau ,
{\vec {\bar \sigma}})}\over {|\vec \sigma -{\vec {\bar \sigma}}|}},
\nonumber \\
\partial_{\tau}{\hat \pi}_{\phi^{*}}(\tau ,\vec \sigma )\, &{\buildrel \circ
\over =}\,& [(\vec \partial -ie{\vec A}_{\perp}(\tau ,\vec \sigma ))^2-m^2]
\hat \phi (\tau ,\vec \sigma )+\nonumber \\
&+&{{ie^2}\over {4\pi}}{\hat \pi}_{\phi^{*}}(\tau ,\vec \sigma )\int d^3\bar
\sigma {{i({\hat \pi}_{\phi^{*}}{\hat \phi}^{*}-{\hat \pi}_{\phi}\hat \phi )
(\tau ,{\vec {\bar \sigma}})}\over {|\vec \sigma -{\vec {\bar \sigma}}|}},
\nonumber \\
&&{}\nonumber \\
\partial_{\tau}A^r_{\perp}(\tau ,\vec \sigma )\, &{\buildrel \circ \over =}\,&
-\pi^r_{\perp}(\tau ,\vec \sigma ),\nonumber \\
\partial_{\tau}\pi^r_{\perp}(\tau ,\vec \sigma )\, &{\buildrel \circ \over =}\,&
\triangle A^r_{\perp}(\tau ,\vec \sigma )+\nonumber \\
&+&ie P^{rs}_{\perp}(\vec \sigma ) [{\hat \phi}^{*}(\partial^s-ieA^s_{\perp})
\hat \phi -\hat \phi (\partial^s+ieA^s_{\perp}){\hat \phi}^{*}](\tau ,\vec
\sigma ).
\label{VIII35}
\end{eqnarray}

\noindent The equations for ${\hat \phi}^{*}$ and $\pi_{\phi}$ are the complex
conjugate of those for $\hat \phi$ and for ${\hat \pi}_{\phi^{*}}$.

By using the results of Ref.\cite{val}, we have the following inversion formula

\begin{eqnarray}
{\hat \pi}_{\phi^{*}}\, &{\buildrel \circ \over =}\,& \partial_{\tau} \hat
\phi +ie^2 \hat \phi {1\over {\triangle}} i({\hat \pi}_{\phi^{*}}{\hat \phi}^{*}
-{\hat \pi}_{\phi}\hat \phi )=\nonumber \\
&=&\partial_{\tau} \hat \phi +ie^2 \hat \phi {1\over {\triangle +2e^2{\hat
\phi}^{*}\hat \phi}} i({\hat \phi}^{*} \partial_{\tau}\hat \phi -\hat \phi
\partial_{\tau} {\hat \phi}^{*}),
\label{VIII36}
\end{eqnarray}

\noindent since we have $i({\hat \phi}^{*}\partial_{\tau}\hat \phi -\hat \phi
\partial_{\tau}{\hat \phi}^{*})=(1+2e^2{\hat \phi}^{*}\hat \phi {1\over
{\triangle}}) i({\hat \pi}_{\phi^{*}}{\hat \phi}^{*}-{\hat \pi}_{\phi}\hat
\phi )$ and where use has been done of the operator identity ${1\over A}{1\over
{1+B{1\over A}}}={1\over A}[1-B{1\over A}+B{1\over A}B{1\over A}-...]={1\over
{A+B}}$ (valid for B a small perturbation of A) for $A=\triangle$ and $B=2e^2
{\hat \phi}^{*}\hat \phi$.

Using this formula, we get the following second order equations of motion

\begin{eqnarray}
&&\{ [\, \partial_{\tau}+ie^2{1\over {\triangle +2e^2{\hat \phi}^{*}\hat \phi}}
\, i({\hat \phi}^{*} \partial_{\tau}\hat \phi -\hat \phi \partial_{\tau}{\hat
\phi}^{*})]^2-(\vec \partial -ie{\vec A}_{\perp})^2+m^2\, \} \hat \phi
{\buildrel \circ \over =}\, 0,\nonumber \\
&&{}\nonumber \\
&&[\partial_{\tau}^2+\triangle ] A^r_{\perp}\, {\buildrel \circ \over =}\,
ie P^{rs}_{\perp}(\vec \sigma ) [{\hat \phi}^{*}(\partial^s -ieA^s_{\perp})\hat
\phi -\hat \phi (\partial^s+ieA^s_{\perp}){\hat \phi}^{*}].
\label{VIII37}
\end{eqnarray}

We see that the non-local velocity-dependent self-energy is formally playing the
role of a scalar potential.

Due to Eqs.(\ref{VIII31}), in the gauge $A_{\tau}(\tau ,\vec \sigma
)=\eta_{em} (\tau ,\vec \sigma )=0$ the Dirac observables $\hat \phi$,
${\hat \pi}_{\phi}$, ${\hat \phi}^{*}$, ${\hat \pi}_{\phi^{*}}$, admit
a Fourier decomposition like in the free case, see Eqs.(\ref{VIII10})
and (\ref{d5}), with Fourier coefficients $\hat a$, $\hat b$ [the
fields ${\hat \phi}_a$, ${\hat \phi}_b$ and the Fourier coefficients
$\hat a$, $\hat b$, tend to the free ones $\phi_a$, $\phi_b$, $a$,
$b$, when we turn off the electromagnetic interaction, switching off
the electric charge]

\begin{eqnarray}
\hat a(\tau ,\vec q)&=&\sqrt{2m\omega (q)} \int d^3\sigma {\hat \phi}_a(\tau
,\vec \sigma )e^{i(\omega (q)\tau -\vec q\cdot \vec \sigma )}=\nonumber \\
&=&\int d^3\sigma [\omega (q) \hat \phi (\tau ,\vec \sigma )+i{\hat \pi}
_{\phi^{*}}(\tau
,\vec \sigma )] e^{i(\omega (q)\tau -\vec q\cdot \vec \sigma )}=\nonumber \\
&=& \sqrt{{\hat I}_a(\tau ,\vec q)}e^{i{\hat \varphi}_a(\tau ,\vec q)},
\nonumber \\
{\hat a}^{*}(\tau ,\vec q)&=&\sqrt{2m\omega (q)} \int d^3\sigma {\hat \phi}
^{*}_a(\tau ,\vec
\sigma )e^{-i(\omega (q)\tau -\vec q\cdot \vec \sigma )}=\nonumber \\
&=&\int d^3\sigma [\omega (q) {\hat \phi}^{*}(\tau ,\vec \sigma )+i{\hat \pi}
_{\phi}(\tau
,\vec \sigma )] e^{-i(\omega (q)\tau -\vec q\cdot \vec \sigma )}=\nonumber \\
&=& \sqrt{{\hat I}_a(\tau ,\vec q)}e^{-i{\hat \varphi}_a(\tau ,\vec q)},
\nonumber \\
\hat b(\tau ,\vec q)&=&\sqrt{2m\omega (q)} \int d^3\sigma {\hat \phi}_b(\tau
,\vec \sigma )e^{i(\omega (q)\tau -\vec q\cdot \vec \sigma )}=\nonumber \\
&=&\int d^3\sigma [\omega (q) \hat \phi(\tau ,\vec \sigma )-i{\hat \pi}
_{\phi^{*}}(\tau
,\vec \sigma )] e^{i(\omega (q)\tau -\vec q\cdot \vec \sigma )}=\nonumber \\
&=&\sqrt{{\hat I}_b(\tau ,\vec q)}e^{i{\hat \varphi}_b(\tau ,\vec q)},
\nonumber \\
{\hat b}^{*}(\tau ,\vec q)&=&\sqrt{2m\omega (q)} \int d^3\sigma {\hat \phi}
^{*}_b(\tau ,\vec
\sigma )e^{-i(\omega (q)\tau -\vec q\cdot \vec \sigma )}=\nonumber \\
&=&\int d^3\sigma [\omega (q) {\hat \phi}^{*}(\tau ,\vec \sigma )+i{\hat \pi}
_{\phi}(\tau
,\vec \sigma )] e^{-i(\omega (q)\tau -\vec q\cdot \vec \sigma )}=\nonumber \\
&=&\sqrt{{\hat I}_b(\tau ,\vec q)}e^{-i{\hat \varphi}_b(\tau ,\vec q)},
\nonumber \\
&&{}\nonumber \\
{\hat \phi}_a(\tau ,\vec \sigma )&=&{1\over {\sqrt{2m\sqrt{m^2+\triangle}}}}
\Big[ \sqrt{m^2+\triangle} \hat \phi (\tau ,\vec \sigma )+i{\hat
\pi}_{\phi^{*}}(\tau ,\vec \sigma )\Big]=\nonumber \\
&=&\sqrt{ {2\over m} } \int d\tilde q \sqrt{\omega (q)} \hat a (\tau ,\vec q)
e^{-i(\omega (q)\tau -\vec q\cdot \vec \sigma)},\nonumber \\
{\hat \phi}_b(\tau ,\vec \sigma )&=&{1\over {\sqrt{2m\sqrt{m^2+\triangle}}}}
\Big[ \sqrt{m^2+\triangle} \hat \phi (\tau ,\vec \sigma )+i{\hat
\pi}_{\phi^{*}}(\tau ,\vec \sigma )\Big]=\nonumber \\
&=&\sqrt{ {2\over m} } \int d\tilde q \sqrt{\omega (q)} \hat b (\tau ,\vec q)
e^{-i(\omega (q)\tau -\vec q\cdot \vec \sigma)},\nonumber \\
&&{}\nonumber \\
\hat \phi (\tau ,\vec \sigma )&=&[{\hat \phi}^{*}(\tau ,\vec \sigma )]^{*}=
\sqrt{ {m\over 2} }{1\over {(m^2+\triangle )^{1/4}}} [{\hat \phi}_a+{\hat \phi}
^{*}_b](\tau ,\vec \sigma ),\nonumber \\
{\hat \pi}_{\phi^{*}}(\tau ,\vec \sigma )&=&[{\hat \pi}_{\phi}(\tau ,\vec
\sigma )]^{*}=-i\sqrt{ {m\over 2} }(m^2+\triangle )^{1/4} [{\hat \phi}_a-{\hat
\phi}^{*}_b](\tau ,\vec \sigma ).
\label{VIII38}
\end{eqnarray}

Let us now assume that the fields satisfy Eq.(\ref{IV29}). Then,
 we can define the canonical
transformations $\hat a, {\hat a}^{*} \mapsto$ ${\hat P}^A_{a\phi}$, ${\hat X}
^A_{a\phi} ({\hat X}^{\tau}_{a\phi}=\tau +{\hat {\tilde X}}^{\tau}_{a\phi})$,
${\hat {\bf H}}_a$, ${\hat {\bf K}}_a$ and $\hat b, {\hat b}^{*} \mapsto$
${\hat P}^A_{b\phi}$, ${\hat X}
^A_{b\phi} ({\hat X}^{\tau}_{b\phi}=\tau +{\hat {\tilde X}}^{\tau}_{b\phi})$,
${\hat {\bf H}}_b$, ${\hat {\bf K}}_b$, and then ${\hat X}^A_{a\phi}$,
${\hat P}^A_{a\phi}$, ${\hat X}^A_{b\phi}$, ${\hat P}^A_{b\phi} \mapsto$
${\hat X}^A_{\phi}$, ${\hat P}^A_{\phi}$, ${\hat R}^A_{\phi}$, ${\hat Q}^A
_{\phi}$. We get

\begin{eqnarray}
{\hat P}^{\tau}_{\phi}&=&m \int d^3\sigma [{\hat \phi}^{*}_a \sqrt{m^2+
\triangle} {\hat \phi}_a+{\hat \phi}^{*}_b \sqrt{m^2+\triangle} {\hat \phi}_b]
(\tau ,\vec \sigma ),\nonumber \\
{\hat {\vec P}}_{\phi}&=&
i m \int d^3\sigma [{\hat \phi}^{*}_a \vec \partial {\hat
\phi}_a+{\hat \phi}^{*}_b \vec \partial {\hat \phi}_b](\tau ,\vec \sigma ),
\nonumber \\
{\hat Q}^{\tau}_{\phi}&=&{m\over 2} \int d^3\sigma [{\hat \phi}^{*}_a \sqrt{m^2+
\triangle} {\hat \phi}_a-{\hat \phi}^{*}_b \sqrt{m^2+\triangle} {\hat \phi}_b]
(\tau ,\vec \sigma ),\nonumber \\
{\hat {\vec Q}}_{\phi}&=&
i{m\over 2} \int d^3\sigma [{\hat \phi}^{*}_a \vec \partial
{\hat \phi}_a-{\hat \phi}^{*}_b \vec \partial {\hat \phi}_b](\tau ,\vec \sigma
),\nonumber \\
{\hat X}^{\tau}_{\phi}&=&\tau +{\hat {\tilde X}}^{\tau}_{\phi}=\tau +
\nonumber \\
&+&{1\over {4\pi i m}} \int d^3q {{e^{-{{4\pi}\over {m^2}} q^2} }\over
{q^2\, \omega (q)}} \nonumber \\
&&\Big( ln\,  {{\int d^3\sigma e^{i\vec q\cdot \vec \sigma} (m^2+\triangle
)^{-1/4}[(\omega (q)+\sqrt{m^2+\triangle}){\hat \phi}_a+(\omega (q)-\sqrt{m^2+
\triangle}){\hat \phi}^{*}_b](\tau ,\vec \sigma )}\over {\int d^3\sigma
e^{-i\vec q\cdot \vec \sigma} (m^2+\triangle)^{-1/4}[(\omega (q)+\sqrt{m^2+
\triangle}){\hat \phi}^{*}_a+(\omega (q)-\sqrt{m^2+\triangle}){\hat \phi}_b]
(\tau ,\vec \sigma )}} +\nonumber \\
&&+ln\, {{\int d^3\sigma e^{i\vec q\cdot \vec \sigma} (m^2+\triangle
)^{-1/4}[(\omega (q)+\sqrt{m^2-\triangle}){\hat \phi}^{*}_a+(\omega (q)+
\sqrt{m^2+\triangle}){\hat \phi}_b](\tau ,\vec \sigma )}\over {\int d^3\sigma
e^{-i\vec q\cdot \vec \sigma} (m^2+\triangle)^{-1/4}[(\omega (q)-\sqrt{m^2+
\triangle}){\hat \phi}_a+(\omega (q)+\sqrt{m^2+\triangle}){\hat \phi}^{*}_b]
(\tau ,\vec \sigma )}} \Big) ,\nonumber \\
{\hat {\vec X}}_{\phi}&=&{{2i}\over {\pi m}} \int d^3q {{\vec q}\over {q^4}}
e^{-{{4\pi}\over {m^2}} q^2}\nonumber \\
&&\Big( ln\, {{\int d^3\sigma e^{i\vec q\cdot \vec \sigma} (m^2+\triangle
)^{-1/4}[(\omega (q)+\sqrt{m^2+\triangle}){\hat \phi}_a+(\omega (q)-\sqrt{m^2+
\triangle}){\hat \phi}^{*}_b](\tau ,\vec \sigma )}\over {\int d^3\sigma
e^{-i\vec q\cdot \vec \sigma} (m^2+\triangle)^{-1/4}[(\omega (q)+\sqrt{m^2+
\triangle}){\hat \phi}^{*}_a+(\omega (q)-\sqrt{m^2+\triangle}){\hat \phi}_b]
(\tau ,\vec \sigma )}} -\nonumber \\
&&-ln\, {{\int d^3\sigma e^{i\vec q\cdot \vec \sigma} (m^2+\triangle
)^{-1/4}[(\omega (q)+\sqrt{m^2-\triangle}){\hat \phi}^{*}_a+(\omega (q)+
\sqrt{m^2+\triangle}){\hat \phi}_b](\tau ,\vec \sigma )}\over {\int d^3\sigma
e^{-i\vec q\cdot \vec \sigma} (m^2+\triangle)^{-1/4}[(\omega (q)-\sqrt{m^2+
\triangle}){\hat \phi}_a+(\omega (q)+\sqrt{m^2+\triangle}){\hat \phi}^{*}_b]
(\tau ,\vec \sigma )}} \Big) ,\nonumber \\
{\hat R}^{\tau}_{\phi}&=&
{1\over {2\pi i m}} \int d^3q {{e^{-{{4\pi}\over {m^2}} q^2} }\over
{q^2\, \omega (q)}} \nonumber \\
&&\Big( ln\, {{\int d^3\sigma e^{i\vec q\cdot \vec \sigma} (m^2+\triangle
)^{-1/4}[(\omega (q)+\sqrt{m^2+\triangle}){\hat \phi}_a+(\omega (q)-\sqrt{m^2+
\triangle}){\hat \phi}^{*}_b](\tau ,\vec \sigma )}\over {\int d^3\sigma
e^{-i\vec q\cdot \vec \sigma} (m^2+\triangle)^{-1/4}[(\omega (q)+\sqrt{m^2+
\triangle}){\hat \phi}^{*}_a+(\omega (q)-\sqrt{m^2+\triangle}){\hat \phi}_b]
(\tau ,\vec \sigma )}} -\nonumber \\
&&-ln\,  {{\int d^3\sigma e^{i\vec q\cdot \vec \sigma} (m^2+\triangle
)^{-1/4}[(\omega (q)+\sqrt{m^2-\triangle}){\hat \phi}^{*}_a+(\omega (q)+
\sqrt{m^2+\triangle}){\hat \phi}_b](\tau ,\vec \sigma )}\over {\int d^3\sigma
e^{-i\vec q\cdot \vec \sigma} (m^2+\triangle)^{-1/4}[(\omega (q)-\sqrt{m^2+
\triangle}){\hat \phi}_a+(\omega (q)+\sqrt{m^2+\triangle}){\hat \phi}^{*}_b]
(\tau ,\vec \sigma )}} \Big) ,\nonumber \\
{\hat {\vec R}}_{\phi}&=&{i\over {\pi m}} \int d^3q {{\vec q}\over {q^4}}
e^{-{{4\pi}\over {m^2}} q^2}\nonumber \\
&&\Big( ln\, {{\int d^3\sigma e^{i\vec q\cdot \vec \sigma} (m^2+\triangle
)^{-1/4}[(\omega (q)+\sqrt{m^2+\triangle}){\hat \phi}_a+(\omega (q)-\sqrt{m^2+
\triangle}){\hat \phi}^{*}_b](\tau ,\vec \sigma )}\over {\int d^3\sigma
e^{-i\vec q\cdot \vec \sigma} (m^2+\triangle)^{-1/4}[(\omega (q)+\sqrt{m^2+
\triangle}){\hat \phi}^{*}_a+(\omega (q)-\sqrt{m^2+\triangle}){\hat \phi}_b]
(\tau ,\vec \sigma )}} +\nonumber \\
&&+ln\, {{\int d^3\sigma e^{i\vec q\cdot \vec \sigma} (m^2+\triangle
)^{-1/4}[(\omega (q)+\sqrt{m^2-\triangle}){\hat \phi}^{*}_a+(\omega (q)+
\sqrt{m^2+\triangle}){\hat \phi}_b](\tau ,\vec \sigma )}\over {\int d^3\sigma
e^{-i\vec q\cdot \vec \sigma} (m^2+\triangle)^{-1/4}[(\omega (q)-\sqrt{m^2+
\triangle}){\hat \phi}_a+(\omega (q)+\sqrt{m^2+\triangle}){\hat \phi}^{*}_b]
(\tau ,\vec \sigma )}} \Big) ,\nonumber \\
&&{}\nonumber \\
{\hat \phi}_a(\tau ,\vec \sigma )&=&\int d\tilde k \sqrt{ {{2\omega (k)}\over
m}} {\bf A}_{a\, \vec k}(\tau ;{1\over 2}{\hat P}^A_{\phi}+{\hat Q}^A_{\phi},
{\hat {\bf B}}_a]\, e^{i(\vec k\cdot \vec \sigma +{\bf B}_{a\, \vec k}(\tau ;
{\hat X}^A_{\phi}+{1\over 2}{\hat R}^A_{\phi},{\hat {\bf K}}_a])},\nonumber \\
{\hat \phi}_b(\tau ,\vec \sigma )&=&\int d\tilde k \sqrt{ {{2\omega (k)}\over
m}} {\bf A}_{b\, \vec k}(\tau ;{1\over 2}{\hat P}^A_{\phi}+{\hat Q}^A_{\phi},
{\hat {\bf B}}_b]\, e^{i(\vec k\cdot \vec \sigma +{\bf B}_{b\, \vec k}(\tau ;
{\hat X}^A_{\phi}+{1\over 2}{\hat R}^A_{\phi},{\hat {\bf K}}_b])}.
\label{VIII39}
\end{eqnarray}

In terms of these variables Eqs.(\ref{VIII34}) and (\ref{VIII33})
become

\begin{eqnarray}
H_R&=&M-\vec \lambda (\tau )\cdot {\vec {\cal H}}
_p(\tau ),\nonumber \\
&&{}\nonumber \\
M&=& {1\over 2} \int d^3\sigma ({\vec \pi}^2_{\perp}+{\vec B}^2)(\tau ,\vec
\sigma )+{\hat P}^{\tau}_{\phi}(\tau )+\nonumber \\
&+&m \int d^3\sigma \Big[ ie {\vec A}_{\perp}(\tau ,\vec \sigma )\cdot {1\over
{\sqrt{m^2+\triangle}}}[{\hat \phi}^{*}_a+{\hat \phi}_b](\tau ,\vec \sigma )
{{\vec \partial}\over {\sqrt{m^2+\triangle}}}[{\hat \phi}_a+{\hat \phi}
_b^{*}](\tau ,\vec \sigma )+\nonumber \\
&+&{{e^2}\over 2}{\vec A}_{\perp}^2(\tau ,\vec \sigma ) {1\over {\sqrt{m^2+
\triangle}}} [{\hat \phi}^{*}_a+{\hat \phi}_b](\tau ,\vec \sigma ) {1\over
{\sqrt{m^2+\triangle}}}[{\hat \phi}_a+{\hat \phi}_b^{*}](\tau ,\vec \sigma )
\Big]-\nonumber \\
&-&{{e^2m^2}\over {32\pi}} \int {{d^3\sigma_1d^3\sigma_2}\over {|{\vec \sigma}
_1-{\vec \sigma}_2|}} \Big[ \sqrt{m^2+\triangle_1}[{\hat \phi}_a-{\hat \phi}_b
^{*}] {1\over {\sqrt{m^2+\triangle_1}}}[{\hat \phi}^{*}_a+{\hat \phi}_b]-
\nonumber \\
&-&\sqrt{m^2+\triangle_1}[{\hat \phi}^{*}_a-{\hat \phi}_b] {1\over
{\sqrt{m^2+\triangle_1}}}[{\hat \phi}_a+{\hat \phi}_b^{*}] \Big]
(\tau ,{\vec \sigma}_1) \nonumber \\
&&\Big[ \sqrt{m^2+\triangle_2}[{\hat \phi}_a-{\hat \phi}_b
^{*}] {1\over {\sqrt{m^2+\triangle_2}}}[{\hat \phi}^{*}_a+{\hat \phi}_b]-
\nonumber \\
&-&\sqrt{m^2+\triangle_2}[{\hat \phi}^{*}_a-{\hat \phi}_b] {1\over
{\sqrt{m^2+\triangle_2}}}[{\hat \phi}_a+{\hat \phi}_b^{*}] \Big]
(\tau ,{\vec \sigma}_2),\nonumber \\
{\vec {\cal H}}_p(\tau )&=&{\hat {\vec P}}_{\phi}(\tau )+\int d^3\sigma
({\vec \pi}_{\perp}\times \vec B)(\tau ,\vec \sigma )\approx 0,\nonumber \\
{\bar S}^{rs}_s&\equiv& {\hat S}^{rs}_{\phi}(\tau )+\int d^3\sigma \Big[
\sigma^r ({\vec \pi}_{\perp}\times \vec B)^s-\sigma^s({\vec \pi}_{\perp}\times
\vec B)^r\Big] .
\label{VIII40}
\end{eqnarray}

In Appendix  Ethere is the expression of the interaction terms in M in
terms of $\hat a$, $\hat b$.

The fields ${\hat \phi}_a$, ${\hat \phi}_b$, are solutions of the coupled
Hamilton equations generated by $H_R$ and these equations cannot be decoupled
when the associated Feshbach-Villars Hamiltonian cannot be diagonalized: in
these cases ${\hat \phi}_a$ [${\hat \phi}_b$] does not correspond to a positive
[negative] energy solution.

Only when we have the collective and relative canonical variables of the
transverse electromagnetic fields ${\vec A}_{\perp}$, ${\vec \pi}_{\perp}$,
we will be able to add the gauge fixings for the constraints ${\vec {\cal H}}
_p\approx 0$, defining the rest frame, and to get the final form of the
physical Hamiltonian for classical scalar electrodynamics in the rest-frame
Wigner-covariant instant form of the dynamics.

\subsection{The energy-momentum tensor.}

The conserved energy-momentum tensor of the isolated system formed by the
Klein-Gordon field plus the electromagnetic field is
[$D^{(A)}_{\mu}\tilde \phi (x)=[\partial_{\mu}-ieA_{\mu}(x)]\tilde \phi (x)$;
$\tilde \phi (z(\tau ,\vec \sigma ))=\phi (\tau ,\vec \sigma )$; ${\tilde A}
_{\mu}(z(\tau ,\vec \sigma ))=z^A_{\mu}(\tau ,\vec \sigma )A_A(\tau ,\vec
\sigma )$; $T^{\mu\nu}_{em}(x)={\tilde F}^{\mu\alpha}(x){\tilde F}_{\alpha}{}
^{\nu}(x)+{1\over 4}\eta^{\mu\nu}{\tilde F}^{\alpha\beta}(x){\tilde F}
_{\alpha\beta}(x)$]

\begin{eqnarray}
\Theta^{\mu\nu}(x)&=&{\tilde F}^{\mu\alpha}(x){\tilde F}_{\alpha}{}^{\nu}(x)+
{1\over 4}\eta^{\mu\nu}{\tilde F}^{\alpha\beta}(x){\tilde F}_{\alpha\beta}(x)
+\nonumber \\
&+&{(D^{(A)\mu}\tilde \phi (x))}^{*}\, D^{(A)\nu}\tilde \phi (x)+{(D^{(A)\nu}
\tilde \phi (x))}^{*}D^{(A)\mu}\tilde \phi (x)-\nonumber \\
&-&\eta^{\mu\nu}[{(D^{(A)\alpha}\tilde \phi (x))}^{*}D^{(A)}{}_{\alpha}
\tilde \phi (x)]=\nonumber \\
&=&\Big[ z^{\mu}_A z^{\nu}_B \Big( [g^{AD}g^{CE}g^{BF}+{1\over 4}g^{AB}g^{CD}
g^{EF}] F_{DE} F_{CF} +g^{AB} m^2 \phi^{*} \phi +\nonumber \\
&+&[g^{AC}g^{BD}+g^{BC}g^{AD}-g^{AB}g^{CD}] (\partial_C+ieA_C)\phi^{*}\,
(\partial_D-ieA_D) \phi \Big) \Big] (\tau ,\vec \sigma )=\nonumber \\
&=&T^{\mu\nu}_{em}(x) + T^{\mu\nu}_{\phi ,A}(x),\nonumber \\
&&{}\nonumber \\
\partial_{\nu}\Theta^{\nu\mu}(x)&{\buildrel \circ \over =}& 0,\nonumber \\
\Rightarrow && \partial_{\nu} T^{\mu\nu}_{\phi ,A}(x)\, {\buildrel \circ \over
=}\, -{\tilde F}^{\mu\nu}(x) J_{\phi \nu}(x),
\label{VIII41}
\end{eqnarray}

\noindent where the conserved electromagnetic current of
the Klein-Gordon field $J^{\mu}_{\phi}$
is [$\partial_{\mu}J^{\mu}_{\phi}\, {\buildrel \circ
\over =}\, 0$ is replaced by ${{dq_{\phi}}\over {dT_s}}\, {\buildrel \circ
\over =}\, 0$ at the Hamiltonian level on the Wigner hyperplanes with $q_{\phi}
=e$ given in Eq.(\ref{VIII11})]

\begin{eqnarray}
J^{\mu}_{\phi}(\tau ,\vec \sigma )
&=&-i[(\partial^{\mu}+ie{\tilde A}^{\mu}){\tilde \phi}^{*}
\tilde \phi -{\tilde \phi}^{*} (\partial^{\mu}-ie{\tilde A}^{\mu})\tilde \phi ]
(z)=\nonumber \\
&=&-i z^{\mu}_A(\tau ,\vec \sigma ) [(\partial^A+ieA^A)\phi^{*} \phi -\phi^{*}
(\partial^A-ieA^A)\phi ](\tau ,\vec \sigma )=\nonumber \\
&=&iz^{\mu}_{\tau}(\tau ,\vec \sigma ) [\pi_{\phi^{*}}\phi^{*}-\pi_{\phi}
\phi ](\tau ,\vec \sigma )+\nonumber \\
&+&iz^{\mu}_r(\tau ,\vec \sigma ) [\phi^{*} (\partial^r-ieA^r)\phi -(\partial^r
+ieA^r)\phi^{*} \phi ](\tau ,\vec \sigma ).
\label{VIII42}
\end{eqnarray}

On Wigner hyperplanes with $T_s=\tau$ in the Coulomb gauge we have

\begin{eqnarray}
{\hat J}^{\mu}_{\phi}(T_s ,\vec \sigma )&=& u^{\mu}(p_s) i[{\hat \pi}_{\phi
^{*}}{\hat \phi}^{*}-{\hat \pi}_{\phi}\hat \phi ](T_s,\vec \sigma )+\nonumber \\
&+&\epsilon^{\mu}_r(u(p_s)) i[{\hat \phi}^{*}(\partial^r-ieA^r_{\perp})\hat
\phi -(\partial^r+ieA^r_{\perp}){\hat \phi}^{*} \hat \phi ](T_s,\vec \sigma )].
\label{VIII43}
\end{eqnarray}

On the Wigner hyperplanes we get the following expression for the
energy-momentum tensor

\begin{eqnarray}
\Theta^{\mu\nu}&&[x^{\beta}_s(T_s)+\epsilon^{\beta}_u(u(p_s)) \sigma^u]=
\nonumber \\
&=&u^{\mu}(p_s) u^{\nu}(p_s) \Big[ {1\over 2}({\vec \pi}^2+{\vec B}^2)+\pi
_{\phi^{*}} \pi_{\phi} +(\vec \partial +ie\vec A)\phi^{*}\cdot (\vec \partial
-ie\vec A)\phi +m^2\phi^{*}\phi \Big](T_s,\vec \sigma )+\nonumber \\
&+&\epsilon^{\mu}_r(u(p_s))\epsilon^{\nu}_s(u(p_s))\Big[ {1\over 2}\delta^{rs}
({\vec \pi}^2+{\vec B}^2)-(\pi^r\pi^s+B^rB^s)+\delta^{rs}(\pi_{\phi^{*}}
\pi_{\phi}-m^2\phi^{*}\phi )+\nonumber \\
&+&(\delta^{ru}\delta^{sv}+\delta^{rv}\delta^{su}-\delta^{rs}\delta^{uv})
(\partial^u+ieA^u)\phi^{*}\, (\partial^v-ieA^v)\phi \Big] (T_s,\vec \sigma )+
\nonumber \\
&+&[u^{\mu}(p_s)\epsilon^{\nu}_r(u(p_s))+u^{\nu}(p_s)\epsilon^{\mu}_r(u(p_s))]
\nonumber \\
&&\Big[ ({\vec \pi}\times \vec B)^r+\pi_{\phi^{*}}(\partial^r+ieA^r)\phi^{*}+
\pi_{\phi} (\partial^r-ieA^r)\phi \Big] (T_s,\vec \sigma ),
\label{VIII44}
\end{eqnarray}

\noindent whose expression in the $A_{\tau}(T_s,\vec \sigma )=\eta_{em}(T_s,
\vec \sigma )=0$ Coulomb gauge, where, from Eq.(\ref{VIII28}), we have $\pi^r
=\pi^r_{\perp}+{e\over {\triangle}} {\hat {\cal Q}}$ with ${\hat {\cal Q}}=i
[{\hat \pi}_{\phi^{*}}{\hat \phi}^{*}-{\hat \pi}_{\phi}\hat \phi ]$
[so that Eq.(\ref{VIII11}) becomes $q_{\phi}=\int d^3\sigma {\hat
{\cal Q}}(\tau ,\vec
\sigma )$], is

\begin{eqnarray}
{\hat \Theta}^{\mu\nu}&&[x^{\beta}_s(T_s)+\epsilon^{\beta}_u(u(p_s)) \sigma^u]=
\nonumber \\
&=&u^{\mu}(p_s) u^{\nu}(p_s) \Big[ {1\over 2}({\vec \pi}^2_{\perp}
+{\vec B}^2)+e {\vec \pi}_{\perp}\cdot {{\vec \partial}\over {\triangle}} {\hat
{\cal Q}}+{{e^2}\over 2} ({{\vec \partial}\over {\triangle}}{\hat {\cal Q}})^2+
\nonumber \\
&+&{\hat \pi}_{\phi^{*}} {\hat \pi}_{\phi} +(\vec \partial +ie{\vec
A}_{\perp}){\hat \phi}^{*}\cdot (\vec \partial -ie{\vec A}_{\perp})\hat \phi
+m^2{\hat \phi}^{*}\hat \phi \Big](T_s,\vec \sigma )+\nonumber \\
&+&\epsilon^{\mu}_r(u(p_s))\epsilon^{\nu}_s(u(p_s))\Big[ {1\over 2}\delta^{rs}
({\vec \pi}^2_{\perp}+{\vec B}^2)-(\pi^r_{\perp}\pi^s_{\perp}+B^rB^s)+
\nonumber \\
&+&e\Big( \delta^{rs}{\vec \pi}_{\perp}\cdot {{\vec \partial}\over {\triangle}}
{\hat {\cal Q}}-[\pi^r_{\perp}{{\partial^s}\over {\triangle}}+\pi^s_{\perp}
{{\partial^r}\over {\triangle}}]{\hat {\cal Q}}\Big) +\nonumber \\
&+&e^2 \Big( {1\over 2}\delta^{rs} ({{\vec \partial}\over {\triangle}}{\hat
{\cal Q}})^2-{{\partial^r}\over {\triangle}}{\hat {\cal Q}}{{\partial^s}\over
{\triangle}}{\hat {\cal Q}}\Big)+\nonumber \\
&+&\delta^{rs}({\hat \pi}_{\phi^{*}}{\hat \pi}_{\phi}-m^2{\hat \phi}^{*}\hat
\phi )+\nonumber \\
&+&(\delta^{ru}\delta^{sv}+\delta^{rv}\delta^{su}-\delta^{rs}\delta^{uv})
(\partial^u+ieA_{\perp}^u){\hat \phi}^{*}\, (\partial^v-ieA^v_{\perp})\hat
\phi \Big] (T_s,\vec \sigma )+\nonumber \\
&+&[u^{\mu}(p_s)\epsilon^{\nu}_r(u(p_s))+u^{\nu}(p_s)\epsilon^{\mu}_r(u(p_s))]
\nonumber \\
&&\Big[ ({\vec \pi}_{\perp}\times \vec B)^r+{\hat \pi}_{\phi^{*}}(\partial^r
+ieA_{\perp}^r){\hat \phi}^{*}+
{\hat \pi}_{\phi} (\partial^r-ieA_{\perp}^r)\hat \phi \Big] (T_s,\vec \sigma ),
\nonumber \\
&&{}\nonumber \\
{\hat \Theta}^{\mu}{}_{\mu}&&[x^{\beta}_s(T_s)+\epsilon^{\beta}_u(u(p_s))
\sigma^u]=\nonumber \\
&=&-2\Big[ {\hat \pi}_{\phi^{*}} {\hat \pi}_{\phi} -(\vec \partial +ie{\vec
A}_{\perp}){\hat \phi}^{*}\cdot (\vec \partial -ie{\vec A}_{\perp})\hat \phi
-2m^2{\hat \phi}^{*}\hat \phi \Big](T_s,\vec \sigma ),\nonumber \\
{\hat P}^{\mu}_{\Theta}&=&\int d^3\sigma {\hat \Theta}^{\mu\nu} u_{\nu}(p_s)=
\nonumber \\
&=& M u^{\mu}(p_s) +{\cal H}_p^r(T_s) \epsilon^{\mu}_r(u(p_s))\approx
M u^{\mu}(p_s),\nonumber \\
{\hat \Theta}^{rs}_S&=&\epsilon^r_{\mu}(u(p_s))\epsilon^s_{\nu}(u(p_s)){\hat
\Theta}^{\mu\nu}=\nonumber \\
&=&\Big[ {1\over 2}\delta^{rs}
({\vec \pi}^2_{\perp}+{\vec B}^2)-(\pi^r_{\perp}\pi^s_{\perp}+B^rB^s)+
\nonumber \\
&+&e\Big( \delta^{rs}{\vec \pi}_{\perp}\cdot {{\vec \partial}\over {\triangle}}
{\hat {\cal Q}}-[\pi^r_{\perp}{{\partial^s}\over {\triangle}}+\pi^s_{\perp}
{{\partial^r}\over {\triangle}}]{\hat {\cal Q}}\Big) +\nonumber \\
&+&e^2 \Big( {1\over 2}\delta^{rs} ({{\vec \partial}\over {\triangle}}{\hat
{\cal Q}})^2-{{\partial^r}\over {\triangle}}{\hat {\cal Q}}{{\partial^s}\over
{\triangle}}{\hat {\cal Q}}\Big)+\nonumber \\
&+&\delta^{rs}({\hat \pi}_{\phi^{*}}{\hat \pi}_{\phi}-m^2{\hat \phi}^{*}\hat
\phi )+\nonumber \\
&+&(\delta^{ru}\delta^{sv}+\delta^{rv}\delta^{su}-\delta^{rs}\delta^{uv})
(\partial^u+ieA_{\perp}^u){\hat \phi}^{*}\, (\partial^v-ieA^v_{\perp})\hat
\phi \Big] (T_s,\vec \sigma ).
\label{VIII45}
\end{eqnarray}

The Dixon multipoles with respect to the origin $x^{\mu}_s(\tau )$ are

\begin{eqnarray}
t^{\mu_1...\mu_n\mu\nu}_{\Theta}(T_s)&=&\int d^3\sigma \delta x^{\mu_1}_s(\vec
\sigma )...\delta x^{\mu_n}_s(\vec \sigma ) {\hat \Theta}^{\mu\nu}[x_s
^{\beta}(T_s)+\epsilon^{\beta}_u(u(p_s))\sigma^u] =\nonumber \\
&=&\epsilon^{\mu_1}_{r_1}(u(p_s))...\epsilon^{\mu_n}_{r_n}(u(p_s))\epsilon^{\mu}_A(u(p_s))
\epsilon^{\nu}_B(u(p_s)) I_{\Theta}^{r_1..r_nAB}(T_s),
\label{VIII46}
\end{eqnarray}

\noindent and all the derived multipoles for the interacting case can be
obtained from these ones following the scheme of Section V.

With the same techniques we can study the multipoles $t_{\phi ,A}^{\mu_1...
\mu_n\mu\nu}(T_s)$ and all the related ones ${\hat P}^{\mu}_{\phi ,A}$,
${\hat S}^{\mu\nu}_{\phi ,A}$, ${\hat J}_{\phi
,A}^{\mu_1...\mu_n\mu\nu}$ of ${\hat T}^{\mu\nu}_{\phi ,A}$. The Dixon
multipoles of the electromagnetic field in the rest-frame instant form
will be studied in a future paper.

Let us remark that in absence of the electromagnetic field, namely in the free
theory of a complex Klein-Gordon field, we have

\begin{eqnarray}
\Theta^{\mu\nu}[x_s^{\beta}(T_s)+\epsilon^{\beta}_u(u(p_s))\sigma^u]&=&
T^{\mu\nu}_{\phi}[x_s^{\beta}(T_s)+\epsilon^{\beta}_u(u(p_s))\sigma^u]=
\nonumber \\
&=&u^{\mu}(p_s)u^{\nu}(p_s) [\pi_{\phi^{*}}\pi_{\phi}+\vec \partial \phi^{*}
\cdot \vec \partial \phi +m^2\phi^{*}\phi ](T_s,\vec \sigma )+\nonumber \\
&+&[u^{\mu}(p_s)\epsilon^{\nu}_r(u(p_s))+u^{\nu}(p_s)\epsilon^{\mu}_r(u(p_s))]
[\pi_{\phi^{*}}\partial^r\phi^{*}+\pi_{\phi}\partial^r\phi ](T_s,\vec \sigma )+
\nonumber \\
&+&\epsilon^{\mu}(u(p_s))\epsilon^{\nu}_s(u(p_s))[\delta^{rs}(\pi_{\phi^{*}}\pi
_{\phi}-m^2\phi^{*}\phi )+\nonumber \\
&+&(\delta^{ru}\delta^{sv}+\delta^{rv}\delta^{su}-
\delta^{rs}\delta^{uv})\partial^u\phi^{*}\partial^v\phi ](T_s,\vec \sigma ),\nonumber \\
 &&{}\nonumber \\
t^{\mu_1...\mu_n\mu\nu}_{T*}(T_s)&=&\int d^3\sigma \delta
x^{\mu_1}_s(\vec
\sigma )...\delta x^{\mu_n}_s(\vec \sigma ) T_{\phi}^{\mu\nu}[x_s
^{\beta}(T_s)+\epsilon^{\beta}_u(u(p_s))\sigma^u] =\nonumber \\
&=&\epsilon^{\mu_1}_{r_1}(u(p_s))...\epsilon^{\mu_n}_{r_n}(u(p_s))\epsilon^{\mu}_A(u(p_s))
\epsilon^{\nu}_B(u(p_s)) I_{T*}^{r_1..r_nAB}(T_s),
\label{VIII47}
\end{eqnarray}

\noindent and, again with the methods of Section V, we can define the Dixon
multipoles [$t_{\phi}^{\mu_1...\mu_n\mu\nu}(T_s)$ and the related
ones] of the complex Klein-Gordon field in absence of interaction. In
particular, from Eq.(\ref{V8}) we get ${\tilde
t}^{\mu_1}_{\phi}(T_s)=\epsilon^{\mu_1}_{r_1}(u(p_s))
I_{T*}^{r_1\tau\tau}(T_s)$ with $I^{r\tau\tau}_{T*}(T_s)\equiv
-P^{\tau}_{\phi} r^r_{\phi} = \int d^3\sigma \,
\sigma^r [\pi_{\phi^{*}}\pi_{\phi} +\vec \partial \phi^{*}\cdot \vec
\partial \phi +m^2\phi^{*}\phi ](T_s,\vec
\sigma )$ and its vanishing identifies Dixon's center of mass (M\"oller
noncanonical center of energy) ${\vec r}_{\phi}$; then the canonical
3-center of mass ${\vec q}_{\phi}$ can be obtained with the methods of Setion
VI.

Then, as in the case of a real Klein-Gordon field, one can look for
the canonical transformation from the center-of-phase canonical basis
$X^A_{\phi}$, $P^A_{\phi}$, $R^A_{\phi}$, $Q^A_{\phi}$, ${\bf
H}_a(\tau ,\vec k)$, ${\bf K}_a(\tau ,\vec k)$, to a center-of-mass
canonical basis $q^{\tau}_{\phi}$, $P^{\tau
{'}}_{\phi}=\sqrt{(P^{\tau}_{\phi})^2-{\vec P}^2_{\phi}}$, ${\vec
q}_{\phi}$, ${\vec P}_{\phi}$, $R^{A {'}}_{\phi}$, $Q^{A {'}}_{\phi}$,
${\bf H}_a^{'}(\tau ,\vec k)$, ${\bf K}_a^{'}(\tau ,\vec k)$.

Now ${\vec R}^{'}_{\phi}$ shouls describe the relative position of the
``centers of mass" for the $``a"$ and $``b"$ modes of the field
configuration. In the free case Eqs.(\ref{d6}) show that the theory
can be reformulated in a non-local way as the sum of two theories, one
for each mode: see Ref.\cite{lam} for the Lagrangian associated to
Eqs.(\ref{d6}) in the framework of pseudo-differential operators. This
implies that the energy momentum tensor can be written as the sum of
two pieces, one for each mode, and that the multipole expansion and
the associated definition of center of mass can be applied to each
piece. Already in the present framework, by using Eqs.(\ref{d5}) and
by making integrations by parts, we get $I^{\tau\tau}_{T*}(T_s)=\int
d^3\sigma [\pi_{\phi^{*}}\pi_{\phi}+\vec
\partial \phi^{*}\cdot \vec \partial \phi +m^2\phi^{*}\phi ](T_s,\vec \sigma )=
m\int d^3\sigma [\phi^{*}_a \sqrt{m^2+\triangle} \phi_a+\phi^{*}_b
\sqrt{m^2+\triangle} \phi_b](T_s,\vec \sigma )=I^{\tau\tau}_{a*}(T_s)
+I^{\tau\tau}_{b*}(T_s)$ and $I^{r\tau\tau}_{T*}(T_s)=\int d^3\sigma
[\pi_{\phi^{*}}\partial^r\phi +\pi_{\phi}\partial^r\phi ](T_s,\vec
\sigma )=im \int d^3\sigma [\phi^{*}_a\partial^r\phi_a+
\phi^{*}_b\partial^r\phi_b](T_s,\vec \sigma )=I^{r\tau\tau}_{a*}(T_s)
+I^{r\tau\tau}_{b*}(T_s)\equiv -P^{\tau}_{a\phi} r^r_{a\phi}-
P^{\tau}_{b\phi} r^r_{b\phi}$. Then we could evaluate ${\vec
q}_{a\phi}$ and ${\vec q}_{b\phi}$ and we expect to have ${\vec
q}_{a\phi}={\vec X}^{'}_{a\phi}={\vec q}_{\phi}+{1\over 2}{\vec
R}^{'}_{\phi}$ and     ${\vec q}_{b\phi}={\vec X}^{'}_{b\phi}={\vec
q}_{\phi}-{1\over 2}{\vec R}^{'}_{\phi}$.

The Dixon multipoles\cite{dixon}  for the electromagnetic current
${\hat J}^{\mu}_{\phi}$ are

\begin{eqnarray}
&& n \geq 0,\nonumber \\
{\hat j}_{\phi}^{\mu_1...\mu_n\mu}(T_s)&=&{\hat j}_{\phi}^{(\mu_1...\mu_n)\mu}
(T_s)=\int d^3\sigma \delta x^{\mu_1}_s(\vec \sigma )...\delta x^{\mu_n}_s(\vec
\sigma ) {\hat J}_{\phi}^{\mu}(T_s,\vec \sigma )=\nonumber \\
&=&\epsilon^{\mu_1}_{r_1}(u(p_s))...\epsilon^{\mu_n}_{r_n}(u(p_s))\nonumber \\
&&\Big[ u^{\mu}(p_s) \int d^3\sigma \sigma^{r_1}...\sigma^{r_n} i[{\hat \pi}
_{\phi^{*}}{\hat \phi}^{*}-{\hat \pi}_{\phi}\hat \phi ](T_s,\vec \sigma )+
\nonumber \\
&+&\epsilon^{\mu}_r(u(p_s)) \int d^3\sigma \sigma^{r_1}...\sigma^{r_n} i[{\hat
\phi}^{*}(\partial^r-ieA^r_{\perp})\hat \phi -(\partial^r+ieA^r_{\perp}){\hat
\phi}^{*} \hat \phi ](T_s,\vec \sigma )\Big] ,\nonumber \\
&&u_{\mu_1}(p_s){\hat j}_{\phi}^{\mu_1...\mu_n\mu}(T_s)=0,\nonumber \\
&&{}\nonumber \\
&& n=0,\nonumber \\
{\hat j}^{\mu}_{\phi}(T_s)&=& q_{\phi} u^{\mu}(p_s) +\epsilon^{\mu}_r(u(p_s))
\nonumber \\
&&\int d^3\sigma i[{\hat
\phi}^{*}(\partial^r-ieA^r_{\perp})\hat \phi -(\partial^r+ieA^r_{\perp}){\hat
\phi}^{*} \hat \phi ](T_s,\vec \sigma )=\nonumber \\
&=&q_{\phi}u^{\mu}(p_s)+\epsilon^{\mu}_r(u(p_s)) im \int d^3\sigma \Big( ({\hat
\phi}^{*}_a+{\hat \phi}_b){{\partial^r}\over {\triangle}}({\hat \phi}_a+{\hat
\phi}^{*}_b)-\nonumber \\
&-&ieA^r_{\perp} [(m^2+\triangle )^{-1/4}({\hat \phi}^{*}_a+{\hat \phi}_b)]
[(m^2+\triangle )^{-1/4}({\hat \phi}_a+{\hat \phi}^{*}_b)]\Big) (T_s,\vec
\sigma ), \nonumber \\
{\hat q}_{\phi}^{\mu_1...\mu_n}(T_s)&=&{\hat q}_{\phi}^{(\mu_1...\mu_n)}(T_s)=
{\hat j}_{\phi}^{\mu_1...\mu_n\mu}(T_s) u_{\mu}(p_s)=
\quad\quad (n > 0)\nonumber \\
&=&\epsilon^{\mu_1}_{r_1}(u(p_s))...\epsilon^{\mu_n}_{r_n}(u(p_s))
\int d^3\sigma \sigma^{r_1}...\sigma^{r_n} i[{\hat \pi}
_{\phi^{*}}{\hat \phi}^{*}-{\hat \pi}_{\phi}\hat \phi ](T_s,\vec \sigma ),
\nonumber \\
&&u_{\mu_1}(p_s) {\hat q}_{\phi}^{\mu_1...\mu_n}(T_s)=0.
\label{VIII48}
\end{eqnarray}

In Ref.\cite{dixon} it is shown how to arrive to the following reconstruction
of the electromagnetic current in terms of the multipoles (if $f(k)$ is
analytic)

\begin{eqnarray}
<{\hat J}^{\mu}_{\phi},f >&=& \int d^4x {\hat J}^{\mu}_{\phi}(x) f(x)=\int dT_s
\int {{d^4k}\over {(2\pi )^4}}\tilde f(k) e^{-ik\cdot x_s(T_s)} \nonumber \\
&&\int d^3\sigma {\hat J}^{\mu}_{\phi}[x_s(T_s)+\delta x_s(\vec \sigma )]
\sum_{n=0}^{\infty} {{(-i)^n}\over {n!}} k_{\mu_1}...k_{\mu_n}{\hat j}_{\phi}
^{\mu_1...\mu_n\mu}(T_s)=\nonumber \\
&=&\int dT_s \sum_{n=0}^{\infty} {1\over {n!}} {\hat j}^{\mu_1...\mu_n\mu}
_{\phi}(T_s){{\partial^nf(x)}\over {\partial x_s^{\mu_1}...\partial
x^{\mu_n}_s}}{|}_{x=x_s(T_s)},\nonumber \\
&&{}\nonumber \\
{\hat J}^{\mu}_{\phi}(x)&=& \sum_{n=0}^{\infty} {{(-)^n}\over {n!}}
{{\partial^n}\over {\partial x_s^{\mu_1}...\partial x^{\mu_n}_s}} \int dT_s
\delta^4(x-x_s(T_s)) {\hat j}_{\phi}^{\mu_1...\mu_n\mu}(T_s).
\label{VIII49}
\end{eqnarray}

It is also shown that if the multipoles ${\hat j}_{\phi}^{\mu_1...\mu_n\mu}(T
_s)$ are known for all $n > N$ for some fixed N, then ${\hat J}_{\phi}^{\mu}$
and the multipoles with $n\leq N$ are completely determined.

See Appendix C for other types of multipoles.

\vfill\eject

\section{Conclusions.}

In this paper we have made a detailed study of the kinematical description of
scalar Klein-Gordon fields on the Wigner hyperplanes of the rest-frame
Wigner-covariant instant form of dynamics. We have considered a Klein-Gordon
field configuration as a relativistic extended object and we utilized both
phase space techniques from relativistic particle mechanics and multipolar
expansions from relativistic fluidodynamics to study aspects of the
Klein-Gordon field which are usually ignored notwithstanding the relevance of
scalar fields in physics: Higgs particles, Bose-Einstein condensation,
Brans-Dicke scalar-tensor general relativity and multipolar expansions for the
theory of gravitational waves, boson stars.

Simultaneously, we have used the Klein-Gordon field as an example to
explore the description of isolated systems on the Wigner hyperplanes
showing how the elusive concept of ``relativistic center of mass" has
to be divided in an ``external" part (with respect to an arbitrary
Lorentz frame) and in an ``internal" part (inside the Wigner
hyperplane). In both cases a canonical noncovariant Newton-Wigner-like
center of mass, a covariant noncanonical Fokker-Pryce center of
inertia and a noncanonical noncovariant M\"oller center of energy may
be defined only in terms of the generators of suitable realizations of
the Poinca\'e algebra. The three ``internal" centers weakly coincide
due to the three first class constraints defining the rest frame of
the isolated system and are, therefore, gauge variables inside the
Wigner hyperplanes. Namely there is the gauge freedom in the choice of
the ``external" timelike worldline to which they have to be attached.
Now, in the description of the Wigner hyperplane with respect to an
arbitrary Lorentz frame in Minkowski spacetime this gauge freedom is
reflected in the arbitrariness of the choice of a timelike worldline
$x^{\mu}_s(\tau )$, with noncanonical 4-coordinates, to be used as
origin of the ``internal" 3-coordinates. The natural gauge fixing for
this gauge freedom is to put the three weakly coinciding ``internal"
centers in this origin: in this way the origin $x^{\mu}_s(\tau )$ is
forced to coincide with the ``external" covariant noncanonical
Fokker-Pryce center of inertia and simultaneously to satisfy the
conditions for being both the Pirani and the Tulczyjew centroid.
Around this worldline there is a noncovariance worldtube (whose finite
extension is measured by the M\"oller radius\cite{moller,dubna})
containing all the pseudoworldlines of the noncovariant ``external"
canonical Newton-Wigner-like center of mass and noncanonical M\"oller
center of energy.

Naturally, one could fix the gauge freedom in a different way by identifying a
different ``internal" collective 3-vector with the origin $x^{\mu}_s$: in such
a case $x^{\mu}_s$ becomes one of the many possible covariant noncanonical
centroids existing in literature and does not coincide with the ``external"
Fokker-Pryce center of inertia. For instance this happens for the Klein-Gordon
field, because we do not yet know its canonical basis containing the
``internal" center of mass 3-vector. In the Longhi-Materassi canonical
decomposition of the Klein-Gordon field in collective and relative variables
this 3-vector is replaced by another 3-vector, which can be named the
``internal" center of phase of the field configuration by its construction.

The other main result of this paper is the identification of the
canonical basis containing the ``internal" center of phase of the
Klein-Gordon field and of its multipoles in the framework of the
rest-frame Wigner-covariant instant form of dynamics.

Therefore, many unrelated kinematical concepts find a well defined setting on
the Wigner hyperplanes of the rest-frame Wigner-covariant instant form of
dynamics, which seems to be the natural tool (like the separation of the
center-of-mass motion in the nonrelativistic case) to be used in relativistic
statistical mechanics and in lattice gauge theories due to its intrinsic
Euclidean signature (without any Wick rotation) and to its associated
description of the evolution by means of the Lorentz scalar rest-frame time
measured by the clock of the decoupled (but noncovariant) point particle
observer defined by the ``external" canonical center of mass.

Even if much work is still needed to clarify all the aspects of the multipolar
expansions of the Klein-Gordon field and of its properties as a relativistic
extended object, especially for the charged fields of scalar electrodynamics,
now we have a well defined framework in which to make further investigations and
a first completely worked out canonical decomposition of the field in
collective and relative variables.

A final comment on quantization. The canonical basis of the Klein-Gordon field
containing the center of phase does not seem a good candidate for the
quantization of the field in the rest-frame instant form of dynamics both due
to the complicated expression of the original fields in terms of the new
variables and to the fact that there is no sound canonical quantization of
phases and angles (different is the case with their exponentials) as can be
seen from the various articles contained in Ref.\cite{phase}. Let us remark
that the problem of the non measurability of absolute phases (see the previous
reference) is connected with the open problem of the measurability of the
``external" decoupled noncovariant canonical center of mass of an isolated
system (the quantization of these degrees of freedom would generate an
equivalent of the ``wave function of the universe") as it is clear from the
Longhi-Materassi canonical basis, in which the collective position variable
is built starting from the phases of the Fourier coefficients of the
Klein-Gordon field.

\vfill\eject

\appendix
\section{Notations for spacelike hypersurfaces.}

In this Appendix we will review the background material from Ref.\cite{lus}
needed in the description of physical systems on spacelike hypersurfaces.

Let $\lbrace \Sigma_{\tau}\rbrace$ be a one-parameter family of spacelike
hypersurfaces foliating Minkowski spacetime $M^4$ and giving a 3+1 decomposition
of it. At fixed $\tau$, let
$z^{\mu}(\tau ,\vec \sigma )$ be the coordinates of the points on $\Sigma
_{\tau }$ in $M^4$, $\lbrace \vec \sigma \rbrace$ a system of coordinates on
$\Sigma_{\tau}$. If $\sigma^{\check A}=(\sigma^{\tau}=\tau ;\vec \sigma
=\lbrace \sigma^{\check r}\rbrace)$ [the notation ${\check A}=(\tau ,
{\check r})$ with ${\check r}=1,2,3$ will be used; note that ${\check A}=
\tau$ and ${\check A}={\check r}=1,2,3$ are Lorentz-scalar indices] and
$\partial_{\check A}=\partial /\partial \sigma^{\check A}$,
one can define the vierbeins

\begin{equation}
z^{\mu}_{\check A}(\tau ,\vec \sigma )=\partial_{\check A}z^{\mu}(\tau ,\vec
\sigma ),\quad\quad
\partial_{\check B}z^{\mu}_{\check A}-\partial_{\check A}z^{\mu}_{\check B}=0,
\label {a1}
\end{equation}

\noindent so that the metric on $\Sigma_{\tau}$ is

\begin{eqnarray}
&&g_{{\check A}{\check B}}(\tau ,\vec \sigma )=z^{\mu}_{\check A}(\tau ,\vec
\sigma )\eta_{\mu\nu}z^{\nu}_{\check B}(\tau ,\vec \sigma ),\quad\quad
g_{\tau\tau}(\tau ,\vec \sigma ) > 0,\nonumber \\
&&g(\tau ,\vec \sigma )=-det\, ||\, g_{{\check A}{\check B}}(\tau ,\vec
\sigma )\, || ={(det\, ||\, z^{\mu}_{\check A}(\tau ,\vec \sigma )\, ||)}^2,
\nonumber \\
&&\gamma (\tau ,\vec \sigma )=-det\, ||\, g_{{\check r}{\check s}}(\tau ,\vec
\sigma )\, ||.
\label{a2}
\end{eqnarray}

If $\gamma^{{\check r}{\check s}}(\tau ,\vec \sigma )$ is the inverse of the
3-metric $g_{{\check r}{\check s}}(\tau ,\vec \sigma )$ [$\gamma^{{\check r}
{\check u}}(\tau ,\vec \sigma )g_{{\check u}{\check s}}(\tau ,\vec
\sigma )=\delta^{\check r}_{\check s}$], the inverse $g^{{\check A}{\check B}}
(\tau ,\vec \sigma )$ of $g_{{\check A}{\check B}}(\tau ,\vec \sigma )$
[$g^{{\check A}{\check C}}(\tau ,\vec \sigma )g_{{\check c}{\check b}}(\tau ,
\vec \sigma )=\delta^{\check A}_{\check B}$] is given by

\begin{eqnarray}
&&g^{\tau\tau}(\tau ,\vec \sigma )={{\gamma (\tau ,\vec \sigma )}\over
{g(\tau ,\vec \sigma )}},\nonumber \\
&&g^{\tau {\check r}}(\tau ,\vec \sigma )=-[{{\gamma}\over g} g_{\tau {\check
u}}\gamma^{{\check u}{\check r}}](\tau ,\vec \sigma ),\nonumber \\
&&g^{{\check r}{\check s}}(\tau ,\vec \sigma )=\gamma^{{\check r}{\check s}}
(\tau ,\vec \sigma )+[{{\gamma}\over g}g_{\tau {\check u}}g_{\tau {\check v}}
\gamma^{{\check u}{\check r}}\gamma^{{\check v}{\check s}}](\tau ,\vec \sigma ),
\label{a3}
\end{eqnarray}

\noindent so that $1=g^{\tau {\check C}}(\tau ,\vec \sigma )g_{{\check C}\tau}
(\tau ,\vec \sigma )$ is equivalent to

\begin{equation}
{{g(\tau ,\vec \sigma )}\over {\gamma (\tau ,\vec \sigma )}}=g_{\tau\tau}
(\tau ,\vec \sigma )-\gamma^{{\check r}{\check s}}(\tau ,\vec \sigma )
g_{\tau {\check r}}(\tau ,\vec \sigma )g_{\tau {\check s}}(\tau ,\vec \sigma ).
\label{a4}
\end{equation}

We have

\begin{equation}
z^{\mu}_{\tau}(\tau ,\vec \sigma )=(\sqrt{ {g\over {\gamma}} }l^{\mu}+
g_{\tau {\check r}}\gamma^{{\check r}{\check s}}z^{\mu}_{\check s})(\tau ,
\vec \sigma ),
\label{a5}
\end{equation}

\noindent and

\begin{eqnarray}
\eta^{\mu\nu}&=&z^{\mu}_{\check A}(\tau ,\vec \sigma )g^{{\check A}{\check B}}
(\tau ,\vec \sigma )z^{\nu}_{\check B}(\tau ,\vec \sigma )=\nonumber \\
&=&(l^{\mu}l^{\nu}+z^{\mu}_{\check r}\gamma^{{\check r}{\check s}}
z^{\nu}_{\check s})(\tau ,\vec \sigma ),
\label{a6}
\end{eqnarray}

\noindent where

\begin{eqnarray}
l^{\mu}(\tau ,\vec \sigma )&=&({1\over {\sqrt{\gamma}} }\epsilon^{\mu}{}_{\alpha
\beta\gamma}z^{\alpha}_{\check 1}z^{\beta}_{\check 2}z^{\gamma}_{\check 3})
(\tau ,\vec \sigma ),\nonumber \\
&&l^2(\tau ,\vec \sigma )=1,\quad\quad l_{\mu}(\tau ,\vec \sigma )z^{\mu}
_{\check r}(\tau ,\vec \sigma )=0,
\label{a7}
\end{eqnarray}

\noindent is the unit (future pointing) normal to $\Sigma_{\tau}$ at
$z^{\mu}(\tau ,\vec \sigma )$.

For the volume element in Minkowski spacetime we have

\begin{eqnarray}
d^4z&=&z^{\mu}_{\tau}(\tau ,\vec \sigma )d\tau d^3\Sigma_{\mu}=d\tau [z^{\mu}
_{\tau}(\tau ,\vec \sigma )l_{\mu}(\tau ,\vec \sigma )]\sqrt{\gamma
(\tau ,\vec \sigma )}d^3\sigma=\nonumber \\
&=&\sqrt{g(\tau ,\vec \sigma )} d\tau d^3\sigma.
\label{a8}
\end{eqnarray}

Let us remark that according to the geometrical approach of
Ref.\cite{kuchar}, one
can use Eq.(\ref{a5}) in the form $z^{\mu}_{\tau}(\tau ,\vec \sigma )=N(\tau ,
\vec \sigma )l^{\mu}(\tau ,\vec \sigma )+N^{\check r}(\tau ,\vec \sigma )
z^{\mu}_{\check r}(\tau ,\vec \sigma )$, where $N=\sqrt{g/\gamma}=\sqrt{g
_{\tau\tau}-\gamma^{{\check r}{\check s}}g_{\tau{\check r}}g_{\tau{\check s}}}$
and $N^{\check r}=g_{\tau \check s}\gamma^{\check s\check r}$ are the
standard lapse and shift functions, so that $g_{\tau \tau}=N^2+
g_{\check r\check s}N^{\check r}N^{\check s}, g_{\tau \check r}=
g_{\check r\check s}N^{\check s},
g^{\tau \tau}=N^{-2}, g^{\tau \check r}=-N^{\check r}/N^2, g^{\check r\check
s}=\gamma^{\check r\check s}+{{N^{\check r}N^{\check s}}\over {N^2}}$,
${{\partial}\over {\partial z^{\mu}_{\tau}}}=l_{\mu}\, {{\partial}\over
{\partial N}}+z_{{\check s}\mu}\gamma^{{\check s}{\check r}} {{\partial}\over
{\partial N^{\check r}}}$, $d^4z=N\sqrt{\gamma}d\tau d^3\sigma$.

The rest frame form of a timelike fourvector $p^{\mu}$ is $\stackrel
{\circ}{p}{}^{\mu}=\eta \sqrt{p^2} (1;\vec 0)= \eta^{\mu o}\eta \sqrt{p^2}$,
$\stackrel{\circ}{p}{}^2=p^2$, where $\eta =sign\, p^o$.
The standard Wigner boost transforming $\stackrel{\circ}{p}{}^{\mu}$ into
$p^{\mu}$ is

\begin{eqnarray}
L^{\mu}{}_{\nu}(p,\stackrel{\circ}{p})&=&\epsilon^{\mu}_{\nu}(u(p))=
\nonumber \\
&=&\eta^{\mu}_{\nu}+2{ {p^{\mu}{\stackrel{\circ}{p}}_{\nu}}\over {p^2}}-
{ {(p^{\mu}+{\stackrel{\circ}{p}}^{\mu})(p_{\nu}+{\stackrel{\circ}{p}}_{\nu})}
\over {p\cdot \stackrel{\circ}{p} +p^2} }=\nonumber \\
&=&\eta^{\mu}_{\nu}+2u^{\mu}(p)u_{\nu}(\stackrel{\circ}{p})-{ {(u^{\mu}(p)+
u^{\mu}(\stackrel{\circ}{p}))(u_{\nu}(p)+u_{\nu}(\stackrel{\circ}{p}))}
\over {1+u^o(p)} },\nonumber \\
&&{} \nonumber \\
\nu =0 &&\epsilon^{\mu}_o(u(p))=u^{\mu}(p)=p^{\mu}/\eta \sqrt{p^2}, \nonumber \\
\nu =r &&\epsilon^{\mu}_r(u(p))=(-u_r(p); \delta^i_r-{ {u^i(p)u_r(p)}\over
{1+u^o(p)} }).
\label{a9}
\end{eqnarray}

The inverse of $L^{\mu}{}_{\nu}(p,\stackrel{\circ}{p})$ is $L^{\mu}{}_{\nu}
(\stackrel{\circ}{p},p)$, the standard boost to the rest frame, defined by

\begin{equation}
L^{\mu}{}_{\nu}(\stackrel{\circ}{p},p)=L_{\nu}{}^{\mu}(p,\stackrel{\circ}{p})=
L^{\mu}{}_{\nu}(p,\stackrel{\circ}{p}){|}_{\vec p\rightarrow -\vec p}.
\label{a10}
\end{equation}

Therefore, we can define the following vierbeins [the $\epsilon^{\mu}_r(u(p))$'s
are also called polarization vectors; the indices r, s will be used for A=1,2,3
and $\bar o$ for A=0]

\begin{eqnarray}
&&\epsilon^{\mu}_A(u(p))=L^{\mu}{}_A(p,\stackrel{\circ}{p}),\nonumber \\
&&\epsilon^A_{\mu}(u(p))=L^A{}_{\mu}(\stackrel{\circ}{p},p)=\eta^{AB}\eta
_{\mu\nu}\epsilon^{\nu}_B(u(p)),\nonumber \\
&&{} \nonumber \\
&&\epsilon^{\bar o}_{\mu}(u(p))=\eta_{\mu\nu}\epsilon^{\nu}_o(u(p))=u_{\mu}(p),
\nonumber \\
&&\epsilon^r_{\mu}(u(p))=-\delta^{rs}\eta_{\mu\nu}\epsilon^{\nu}_r(u(p))=
(\delta^{rs}u_s(p);\delta^r_j-\delta^{rs}\delta_{jh}{{u^h(p)u_s(p)}\over
{1+u^o(p)} }),\nonumber \\
&&\epsilon^A_o(u(p))=u_A(p),
\label{a11}
\end{eqnarray}

\noindent which satisfy

\begin{eqnarray}
&&\epsilon^A_{\mu}(u(p))\epsilon^{\nu}_A(u(p))=\eta^{\mu}_{\nu},\nonumber \\
&&\epsilon^A_{\mu}(u(p))\epsilon^{\mu}_B(u(p))=\eta^A_B,\nonumber \\
&&\eta^{\mu\nu}=\epsilon^{\mu}_A(u(p))\eta^{AB}\epsilon^{\nu}_B(u(p))=u^{\mu}
(p)u^{\nu}(p)-\sum_{r=1}^3\epsilon^{\mu}_r(u(p))\epsilon^{\nu}_r(u(p)),
\nonumber \\
&&\eta_{AB}=\epsilon^{\mu}_A(u(p))\eta_{\mu\nu}\epsilon^{\nu}_B(u(p)),\nonumber
\\
&&p_{\alpha}{{\partial}\over {\partial p_{\alpha}} }\epsilon^{\mu}_A(u(p))=
p_{\alpha}{{\partial}\over {\partial p_{\alpha}} }\epsilon^A_{\mu}(u(p))
=0.
\label{a12}
\end{eqnarray}

The Wigner rotation corresponding to the Lorentz transformation $\Lambda$ is

\begin{eqnarray}
R^{\mu}{}_{\nu}(\Lambda ,p)&=&{[L(\stackrel{\circ}{p},p)\Lambda^{-1}L(\Lambda
p,\stackrel{\circ}{p})]}^{\mu}{}_{\nu}=\left(
\begin{array}{cc}
1 & 0 \\
0 & R^i{}_j(\Lambda ,p)
\end{array}
\right) ,\nonumber \\
{} && {}\nonumber \\
R^i{}_j(\Lambda ,p)&=&{(\Lambda^{-1})}^i{}_j-{ {(\Lambda^{-1})^i{}_op_{\beta}
(\Lambda^{-1})^{\beta}{}_j}\over {p_{\rho}(\Lambda^{-1})^{\rho}{}_o+\eta
\sqrt{p^2}} }-\nonumber \\
&-&{{p^i}\over {p^o+\eta \sqrt{p^2}} }[(\Lambda^{-1})^o{}_j- { {((\Lambda^{-1})^o
{}_o-1)p_{\beta}(\Lambda^{-1})^{\beta}{}_j}\over {p_{\rho}(\Lambda^{-1})^{\rho}
{}_o+\eta \sqrt{p^2}} }].
\label{a13}
\end{eqnarray}

The polarization vectors transform under the
Poincar\'e transformations $(a,\Lambda )$ in the following way

\begin{equation}
\epsilon^{\mu}_r(u(\Lambda p))=(R^{-1})_r{}^s\, \Lambda^{\mu}{}_{\nu}\,
\epsilon^{\nu}_s(u(p)).
\label{a14}
\end{equation}

\vfill\eject

\section{The Klein-Gordon field.}

Let us review the main results of Ref.\cite{lon}. For the sake of simplicity
we shall denote $\phi (x)$ the Klein-Gordon field $\tilde \phi (x)$ of
Section III.

Given a real field, solution of the Klein-Gordon equation, $\phi (x)=\phi^{*}(x)
=\phi (x^o,\vec x)= \int d\tilde k [a(\vec k)e^{-ik\cdot x}+
a^{*}(\vec k)e^{ik\cdot x}]$ [with $d\tilde k=d^3k/\Omega (k)$, $\Omega (k)=
(2\pi )^3 \omega (k)$, $\omega (k)=k^o=\sqrt{m^2+k^2}$, $k=|\vec k|$; $a(\vec
k)$ and $\partial a(\vec k)/\partial k^i$ are assumed to be in $L_2(d\tilde
k)$; $a(\vec k)$ is $x^o$-independent in the free case without self
interactions $V(\phi )$],
its conjugate momentum is $\pi (x)={{\partial \phi (x)}\over {\partial
x^o}}= -i \int d\tilde k \omega (k)[a(\vec k)e^{-ik\cdot x}-a^{*}(\vec
k)e^{ik\cdot x}]$. The 10 Poincar\'e generators

\begin{eqnarray}
P^{\mu}_{\phi}&=&{1\over 2}
\int d^3x \Big[\pi \partial^{\mu} \phi -{1\over 2}\eta^{o\mu}[\pi^2 -(\vec
\partial \phi )^2-m^2\phi^2]\Big] (x^o,\vec x)=\nonumber \\
&=&\int d\tilde k\, k^{\mu}
a^{*}(\vec k)a(\vec k),\nonumber \\
J^{\mu\nu}_{\phi}&=&\int d^3x \Big[ \Big( \pi (x^{\mu}\partial^{\nu}-
x^{\nu}\partial^{\mu}) \phi \Big) (x^o,\vec x)-{1\over 2}(x^{\mu}\eta^{\nu o}-
x^{\nu}\eta^{\mu o})(\pi^2-(\vec \partial \phi )^2-m^2\phi^2)\Big] (x^o,\vec x)
,\nonumber \\
&&{}\nonumber \\
J^{ij}_{\phi}&=&\int d^3x \Big[ \pi (x^i\partial^j-x^j\partial^i) \phi \Big]
(x^o,\vec x)=\nonumber \\
&=&-i\int d\tilde k a^{*}(\vec k) (k^i{{\partial}\over {\partial
k^j}}-k^j{{\partial}\over {\partial k^i}}) a(\vec k),\nonumber \\
J^{oi}_{\phi}&=&x^o \int d^3x (\pi \partial^i \phi )(x^o,\vec x)-{1\over 2}
\int d^3x x^i\, [\pi^2+(\vec \partial \phi )^2+m^2\phi^2](x^o,\vec x)=
\nonumber \\
&=&x^oP^i
_{\phi}-i\int d\tilde k a^{*}(\vec k) \omega (k){{\partial}\over {\partial k^i}}
a(\vec k),\nonumber \\
&&{}\nonumber \\
&&\{ P^{\mu}_{\phi},P^{\nu}_{\phi} \} =0,\quad\quad \{ J^{\mu\nu}_{\phi},p
^{\rho}_{\phi} \} =P^{\mu}_{\phi}\eta^{\nu\rho}-P^{\nu}_{\phi}\eta^{\mu\rho},
\nonumber \\
&&\{ J^{\mu\nu}_{\phi},J^{\rho\sigma}_{\phi} \} =\eta^{\mu\sigma}J^{\nu\rho}
_{\phi}+\eta^{\nu\rho}J^{\mu\sigma}_{\phi}-\eta^{\mu\rho}J^{\nu\sigma}_{\phi}-
\eta^{\nu\sigma}J^{\mu\rho}_{\phi},
\label{bb1}
\end{eqnarray}

\noindent are finite if $|a(\vec k)|{\rightarrow}_{k >> m}\, k^{-{3\over 2}
-\sigma}$, $\sigma > 0$, and $|a(\vec k)|\, {\rightarrow}_{k << m}\,
k^{-{3\over 2}+\epsilon}$, $\epsilon > 0$.

By putting $a(\vec k)=\sqrt{I(\vec k)} e^{i\varphi (\vec k)}$, one gets
\hfill\break
\hfill\break
$P^{\mu}_{\phi}=\int d\tilde k\, k^{\mu} I(\vec k)$,\hfill\break
 $J^{ij}_{\phi}=\int
d\tilde k I(\vec k)(k^i{{\partial}\over {\partial k^j}}-k^j{{\partial}\over
{\partial k^i}})\varphi (\vec k)$,\hfill\break
 $J^{oi}_{\phi}=x^oP^i_{\phi}+\int d\tilde k
I(\vec k) \omega (k) {{\partial}\over {\partial k^i}} \varphi (\vec k)$,
\hfill\break
\hfill\break
and they are finite if: a) for $k \rightarrow \infty$ one has $|I(\vec k)|
\rightarrow k^{-3-\sigma}$, $\sigma > 0$, $|\varphi (\vec k)| \rightarrow
k$; b) for $k \rightarrow 0$ one has $|I(\vec k)| \rightarrow k^{-3+\epsilon}$,
$\epsilon > 0$, $|\varphi (\vec k)| \rightarrow k^{\eta}$, $\eta > -\epsilon$.

The non vanishing Poisson brackets of these 3 canonical bases are: $\{ \phi
(x^o,\vec x), \pi (x^o,\vec y) \} =\delta^3(\vec x-\vec y)$, $\{ a(\vec k),
a^{*}(\vec q) \} =-i \Omega (k) \delta^3(\vec k-\vec q)$, $\{ I(\vec k),
\varphi (\vec q) \} = \Omega (k) \delta^3(\vec k-\vec q)$.

Let $F(P_{\phi},k)=F(P_{\phi}\cdot k, P^2_{\phi})$ be a real scalar weight
function, arbitrary apart the normalization
$\int d\tilde k k^{\mu} F(P_{\phi},k)=
P^{\mu}_{\phi}$; a simple choice for $F$ is $F \approx e^{-P_{\phi}\cdot k}$.
One assumes a behaviour of $F$ for $k \rightarrow \infty$ and $k \rightarrow 0$
such that all the integrals, which depend on $F$, are finite.

Then one can define the following canonical basis

\begin{eqnarray}
X^{\mu}_{\phi} &=& \int d\tilde k\, \varphi (\vec k) {{\partial}\over {\partial
P_{\phi \, \mu} }} F(P_{\phi},k),\nonumber \\
P^{\mu}_{\phi} &=& \int d\tilde k\, k^{\mu} I(\vec k) = \int d\tilde k\, k^{\mu}
F(P_{\phi},k),\nonumber \\
{\cal H}(\vec k) &=& \int d\tilde k^{'}\, G(\vec k,{\vec k}^{'}) [I({\vec k}
^{'})-F(P_{\phi},k^{'})]\nonumber \\
&&{\rightarrow}_{k\rightarrow \infty}\, k^{-3-\sigma},\quad \sigma > 0;
\quad\quad
{\rightarrow}_{k\rightarrow 0}\, k^{-1+\epsilon},\quad \epsilon > 0,
\nonumber \\
{\cal K}(\vec k) &=& {\cal D} \varphi (\vec k)\, {\rightarrow}_{k\rightarrow
\infty}\, k^{1-\epsilon},\quad \epsilon > 0;\quad\quad
{\rightarrow}_{k\rightarrow 0}\, k^{\eta -2},\quad \eta > -\epsilon ,
\nonumber \\
&&{}\nonumber \\
&&\{ X^{\mu}_{\phi},P^{\nu}_{\phi} \} =-\eta^{\mu\nu},\quad\quad
\{ {\cal H}(\vec k),{\cal K}(\vec q) \} =\Omega (k) \delta^3(\vec k-\vec q),
\nonumber \\
&&{}\nonumber \\
I(\vec k) &=& {\cal D} {\cal H}(\vec k) + F(P_{\phi},k),\nonumber \\
\varphi (\vec k) &=& k\cdot X_{\phi} +\int d{\tilde k}^{'} \int d{\tilde k}^{"}
\, {\cal K}({\vec k}^{'}) G({\vec k}^{'},{\vec k}^{"}) \triangle ({\vec k}^{"},
\vec k).
\label{b1}
\end{eqnarray}

\noindent Here, ${\cal D}$ is the non compact operator ${\cal D}=3-m^2\triangle
_{LB}$, with $\triangle_{LB}$ being the Laplace-Beltrami operator on the mass
shell submanifold $H^1_3$ defined by $k^{\mu}k_{\mu}=m^2$ and $k^o > 0$. The
Green function $G$ satisfies ${\cal D}_{\vec k} G(\vec k,\vec q)=\Omega (k)
\delta^3(\vec k-\vec q)$. Finally, one has $\triangle (\vec k,\vec q)=\Omega (k)
\delta^3(\vec k-\vec q) -q^{\mu} {{\partial}\over {\partial P^{\mu}_{\phi} }}
F(P_{\phi},k)$.

The Poincar\'e generators are given by

\begin{eqnarray}
P^{\mu}_{\phi},&&\nonumber \\ J^{ij}_{\phi} &=&
X^i_{\phi}P^j_{\phi}-X^j_{\phi}P^i_{\phi} + \int d\tilde k\, {\cal
H}(\vec k) (k^i{{\partial}\over {\partial k^j}}-k^j{{\partial}\over
{\partial k^i}}) {\cal K}(\vec k),\nonumber \\ J^{oi}_{\phi} &=&
(x^o+X^o_{\phi})P^i_{\phi}-X^i_{\phi}P^o_{\phi} +
\int d\tilde k\, {\cal H}(\vec k) \omega (k) {{\partial}\over {\partial k^i}}
{\cal K}(\vec k).
\label{b2}
\end{eqnarray}

This canonical transformation exists (without problems from the
existence of the zero modes of the operator ${\cal D}$) only if the field
configurations $\phi (x)$, solution of the Klein-Gordon equation, with
$\pi (x)=\partial \phi (x)/\partial x^o$, satisfy the conditions:\hfill\break
A) $P_{lm}=\int d\tilde k I(\vec k) v^{(o)}_{1,-3,lm}(\vec k) \equiv
\int d\tilde k F(P_{\phi},k) v^{(o)}_{1,-3,lm}(\vec k) =0,$\hfill\break
and, in the case that $I(\vec k)\, {\rightarrow}_{k\rightarrow 0}\,
k^{-3+\eta}$ with $\eta > l+1$, also\hfill\break
B) $Q_{lm}=\int d\tilde k I(\vec k) v^{(o)}_{2,-3,lm}(\vec k) \equiv
\int d\tilde k F(P_{\phi},k) v^{(o)}_{2,-3,lm}(\vec k) =0$,\hfill\break
where the $v^{(o)}$'s are zero modes of ${\cal D}$. The conditions A)
are non void only for $l \geq 2$ and are called ``no supertranslations
conditions" in Ref.\cite{lon1}: when they are satisfied, the constants of
motion $P_{lm}$ are functionally dependent on the momenta of the new
canonical basis [when they are not satisfied, the $P_{lm}$ are independent
constants of motion allowing the definition of a BMS algebra and this destroys
the possibility of definining a unique Poincar\'e algebra in a non ambigous
way].

One has

\begin{eqnarray}
\phi (x) &=& \int d\tilde k \,\, \sqrt{F(P_{\phi},k) + {\cal D}{\cal H}(\vec k)}
\nonumber \\
&&\Big[ e^{ik\cdot (X_{\phi}-x)+i\int d{\tilde k}^{'} \int d{\tilde k}^{"}
{\cal K}({\vec k}^{'}) G({\vec k}^{'},{\vec k}^{"})
\triangle ({\vec k}^{"},\vec k)} + c.c. \Big] ,\nonumber \\
\pi (x) &=& -i \int d\tilde k\,\, \omega (k)\,
\sqrt{F(P_{\phi},k) + {\cal D}{\cal H}(\vec k)}
\nonumber \\
&&\Big[ e^{ik\cdot (X_{\phi}-x)+i\int d{\tilde k}^{'} \int d{\tilde k}^{"}
{\cal K}({\vec k}^{'}) G({\vec k}^{'},{\vec k}^{"})
\triangle ({\vec k}^{"},\vec k)} - c.c. \Big] .
\label{b3}
\end{eqnarray}

Let us remark that on the Wigner hyperplane $F(P_{\phi},k)$ may be replaced
with two scalar functions ${\tilde F}^{\tau}(P^{\tau}_{\phi},q)$,
$\tilde F({\vec P}_{\phi},\vec q)$ [$F \mapsto {\tilde {\cal F}}(P_{\phi},q)=
{\tilde F}^{\tau}(P^{\tau}_{\phi},|\vec q|)-\tilde F({\vec P}_{\phi},\vec q)$],
so that the center-of-mass variables would be

\begin{eqnarray}
X^{\tau}_{\phi}&=&
\int d\tilde q \,\, \varphi (\tau ,\vec q)\,
{{\partial {\tilde F}^{\tau}(P^{\tau}_{\phi},q)}
\over {\partial P^{\tau}_{\phi}}} ,\nonumber\\
P^{\tau}_{\phi}&=&\int d\tilde q\,\, \omega (q)\, I(\tau ,\vec q)=\int d\tilde q
\,\, \omega (q)\,  {\tilde F}^{\tau}(P^{\tau}_{\phi},q),\nonumber \\
{\vec X}_{\phi}&=&\int d\tilde q\,\, \varphi (\tau ,\vec q)\,
{{\partial \tilde F({\vec P}_{\phi},\vec q)}
\over {\partial {\vec P}_{\phi}}} ,\nonumber \\
{\vec P}_{\phi}&=&\int d\tilde q\,\, \vec q\, I(\tau ,\vec q)=\int d\tilde q\,
\, \vec q\, \tilde F({\vec P}_{\phi},\vec q).
\label{b4}
\end{eqnarray}

Instead of this generically non polynomial weight functions, we use
in this paper weight functions linear in the Poincar\'e momenta:
${\tilde F}^{\tau}(P^{\tau}_{\phi},q) = P^{\tau}_{\phi}\, F^{\tau}(q)$,
$\tilde F({\vec P}_{\phi},\vec q)={\vec P}_{\phi}\cdot \vec q\, F(q)$.

\vfill\eject

\section{More on Dixon's Multipoles.}

Let us add other forms of the Dixon multipoles.

In  the case of the real Klein-Gordon field treated in Section V, the
Hamilton equations (\ref{IV40}) imply [in Ref.\cite{dixon} this is a
consequence of $\partial_{\mu}T^{\mu\nu}\, {\buildrel \circ
\over
=}\, 0$]

\begin{eqnarray}
{{dp_T^{\mu}(T_s)}\over {dT_s}}\, &{\buildrel \circ \over =}\,& 0,\quad
for\, n=0,\nonumber \\
{{d p_T^{\mu_1...\mu_n\mu}(T_s)}\over {dT_s}}\, &{\buildrel \circ \over =}\,&
-nu^{(\mu_1}(p_s) p_T^{\mu_2...\mu_n)\mu}(T_s)+n t_T^{(\mu_1...\mu_n)\mu}(T_s),
\quad n\geq 1.
\label{c1}
\end{eqnarray}

Let us define for $n \geq 1$

\begin{eqnarray}
b_T^{\mu_1...\mu_n\mu}(T_s)&=&p_T^{(\mu_1...\mu_n\mu )}(T_s)=\nonumber
\\
&=&\epsilon^{(\mu_1}_{r_1}(u(p_s))...\epsilon^{\mu_n}_{r_n}(u(p_s))\epsilon^{\mu
)}_A(u(p_s)) I_T^{r_1..r_nA\tau}(T_s),\nonumber \\
 &&{}\nonumber \\
c_T^{\mu_1...\mu_n\mu}(T_s)&=&c_T^{(\mu_1...\mu_n
)\mu}(T_s)=p_T^{\mu_1...
\mu_n\mu}(T_s)-p_T^{(\mu_1...\mu_n\mu )}(T_s)=\nonumber \\
&=&[\epsilon^{\mu_1}_{r_1}(u(p_s))...\epsilon^{\mu_n}_{r_n}\epsilon^{\mu}_A(u(p_s))-
\epsilon^{(\mu_1}_{r_1}(u(p_s))...\epsilon^{\mu_n}_{r_n}(u(p_s))\epsilon^{\mu )}_A(u(p_s))]
I_T^{r_1..r_nA\tau}(T_s),\nonumber \\
 &&{}\nonumber \\
&&c_T^{\mu_1...\mu_n\mu}(T_s)=0,
\label{c2}
\end{eqnarray}

\noindent and then for $n\geq 2$

\begin{eqnarray}
d_T^{\mu_1...\mu_n\mu\nu}(T_s)&=&d_T^{(\mu_1...\mu_n)(\mu\nu )}(T_s)=
t_T^{\mu_1...\mu_n\mu\nu}(T_s)-\nonumber \\ &-&{{n+1}\over
n}[t_T^{(\mu_1...\mu_n\mu )\nu}(T_s)+t_T^{(\mu_1...\mu_n\nu )\mu}
(T_s)]+{{n+2}\over n}t_T^{(\mu_1...\mu_n\mu\nu )}(T_s)=\nonumber \\
 &=&\Big[ \epsilon^{\mu_1}_{r_1} ... \epsilon^{\mu_n}_{r_n} \epsilon^{\mu}_A \epsilon^{\nu}_B-
 {{n+1}\over n}\Big( \epsilon^{(\mu_1}_{r_1} ... \epsilon^{\mu_n}_{r_n} \epsilon^{\mu )}_A
 \epsilon^{\nu}_B+\nonumber \\
 &+&\epsilon^{(\mu_1}_{r_1} ... \epsilon^{\mu_n}_{r_n} \epsilon^{\nu )}_B
 \epsilon^{\mu}_A\Big) +{{n+2}\over n} \epsilon^{(\mu_1}_{r_1} .. \epsilon^{\mu_n}_{r_n}
 \epsilon^{\mu}_A \epsilon^{\nu}_B\Big] (u(p_s)) I_T^{r_1..r_nAB}(T_s),
 \nonumber \\
 &&{}\nonumber \\
&&d_T^{(\mu_1...\mu_n\mu )\nu}(T_s)=0.
\label{c3}
\end{eqnarray}

Then Eqs.(\ref{c1}) may be rewritten in the form

\begin{eqnarray}
&&1)\quad n=1\nonumber \\
 &&{}\nonumber \\
t^{\mu\nu}_T(T_s)&=&t^{(\mu\nu )}_T(T_s)\, {\buildrel \circ \over =}\,
p_T^{\mu}(T_s)u^{\nu }(p_s)+{1\over 2}{d\over
{dT_s}}(S^{\mu\nu}_T(T_s)[\phi ]+2b_T
^{\mu\nu}(T_s)),\nonumber \\
&&\Downarrow \nonumber \\
 t^{\mu\nu}_T(T_s)\, &{\buildrel \circ \over
=}\,&p_T^{(\mu}(T_s)u^{\nu )}(p_s) +{d\over
{dT_s}}b_T^{\mu\nu}(T_s)=P^{\tau}_{\phi}u^{\mu}(p_s)u^{\nu}(p_s)+
P^r_{\phi}u^{(\mu}(p_s)\epsilon^{\nu )}_r(u(p_s))+\nonumber \\
&+&\epsilon^{(\mu}_r(u(p_s))u^{\nu )}(p_s) \int d^3\sigma \sigma^r
{1\over 2} [\pi^2+(\vec \partial \phi )^2+m^2\phi^2](T_s,\vec \sigma
)+\nonumber \\
 &+&\epsilon^{(\mu}_r(u(p_s))\epsilon^{\nu )}_s(u(p_s))
\int d^3\sigma \sigma^r [\pi \partial^s\phi ](T_s,\vec \sigma
),\nonumber \\
 {d\over {dT_s}}S^{\mu\nu}_T(T_s)[\phi ]\, &{\buildrel
\circ \over =}\,&2p_T^{[\mu}(T_s) u^{\nu
]}(p_s)=2P^r_{\phi}\epsilon^{[\mu}_r(u(p_s))u^{\nu ]}(p_s) \approx
0,\nonumber \\
 &&{}\nonumber \\
 &&2)\quad n=2\quad [identity\,\,
t_T^{\rho\mu\nu}=t_T^{(\rho\mu )\nu}+t_T^{(\rho
\nu )\mu}+t_T^{(\mu\nu )\rho}]\nonumber \\
 &&{}\nonumber \\
2t_T^{(\rho\mu )\nu}(T_s)\, &{\buildrel \circ \over =}\,& 2u^{(\rho}(p_s)b_T
^{\mu )\nu}(T_s)+u^{(\rho}(p_s)S_T^{\mu )\nu}(T_s)[\phi ]
+{d\over {dT_s}}(b_T^{\rho\mu\nu}
(T_s)+c_T^{\rho\mu\nu}(T_s)),\nonumber \\
 &&\Downarrow \nonumber \\
t_T^{\rho\mu\nu}(T_s)\, &{\buildrel \circ \over =}\,&u^{\rho}(p_s)b_T
^{\mu\nu}(T_s)+S_T^{\rho (\mu}(T_s)[\phi ]
u^{\nu )}(p_s)+{d\over {dT_s}}({1\over 2}b_T
^{\rho\mu\nu}(T_s)-c_T^{\rho\mu\nu}(T_s)),\nonumber \\
&&{}\nonumber \\
&&3) \quad n \geq 3 \nonumber \\
&&{}\nonumber \\
t_T^{\mu_1...\mu_n\mu\nu}(T_s)\, &{\buildrel \circ \over =}\,& d_T^{\mu_1...
\mu_n\mu\nu}(T_s)+u^{(\mu_1}(p_s)b_T^{\mu_2...\mu_n)\mu\nu}(T_s)+2u^{(\mu
_1}(p_s)c_T^{\mu_2...\mu_n)(\mu\nu )}(T_s)+\nonumber \\
&=&{2\over n}c_T^{\mu_1...\mu_n(\mu}(T_s)u^{\nu )}(p_s)+{d\over {dT_s}}
[{1\over {n+1}}b_T^{\mu_1...\mu_n\mu\nu}(T_s)+{2\over n}c_T^{\mu_1...\mu
_n(\mu\nu )}(T_s)],
\label{c4}
\end{eqnarray}

This allows \cite{dixon} to rewrite $< T^{\mu\nu},f >$ in the following form

\begin{eqnarray}
< T^{\mu\nu},f > &=&\int dT_s \int {{d^4k}\over {(2\pi )^4}} \tilde
f(k) e^{-ik\cdot x_s(T_s)} \Big[ u^{(\mu}(p_s)p_T^{\nu
)}(T_s)-ik_{\rho}S^{\rho (\mu}_T(T_s)[\phi  ]u^{\nu )}(p_s)+\nonumber
\\ &+&\sum_{n=2}^{\infty}{{(-i)^n}\over {n!}} k_{\rho_1}...k_{\rho_n}
{\cal I}_T^{\rho_1...\rho_n\mu\nu}(T_s)\Big],
\label{c5}
\end{eqnarray}

\noindent with

\begin{eqnarray}
{\cal I}_T^{\mu_1...\mu_n\mu\nu}(T_s)&=&{\cal I}_T^{(\mu_1...\mu_n
)(\mu\nu )}(T_s)= d_T^{\mu_1...\mu_n\mu\nu}(T_s)-\nonumber \\
&-&{2\over {n-1}}u^{(\mu_1}(p_s) c_T^{\mu_2...\mu_n )(\mu\nu
)}(T_s)+{2\over n} c_T^{\mu_1...\mu_n(\mu}(T_s)u^{\nu
)}(p_s)=\nonumber \\
 &=&\Big[ \epsilon^{\mu_1}_{r_1}...\epsilon^{\mu_n}_{r_n} \epsilon^{\mu}_A \epsilon^{\nu}_B-
 {{n+1}\over n}\Big( \epsilon^{(\mu_1}_{r_1}...\epsilon^{\mu_n}_{r_n} \epsilon^{\mu )}_A
 \epsilon^{\nu}_B+\nonumber \\
 &+&\epsilon^{(\mu_1}_{r_1}...\epsilon^{\mu_n}_{r_n} \epsilon^{\nu )}_B \epsilon^{\mu}_A\Big) +
 {{n+2}\over n} \epsilon^{(\mu_1}_{r_1}...\epsilon^{\mu_n}_{r_n}
 \epsilon^{\mu}_A \epsilon^{\nu )}_B\Big] (u(p_s)0 I_T^{r_1..r_nAB}(T_s)-\nonumber \\
 &-&\Big[ {2\over {n-1}} u^{(\mu_1}(p_s) \Big( \epsilon^{\mu_2}_{r_1}...
 \epsilon^{\mu_n)}_{r_{n-1}} \epsilon^{(\mu}_{r_n} \epsilon^{\nu )}_A-
 \epsilon^{(\mu_2}_{r_1}...\epsilon^{\mu_n)}_{r_{n-1}} \epsilon^{(\mu}_{r_n}
 \epsilon^{\nu ))}_A\Big)-\nonumber \\
 &-&{2\over n} \Big( \epsilon^{\mu_1}_{r_1}...\epsilon^{\mu_n}_{r_n} \epsilon^{(\mu}_A-
 \epsilon^{(\mu_1}_{r_1}... \epsilon^{\mu_n}_{r_n} \epsilon^{(\mu )}_A u^{\nu )}(p_s)
 \big] (u(p_s)0 I_T^{r_1..r_nA\tau}(T_s),\nonumber \\
 &&{}\nonumber \\
 &&{\cal I}_T^{(\mu_1...\mu_n\mu )\nu}(T_s)=0.
\label{c6}
\end{eqnarray}

Finally, a set of multipoles equivalent to the ${\cal
I}_T^{\mu_1...\mu_n\mu\nu}$ is

\begin{eqnarray}
&&n \geq 0\nonumber \\
 &&{}\nonumber \\
J_T^{\mu_1...\mu_n\mu\nu\rho\sigma}(T_s)&=&J_T^{(\mu_1...\mu_n)[\mu\nu ][\rho
\sigma ]}(T_s)= {\cal I}_T^{\mu_1...\mu_n[\mu [\rho\nu ]\sigma ]}(T_s)=\nonumber \\
&=&t_T^{\mu_1...\mu_n[\mu [\rho\nu ]\sigma ]}(T_s)-{1\over {n+1}}\Big[
u^{[\mu}(p_s)p_T^{\nu ]\mu_1...\mu_n[\rho\sigma ]}(T_s)+\nonumber \\
 &+&u^{[\rho}(p_s)p_T
^{\sigma ]\mu_1...\mu_n[\mu\nu ]}(T_s)\Big] =\nonumber \\
 &=&\Big[ \epsilon^{\mu_1}_{r_1} .. \epsilon^{\mu_n}_{r_n} \epsilon^{[\mu}_r
 \epsilon^{[\rho}_s \epsilon^{\nu ]}_A \epsilon^{\sigma ]}_B\Big] (u(p_s))
 I_T^{r_1..r_nAB}(T_s)-\nonumber \\
  &-&{1\over {n+1}}\Big[ u^{[\mu}(p_s) \epsilon^{\nu ]}_r(u(p_s)) \epsilon^{[\rho}_s(u(p_s))
  \epsilon^{\sigma ]}_A(u(p_s))+\nonumber \\
   &+&u^{[\rho}(p_s) \epsilon^{\sigma ]}_r(u(p_s))
  \epsilon^{[\mu}_s(u(p_s)) \epsilon^{\nu ]}_A(u(p_s))\Big]\nonumber \\
  &&\epsilon^{\mu_1}_{r_1}(u(p_s))...\epsilon^{\mu_n}_{r_n}(u(p_s))
  I_T^{rr_1..r_nsA\tau}(T_s),\nonumber \\
 &&{}\nonumber \\
 &&[ (n+4)(3n+5)\, linearly\, independent\, components],\nonumber \\
&&{}\nonumber \\
 && n \geq 1\nonumber \\
  &&{}\nonumber \\
u_{\mu_1}(p_s)&&J_T^{\mu_1...\mu_n\mu\nu\rho\sigma}(T_s)=
J_T^{\mu_1...\mu_{n-1}(\mu_n\mu\nu )\rho\sigma}(T_s)=0,\nonumber \\
 &&{}\nonumber \\
 && n \geq 2 \nonumber \\
 &&{}\nonumber \\
{\cal I}_T^{\mu_1...\mu_n\mu\nu}(T_s)&=&{{4(n-1)}\over {n+1}}
J_T^{(\mu_1...\mu_{n-1}|
\mu |\mu_n)\nu}(T_s).
\label{c7}
\end{eqnarray}

The $J_T^{\mu_1...\mu_n\mu\nu\rho\sigma}$ are the Dixon
``$2^{n+2}$-pole inertial moment tensors" of the extended system: they
[or equivalently the ${\cal I}_T^{\mu_1...
\mu_n\mu\nu}$'s] determine its
energy-momentum tensor together with the monopole $p^{\mu}_T$ and the
spin dipole $S^{\mu\nu}_T$. The equations $\partial_{\mu} T^{\mu\nu}\,
{\buildrel \circ \over =}\, 0$ are satisfied due to the equations of
motion (\ref{c4}) for $P^{\mu}_T$ and $S^{\mu\nu}_T$ [the so called
Papapetrou-Dixon-Souriau equations given in Eqs.(\ref{V16})] without
the need of the equations of motion for the
$J_T^{\mu_1...\mu_n\mu\nu\rho\sigma}$. When all the multipoles
$J_T^{\mu_1...\mu_n\mu\nu\rho\sigma}$ are zero [or negligible] one
speaks of a pole-dipole field configuration of the real Klein-Gordon
field.

For the electromagnatic current of Section VIII
the Hamilton equations (\ref{VIII35}) [equivalent to
$\partial_{\mu}{\hat J}^{\mu}
_{\phi}\, {\buildrel \circ \over =}\, 0$] give

\begin{eqnarray}
{{d {\hat q}_{\phi}}\over {dT_s}}\,&{\buildrel \circ \over =}\,& 0,\nonumber \\
{{d{\hat q}_{\phi}^{\mu_1...\mu_n}(T_s)}\over {dT_s}}\,&{\buildrel \circ \over
=}\,& n {\hat j}_{\phi}^{(\mu_1...\mu_n)}(T_s)-n u^{(\mu_1}(p_s) {\hat q}_{\phi}
^{\mu_2...\mu_n)}(T_s),\quad\quad n \geq 1,
\label{c8}
\end{eqnarray}

By introducing the multipoles

\begin{equation}
{\hat a}_{\phi}^{\mu_1...\mu_n\mu}(T_s)={\hat j}_{\phi}^{\mu_1...\mu_n\mu}(T_s)
-{\hat j}_{\phi}^{(\mu_1...\mu_n\mu )}(T_s).
\label{c9}
\end{equation}

\noindent the previous equations become

\begin{eqnarray}
{\hat j}^{\mu}_{\phi}(T_s)
&\,{\buildrel \circ \over =}\,& {\hat q}_{\phi}u^{\mu}(p_s)+{{d{\hat q}_{\phi}
^{\mu}(T_s)}\over {dT_s}},\quad\quad n=0,\nonumber \\
{\hat j}^{\mu_1...\mu_n\mu}_{\phi}(T_s)
&\,{\buildrel \circ \over =}\,& {\hat a}_{\phi}^{\mu_1...\mu_n\mu}(T_s)+{\hat
q}_{\phi}^{(\mu_1...\mu_n}(T_s) u^{\mu )}(p_s)+\nonumber \\
&+&{1\over {n+1}} {{d{\hat q}_{\phi}^{\mu_1...\mu_n\mu}(T_s)}\over {dT_s}}
,\quad\quad n \geq 1.
\label{c10}
\end{eqnarray}

Let us introduce the new multipoles

\begin{eqnarray}
{\hat m}_{\phi}^{\mu_1...\mu_n\mu}(T_s)&=&{\hat m}_{\phi}^{(\mu_1...\mu_n)
\mu}(T_s)=\nonumber \\
&=&{\hat a}_{\phi}^{\mu_1...\mu_n\mu}(T_s)+{\hat q}_{\phi}^{(\mu_1...\mu_n}(T_s)
u^{\mu )}(p_s)-u^{(\mu_1}(p_s) {\hat q}_{\phi}^{\mu_2...\mu_n)\mu}(T_s)
=\nonumber \\
&=&{\hat j}_{\phi}^{\mu_1...\mu_n\mu}(T_s)-u^{(\mu_1}(p_s) {\hat q}_{\phi}
^{\mu_2...\mu_n)\mu}(T_s)-{1\over {n+1}}{{d{\hat q}_{\phi}^{\mu_1...\mu_n\mu}(T
_s)}\over {dT_s}},\quad\quad n \geq 1,\nonumber \\
&&{\hat m}_{\phi}^{(\mu_1...\mu_n\mu )}(T_s)=0,\nonumber \\
{\hat q}_{\phi}^{\mu_1...\mu_n}(T_s)&=&-nu_{\alpha}(p_s) {\hat m}_{\phi}
^{\alpha \mu_1...\mu_n}(T_s)=-n u_{\alpha}(p_s) {\hat m}_{\phi}^{\alpha (\mu_1
...\mu_n)}(T_s).
\label{c11}
\end{eqnarray}

In Ref.\cite{dixon} it is shown that we have

\begin{eqnarray}
< {\hat J}_{\phi}^{\mu}, f > &=& \int dT_s \int {{d^4k}\over {(2\pi )^4}}
\tilde f(k) Q^{\mu}_{\phi}(k,T_s),\nonumber \\
Q^{\mu}_{\phi}(k,T_s) &=& e^{-ik\cdot x_s(T_s)}\Big[ {\hat q}_{\phi}u^{\mu}(p_s)
+\sum_{n=1}^{\infty} {{(-i)^n}\over {n!}} k_{\mu_1}...k_{\mu_n} {\hat m}_{\phi}
^{\mu_1...\mu_n\mu}(T_s) \Big] =\nonumber \\
&=&e^{-ik\cdot x_s(T_s)}\Big[ {\hat q}_{\phi}u^{\mu}(p_s)
+\sum_{n=1}^{\infty} {{(-i)^n 2n}\over {(n+1)!}} k_{\mu_1}...k_{\mu_n} {\hat Q}
_{\phi}^{(\mu_1...\mu_n)\mu}(T_s) \Big] ,
\label{c12}
\end{eqnarray}

\noindent where the following new multipoles have been introduced
[for n=0 we get $Q^{[\mu\nu ]}_{\phi}={\hat j}^{[\mu\nu ]}_{\phi}+{\hat q}
^{[\mu}_{\phi}\, u^{\nu ]}(p_s)$]

\begin{eqnarray}
{\hat Q}^{\mu_1...\mu_n\mu\nu}_{\phi}(T_s)&=&{\hat Q}_{\phi}^{(\mu_1...
\mu_n)[\mu\nu ]}(T_s)={\hat m}_{\phi}^{\mu_1...\mu_n[\mu\nu ]}(T_s)=\nonumber \\
&=& {\hat j}_{\phi}^{\mu_1...\mu_n[\mu\nu ]}(T_s)+{1\over {n+1}}{\hat q}_{\phi}
^{\mu_1...\mu_n[\mu}(T_s) u^{\nu ]}(p_s),\quad\quad n \geq 0,\nonumber \\
&&{\hat Q}_{\phi}^{\mu_1...\mu_{n-1}[\mu_n\mu\nu ]}(T_s)=0,\quad\quad n \geq 1,
\nonumber \\
&&u_{\mu_1}(p_s) {\hat Q}_{\phi}^{\mu_1...\mu_n\mu\nu}(T_s)=0,\nonumber \\
{\hat m}_{\phi}^{\mu_1...\mu_n\mu}(T_s)&=&{{2n}\over {n+1}} {\hat Q}_{\phi}
^{(\mu_1...\mu_n)\mu}(T_s).
\label{c13}
\end{eqnarray}

The multipole ${\hat Q}_{\phi}^{\mu_1...\mu_n\mu\nu}(T_s)$ is called the Dixon
``$2^{n+1}$-pole electromagnetic moment tensor". The charge $q_{\phi}$ and these
multipoles determine the electromagnetic current ${\hat J}_{\phi}^{\mu}$,
whose conservation is simply equivalent to ${{dq_{\phi}}\over {dT_s}}\,
{\buildrel \circ \over =}\, 0$.

Therefore the complex Klein-Gordon field interacting with the electromagnetic
field is an extended object defined by the inertial multipoles ${\hat P}^{\mu}
_{\phi ,A}$, ${\hat S}^{\mu\nu}_{\phi ,A}$, ${\hat J}_{\phi ,A}^{\mu_1...
\mu_n\mu\nu}$ and by the electromagnetic multipoles $q_{\phi}$, ${\hat Q}
_{\phi}^{\mu_1...\mu_n\mu}$.

Then in Ref.\cite{dixon} there is a study of the Lorentz force ${\hat F}^{\mu}
=-{\hat F}^{\mu\nu}{\hat J}_{\phi \nu}$ and it is shown that the equations
$\partial_{\nu}{\hat T}^{\mu\nu}_{\phi ,A}\, {\buildrel \circ \over =}\,
-{\hat F}^{\mu}$ and $\partial_{\mu}{\hat J}^{\mu}_{\phi}\, {\buildrel \circ
\over =}\, 0$ imply only the relations ${{dq_{\phi}}\over {dT_s}}
\, {\buildrel \circ \over =}\, 0$, ${{d{\hat P}^{\mu}_{\phi ,A}}\over {dT_s}}
\, {\buildrel \circ \over =}\, {\cal F}^{\mu}({\hat F}^{\alpha\beta}, {\hat Q}
_{\phi}^{\nu_1...\nu_n\rho\sigma})$, ${{d{\hat S}^{\mu\nu}_{\phi ,A}}\over
{dT_s}}\, {\buildrel \circ \over =}\, {\cal F}^{\mu\nu}({\hat F}^{\alpha\beta},
{\hat Q}_{\phi}^{\nu_1...\nu_n\rho\sigma})$ [see Dixon's paper for the
criticism of the standard pole-dipole approximation].

\vfill\eject

\section{The Feshbach-Villars formalism.}

As in Ref.\cite{albad}, let us consider the two-component Feshbach-Villars
formalism for the complex Klein-Gordon field \cite{fv} [see also Ref.
\cite{cs,gross}]. If we put ($\tau_i$ are the Pauli matrices)

\begin{eqnarray}
\phi (\tau ,\vec \sigma ) &=&{1\over {\sqrt{2}}}[\varphi +\chi ](\tau ,\vec
\sigma )={1\over {\sqrt{2}}} (\phi_1+i\phi_2)(\tau ,\vec \sigma )=\nonumber \\
&&=\int d\tilde q [a(\tau ,\vec q) e^{-i(\omega (q)\tau -\vec q\cdot \vec
\sigma )} +b^{*}(\tau ,\vec q) e^{+i(\omega (q)\tau -\vec q\cdot \vec \sigma )}]
,\nonumber \\
{i\over m}\pi_{\phi^{*}}(\tau ,\vec \sigma )&=&{1\over {\sqrt{2}}}[\varphi -
\chi ](\tau ,\vec \sigma )={i\over {\sqrt{2} m}} (\pi_1+i\pi_2)(\tau ,\vec
\sigma )=\nonumber \\
&&=\int d\tilde q {{\omega (q)}\over m}[a(\tau ,\vec q) e^{-i(\omega (q)\tau
-\vec q\cdot \vec \sigma )} -b^{*}(\tau ,\vec q) e^{+i(\omega (q)\tau
-\vec q\cdot \vec \sigma )}],\nonumber \\
&&{}\nonumber \\
\varphi (\tau ,\vec \sigma )&=& {1\over {\sqrt{2}}} [\phi +{i\over m}\pi
_{\phi^{*}}](\tau ,\vec \sigma )={1\over 2}[\phi_1+{i\over m}\pi_1+i(\phi_2+
{i\over m}\pi_2)](\tau ,\vec \sigma )=\nonumber \\
&&={1\over {\sqrt{2}}} \int d\tilde q [{{m+\omega (q)}\over m} a(\tau ,\vec q)
e^{-i(\omega (q)\tau -\vec q\cdot \vec \sigma )}+{{m-\omega (q)}\over m}
b^{*}(\tau ,\vec q) e^{+i(\omega (q)\tau -\vec q\cdot \vec \sigma )}],
\nonumber \\
\chi (\tau ,\vec \sigma )&=& {1\over {\sqrt{2}}} [\phi -{i\over m}\pi
_{\phi^{*}}](\tau ,\vec \sigma )={1\over 2}[\phi_1-{i\over m}\pi_1+i(\phi_2-
{i\over m}\pi_2)](\tau ,\vec \sigma )=\nonumber \\
&&={1\over {\sqrt{2}}} \int d\tilde q [{{m-\omega (q)}\over m} a(\tau ,\vec q)
e^{-i(\omega (q)\tau -\vec q\cdot \vec \sigma )}+{{m+\omega (q)}\over m}
b^{*}(\tau ,\vec q) e^{+i(\omega (q)\tau -\vec q\cdot \vec \sigma )}],
\nonumber \\
&&{}\nonumber \\
\Phi (\tau ,\vec \sigma )&=&\left( \begin{array}{c} \varphi \\ \chi
\end{array} \right) (\tau ,\vec \sigma )=\int d\tilde q \, e^{-i\vec q\cdot
\vec \sigma} \tilde \Phi (\tau ,\vec q)=\int d\tilde q\, e^{-i\vec q\cdot \vec
\sigma} \left( \begin{array}{c} \tilde \varphi \\ \tilde \chi \end{array}
\right) (\tau ,\vec q)=\nonumber \\
&&={1\over {\sqrt{2}}}\int d\tilde q [\left( \begin{array}{c} {{m+\omega (q)}
\over m}\\{{m-\omega (q)}\over m} \end{array} \right) a(\tau ,\vec q)
e^{-i(\omega (q)\tau -\vec q\cdot \vec \sigma )}+\nonumber \\
&&+\left( \begin{array}{c}
{{m-\omega (q)}\over m} \\ {{m+\omega (q)}\over m} \end{array} \right)
b^{*}(\tau ,\vec q) e^{+i(\omega (q)\tau -\vec q\cdot \vec \sigma )}],
\label{d1}
\end{eqnarray}

\noindent the Hamilton equations for the Klein-Gordon field become

\begin{eqnarray}
&&i\partial_{\tau}\varphi (\tau ,\vec \sigma )\,{\buildrel \circ \over =}\,
[{1\over {2m}}{(-i\vec \partial )}^2(\varphi +\chi )+m\varphi ](\tau
,\vec \sigma ) ,\nonumber \\
&&i\partial_{\tau}\chi (\tau ,\vec \sigma )\, {\buildrel \circ \over  =}\,
[-{1\over {2m}}{(-i\vec \partial
)}^2(\varphi +\chi )-m\chi ](\tau ,\vec \sigma ) ,
\label{d2}
\end{eqnarray}

\noindent In the $2\times 2$ matrix formalism we have

\begin{eqnarray}
i\partial_{\tau} \Phi (\tau ,\vec \sigma )&=& [{1\over {2m}}{(-i\vec \partial
)}^2\, (\tau_3+i\tau_2)+m\tau_3 ]\Phi (\tau ,\vec \sigma ) =\nonumber \\
&=&\hat H \Phi (\tau ,\vec \sigma ) .
\label{d3}
\end{eqnarray}

Since $\rho ={i\over m}\Phi^{*}\tau_3\Phi={i\over m}(\varphi^{*}\varphi -\chi
^{*}\chi )={i\over m}({\hat \pi}_{\phi^{*}}{\hat \phi}^{*}-{\hat \pi}_{\phi}
\hat \phi )$ is the density of the conserved charge
${e\over m}$ (see the Gauss law),
the normalization of $\Phi$ can be taken $\int d^3\sigma (\Phi^{*}\tau_3\Phi )
(\tau ,\vec \sigma )={{q_{\phi}}\over m}={{N_{a \phi}-N_{b \phi}}\over m}=
{e\over m}$.

As shown in Ref.\cite{fv}, the free Klein-
Gordon field has the Hamiltonian  $H_o={ {{\vec p}^2}\over {2m}}(\tau_3+
i\tau_2)+m\tau_3$ in the momentum representation and this Hamiltonian can be
diagonalized ($\omega (p)=+\sqrt{m^2+{\vec p}^2}$) with a $\tau_3$-unitary
matrix $U(\omega (p))$ [${}U^{-1}=\tau_3 U^{\dagger} \tau_3$; $H_o$ is
$\tau_3$-hermitean, $H_o=\tau_3 H^{\dagger} \tau_3$]

\begin{eqnarray}
H_{o,U}&=&U^{-1}(\omega (p))H_oU(\omega (p))=\omega (p)\, \tau_3=\left(
\begin{array}{cc}
\sqrt{m^2+{\vec p}^2}&0\\ 0&-\sqrt{m^2+{\vec p}^2} \end{array} \right) ,
\nonumber \\
{\tilde \Phi}_U(\tau ,\vec p)&=&U^{-1}(\omega (p))\tilde \Phi (\tau ,\vec p)=
\left( \begin{array}{c} {\tilde \varphi}_U\\ {\tilde \chi}_U \end{array}
\right) (\tau ,\vec p)=\nonumber \\
&&={{\sqrt{\omega (p)}}\over {m\sqrt{2m}}} [\left( \begin{array}{c} 1\\0
\end{array} \right) a(\tau ,\vec p) e^{-i\omega (p)\tau}+
\left( \begin{array}{c} 0\\1 \end{array} \right) b^{*}(\tau ,\vec p)
e^{+i\omega (p)\tau}],\nonumber \\
i\partial_{\tau}{\tilde \Phi}_U(\tau ,\vec p)&=&\left( \begin{array}{c}
i\partial_{\tau} {\tilde \varphi}_U(\tau ,\vec p)\\ i\partial_{\tau}{\tilde
\chi}_U(\tau ,\vec p) \end{array} \right) =H_{o,U}{\tilde \Phi}_U(\tau
,\vec p)=\left( \begin{array}{c} \omega (p){\tilde \varphi}_U(\tau ,\vec p)\\
-\omega (p){\tilde \chi}_U(\tau ,\vec p) \end{array}\right) ,\nonumber \\
&&{}\nonumber \\
&&U(\omega (p))={1\over {2\sqrt{m\omega (p)}}}[(m+\omega (p)) 1+(m-
\omega (p))\tau_1],\nonumber \\
&&U^{-1}(\omega (p))={1\over {2\sqrt{m\omega (p)}}}[(m+\omega (p)) 1-(m-
\omega (p))\tau_1],\nonumber \\
&&{}\nonumber \\
\left( \begin{array}{c} {\tilde \varphi}_U(\tau ,\vec p)\\ {\tilde \chi}
_U(\tau ,\vec p) \end{array} \right)&=&{1\over {2\sqrt{m\omega (p)}}} \left(
\begin{array}{c} (m+\omega (p)){\tilde \varphi}(\tau ,\vec p)-(m-\omega (p))
\tilde \chi (\tau ,\vec p)\\ (m+\omega (p))\tilde \chi (\tau ,\vec p)-(m-
\omega (p))\tilde \varphi (\tau ,\vec p) \end{array} \right)=\nonumber \\
&=&{1\over {\sqrt{2m}}}
\left( \begin{array}{c} \sqrt{\omega (p)}\tilde \phi (\tau ,\vec p)+{i\over
{\sqrt{\omega (p)}}}{\tilde \pi}_{\phi^{*}}(\tau ,\vec p)\\ \sqrt{\omega (p)}
\tilde \phi (\tau ,\vec p)-{i\over {\sqrt{\omega (p)}}}{\tilde \pi}_{\phi^{*}}
(\tau ,\vec p) \end{array} \right) =\left( \begin{array}{c} {\tilde \phi}_a(\tau
,\vec p) \\ {\tilde \phi}^{*}_b(\tau ,\vec p) \end{array} \right) ,\nonumber \\
\left( \begin{array}{c} {\tilde \varphi}(\tau ,\vec p)\\ {\tilde \chi}
(\tau ,\vec p) \end{array} \right)&=& U(\omega (p))
\left( \begin{array}{c} {\tilde \phi}_a(\tau ,\vec p)\\ {\tilde \phi}^{*}_b
(\tau ,\vec p) \end{array} \right) =\nonumber \\
&=&{1\over {2\sqrt{m\omega (p)}}}
\left( \begin{array}{c} (m+\omega (p)){\tilde \phi}_a(\tau ,\vec p)+(m-\omega
(p)){\tilde \phi}^{*}_b(\tau ,\vec p)\\ (m+\omega (p)){\tilde \phi}^{*}_b
(\tau ,\vec p) +(m-\omega (p)){\tilde \phi}_a(\tau ,\vec p)
\end{array} \right) .
\label{d4}
\end{eqnarray}

Therefore, we can introduce the following two Klein-Gordon fields

\begin{eqnarray}
\phi_a(\tau ,\vec \sigma )&=&\varphi_U(\tau ,\vec \sigma )=
{1\over {2\sqrt{m\sqrt{m^2+\triangle}}}}[(m+\sqrt{m^2+\triangle}) \varphi -
(m-\sqrt{m^2+\triangle})\chi ](\tau ,\vec \sigma )=\nonumber \\
&=&{1\over {\sqrt{2m\sqrt{m^2+\triangle}}}}[\sqrt{m^2+\triangle}\phi +i\pi
_{\phi^{*}}](\tau ,\vec \sigma )=\nonumber \\
&=&\sqrt{{2\over m}} \int d\tilde q \sqrt{\omega (q)} a(\tau ,\vec q)
e^{-i(\omega (q)\tau -\vec q\cdot \vec \sigma )}=\nonumber \\
&=&\sqrt{{2\over m}} \int d\tilde q \sqrt{\omega (q)}
\sqrt{F^{\tau}(q)\omega (q)P^{\tau}
_{a \phi}-F(q)\vec q\cdot {\vec P}_{a \phi}+{\cal D}_q {\bf H}_a(\tau ,
\vec q) }\nonumber \\
&&e^{i\int d\tilde k\int d\tilde k^{\prime }\, {\bf K}_a( \tau ,\vec k)
{\cal G}\left( \vec k,\vec k^{\prime }\right) \Delta \left( \vec k^{\prime }
,\vec q\right) -i\omega (q) \left( \tau -X_{a \phi}^\tau \right) +
i\vec q\cdot \left(\vec \sigma -{\vec X}_{a \phi}\right) }=\nonumber \\
&=&\sqrt{{2\over m}} \int d\tilde q \sqrt{\omega (q)}
\sqrt{F^{\tau}(q)\omega (q)[{1\over 2}P^{\tau}_{\phi}+Q^{\tau}_{\phi}]
-F(q)\vec q\cdot [{1\over 2}{\vec P}_{\phi}+{\vec Q}_{\phi}]+
{\cal D}_q {\bf H}_a(\tau , \vec q) }\nonumber \\
&&e^{+i\int d\tilde k\int d\tilde k^{\prime }\, {\bf K}_a( \tau ,\vec k)
{\cal G}\left( \vec k,\vec k^{\prime }\right) \Delta \left( \vec k^{\prime }
,\vec q\right) -i\omega (q) \left( \tau -[X_{\phi}^\tau +{1\over 2}R^{\tau}
_{\phi}] \right) +i\vec q\cdot \left(\vec \sigma -[{\vec X}_{\phi}+
{1\over 2}{\vec R}_{\phi}]\right) },\nonumber \\
\phi_b(\tau ,\vec \sigma )&=&\chi^{*}_U(\tau ,\vec \sigma )=
{1\over {2\sqrt{m\sqrt{m^2+\triangle}}}}
[(m+\sqrt{m^2+\triangle}) \chi^{*} -(m-\sqrt{m^2+\triangle})\varphi^{*}](\tau
,\vec \sigma )=\nonumber \\
&=&{1\over {\sqrt{2m\sqrt{m^2+\triangle}}}} [\sqrt{m^2+\triangle} \phi^{*} +i
\pi ](\tau ,\vec \sigma )=\nonumber \\
&=&\sqrt{{2\over m}} \int d\tilde q \sqrt{\omega (q)} b(\tau ,\vec q)
e^{-i(\omega (q)\tau -\vec q\cdot \vec \sigma )}=\nonumber \\
&=&\sqrt{{2\over m}} \int d\tilde q \sqrt{\omega (q)}
\sqrt{F^{\tau}(q)\omega (q)P^{\tau}
_{b \phi}-F(q)\vec q\cdot {\vec P}_{b \phi}+{\cal D}_q {\bf H}_b(\tau ,
\vec q) }\nonumber \\
&&e^{i\int d\tilde k\int d\tilde k^{\prime }\, {\bf K}_b( \tau ,\vec k)
{\cal G}\left( \vec k,\vec k^{\prime }\right) \Delta \left( \vec k^{\prime }
,\vec q\right) -i\omega (q) \left( \tau -X_{b \phi}^\tau \right) +
i\vec q\cdot \left(\vec \sigma -{\vec X}_{b \phi}\right) }=\nonumber \\
&=&\sqrt{{2\over m}} \int d\tilde q \sqrt{\omega (q)}
\sqrt{F^{\tau}(q)\omega (q)[{1\over 2}P^{\tau}_{\phi}-Q^{\tau}_{\phi}]
-F(q)\vec q\cdot [{1\over 2}{\vec P}_{\phi}-{\vec Q}_{\phi}]+
{\cal D}_q {\bf H}_b(\tau , \vec q) }\nonumber \\
&&e^{+i\int d\tilde k\int d\tilde k^{\prime }\, {\bf K}_b( \tau ,\vec k)
{\cal G}\left( \vec k,\vec k^{\prime }\right) \Delta \left( \vec k^{\prime }
,\vec q\right) -i\omega (q) \left( \tau -[X_{\phi}^\tau -{1\over 2}R^{\tau}
_{\phi}] \right) +i\vec q\cdot \left(\vec \sigma -[{\vec X}_{\phi}-
{1\over 2}{\vec R}_{\phi}]\right) },\nonumber \\
&&{}\nonumber \\
\phi (\tau ,\vec \sigma )&=&\sqrt{{m\over 2}} {1\over {(m^2+\triangle )^{1/4}}}
[\phi_a(\tau ,\vec \sigma )+\phi_b^{*}(\tau ,\vec \sigma )]=
[\phi^{*}(\tau ,\vec \sigma )]^{*},\nonumber \\
\pi_{\phi^{*}}(\tau ,\vec \sigma )&=&\dot \phi (\tau ,\vec \sigma )=-i
\sqrt{{m\over 2}} (m^2+\triangle )^{1/4} [\phi_a(\tau ,\vec \sigma )-\phi_b^{*}
(\tau ,\vec \sigma )]=[\pi_{\phi}(\tau ,\vec \sigma )]^{*},\nonumber \\
&&{}\nonumber \\
a(\tau ,\vec q)&=&\sqrt{2m\omega (q)} \int d^3\sigma \phi_a(\tau ,\vec \sigma )
e^{i(\omega (q)\tau -\vec q\cdot \vec \sigma )}=\nonumber \\
&=&\int d^3\sigma [\omega (q) \phi (\tau ,\vec \sigma )+i\pi_{\phi^{*}}(\tau
,\vec \sigma )] e^{i(\omega (q)\tau -\vec q\cdot \vec \sigma )},\nonumber \\
a^{*}(\tau ,\vec q)&=&\sqrt{2m\omega (q)} \int d^3\sigma \phi^{*}_a(\tau ,\vec
\sigma )e^{-i(\omega (q)\tau -\vec q\cdot \vec \sigma )}=\nonumber \\
&=&\int d^3\sigma [\omega (q) \phi^{*}(\tau ,\vec \sigma )+i\pi_{\phi}(\tau
,\vec \sigma )] e^{-i(\omega (q)\tau -\vec q\cdot \vec \sigma )},\nonumber \\
b(\tau ,\vec q)&=&\sqrt{2m\omega (q)} \int d^3\sigma \phi_b(\tau ,\vec \sigma )
e^{i(\omega (q)\tau -\vec q\cdot \vec \sigma )}=\nonumber \\
&=&\int d^3\sigma [\omega (q) \phi(\tau ,\vec \sigma )-i\pi_{\phi^{*}}(\tau
,\vec \sigma )] e^{i(\omega (q)\tau -\vec q\cdot \vec \sigma )},\nonumber \\
b^{*}(\tau ,\vec q)&=&\sqrt{2m\omega (q)} \int d^3\sigma \phi^{*}_b(\tau ,\vec
\sigma )e^{-i(\omega (q)\tau -\vec q\cdot \vec \sigma )}=\nonumber \\
&=&\int d^3\sigma [\omega (q) \phi^{*}(\tau ,\vec \sigma )+i\pi_{\phi}(\tau
,\vec \sigma )] e^{-i(\omega (q)\tau -\vec q\cdot \vec \sigma )}.
\label{d5}
\end{eqnarray}

They are solutions of the square-root Klein-Gordon equation [see Ref.\cite{lam}
for the study of this equation by using the theory of pseudodifferential
operators] with both the energy and the electric charge of the same sign:
positive for $\phi_a$ and negative for $\phi_b$.

\begin{eqnarray}
i\partial_{\tau}\phi_a(\tau ,\vec \sigma )\, &{\buildrel \circ \over =}\,&
\sqrt{m^2+\triangle} \phi_a(\tau ,\vec \sigma ),\quad\quad q_{\phi}=N_{a\phi}
,\nonumber \\
i\partial_{\tau}\phi_b(\tau ,\vec \sigma )\, &{\buildrel \circ \over =}\,&
-\sqrt{m^2+\triangle} \phi_b(\tau ,\vec \sigma ),\quad\quad
q_{\phi}=-N_{b\phi}.
\label{d6}
\end{eqnarray}

Like in the case of the Foldy-Wouthuysen transformation for particles
of spin 1/2, also in the spin 0 case the exact diagonalization of the
Hamiltonian cannot be achieved in presence of an arbitrary external
electromagnetic field \cite{fv}, when Eqs.(\ref{d2}) become [see
Ref.\cite{albad}]

\begin{eqnarray}
&&i\partial_{\tau}\varphi (\tau ,\vec \sigma )
\,{\buildrel \circ \over =}\, \Big[ {1\over {2m}}{(-i\vec
\partial -e{\vec A}_{\perp})}^2(\varphi +\chi )+(m+K)\varphi
\Big] (\tau ,\vec \sigma ),\nonumber \\
&&i\partial_{\tau}\chi (\tau ,\vec \sigma )\, {\buildrel \circ \over =}\,
\Big[ -{1\over {2m}}{(-i\vec \partial -e{\vec A}_{\perp})}^2
(\varphi +\chi )+(-m+K)\chi \Big] (\tau ,\vec \sigma ),\nonumber \\
&&{}\nonumber \\
&&K(\tau ,\vec \sigma )=-{{e^2}\over {4\pi}} \int d^3\sigma_1 {{i({\hat \pi}
_{\phi^{*}}\, {\hat \phi}^{*}-{\hat \pi}_{\phi}\, {\hat \phi})
(\tau ,{\vec \sigma}_1)}\over {|\vec \sigma -{\vec
\sigma}_1|}}=\nonumber \\
&&=-{{me^2}\over {4\pi}} \int d^3\sigma_1 {{(\varphi^{*}
\varphi -\chi^{*}\chi )(\tau ,{\vec \sigma}_1)}\over {|\vec \sigma -{\vec
\sigma}_1|}}=\nonumber \\
&&=-{{me^2}\over {4\pi}}\int d^3\sigma_1 {{(\Phi^{*}\tau_3\Phi )(\tau ,{\vec
\sigma}_1)}\over {|\vec \sigma -{\vec \sigma}_1|}} =\int d\tilde k K(\tau
,\vec k) e^{-i\vec k\cdot \vec \sigma }.
\label{d7}
\end{eqnarray}

In this case, Eq.(\ref{d3}) assumes the following form after Fourier
transform

\begin{eqnarray}
i\partial_{\tau} \tilde \Phi (\tau ,\vec p)&=& \tilde H \tilde \Phi (\tau
,\vec p),\nonumber \\
\tilde H&=&{1\over {2m}} [\vec p-e\int d^3k {\vec A}_{\perp}(\tau ,\vec k)e^{-
\vec k\cdot \vec \partial}\, ]^2(\tau_3+i\tau_2)+m\tau_3+\nonumber \\
&+&\int d^3k K(\tau ,\vec k) e^{-\vec k\cdot \vec \partial}\, \openone .
\label{d8}
\end{eqnarray}

If we put $\tilde \Phi(\tau ,\vec p)=U(\omega (p)) {\tilde \Phi}_U(\tau
,\vec p)$ with the same $U(\vec p)$ of the free case, we get [see Ref.\cite{fv}]

\begin{eqnarray}
i\partial_{\tau} \tilde \Phi_U(\tau ,\vec p)&=& \sqrt{m^2+{\vec p}^2} \tau_3
{\tilde \Phi}_U(\tau ,\vec p)+\nonumber \\
&+&\int d^3k K(\tau ,\vec p-\vec k){{(\sqrt{m^2+{\vec p}^2}+\sqrt{m^2+{\vec
k}^2})\openone +(\sqrt{m^2+{\vec p}^2}-\sqrt{m^2+{\vec k}^2})\tau_1}\over
{2\sqrt{ \sqrt{m^2+{\vec p}^2} \sqrt{m^2+{\vec k}^2} } }} \nonumber \\
&&\openone {\tilde \Phi}_U(\tau ,\vec k)+
\nonumber \\
&+&\int d^3k {m\over {2\sqrt{ \sqrt{m^2+{\vec p}^2} \sqrt{m^2+{\vec k}^2} }}}
[-{e\over m}\vec k\cdot {\vec A}_{\perp}(\tau ,\vec p-\vec k)+{{e^2}\over {2m}}
({\vec A}^2_{\perp})(\tau ,\vec p-\vec k)] \nonumber \\
&&(\openone +\tau_1) {\tilde \Phi}_U(\tau
,\vec k),
\label{d9}
\end{eqnarray}

\noindent where $({\vec A}^2_{\perp})(\tau ,\vec p)$ means the Fourier
transform of ${\vec A}^2_{\perp}(\tau ,\vec \sigma )$.

In ref.\cite{fv}, it is shown that this Hamiltonian cannot be diagonalized,
because the separation of positive and negative energies is inhibited by effects
which (in a second quantized formalism) can be ascribed to the vacuum
polarization, namely to the pair production. This (i.e. the nonseparability
of positive and negative energies) is also the source of the
zitterbewegung effects for localized Klein-Gordon wave packets as discussed in
Ref.\cite{fv}.

With an integration by parts, the constraint ${\cal H}(\tau )$ of
Eq.(\ref{VIII32}) can be rewritten as

\begin{eqnarray}
\epsilon_s&-&{1\over 2}\int d^3\sigma ({\vec \pi}_{\perp}^2+{\vec B}^2)(\tau ,
\vec \sigma ) -\nonumber \\
&-&\int d^3\sigma \Phi^{*}(\tau ,\vec \sigma )\tau_3 [{1\over 2}(-i\vec
\partial -e{\vec A}_{\perp}(\tau ,\vec \sigma ))^2 (\tau_3+i\tau_2)+m^2\tau_3]
\Phi (\tau ,\vec \sigma )+\nonumber \\
&+&\int d^3\sigma \Phi^{*}(\tau ,\vec \sigma )\tau_3[{{e^2m^2}\over {8\pi}}
\int d^3\sigma_1 {{(\Phi^{*}\tau_3\Phi )(\tau ,{\vec \sigma}_1)}\over
{|\vec \sigma -{\vec \sigma}_1|}} \openone ] \Phi (\tau ,\vec \sigma )\approx 0.
\label{d10}
\end{eqnarray}

If we suppose that $\Phi (\tau ,\vec \sigma )$ is normalized to $\int d^3\sigma
\Phi^{*}(\tau ,\vec \sigma )\tau_3 \Phi (\tau ,\vec \sigma )=1/m$ [this is a
charge normalization compatible with the nonlinear equations of motion, because
the electric charge is conserved], we can rewrite the previous formula as

\begin{eqnarray}
&&\int d^3\sigma \Phi^{*}(\tau ,\vec \sigma )\tau_3 \{ \, [
\epsilon_s-{1\over 2}\int d^3\sigma_1 ({\vec \pi}_{\perp}^2+{\vec B}^2)(\tau ,
{\vec \sigma}_1) +\nonumber \\
&+&{{e^2m}\over {8\pi}}
\int d^3\sigma_1 {{(\Phi^{*}\tau_3\Phi )(\tau ,{\vec \sigma}_1)}\over
{|\vec \sigma -{\vec \sigma}_1|}} ] \openone -\nonumber \\
&-&[{1\over {2m}}(-i\vec
\partial -e{\vec A}_{\perp}(\tau ,\vec \sigma ))^2 (\tau_3+i\tau_2)+m\tau_3]
\} \Phi (\tau ,\vec \sigma ) \approx 0.
\label{d11}
\end{eqnarray}

If we assume that the nonlinear equations for the reduced Klein-Gordon field
have solutions of the form $\Phi (\tau ,\vec \sigma )=\Phi (\tau ,\vec p)
e^{i\vec p\cdot \vec \sigma}+\Phi_1(\tau ,\vec \sigma )$ with $\Phi_1$
negligible, namely that the global form of the nonlinear wave admits a sensible
eikonal approximation, then, neglecting $\Phi_1$, we get approximately

\begin{eqnarray}
&&\int d^3\sigma \Phi^{*}(\tau ,\vec \sigma )\tau_3 \{ \, [
\epsilon_s-{1\over 2}\int d^3\sigma ({\vec \pi}_{\perp}^2+{\vec B}^2)(\tau ,
\vec \sigma ) +\nonumber \\
&+&{{e^2m}\over {8\pi}}
\int d^3\sigma_1 {{(\Phi^{*}\tau_3\Phi )(\tau ,{\vec \sigma}_1)}\over
{|\vec \sigma -{\vec \sigma}_1|}} ] \openone -\nonumber \\
&-&[{1\over {2m}}(\vec p
-e{\vec A}_{\perp}(\tau ,\vec \sigma ))^2 (\tau_3+i\tau_2)+m\tau_3]
\} \Phi (\tau ,\vec \sigma ) \approx 0.
\label{d12}
\end{eqnarray}

If we now redefine $\Phi_U(\tau ,\vec \sigma )=U^{-1}(\sqrt{m^2+ (\vec
p-e{\vec A}_{\perp} (\tau ,\vec \sigma ))^2}) \Phi (\tau ,\vec \sigma
)$ with the same U of Eq.(\ref{d4}), we get

\begin{eqnarray}
&&\int d^3\sigma \Phi_U^{*}(\tau ,\vec \sigma ) \tau_3
\left( \begin{array}{cc}{\cal H}_{+}(\tau ,\vec \sigma )&0\\
0&{\cal H}_{-}(\tau ,\vec \sigma )\end{array} \right) \,
\Phi_U(\tau ,\vec \sigma ) =\nonumber \\
&=&\int d^3\sigma \Big[ \phi^{*}_a {\cal H}_{+} \phi_a - \phi^{*}_b {\cal
H}_{-} \phi_b \Big] (\tau ,\vec \sigma ) \approx 0,\nonumber \\
&&{}\nonumber \\
&&{\cal H}_{\pm}(\tau ,\vec \sigma )=\epsilon_s\mp
\sqrt{m^2+({\vec p}\mp |e|{\vec A}_{\perp}(\tau ,\vec \sigma ))^2}+\nonumber \\
&+&{{e^2m}\over {8\pi}} \int d^3\sigma_1
{{(\Phi^{*}\tau_3\Phi )(\tau ,{\vec \sigma}_1)}\over
{|\vec \sigma -{\vec \sigma}_1|}}
-{1\over 2}\int d^3\sigma_1({\vec \pi}^2_{\perp}+{\vec B}^2)(\tau
,{\vec \sigma}_1).
\label{d13}
\end{eqnarray}

\noindent where ${\cal H}_{\pm}(\tau ,\vec \sigma )\approx 0$ are the form
for N=1  of the constraints

\begin{eqnarray}
{\cal H}(\tau )&=&\epsilon_s-\{
\sum_{i=1}^N\eta_{i}\sqrt{m_{i}^{2}+(\check{\vec{\kappa}}_{i}(\tau)
-Q_{i}\vec{A}_{\perp}(\tau,\vec{\eta}_{i}(\tau )))^{2}}
+\nonumber\\
&+&\sum_{i\neq j}\frac{Q_{i}Q_{j}}
{4\pi\mid\vec{\eta}_{i}(\tau )-\vec{\eta}_{j}(\tau )\mid}+
\int d^{3}\sigma\:
{1\over 2}[\check{\vec{\pi}}^{2}_{\perp}(\tau,\vec{\sigma})
+\check{\vec{B}}^{2}(\tau,\vec{\sigma})]\}\approx 0,\nonumber \\
{\vec {\cal H}}_p(\tau )&=&{\check {\vec \kappa}}_{+}(\tau )+\int d^3\sigma
[{\check {\vec \pi}}_{\perp}\times {\check {\vec B}}](\tau ,\vec \sigma
)\approx 0,
\label{d14}
\end{eqnarray}

\noindent given in Ref.\cite{albad}
for the invariant mass of charged scalar particles plus the
electromagnetic field, which was determined in Ref.\cite{lus} for the
two possible signs of the energy $\eta =\pm$ [with the Grassmann
charges $Q_i$ replaced with $\pm |e|$], when evaluated at $\vec \sigma
={\vec
\eta}_i(\tau )$. Under the square root there is only a magnetic
coupling to ${\vec A}_{\perp} $: in this way the quoted problems of
the interconnession of electric fields with pair production are
avoided. The Klein-Gordon self-energy should go in the particle limit
(eikonal approximation of field theory) in the Coulomb self-energy of
the classical particle, which is absent in Ref.\cite{lus} because it
is regularized by assuming that the particle electric charge $Q$ is
pseudoclassically described by Grassmann variables so that $Q^2=0$.
Therefore, the particle description of Ref.\cite{lus} is valid only
when one disregards the effects induced by vacuum polarization and
pair production and uses  a strong eikonal approximation neglecting
diffractive effects.

\vfill\eject

\section{The invariant mass in terms of $\hat a$, $\hat b$.}

The invariant mass $M$ of Eq.(\ref{VIII40}) has the following form in
terms of the Fourier coefficients $\hat a$ and $\hat b$ [see
Eqs.(\ref{d5}) for their expression in terms of the final canonical
variables]:

\begin{eqnarray}
M
&=& {1\over 2} \int d^3\sigma ({\vec \pi}^2_{\perp}+{\vec B}^2)(\tau ,\vec
\sigma )+{\hat P}^{\tau}_{\phi}(\tau )+\nonumber \\
&+&m \int d{\tilde q}_1d{\tilde q}_2 \int d^3\sigma \Big[ e^{i(\omega (q_1)\tau
 -{\vec q}_1\cdot \vec \sigma )}{\hat a}^{*}(\tau ,{\vec q}_1)+e^{i(\omega (q_1)
\tau -{\vec q}_1\cdot \vec \sigma )}\hat b(\tau ,{\vec q}_1)\Big] \nonumber \\
&&\Big( -2e{\vec q}_2\cdot {\vec A}_{\perp}(\tau ,\vec \sigma )\Big[ e^{-i(
\omega (q_2)\tau -{\vec q}_2\cdot \vec \sigma )}\hat a(\tau ,{\vec q}_2)-
e^{i(\omega (q_2)\tau -{\vec q}_2\cdot \vec \sigma )}{\hat b}^{*}(\tau
,{\vec q}_2)\Big] +\nonumber \\
&+& e^2 {\vec A}_{\perp}^2(\tau ,\vec \sigma ) \Big[ e^{-i(\omega (q_2)\tau -
{\vec q}_2\cdot \vec \sigma )}\hat a(\tau ,{\vec q}_2)+e^{i(\omega (q_2)\tau -
{\vec q}_2\cdot \vec \sigma )}{\hat b}^{*}(\tau ,{\vec q}_2) \Big] \Big)
-\nonumber \\
&-& {{e^2m^2}\over 8} (2\pi )^3 \int d{\tilde q}_1d{\tilde q}_2d{\tilde q}
_3d{\tilde q}_4 \sqrt{ {{\omega (q_1)\omega (q_3)}\over {\omega (q_2)\omega
(q_4)}} }\nonumber \\
&&\Big( {1\over {({\vec q}_1-{\vec q}_2)^2}}\nonumber \\
&&\Big[ \delta^3({\vec q}_1-{\vec q}_2+{\vec q}_3-{\vec q}_4) \Big(
e^{-i[\omega (q_1)-\omega (q_2)+\omega (q_3)-\omega (q_4)]\tau}\nonumber \\
&&({\hat a}(\tau ,{\vec q}_1){\hat a}^{*}(\tau ,{\vec q}_2)+
{\hat b}(\tau ,{\vec q}_1){\hat b}^{*}(\tau ,{\vec q}_2))
({\hat a}(\tau ,{\vec q}_3){\hat a}^{*}(\tau ,{\vec q}_4)+
{\hat b}(\tau ,{\vec q}_3){\hat b}^{*}(\tau ,{\vec q}_4))+\nonumber \\
&+&e^{i[\omega (q_1)-\omega (q_2)+\omega (q_3)-\omega (q_4)]\tau}\nonumber \\
&&({\hat a}^{*}(\tau ,{\vec q}_1){\hat a}(\tau ,{\vec q}_2)+
{\hat b}^{*}(\tau ,{\vec q}_1){\hat b}(\tau ,{\vec q}_2))
({\hat a}^{*}(\tau ,{\vec q}_3){\hat a}(\tau ,{\vec q}_4)+
{\hat b}^{*}(\tau ,{\vec q}_3){\hat b}(\tau ,{\vec q}_4)) \Big) -\nonumber \\
&-& \delta^3({\vec q}_1-{\vec q}_2-{\vec q}_3+{\vec q}_4) \Big(
e^{-i[\omega (q_1)-\omega (q_2)-\omega (q_3)+\omega (q_4)]\tau}\nonumber \\
&&({\hat a}(\tau ,{\vec q}_1){\hat a}^{*}(\tau ,{\vec q}_2)+
{\hat b}(\tau ,{\vec q}_1){\hat b}^{*}(\tau ,{\vec q}_2))
({\hat a}^{*}(\tau ,{\vec q}_3){\hat a}(\tau ,{\vec q}_4)+
{\hat b}^{*}(\tau ,{\vec q}_3){\hat b}(\tau ,{\vec q}_4))+\nonumber \\
&+&e^{i[\omega (q_1)-\omega (q_2)-\omega (q_3)+\omega (q_4)]\tau}\nonumber \\
&&({\hat a}^{*}(\tau ,{\vec q}_1){\hat a}(\tau ,{\vec q}_2)+
{\hat b}^{*}(\tau ,{\vec q}_1){\hat b}(\tau ,{\vec q}_2))
({\hat a}(\tau ,{\vec q}_3){\hat a}^{*}(\tau ,{\vec q}_4)+
{\hat b}(\tau ,{\vec q}_3){\hat b}^{*}(\tau ,{\vec q}_4)) \Big) +\nonumber \\
&+& \delta^3({\vec q}_1-{\vec q}_2+{\vec q}_3+{\vec q}_4) \Big(
e^{-i[\omega (q_1)-\omega (q_2)+\omega (q_3)+\omega (q_4)]\tau}\nonumber \\
&&({\hat a}(\tau ,{\vec q}_1){\hat a}^{*}(\tau ,{\vec q}_2)+
{\hat b}(\tau ,{\vec q}_1){\hat b}^{*}(\tau ,{\vec q}_2))
({\hat a}(\tau ,{\vec q}_3){\hat b}(\tau ,{\vec q}_4)+
{\hat b}(\tau ,{\vec q}_3){\hat a}(\tau ,{\vec q}_4))+\nonumber \\
&+&e^{i[\omega (q_1)-\omega (q_2)+\omega (q_3)+\omega (q_4)]\tau}\nonumber \\
&&({\hat a}^{*}(\tau ,{\vec q}_1){\hat a}(\tau ,{\vec q}_2)+
{\hat b}^{*}(\tau ,{\vec q}_1){\hat b}(\tau ,{\vec q}_2))
({\hat a}^{*}(\tau ,{\vec q}_3){\hat b}^{*}(\tau ,{\vec q}_4)+
{\hat b}^{*}(\tau ,{\vec q}_3){\hat a}^{*}(\tau ,{\vec q}_4)) \Big)
-\nonumber \\
&-& \delta^3({\vec q}_1-{\vec q}_2-{\vec q}_3-{\vec q}_4) \Big(
e^{-i[\omega (q_1)-\omega (q_2)-\omega (q_3)-\omega (q_4)]\tau}\nonumber \\
&&({\hat a}(\tau ,{\vec q}_1){\hat a}^{*}(\tau ,{\vec q}_2)+
{\hat b}(\tau ,{\vec q}_1){\hat b}^{*}(\tau ,{\vec q}_2))
({\hat a}^{*}(\tau ,{\vec q}_3){\hat b}^{*}(\tau ,{\vec q}_4)+
{\hat b}^{*}(\tau ,{\vec q}_3){\hat a}^{*}(\tau ,{\vec q}_4))+\nonumber \\
&+&e^{i[\omega (q_1)-\omega (q_2)-\omega (q_3)-\omega (q_4)]\tau}\nonumber \\
&&({\hat a}^{*}(\tau ,{\vec q}_1){\hat a}(\tau ,{\vec q}_2)+
{\hat b}^{*}(\tau ,{\vec q}_1){\hat b}(\tau ,{\vec q}_2))
({\hat a}(\tau ,{\vec q}_3){\hat b}(\tau ,{\vec q}_4)+
{\hat b}(\tau ,{\vec q}_3){\hat a}(\tau ,{\vec q}_4)) \Big) \Big] +\nonumber \\
&+&{1\over {({\vec q}_1+{\vec q}_2)^2}}\nonumber \\
&&\Big[ \delta^3({\vec q}_1+{\vec q}_2+{\vec q}_3-{\vec q}_4) \Big(
e^{-i[\omega (q_1)+\omega (q_2)+\omega (q_3)-\omega (q_4)]\tau}\nonumber \\
&&({\hat a}(\tau ,{\vec q}_1){\hat b}(\tau ,{\vec q}_2)+
{\hat b}(\tau ,{\vec q}_1){\hat a}(\tau ,{\vec q}_2))
({\hat a}(\tau ,{\vec q}_3){\hat a}^{*}(\tau ,{\vec q}_4)+
{\hat b}(\tau ,{\vec q}_3){\hat b}^{*}(\tau ,{\vec q}_4))+\nonumber \\
&+&e^{i[\omega (q_1)+\omega (q_2)+\omega (q_3)-\omega (q_4)]\tau}\nonumber \\
&&({\hat a}^{*}(\tau ,{\vec q}_1){\hat b}^{*}(\tau ,{\vec q}_2)+
{\hat b}^{*}(\tau ,{\vec q}_1){\hat a}^{*}(\tau ,{\vec q}_2))
({\hat a}^{*}(\tau ,{\vec q}_3){\hat a}(\tau ,{\vec q}_4)+
{\hat b}^{*}(\tau ,{\vec q}_3){\hat b}(\tau ,{\vec q}_4)) \Big) -\nonumber \\
&-&\delta^3({\vec q}_1+{\vec q}_2-{\vec q}_3+{\vec q}_4) \Big(
e^{-i[\omega (q_1)+\omega (q_2)-\omega (q_3)+\omega (q_4)]\tau}\nonumber \\
&&({\hat a}(\tau ,{\vec q}_1){\hat b}(\tau ,{\vec q}_2)+
{\hat b}(\tau ,{\vec q}_1){\hat a}(\tau ,{\vec q}_2))
({\hat a}^{*}(\tau ,{\vec q}_3){\hat a}(\tau ,{\vec q}_4)+
{\hat b}^{*}(\tau ,{\vec q}_3){\hat b}(\tau ,{\vec q}_4))+\nonumber \\
&+&e^{i[\omega (q_1)+\omega (q_2)-\omega (q_3)+\omega (q_4)]\tau}\nonumber \\
&&({\hat a}^{*}(\tau ,{\vec q}_1){\hat b}^{*}(\tau ,{\vec q}_2)+
{\hat b}^{*}(\tau ,{\vec q}_1){\hat a}^{*}(\tau ,{\vec q}_2))
({\hat a}(\tau ,{\vec q}_3){\hat a}^{*}(\tau ,{\vec q}_4)+
{\hat b}(\tau ,{\vec q}_3){\hat b}^{*}(\tau ,{\vec q}_4)) \Big) +\nonumber \\
&+&\delta^3({\vec q}_1+{\vec q}_2+{\vec q}_3+{\vec q}_4) \Big(
e^{-i[\omega (q_1)+\omega (q_2)+\omega (q_3)+\omega (q_4)]\tau}\nonumber \\
&&({\hat a}(\tau ,{\vec q}_1){\hat b}(\tau ,{\vec q}_2)+
{\hat b}(\tau ,{\vec q}_1){\hat a}(\tau ,{\vec q}_2))
({\hat a}(\tau ,{\vec q}_3){\hat b}(\tau ,{\vec q}_4)+
{\hat b}(\tau ,{\vec q}_3){\hat a}(\tau ,{\vec q}_4))+\nonumber \\
&+&e^{i[\omega (q_1)+\omega (q_2)+\omega (q_3)+\omega (q_4)]\tau}\nonumber \\
&&({\hat a}^{*}(\tau ,{\vec q}_1){\hat b}^{*}(\tau ,{\vec q}_2)+
{\hat b}^{*}(\tau ,{\vec q}_1){\hat a}^{*}(\tau ,{\vec q}_2))
({\hat a}^{*}(\tau ,{\vec q}_3){\hat b}^{*}(\tau ,{\vec q}_4)+
{\hat b}^{*}(\tau ,{\vec q}_3){\hat a}^{*}(\tau ,{\vec q}_4)) \Big)
-\nonumber \\
&-&\delta^3({\vec q}_1+{\vec q}_2-{\vec q}_3-{\vec q}_4) \Big(
e^{-i[\omega (q_1)+\omega (q_2)-\omega (q_3)-\omega (q_4)]\tau}\nonumber \\
&&({\hat a}(\tau ,{\vec q}_1){\hat b}(\tau ,{\vec q}_2)+
{\hat b}(\tau ,{\vec q}_1){\hat a}(\tau ,{\vec q}_2))
({\hat a}^{*}(\tau ,{\vec q}_3){\hat b}^{*}(\tau ,{\vec q}_4)+
{\hat b}^{*}(\tau ,{\vec q}_3){\hat a}^{*}(\tau ,{\vec q}_4))+\nonumber \\
&+&e^{i[\omega (q_1)+\omega (q_2)-\omega (q_3)-\omega (q_4)]\tau}\nonumber \\
&&({\hat a}^{*}(\tau ,{\vec q}_1){\hat b}^{*}(\tau ,{\vec q}_2)+
{\hat b}^{*}(\tau ,{\vec q}_1){\hat a}^{*}(\tau ,{\vec q}_2))
({\hat a}(\tau ,{\vec q}_3){\hat b}(\tau ,{\vec q}_4)+
{\hat b}(\tau ,{\vec q}_3){\hat a}(\tau ,{\vec q}_4)) \Big) \Big] \Big) .
\nonumber \\
&&
\label{e1}
\end{eqnarray}

\vfill\eject

\end{document}